\pdfoutput=1 
\newcommand{\beq}{\begin{equation}}
\newcommand{\eeq}{\end{equation}}
\newcommand{\beqs}{\begin{equation*}}
\newcommand{\eeqs}{\end{equation*}}
\newcommand{\ssty}{\scriptstyle}
\newcommand{\sssty}{\scriptscriptstyle}
\newcommand{\ssiz}{\scriptsize}
\newcommand{\sssiz}{\tiny}

\newcommand{\qqquad}{\qquad \quad}
\newcommand{\qqqquad}{\qquad \qquad}
\newcommand{\ADS}{^{\sssty \!A\!d\!S}}
\newcommand{\ads}{_{\sssty \!A\!d\!S}}
\newcommand{\s}{_{\sssty \!S}}

\newcommand{\axs}{AdS$\times$S \,}
\newcommand{\DS}{^{\sssty d\!S}}
\newcommand{\ds}{_{\sssty \!d\!S}}
\newcommand{\dvb}{_{\sssty \!D\!v\!B}}
\newcommand{\del}{\partial}
\newcommand{\degree}{^\circ}
\newcommand{\snake}{\tilde}
\newcommand{\cross}{\times}
\newcommand{\limepszero}
						{\underset{\epsilon \rightarrow 0}{\lim} \,}
\newcommand{\sDelta}{{\ssty \Delta}}
\newcommand{\sign}{\text{sign}}
\newcommand{\diag}{\text{diag}}

\newcommand{\elof}{{\ssty \in}}
\newcommand{\phii}{\varphi}

\newcommand{\rhobar}{\ovl{\rho}}
\newcommand{\taubar}{\ovl{\tau}}
\newcommand{\ovl}{\overline}
\newcommand{\os}{\overset}
\newcommand{\ob}{\overbrace}

\newcommand{\ub}{\underbrace}
\newcommand{\igx}{\includegraphics}
\documentclass[a4paper,10pt]{book}
\usepackage {amsmath}
\usepackage {amssymb}
\usepackage {amsxtra}
\usepackage [english]{babel}
\usepackage {bbm}
\usepackage {cancel}
\usepackage {dsfont}
\usepackage {graphicx}
\usepackage {esint}
\usepackage {float}
\usepackage {latexsym}
\usepackage {mathbbol}
\usepackage {slashed}
\usepackage {subfigure}
\usepackage {verbatim}
\usepackage {hyperref}
\begin{document}
%%___________________Prefix_________________
\setlength{\parindent}{0cm}
\numberwithin{equation}{chapter}
%% Equation numbers are in format (#chapter.#equation)
\setcounter{tocdepth}{3}
%%
%%
%% diploma thesis, 113 pages, 35 figures, PDF-LaTeX
\begin{center} \LARGE \bfseries
Configuration Space Methods and Time Ordering for %%@
Scalar Propagators
in (Anti) de Sitter Spacetimes
\end{center}
\vspace{-0.2cm}
\begin{center}
	Diploma Thesis (Corrected Version) \\
\end{center}
\vspace{0.1cm}
\begin{figure} [H]
	\begin{center}
	\igx[width=3.5cm]{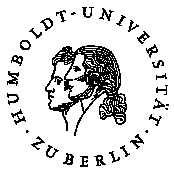} 
	\end{center}
\end{figure}
\vspace{-0.9cm}
\begin{center}
	Humboldt University at Berlin \\
	Faculty of Mathematical and Natural Sciences I \\
\end{center}
\vspace{0.2cm}
\vspace{-0.8cm}
\begin{center}
	Institute of Physics \\
	Quantum Field Theory and String Theory Group \\
\end{center}
\vspace{0.8cm}
\begin{center}
	handed in by Max Dohse,
	\footnote{ email: max.dohse@physik.hu-berlin.de} \\
	born $11^{\text th}$ of December 1979\\
	in K\"onigs Wusterhausen, Germany
\end{center}
\vspace{1cm}
\begin{align*}
	\text{Referees: } & \text{ Dr. habil. Harald Dorn} \\
								& \text{ Prof. Dr. Jan Plefka}
	\qqqquad \qqqquad \qqqquad \qqqquad \qquad \;\,
\end{align*}
\vspace{1.3cm}
\hspace{0.1cm}
Berlin, March 2007 \\
\vspace{0.5cm}
\thispagestyle{empty}
\pagebreak
\vspace{8cm}
\section*{Abstract}
In this diploma thesis a configuration space method 
presented by C. Dullemond and E. van Beveren
for computing all propagators of a scalar field
(Wightman, Hadamard and Schwinger functions,
retarded, advanced and Feynman propagator) 
is reviewed for four-dimensional Minkowski
and Anti de Sitter spacetime AdS$_4$.
This method is then applied for AdS$_d$ 
as well as de Sitter spacetime dS$_d$ of arbitrary dimension $d$,
obtaining results in agreement with the literature.
\\
The advantages of the method are that it
needs neither mode summation
nor analytic continuation from euclidean time,
while delivering the propagators above
including $i\epsilon$-prescription,
plus as a nice bonus the conformal dimension
of a corresponding CFT field.
\\ 
General properties of the considered spacetimes
(namely various coordinate systems and their metrics,
chordal distances, relations between conformal dimensions $\Delta$
and the mass $m$ of the scalar field,
geodesics and the invariance of time ordering)
are also examined and compiled from various sources,
providing an overview of geometrical properties of AdS and dS %%@
spacetimes.  
\\
\vspace{2cm}
\\

Writing a thesis is compulsory in Germany in order to obtain
the academic degree of Diploma Physicist which is equivalent to a Master degree.
This diploma thesis was written in the QFT and String Theory Group of Prof. Jan Plefka
at the Institute of Physics of the Humboldt University at Berlin
under the kind supervision of Dr. Harald Dorn, to whom I
therefore wish to express my deepest gratitude.
\\
A thesis naturally contains calculations down to a certain level of detail.
The inclined reader thus has the choice between following the calculations
or just looking up and enjoying the results.  
\vspace{6cm}
\thispagestyle{plain}
\pagebreak
\tableofcontents
\clearpage
{\LARGE \bfseries  Notation and Conventions}
\\
\begin{align*}
	\text{Re} \, (z), \; \text{Im} \, (z) \; 
	& = \; \text{the real and the imaginary part of the complex number } z \\
	z^* \;
	& = \; \text{the complex conjugate of } z \\
	M^T \;
	& = \; \text{the transposed of the vector or matrix } M \\
	U^\dagger \;
	& = \; \text{the hermitian adjoint of } U \\
	\\
	[a,b] \;
	& = \; \text{the closed interval from } a \text{ to } b \\
	]a,b[ \;
	& = \; \text{the open interval } (a,b \in \mathbb{R})\\
	\\
	[A,B \, ] \;
	& = \; AB-BA \quad \text{the commutator of A and B} \\
	\{ A,B \} \,
	& = \; AB+BA \quad \text{the anticommutator} \\
	\\
	\vec{x}^{\, 2} & = \;\;\; x^k x^k \;\; = \sum_{k=1}^{m} \; \bigl( x^k %%@
\bigr)^2 
	\qquad \;\;
		\begin{matrix}
			\text{an arrow indicates a euclidean sum of squares} \\
			\quad \vec{x} = (x^1,\ldots,x^m) \\
		\end{matrix}
		\\
	x^2 & = g_{jk} \, x^j x^k = \!\! \sum_{j,k=1}^{n} g_{jk} \, x^j x^k
	\quad \; 
		\begin{matrix}
			\text{no arrow indicates the use of the metrical tensor} \, g \\
			\!\!\! x = (x^1,\ldots,x^n) \\
		\end{matrix}
		\\
	\\
	(\del_j)^m
	& = \; \biggl( \! \frac{\del}{\del x^j} \! \biggr)^{\!\! m}
	\quad \text{partial derivatives for contravariant } \ldots \\
	(\del^k)^n
	& = \; \biggl( \! \frac{\del}{\del x_k} \! \biggr)^{\!\! n}
	\quad \ldots \text{ and for covariant coordinates} \\
	\\
	T^M_X
	& = \; \text{tangent space of a point $X$ in the manifold } M \\
	\nabla_{ \! V}  X
	& = \; \text{covariant derivative of } X \text{ in the direction of the vector } %%@
V \\
	\\
	P(f) \;
	& = \; \text{Cauchy's principal value of the function } f \\
	\\
	\theta {\scriptstyle (\!x\!)}
	& =
	\begin{cases}
		\;\;\: 1              & {\ssty x \, > \, 0} \\
		\;\;\: \! \frac{1}{2} & {\ssty x \, = \, 0}
					\qquad \quad \text{Heaviside's } \theta \text{ step function}\\
		\;\;\: 0              & {\ssty x \, < \, 0}
	\end{cases}
	\\
	\epsilon {\scriptstyle (\!x\!)}
	= \, \theta {\scriptstyle (\!x\!)} \! - \! \theta {\scriptstyle (\!-x\!)}
	& =
	\begin{cases}
		      +1 & {\ssty x \, > \, 0} \\
		\;\;\: 0 & {\ssty x \, = \, 0}
				\qquad \quad \epsilon \text{ step function = sign function}\\
		      -1 & {\ssty x \, < \, 0}
	\end{cases} 
\end{align*}
\\
Einstein's Sum Convention: \\
Whenever not indicated otherwise, a summation is understood
over all indices appearing exactly twice in a product,
no matter whether they are lower or upper case, greek or latin,
co- or contravariant.
\thispagestyle{empty}
\pagebreak

%%
%%___________________Main_Part__________________
%%
\chapter{Introduction}
\label{chap:intro}
	\section{Motivation and general introduction}
		\subsection{General Introduction}
The physics of today knows four fundamental forces
acting on elementary particles.
Three of these forces, namely the strong color interaction
and the electroweak interaction
(which unifies the electromagnetic with the weak interaction)
can be well described by the quantum field theory (QFT)
known as the Standard Model (SM) of particle physics.  
The fourth force in league is gravitation,
which is described classically (nonquantized)
by the theory of General Relativity (GR).
\\
The quantum field theories of the Standard Model are gauge field theories.
Gauge fields (e.g. the electromagnetic four potential $A^\mu$)
are needed in order to construct gauge covariant derivatives
(e.g. $D_\mu = \del_\mu + i e A_\mu$)
rendering the Lagrangian of the QFT invariant
under the action of local symmetry transformations of the matter fields.
These gauge transformations form groups.
\\
In contrast to the gauge group U(1)
of quantum electrodynamics (QED),
the gauge group SU(N=3)$_C$ of quantum chromodynamics (QCD) is non-abelian.
This leads to an interaction between the QCD gauge bosons (gluons).
At low energies the coupling constant $g_{Y \! M}$
for the interaction between quarks and gluons is large,
which leads to quark confinement,
while it is small for high energies,
leading to an asymptotic freedom of quarks.
\\
One standard tool used in QFT is perturbation theory
which is applicable for small perturbations of the free system
i.e. small coupling constants.
In the perturbative calculations of Yang-Mills (YM) theories 
one usually uses expansions
in powers of the YM coupling constant $g_{Y \! M}$.
So far the Standard Model works well at small couplings,
while the strong coupling behaviour is understood less well,
accessible non-perturbatively today only via numeric computations
on a discretised spacetime lattice.
\\
The situation is different for gravitation.
Being a classical theory, GR works well at large length scales
corresponding to low energies.
Yet a microscopical description of spacetime at lengths
near the Planck scale or energies near the Planck energy
requires a quantum theory of gravity.
The most famous example of large gravitational energies
are the spacetime regions near the horizons of black holes.
\\
Several candidates for theories unifying the Standard Model and gravity
with or without supersymmetry 
are provided by different string theories.
While in the Standard Model elementary particles
are considered  to be pointlike and to interact locally,
string theories considers them as one-dimensional extended objects: strings.
These strings can oscillate
and different modes of oscillation correspond to different particles.
String theories also make use of perturbation theory
and thus work well if the coupling constant $g_S$ 
for interactions between strings is small.
Supersymmetric string theories live in ten dimensions,
from which four-dimensional spacetime can be obtained
via compactifying six dimensions.
\\
We see, that in both Yang-Mills and string theories
perturbative expansions are usually done
in powers of the coupling constant $g_{Y \! M}$ respectively $g_S$.
\\
However, as discovered by 't Hooft \cite{uthooft},
in YM-theories one can also perform
expansions in powers of $\lambda=g^2_{Y \! M}N$.
He hoped that one could solve the theories for $N=\infty$
and then perform an expansion in $^1 \!\! / \! _N$
with the value $N=3$ of the standard model.
In addition to this expansion Feynman graphs
can be classified in powers of $^1 \!\! / \! _{N^2}$.
For YM-theories a perturbative expansion in $^1 \!\! / \! _{N^2}$ is possible
in the 't Hooft limit where $N \rightarrow \infty$
with $\lambda$ kept fixed.
In this limit only the Feynman graphs of topological genus zero survive
(i.e. those which can be drawn on a sphere without crossing lines
in double line notation).
\\
The form of the $^1 \!\! / \! _{N^2}$ expansion series of the YM-theories
is the same as one finds in a perturbative theory with closed oriented strings
if one identifies $g_S$ with $^1 \!\! / \! _N$.
Because of this resemblance
a relation between YM and string theories was conjectured.
\\
The AdS/CFT correspondence is a candidate for such a relation 
and thus an example of the phenomenon named duality,
which states that a theory can possess (usually two) different descriptions.
This can be very useful
if one description is weakly coupled in a regime
where the other description is strongly coupled,
because then one can apply perturbation theory
for both low and high energies. 
The AdS/CFT duality was conjectured by Maldacena \cite{maldacena}
and states that a four-dimensional supersymmetric SU(N)
$(\mathcal{N} \! =4)$-Yang-Mills theory
describes the same physics as (type IIB) string theory
on an AdS$_5 \times$S$^5$ background 
(which is shortly referred to as AdS in this section).
$\mathcal{N}$ is the number of certain charges of the YM theory,
namely the four-spinor supercharges,
while $N$ denotes the number of colors in the SU(N) Yang-Mills theory,
which we had found to be three for the Standard Model.
This YM theory is a conformally invariant field theory (CFT). 
\\
The contents of AdS/CFT is that each supergravity field 
(which is derived from string theory) propagating on AdS $\cross$ S
is in a one to one correspondence  with an operator in the CFT
on the conformal boundary of AdS $\cross$ S.
\\
The AdS/CFT correspondence relates the masses of supergravity fields
with the conformal dimensions of the CFT operators.
A field $\phi{\ssty(x)}$ of a CFT has the scaling dimension $\Delta$
if under spacetime scalings $x^\mu \rightarrow x'^\mu= \lambda x^\mu$
it transforms as
$\phi{\ssty(x)} \rightarrow \phi'{\ssty(x)} = \lambda^\Delta \phi{\ssty(x')}$.
CFT operators $\mathcal{O}$ are constructed from traces of the CFT fields
$\phi_k$ and their derivatives:
\beqs
	\mathcal{O} = Tr \, (\phi_{k_1}\phi_{k_2}\phi_{k_3} \ldots)
\eeqs
The conformal dimension of a CFT operator is the sum of the scaling dimensions
of its constituting fields plus quantum corrections
and the two-point function of operators behaves like
\beqs
	\bigl< \mathcal{O}_1{\ssty(x_1 \!)} \, \mathcal{O}_2 {\ssty(x_2 \!)} \bigr> \;
	\sim \; \frac{\delta_{\Delta_1, \Delta_2} \;}
		{\, \mid \! x_1 \! - \!x_2 \! \mid ^{\, 2 \Delta_1}}
\eeqs
In \cite{arx9905111} one finds
that the geometry of the string theory background
(in a near horizon limit) is determined by the metric
\begin{align}
	\label{motiv_metric_ads_x_s_01}
	ds^2 = \, \frac{r^2}{R^2} \,
	\Bigl( -dt^{^2} \!\! + dx^{1^2} \!\! + dx^{2^2} \!\!
	+ dx^{3^2} \Bigr) + \, \frac{R^2}{r^2} \, dr^2 + R^2 \, d\Omega_5^2
\end{align}
which via the substitution $(x^4=R^2/r)$
taken from \cite{sieg} transforms into:
\begin{align}
	\label{motiv_metric_ads_x_s_02}
	ds^2 = \, \frac{R^2}{x^{4^2}} \,
	\Bigl( -dt^{^2} \!\! + dx^{1^2} \!\! + dx^{2^2} \!\!
	+ dx^{3^2}  \!\!+ dx^{4^2} \Bigr) + R^2 \, d\Omega_5^2
\end{align}
This is the metric of AdS$_5 \times$S$^5$ spacetime
in Poincar\'e coordinates with the radii of curvature
$R\ads \!=R_S \!=R$.
Moreover in \cite{arx9905111} one finds a relation
between $N,R$ and the string length $l_S$:
\begin{align*}
	R^4 \! \sim l_S^4 g_S N
\end{align*}
\\
Herein $g_S$ is the string coupling which is not a constant as in QFT
but a dynamical variable depending on the expectation value
of the scalar dilaton field $\phi{\ssty (x)}$
which is contained in all 10-dimensional String Theories:
\beqs
	g_S{\ssty (x)} = \exp \bigl( \phi{\ssty (x)} \bigr)
\eeqs
Now the perturbative analysis of the Yang-Mills theory is valid
in the case of small couplings
\begin{align*}
	\lambda \, = \, g^2_{Y \! M} N
	\sim \, g_S N \sim \, \frac{R^4}{l^4_S} \, \ll \, 1
\end{align*}
while the supergravity description via type IIB string theory
is valid for the opposite case
of small curvature corresponding to a large radius of AdS and the sphere:
\begin{align*}
	\lambda \, = \, g^2_{Y \! M} N
	\sim \, g_S N \sim \, \frac{R^4}{l^4_S} \, \gg \, 1
\end{align*}
Thus both descriptions complement each other.
Each one works fine in its perturbative regime
and can be used as a tool
in order to research the regime of strong coupling
in the other description.
\\
The AdS/CFT correspondence is supposed to hold
on both string and field side for all values of $\lambda$,
but up to now only the duality between large $\lambda$
on the string side and small $\lambda$ on the field side
can be tested via perturbative methods. 
\\
More detailed overviews and introductions
for the AdS/CFT correspondence
can be found in \cite{arx9905111} in chapters 1.1, 1.2 and 3.1
and in chapters 1 and 1.2 of \cite{sieg}.
Above the main points discussed therein are outlined briefly.

		\pagebreak
		\subsection{Motivation}
\begin{figure} [H]
	\begin{center}
		\igx[height=4.6cm]{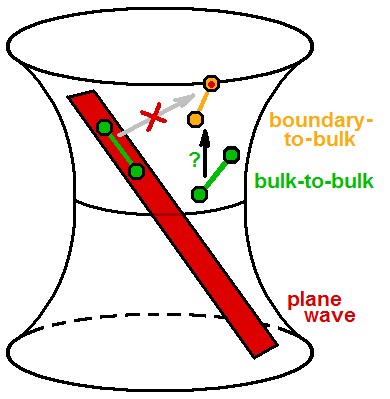}
	\end{center}
	\caption{Types of propagators}
	\label{fig:btb}
\end{figure}

On the string side of the correspondence
supergravity fields are propagating on the \axs background
and therefore their propagator functions are an essential tool. 
The bulk-to-bulk propagator of the complete \axs background
is a subject of interest because in pure AdS
the boundary-to-bulk propagator can be derived from it
and this may also hold for \axs.
The bulk-to-bulk propagator for \axs has not yet been constructed
for arbitrary values of mass.
It has however been constructed in the limit of a plane wave background
but therefrom the boundary-to-bulk propagator cannot be derived
because the boundary is not lying within the region
which converges to the plane wave.
This situation is sketched in figure \ref{fig:btb}.
A plane wave background is a metric 
which is a solution of Einstein's equation without matter
representing plane gravitational waves \cite{bondi}.
From the AdS background it can be obtained via a process called Penrose limit. 
In $d$-dimensional spacetime a plane wave background
has $(d+1)$ continuous symmetries: $(d-1)$ translations and 2 rotations.
\\
The organization of the thesis is as follows:
in subsection \ref{sec:props_mink_gen} we compile
the well known standard relations between the various propagator functions
and in subsection \ref{sec:minkfou} we review in Minkowski spacetime
the well known standard procedure
for finding the propagator functions in momentum space.
\\
Dorn, Salizzoni and Sieg in \cite{arx0307229} 
calculate scalar bulk-to-bulk propagators for AdS$_d \times$S$^{d'}$
up to a time dependent $i\epsilon$-prescription.
Therefore in subsection \ref{sec:dvb_mink}
we review in fourdimensional Minkowski spacetime
a configuration space method
presented by Dullemond and van Beveren (DvB) in \cite{dvb}.
This method can be applied
in various constantly curved spacetimes
in order to find all propagators including the appropriate $i\epsilon$-term.
In subsection \ref{sec:props_ads}
we review their method for AdS$_4$
and extend it slightly to $d$-dimensional AdS$_d$.
\\
Since also a dS/CFT duality has been conjectured
by Strominger in \cite{strominger} in subsection \ref{sec:props_ds}
we also apply the DvB method to de Sitter spacetime dS$_d$
of arbitrary dimension $d$.
\\
In sections \ref{sec:basprop_ads} and \ref{sec:basprop_ds}
we also review and examine basic geometric properties
of AdS and dS spacetime:
different coordinate systems and their metrics,
chordal distances, relations between conformal dimensions $\Delta$
and the mass $m$ of the scalar field,
geodesics and the invariance of time ordering
are examined and compiled from various sources.
\vspace{0.2cm}
	\clearpage
	\section{Basic properties of Anti de Sitter spacetime}
	\label{sec:basprop_ads}
		\subsection{Coordinate systems}
		\label{sec:ads_coord}
\begin{figure} [H]
	\centering
		\subfigure[]{\igx[width = 6cm]{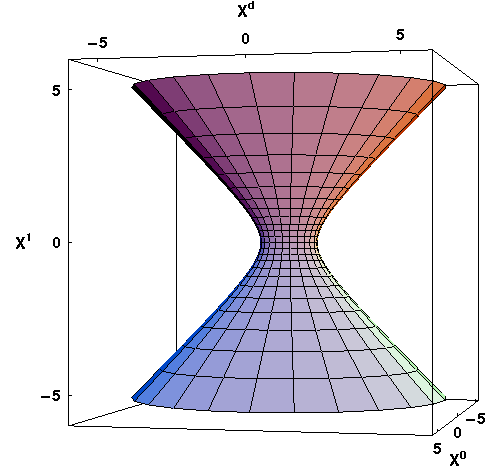}}
		\quad
		\subfigure[]{\igx[width = 5.5cm]{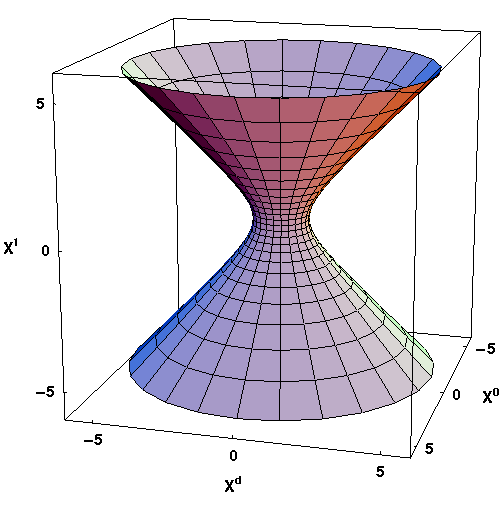}}
	\caption{AdS hyperboloid}
	\label{fig:ads_hyp_global}
\end{figure}

An Anti de Sitter space AdS is a homogeneous isotropic space
with constant negative curvature. $d$-dimensional Anti de Sitter space AdS$_d$
can be realized as a hyperboloid in a $(d+1)$-dimensional embedding space
with the indefinite metric
\begin{align}
	\label{eq:ads_coord01}
	\eta = \sigma \, \text{diag} \, (+,-,\ldots,-,+) \qquad \qquad \sigma = \pm 1
\end{align}
The overall sign $\sigma$ is a purely conventional choice
but is maintained for the convenience of the reader (and the author).
Doing so shall facilitate the comparison of different publications
and mark where the choice of $\sigma$ leads to sign changes and where not.
Cartesian coordinates for a point X in the embedding space are given by
\begin{align}
	X = (X^0 \! , \vec{X}, X^d) \qquad \vec{X} = (X^1,\ldots,X^{d-1})
\end{align}
and the $(d\!+\!1)$-dimensional d'Alembertian in embedding space is
\begin{align}
	\Box_{bulk} = \, \eta^{MN} \del_M \del_N 
	= \, \sigma \, (\del_{0}^2 \! + \del_{d}^2-\vec{\del} \, ^2)
\end{align}
with indices in Latin upper cases running within $(0,\ldots,d)$.
AdS$_d$ then corresponds to a hyperboloid in embedding space with
\beq
	\label{ads_coord04}
	X^{^2} =  \, \eta_{MN} X^M X^N
	= \, \sigma \Bigl( X^{0^2} \!\! +X^{d^2} \! -\vec{X}^{^2} \Bigr) 
	= \, \sigma R^{^2} \ads = \, const.
\eeq
Figure \ref{fig:ads_hyp_global} shows such a hyperboloid
for the case of AdS$_{d=2}$.
For the embedding space we can introduce
several systems of radial coordinates,
from which AdS emerges by demoting the radial coordinate $R$
to a constant parameter $R \ads$.
\begin{align}
	X^0 & \; = \; R \cosh \rho \, \sin t \: = \; R \; \frac{\sin t}{\cos \overline{\rho}}
	              \quad \;\; = \; R \, \sqrt{1+\vec{x} \,^2} \, \sin t 
				\qquad \quad {\ssty 0 \leq \rho < \infty}
	 			\notag \\
	X^d & \; = \; R \cosh \rho \, \cos t \, = \; R \; \frac{\cos t}{\cos \overline{\rho}}
	              \quad \;\; = \; R \, \sqrt{1+\vec{x}\,^2} \, \cos t 
	 			\qquad \quad {\ssty 0 \leq \overline{\rho} < \frac{\pi}{2}}
				\notag \\
	\label{ads_coord05}
	\vec{X} & \; = \; R \sinh \rho \;\;\; \vec{\xi} {\ssty \! (\!\vec{\varphi})} \;
				= \; R \, \vec{\xi}{\ssty \! (\!\vec{\varphi})}
				\, \tan \overline{\rho} \, = \;\; R \, \vec{x}
				 \qquad \qquad \qquad \qquad \:
				 {\ssty \vec{\xi}\, ^2=1 }				  
\end{align}
In figure \ref{fig:ads_hyp_global} the circles
running around the hyperboloid
are lines of constant $\rho, \rhobar$ and $\vec{x}$
and the hyperbolae are lines of constant time $t$.
We can read off, that the topology of AdS$_d$ is 
$S^1_{\text{time}} \cross \mathbb{R}^{d-1}_{\text{space}}$ \cite{avis}.
Therefore we are faced with closed timelike curves.
\\
The point $X_O=(0,\vec{0},R)$ with $n=0$
in the embedding space is the origin
$\; t=\rho=\overline{\rho}=\vec{\varphi}=\vec{x} =0\;$ of AdS
($n$ is a winding number to be defined soon).
The antipodal point $\tilde{X}$ of a point $X$ on the hyperboloid
in embedding space is unique: $\tilde{X}=-X$.
Yet on there exist two possibilities of traveling
to the antipodal point in direction of either increasing or decreasing time $t$:
\begin{align}
	X \, & = \, (R, \, t, \; \vec{x},\; n) \notag \\
	\label{eq:ads_coord10_5}
	\rightarrow \quad
	\snake{X}_\pm \! & = \, (R,t \! \pm \! \pi, -\vec{x}, n \! \pm \! 1)
\end{align}
All of the coordinate sets given above are global coordinates,
because at constant radius they cover the whole AdS hyperboloid.
For this purpose it is sufficient to have $-\pi < t \leq +\pi$
with the points at $t=-\pi$ and $t=+\pi$ identified.
\begin{figure} [H]
	\begin{center}
		\igx[width = 11cm]{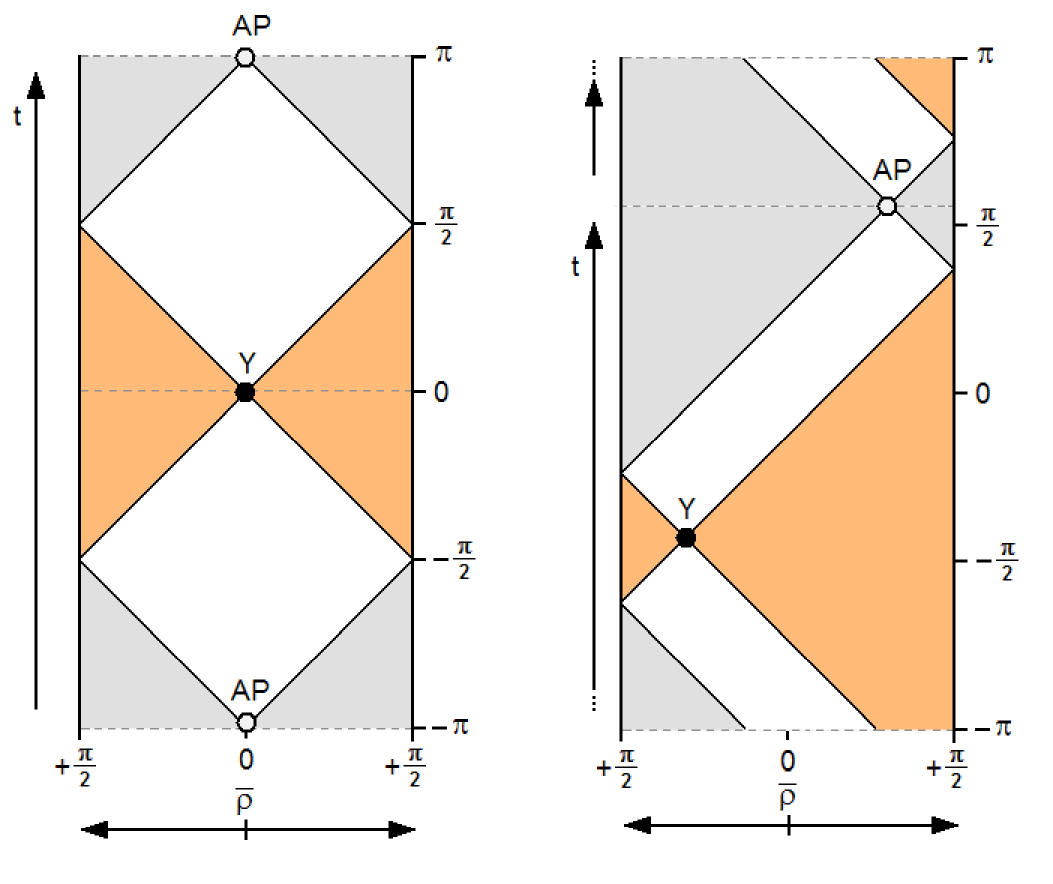}
	\end{center}
	\caption{Penrose diagram of the AdS hyperboloid}
	\label{fig:penrose_ads_hyp}
\end{figure}
Figure \ref{fig:penrose_ads_hyp} is the Penrose diagram
of this AdS hyperboloid in the conformal $(t,\rhobar)$ coordinates.
The dashed grey lines at $t=\pm \pi$ indicate
the identification of these points.
The coordinate $\rhobar$ is nonnegative by definition,
the left half of the diagrams with $\rhobar$ increasing leftward
shows the half of the hyperboloid with $X^1<0$
and the part with $\rhobar$ increasing rightward
shows the other one.
In dimensions $d>2$ each point on the hyperboloid
represents a sphere S$^{d-2}$ of radius $\tan \rhobar$
and the halves of the diagram correspond
to the southern and northern hemisphere.
\\
The left diagram shows the lightcone of the reference point $Y$,
situated at the origin of AdS, and its antipode point $AP$.
The diagram on the right is drawn for a arbitrary reference point
sitting somewhere else.
Lightrays travel at angles of $45\degree$ and 
the orange regions are the spacelike part of the lightcone
while all other parts are timelike regions.
However we will see in subsection \ref{sec:ads_geod}
that only the white yet not the grey parts
can be reached from $Y$ via a geodesic.
This illustrates that AdS is not geodesically convex \cite{fronsdal}
i.e. geodesics originating at an arbitrary point
do not cover the entire spacetime.
(Geodesically convex is meant in the sense that any two points
of a manifold can be connected by a geodesic.
This is sometimes also called a geodesically complete manifold,
however there exists a different definition of geodesical completeness
and therefore we will use the term geodesically convex.)
\\
As we will see in \ref{sec:ads_timord},
due to the fact that the hyperboloid is closed in direction of $t$,
and one encounters closed timelike curves,
we therefor have to introduce a special way of time ordering.
This leads to the somewhat awkward picture
that in the diagram on the right
points with $t \gtrapprox \pi \!/2 \,$ 'behind' the antipode point $AP$ 
need to be considered to be earlier in time than $Y$ itself. 
\\
We also see another apparent feature of AdS
presented by Avis, Isham and Storey in \cite{avis}
wherein also the Penrose diagrams can be found:
lightlike geodesics starting at $Y$
reach the conformal boundary $\rhobar = \pi \!/2$
representing spatial infinity
within finite coordinate time $\Delta t \leq \pi$.
Thus initial data from a spacelike hypersurface
can propagate to spatial infinity in finite time.
Vice versa data from the timelike conformal boundary at spatial infinity
can influence the interior of AdS.
This illustrates that AdS is not globally hyperbolic:
(the intersection of an arbitrary point's future
with another arbitrary point's past
is not necessarily compact, which implies that)
the causal future of a Cauchy surface
is not entirely determined by the equations of motion
plus the initial data on the Cauchy surface.
\\
Whenever not otherwise indicated, we refer to AdS
as the universal covering space of the hyperboloid
obtained by 'unwrapping' the circle $S^1_{\text{time}}$ 
i.e. extending the range of $t$ to $-\infty < t < +\infty$.
Then $t$ is many valued in embedding space
and serves as time variable in AdS spacetime.
The topology of universal covering AdS$_d$ is
the same as for $d$-dimensional Minkowski spacetime:
$\mathbb{R}_{\text{time}} \cross \mathbb{R}^{d-1}_{\text{space}}$.
\begin{figure} [H]
	\begin{center}
		\igx[height=12cm]{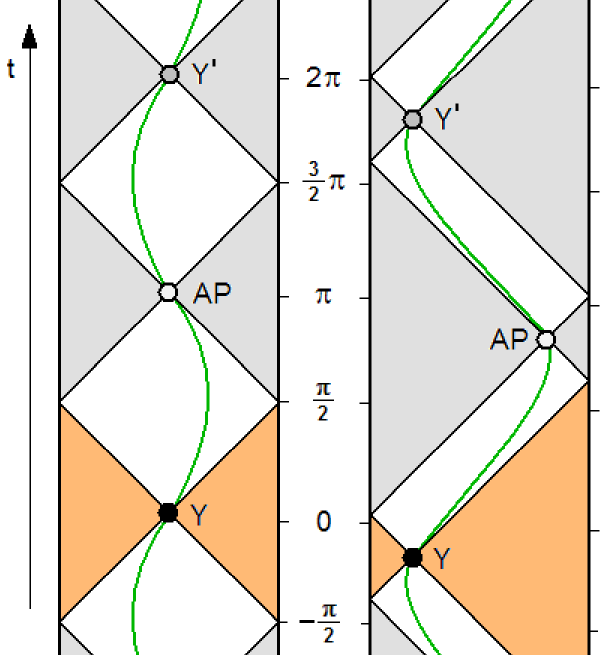}
	\end{center}
	\caption{Penrose diagram of universal covering AdS}
	\label{fig:penrose_ads_uc}
\end{figure}
The conformal boundary of AdS$_d$ is formed by all points with
$\rhobar= \pi / 2$ and has the topology of
$S^1 \! \cross S^{d-2}$ for the hyperboloid
and $\mathbb{R} \cross S^{d-2}$ for the universal cover.
Figure \ref{fig:penrose_ads_uc} shows the Penrose diagram
of universal covering AdS.
The green curve represents a timelike geodesic.
All timelike geodesics starting from a reference point $Y$
intersect again at all points $Y'$ and $AP$ of AdS,
which by the coordinate systems \eqref{ads_coord05}
are mapped to either the point Y or its antipode point
on the hyperboloid in embedding space
(see section \ref{sec:ads_geod}).
\\
Figure \ref{fig:ads_hyp_lightcones} illustrates
the form of the lightcones one the AdS hyperboloid.
Subfigures (a), (c) and (e) on the left show the lightcones of the AdS origin
while (b), (d) and (f) one the right are plotted for an arbitrary reference point.
The spacelike regions are kept in sunny orange again
and the timelike ones in emerald green.
The lightcones are cut out of the hyperboloid
by a plane in embedding space containing our reference point
and being perpendicular (with respect to the metric of embedding space)
to the vector pointing from embedding space origin
towards the reference point.
\begin{figure} [H]
	\centering
		\subfigure[]{\igx[width = 5cm]{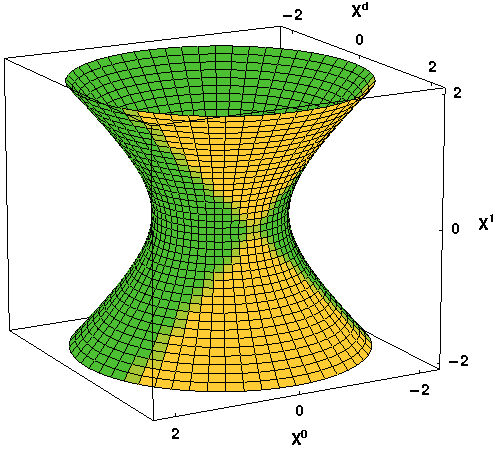}}
		\subfigure[]{\igx[width = 5cm]{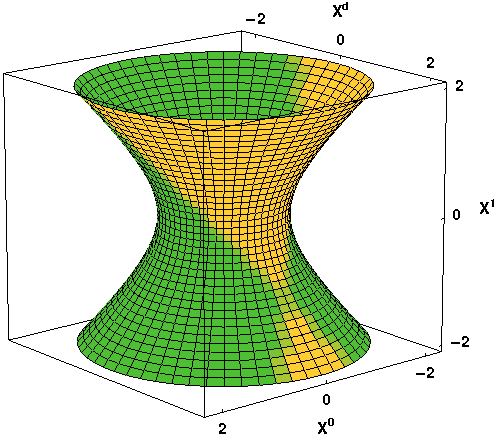}}
		\\
		\vspace{-0.4cm}
		\subfigure[]{\igx[width = 5cm]{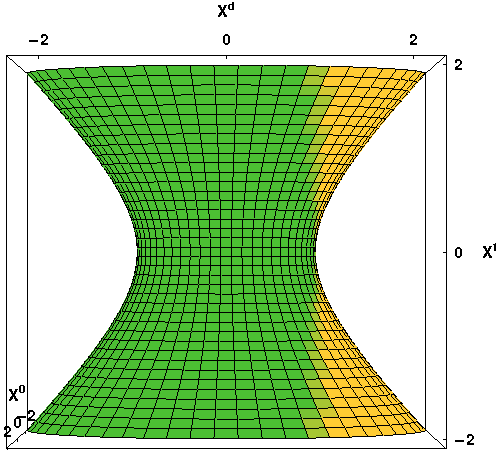}}
		\subfigure[]{\igx[width = 5cm]{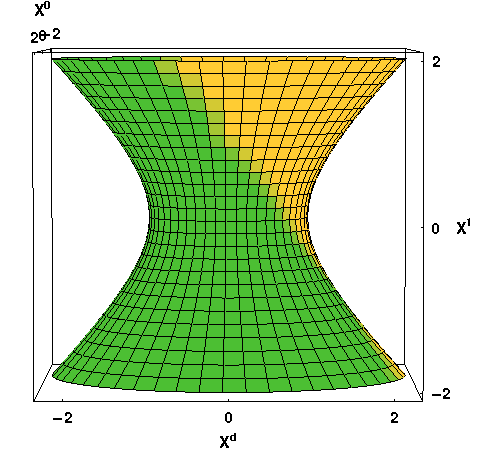}}
		\\
		\vspace{-0.4cm}
		\subfigure[]{\igx[width = 5cm]{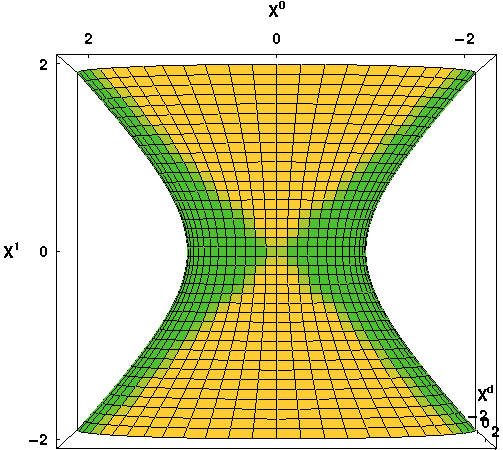}}
		\subfigure[]{\igx[width = 5cm]{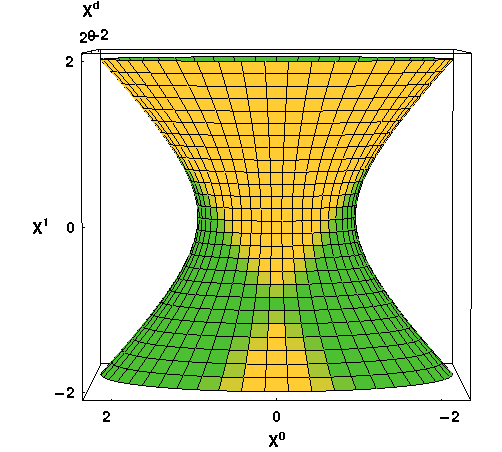}}
		\vspace{-0.2cm}
	\caption{AdS hyperboloid: lightcones}
	\label{fig:ads_hyp_lightcones}
\end{figure}
\vspace{-1cm}
$(\rho,\vec{\varphi}), \; (\overline{\rho},\vec{\varphi})$ and ${\vec{x}}$
are sets of spatial coordinates in AdS.
For the $\vec{\xi}{\ssty \! (\!\vec{\varphi})}$
and $\vec{\varphi}=(\varphi^1,\ldots,\varphi^{d-2})$ coordinates
we choose standard spherical coordinates
for the $(d\!-\!2)$-sphere $S^{d-2}$
which e.g. can be found in \cite{leshouches} and \cite{sieg}:
\begin{align}
	\xi^p \, & = \, \cos \varphi^p \, \prod_{j=1}^{p-1} \sin \varphi^j
	 	& & 		^{0 \, \leq \; \varphi^k \;\;\;\; \leq \; \pi}
		_{0 \, \leq \, \varphi^{d-2} \, < \, 2\pi}
		\notag \\
	\xi^{d-1} & = \, \sin \varphi^{d-2} \, \prod_{j=1}^{d-3} \sin \varphi^j
	 	& & ^{p \, = \, 1, \, \ldots, \, (d-2)}
		_{k \, = \, 1, \, \ldots, \, (d-3)}
		\notag
\end{align}
The inverted relations are:
\begin{align}
	R & = \sqrt{X^{0^2} \!\! +X^{d^2} \! -\vec{X}^{^2}} \phantom{\biggl(  \biggr)}
	 \\
	\sinh \rho = \tan \overline{\rho} 
	    & = \sqrt{\vec{X}^{^2}/(X^{0^2} \!\! +X^{d^2} \! -\vec{X}^{^2})} \\
	\vec{x} & = \frac{\vec{X}}{\sqrt{X^{0^2} \!\! +X^{d^2} \! -\vec{X}^{^2}}} \\
	t & = \arctan \bigl( X^0 \! / X^{d} \bigr) + \, n \pi \phantom{\biggl(  \biggr)} \\
	n & = \bigl( t - \arctan (\tan t) \bigr) / \, \pi
\end{align}
wherein we finally have introduced the coordinate winding number $n$.
Half a turn around the hyperboloid in the direction of
increasing(decreasing) $t$ increases(decreases) $n$ by $1$.
So every point of AdS is uniquely defined by its coordinates
in embedding space and its coordinate winding number.
\\
Moreover there also exists a patchwise coordinate system
called Poincar\'e patch, in which the metric takes a particularly simple form:
\begin{align}
	X^0 & \; = \; R \, \frac{\,\ovl{\tau}}{a^1}
				&
				{\ssty -\infty \, < \; \ovl{\tau},a^1,a^k \, < \; +\infty} &
	 			\notag \\
	X^d & \; = \; \frac{R}{2a^1} \,
					\Bigl(- \ovl{\tau}^{\,2} \! + a^{1^2} \!\! + \vec{a}^{\,2} \! + 1 \Bigr)
					&
					{\ssty a^1 \, \neq \, 0} &
				\notag \\
	X^1 & \; = \; \frac{R}{2a^1} \,
					\Bigl(- \ovl{\tau}^{\,2} \! + a^{1^2} \!\! + \vec{a}^{\,2} \! - 1 \Bigr)
					&
					{\ssty \vec{a} \, = \, (a^2,\ldots,a^{d-1})} &
				\notag \\
	X^k & \; =  \; R \, \frac{\,a^k}{a^1}
				&
				{\ssty k \, = \, 2, \, \ldots, \, (d-1)} &
				\notag				  
\end{align}
The inverted relations are:
\begin{align}
	a^1 \, & = \, \frac{(X^{0^2} \!\! + \! X^{d^2} \!\! - \! \vec{X}^{^2})}
			{(X^d \! - \! X^1)}
			\notag \\
	\ovl{\tau} \, & = \; \frac{X^0}{(X^d \! - \! X^1)} \;
							\sqrt{X^{0^2} \!\! + \! X^{d^2} \!\! - \!\vec{X}^{^2}}
			\notag \\
	a^k & = \; \frac{X^k}{(X^d \! - \! X^1)} \;
							\sqrt{X^{0^2} \!\! + \! X^{d^2} \!\! - \!\vec{X}^{^2}}
			\notag
\end{align}
We see that the Poincar\'e coordinates are not well defined
for $X^d \!\!=\! X^1$ .
The patch with $a^1 > 0$
covers the one half of the hyperboloid with $X^d \! > \! X^1$
and the patch with $a^1 < 0$ covers the other half $X^d \! < X^1$.
Poincar\'e coordinates (nearly) cover the hyperboloid once.
Thus in order to avoid closed timelike curves
again we need to introduce a universal covering
(e.g. by introducing a coordinate winding number in some way). 
\begin{figure} [H]
	\centering
		\subfigure[]{\igx[width = 7cm]{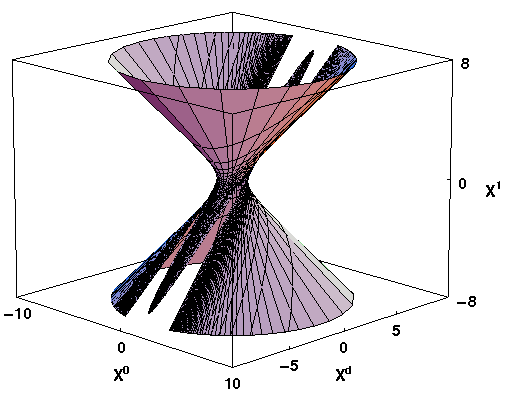}}
		\\
		\vspace{-0.4cm}
		\subfigure[]{\igx[width = 7.5cm]{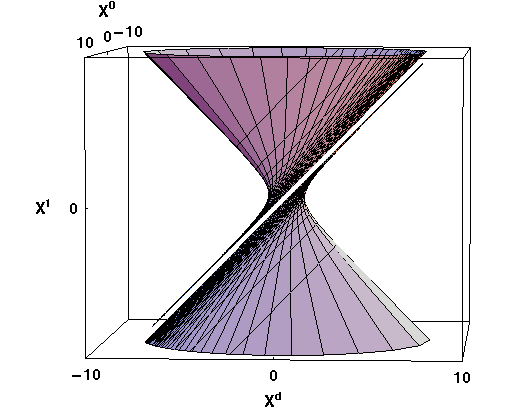}}
		\\
		\vspace{-0.2cm}
		\subfigure[]{\igx[width = 6cm]{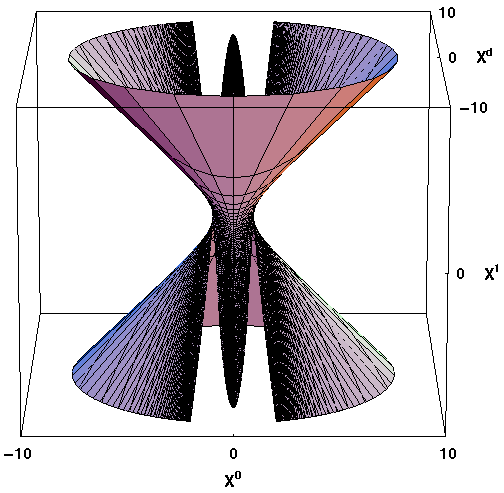}}
		\vspace{-0.1cm}
	\caption{AdS hyperboloid: Poincar\'e coordinates}
	\label{fig:ads_hyp_poincare}
\end{figure}
\clearpage
Figure \ref{fig:ads_hyp_poincare} shows the AdS hyperboloid
in Poincar\'e coordinates.
\ref{fig:ads_hyp_poincare} (b) demonstrates (drawn slightly exaggerated)
how the $(X^d \! = X^1)$ - plane separates the two patches.
\begin{figure} [H]
	\begin{center}
		\igx[height = 10cm]{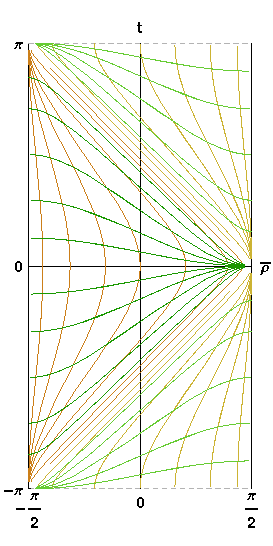}
	\end{center}
	\caption{Penrose diagram of AdS: Poincar\'e coordinates}
	\label{fig:penrose_ads_poincare}
\end{figure}
Figure \ref{fig:penrose_ads_poincare} is the Penrose diagram
of AdS showing the coordinate lines of Poincar\'e coordinates
for the two patches.
The green curves are lines of constant $\taubar$
and the orange ones of constant $a^1$.
If in our mind we vertically roll up the diagram as a cylinder
gluing together the lines of $t=-\pi$ and $t=+\pi$,
then we obtain a flattened version
of the AdS hyperboloid covered by Poincar\'e coordinates. 
The negative value of $\rhobar$ is symbolically only
and indicates negative values of $X^1$.
\\
In the Poincar\'e coordinates there is no origin because of $a^1 \neq 0$.
We just find that $X_O$ corresponds to $\ovl{\tau}=\vec{a}=0$ with $a^1=1$.
\\
In Poincar\'e coordinates we find for the antipodal point:
\begin{align}
	X \, & = \, (R,\ovl{\tau}, \; a^1,\vec{a}) \notag \\
			\rightarrow \quad
	\snake{X} \, & = \, (R,\ovl{\tau},-a^1,\vec{a})
			 \qquad \notag
\end{align}
In the following considerations we will often switch
between regarding AdS as a self-contained spacetime
and using viewpoints of embedding space.

			\subsubsection{Chordal distance}
The chordal distance $u$ of two points $x$ and $y$ in AdS
is defined as the squared distance
of their corresponding points $X$ and $Y$
on the hyperboloid in embedding space:
\begin{align}
	u {\scriptstyle (X\!,Y)} & \equiv \bigl( X-Y \bigr)^2 \\
	\tilde{u} {\scriptstyle (X\!,Y)}
	\equiv \, u {\scriptstyle (X\!,\tilde{Y})} & = \bigl( X+Y \bigr)^2
\end{align}
In some publications these definition include a factor of $^1 \!\! / \! _2$
on the right hand side which we avoid.
$\snake{u}$  is called antipodal chordal distance.
Both chordal distances are unique on the hyperboloid,
periodic on the universal cover
and SO${\ssty (2,d-1)}$-invariant.
\\
When written without arguments, the chordal distances
are meant to refer to the arguments $(X,Y)$ as above.
There exist a few simple relations \cite{arx0307229}
between them on AdS:
\begin{align}
	\label{eq:ads_chordal03}
	u & = 2 \sigma R^2\ads \! - 2XY
	& u \! + \snake{u} & = 4 \sigma R^2\ads \\
	\label{eq:ads_chordal04}
	\snake{u} & = 2 \sigma R^2\ads \! + 2XY
	& u \snake{u} & = 4R^4\ads - 4 (XY)^2
\end{align}

		\subsection{AdS metrics}
		\label{sec:ads_metrics}
The squared infinitesimal length element is
%%
%%____________________equation_ads_coord12_____________
\begin{align}
	ds^2 \ads & = \, \Bigl[ \eta_{MN}dX^MdX^N \Bigr] _{ \! R=R \ads=const.}
			\notag \\
	& = \, \sigma R^2 \ads \, \Bigl( \cosh^2 \!\! \rho \, dt^2 - d \rho^2 
	        - \sinh^2 \!\! \rho \; d \Omega^2_{d-2} \Bigr) \\
	\label{eq:ads_metrics17}
	& = \, \frac{\sigma R^2 \ads}{\cos^2 \overline{\rho}} \,
	     	\Bigl( dt^2 - d \overline{\rho}^2
		 	- \sin^2 \! \overline{\rho} \; d \Omega^2_{d-2} \Bigr) \\
	& = \, \sigma R^2 \ads \Biggl[ (1\!+\!\vec{x}\, ^2) \, dt^2 + \biggl( \!
	       	\frac{x^i x^j}{\,(1+\vec{x}\, ^2)}-\delta^{ij} \! \biggr) dx^i dx^j %%@
\Biggr] \\
	\label{eq:ads_metrics17_1}
	& = \, \sigma \frac{R\ads^2}{a^{1^2}} \,
			\Bigl( d\ovl{\tau}^2 - da^{1^2} \! - d\vec{a}^{\,2} \Bigr)	
\end{align}
wherein indices in Latin lower cases are running within $(1,\ldots,d\!-\!1)$
and $d \Omega _{d-2}$ is the infinitesimal line element
on the unit sphere $S^{d-2}$.
In standard spherical coordinates $d \Omega^2_{d-2}$ reads:
\beqs
	d \Omega^2_{d-2} 
		= d\varphi^{1^2} \!\! + \sin^2 \! \varphi^1
		\biggl( \!d\varphi^{2^2} \!\! + \sin^2 \! \varphi^2
		\Bigl( \ldots \! \bigl( d\varphi^{d-3^2} \!\! + \sin^2 \! \varphi^{d-3} %%@
d\varphi^{d-2^2}
		\bigr) \! \ldots \Bigr) \! \biggr)
\eeqs
Metric (\ref{eq:ads_metrics17}) is conformally equivalent
to the metric of the spacetime known as Einstein static universe (ESU).
Therefore we refer to $(t, \overline{\rho}, \vec{\varphi} \,)$ 
as conformal coordinates.
Thus the time $t$ used in the first three coordinate systems
is also called conformal time variable.
\\
The $d$-dimensional Einstein static universe ESU$_d$
has the topological structure
$\mathbb{R}_{\text{time}} \cross S^{d-1}$
(full real axis and complete sphere).
Since for AdS we only have $\rhobar \in [0,\pi/2[$
(and not $\rhobar \in [0,\pi]$ as for the ESU)
we see that AdS$_d$ can be conformally mapped
into one half of ESU$_d$ \cite{avis}. 
\\
The metric (\ref{eq:ads_metrics17_1}) in Poincar\'e coordinates
is conformally flat and thus $\ovl{\tau}$ is a conformal time variable.
\\
The metrics of all four coordinate systems are static.
This is the reason why both vectors $\del_t$
and $\del_{\ovl{\tau}}$ are killing vectors.
This is important because Killing vectors of time variables
generate a symmetry in time translation
and thus can be used for defining the Hamiltonian.
A vector $X$ is a Killing vector if it fulfills the Killing equation  
which we can find e.g. in \cite{nakahara} by Nakahara:
\beq
	\label{eq:ads_metric_killing}
	0 \, = \, 
	g_{\alpha \mu} \del_\nu X^\alpha + g_{\alpha \nu} \del_\mu X^\alpha
	+ (\del_\lambda g_{\mu \nu}) X^\lambda
	\qqqquad {\ssty \forall \, \mu,\nu}  
\eeq
For the induced metric of the $(t,\vec{x})$ coordinate set
we find with indices in greek lower cases
running within $(0,\ldots,d\!-\!1)$ and $x^0 \! = t$:
%%
%%____________________equation_ads_coord16_____________
\begin{align}
	\label{ads_metrics_01}
	g_{\mu\nu} \! & = \, \sigma R^2 \ads \!
		\begin{pmatrix}
			(1+\vec{x}\,^2) & 0 & 0 & 0 & \cdots \quad \\
			0 & \frac{x^1x^1}{(1+\vec{x}\,^2)}-1
			   & \frac{x^1x^2}{(1+\vec{x}\,^2)}
			   & \frac{x^1x^3}{(1+\vec{x}\,^2)}	& \cdots \quad \\
			0 & \frac{x^2x^1}{(1+\vec{x}\,^2)}
			   & \frac{x^2x^2}{(1+\vec{x}\,^2)}-1
			   & \frac{x^2x^3}{(1+\vec{x}\,^2)}	& \cdots \quad \\
			0 & \frac{x^3x^1}{(1+\vec{x}\,^2)}
			   & \frac{x^3x^2}{(1+\vec{x}\,^2)}
			   & \frac{x^3x^3}{(1+\vec{x}\,^2)}-1 & \cdots \quad \\
			\vdots & \vdots & \vdots & \vdots & \ddots \quad
		\end{pmatrix} \\
		\notag \\
	g^{\mu\nu} \! & = \, \frac{-\sigma}{R^2 \ads}
		\begin{pmatrix}
			\; \, \frac{-1}{(1+\vec{x}\,^2)} \;\, & 0 & 0 & 0 & \cdots \quad \\
			0 & x^1 x^1 \! +1 \quad
			   & x^1x^2
			   & x^1x^3	& \cdots \quad \\
			0 & x^2x^1
			   & x^2x^2 \! +1 \quad
			   & x^2x^3	& \cdots \quad \\
			0 & x^3x^1
			   & x^3x^2
			   & x^3x^3 \! +1	\quad & \cdots \quad \\
			\vdots & \vdots & \vdots & \vdots & \ddots \quad
		\end{pmatrix}
\end{align}
%%
%%____________________equation_ads_coord19_____________
\begin{align}
	g_{ij} \, & = \sigma R^2 \ads \biggl[ \frac{x^ix^j}{(1+\vec{x}\,^2)}
				- \delta^{ij} \biggr] \\
	g^{ij} \, & = \, \frac{-\sigma}{R^2 \ads} \,
						\biggl[ \;\; x^ix^j \;\; + \;\; \delta^{ij} \; \biggr] \\
	\notag \\
	\det g_{ij} \,
		& = \, (-\sigma R\ads^2)^{d-\!1} / \, (1 \! + \! \vec{x}\,^2) \\
	\det g_{\mu \nu}  & = \, -(-\sigma R\ads^2)^d
\end{align}
The d'Alembertian on AdS then is
%%
%%____________________equation_ads_coord23_____________
\begin{align}
	\Box \ads \, & = \, \frac{1}{\sqrt{g}} \, \del _\mu\sqrt{g} \, g^{\mu \nu}
	\del _\nu
			\qquad \qquad \qquad \qquad \qquad
			{\scriptstyle g \, = \, \mid \det g_{\mu \nu} \mid \, = \, R \ads^{2d}}
			\notag \\
	\label{eq:ads_coord23}
	& = \, \frac{\sigma}{R^2 \ads} \, \biggl( \frac{1}{\,(1+\vec{x}\,^2)} \, \del %%@
_t^2
			\underbrace {- \, \del _k \del _k - x^j \! x^k \del _j \del _k
								-2 x^k \! \del _k} \biggr) \\
	& \qquad \qquad \qquad \qquad \qquad \qquad \qquad \quad
	^{= \, \sigma R^2 \ads \Box_{\vec{x}}} \notag
\end{align}
and the Christoffel symbols (independent of $\sigma$) read
%%
%%____________________equation_ads_coord24_____________
\begin{align}
	\Gamma ^\lambda _{\alpha \beta} & \equiv \frac{1}{2} \, g^{\lambda \mu}
		\bigl( \del _\alpha g_{\beta \mu} + \del _\beta g_{\mu \alpha}
			- \del _\mu g_{\alpha \beta} \bigr) \\
	\Gamma ^t _{tt} & = \Gamma ^t _{\!jk} = \Gamma ^j _{tk}
		=  \Gamma ^j _{kt} = \, 0 \\
	\Gamma ^t _{tk} & =  \Gamma ^t _{kt} =
	 	\frac{x^k}{(1+\vec{x}\,^2)} \\
	\Gamma ^k _{tt} & = \qquad x^k (1+\vec{x}\,^2)
		\qquad = \, \sigma \, g_{tt} \, \frac{x^k}{R \ads ^2} \\
	\Gamma ^k _{ab} & = \, x^k \biggl( \! \frac{x^a x^b}{\, (1+\vec{x}\,^2)}
															-\delta^{ab} \! \biggr) 
		= \, \sigma \, g_{ab} \, \frac{x^k}{R \ads ^2}		
\end{align}

		\subsection{Conformal dimension and mass term in AdS}
		\label{sec:ads_breitfreed}
In \cite{mezincescu} Mezincescu and Townsend show for AdS$_{d}$ %%@
spacetime 
that the squared mass $m^2$ of a scalar field fulfills the equation:
\beq
	\label{eq:breitfreed01}
	\Delta\!^2 -(d\!-\!\!1)\Delta - m^2 \! R \ads^2 = \, 0
\eeq 
with the two solutions
\beq
	\label{eq:breitfreed02}
	\Delta_\pm = \: \frac{d\!-\!\!1}{2} \pm
							\sqrt{\biggl( \frac{d\!-\!\!1}{2} \biggr)^{\!\! 2} \! 
								 + m^2 \! R \ads^2 \;}
\eeq 
This $\Delta_\pm$ is the conformal dimension (conformal weight)
of a CFT field that via AdS/CFT correspondence is related
to a scalar field with mass $m^2$ on AdS,
see \cite{arx0201253} eq.(6.8), \cite{arx0307229} eq.(9). \\
In order to obtain a positive definite energy for our scalar field,
$\Delta_\pm$ needs to be real. This yields the Breitenlohner-Freedman %%@
bound:
\beq
	\label{eq:breitfreed03}
	m^2 \! R \ads^2 \, \geq \, - \biggl( \frac{d\!-\!\!1}{2} \biggr)^{\!\! 2}
\eeq 
When working with $\Delta_-$, one has to use an alternative improved
definition of the energy, which differs from the conventional definition
by a surface term \cite{mezincescu}. \\
Moreover the conformal weight is bounded from below by the unitary bound
\beq
	\label{eq:breitfreed04}
	\Delta > \biggl( \frac{d\!-\!\!3}{2} \biggr)
\eeq 
For values below the unitary bound the corresponding representation
of the symmetry group $SO(2,d \! - \! 1)$ ceases to be unitary \cite{breit_a}.
This bound is also necessary for the conservation of energy.
Hence we can easily check \cite{arx0307229} that both $\Delta_+$ and %%@
$\Delta_-$
are allowed for mass values
\beq
	-\biggl( \frac{d\!-\!\!1}{2} \biggr) \, \leq \, m^2 \! R \ads^2 \, 
		< \, -\biggl( \frac{d\!-\!\!3}{2} \biggr)
\eeq
but only $\Delta_+$ is allowed for masses
\beq
	m^2 \! R \ads^2 \, \geq \, -\biggl( \frac{d\!-\!\!3}{2} \biggr)
\eeq
In \cite{breit_a} Breitenlohner and Freedman find (explicitly for the case of %%@
$d=4$)
that the regular definition of the energy functional converges only if the field
vanishes at spatial infinity where $\overline{\rho} = \pi \! /2$ faster than
\beq
	\label{eq:breitfreed07}
	(\cos \overline{\rho})^{\frac{d-1}{2}}
\eeq
and that the improved version converges if the field vanishes faster than
\beq
	\label{eq:breitfreed08}
	(\cos \overline{\rho})^{\frac{d-3}{2}}
\eeq
\\
If the scalar is propagated by the conformal wave operator,
then the conformal mass value is
\beq
	\label{eq:breitfreed09}
	m_c^2 R \ads^2 \, = \, -\frac{d}{2}\biggl( \frac{d\!-\!\!2}{2} \biggr)
	\qquad \longrightarrow \qquad
	^{\Delta_{c+} \, = \, d/2} _{\Delta_{c-} \, = \, d/2 \, -1}
\eeq
which for $d \geq 4$ is negative.

		\subsection{Geodesics in AdS}
		\label{sec:ads_geod}
\begin{figure} [H]
	\centering
		\subfigure[]{\igx[width = 5cm]{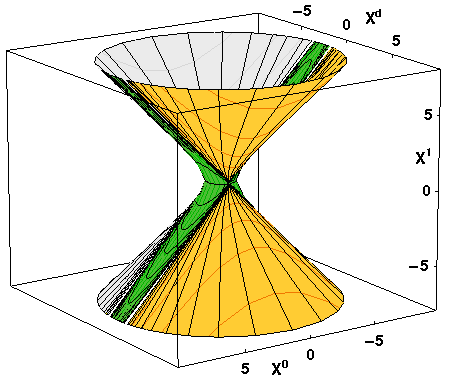}}
		\subfigure[]{\igx[width = 5cm]{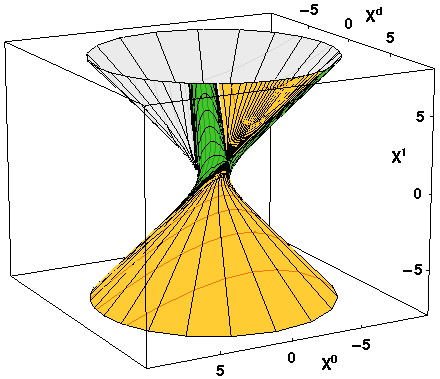}}
		\\
		\vspace{-0.2cm}
		\subfigure[]{\igx[width = 5cm]{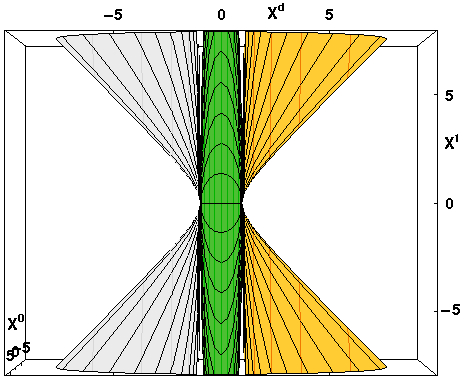}}
		\subfigure[]{\igx[width = 5cm]{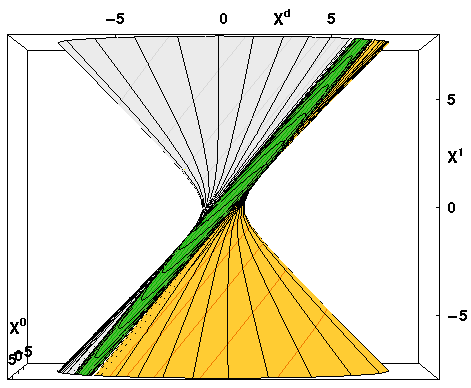}}
		\\
		\vspace{-0.2cm}
		\subfigure[]{\igx[width = 5cm]{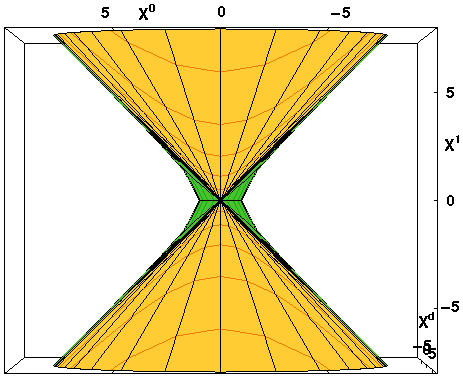}}
		\subfigure[]{\igx[width = 5cm]{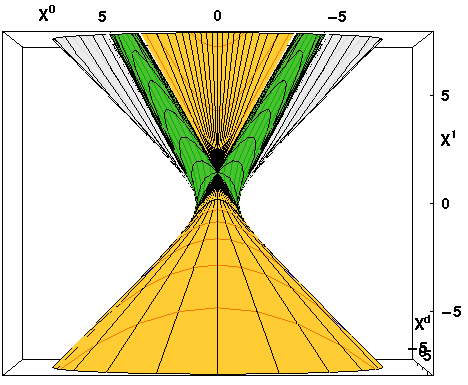}}
		\\
		\vspace{-0.3cm}
		\subfigure[]{\igx[width = 5cm]{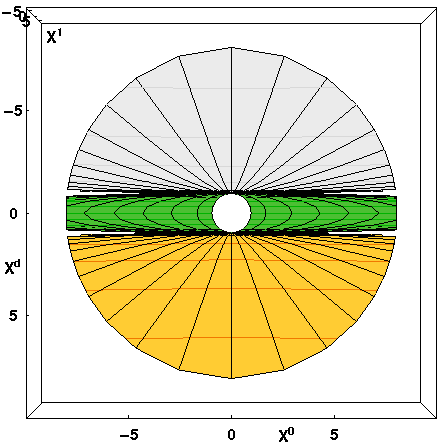}}
		\subfigure[]{\igx[width = 5cm]{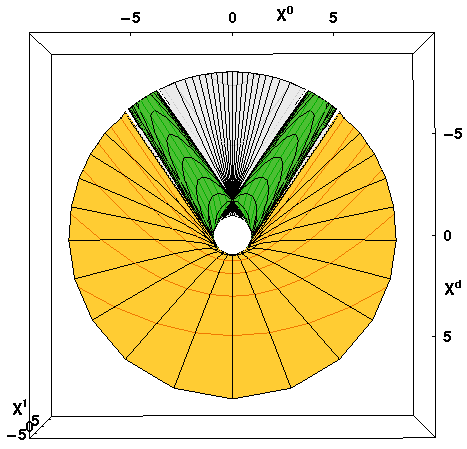}}
	\caption{AdS hyperboloid: geodesics}
	\label{fig:ads_geod}
\end{figure}

In this section we engage in finding geodesics for AdS$_d$ spacetime.
We begin considering an arbitrary curve $X^M \! {\ssty ( \! \lambda \!)}$
with contour parameter $\lambda$ on the hyperboloid (\ref{ads_coord04})
in embedding space:
\begin{align}
	\label{ads_geod00}
	X^M X_M \, & = \, \sigma R \ads ^2 \\
	\label{ads_geod01}
	\rightarrow \!\! \qquad \qquad \qquad \dot{X}^M X_M \, & = \, 0
	\qquad \qquad { \ssty \dot{X}^M = \frac{d}{d \lambda} X^M }
	\\
	\label{ads_geod02}
	\rightarrow \quad \ddot{X}^M X_M + \dot{X}^M \dot{X}_M \, & = \, 0
\end{align}
Since (\ref{ads_geod01}) is true for all curves
passing the point $X$ on the hyperboloid
we know that 
\beq
	\label{ads_geod03}
	X^M V_M = \, 0 \qquad \qquad \forall \;\; V \! \in T_X \ADS
\eeq
holds for all vectors $V$ in the tangent space of the point $X$ of AdS.
By definition a curve in embedding space is called:
\begin{align}
	\text{timelike in } X   \quad
	&  \leftrightarrow \quad \sigma\dot{X}^2 {\ssty \! ( \! X \! )}  > 0 \notag \\
	\text{lightlike in } X  \quad
	&  \leftrightarrow \quad \sigma\dot{X}^2 {\ssty \! ( \! X \! )} = 0 \\
	\text{spacelike in } X \quad
	&  \leftrightarrow \quad \sigma\dot{X}^2 {\ssty \! ( \! X \! )} < 0 \notag	
\end{align}
We can decompose the vector $\dot{X}{\ssty \! ( \! X \! )}$ into
a vector $\dot{X}_\parallel$ living in the tangent space $T_X\ADS$
and another vector $\dot{X}_\perp$ which does not live there:
\begin{align}
	\dot{X} & = \dot{X}_\parallel + \dot{X}_\perp
	\notag \\
	\rightarrow \quad \dot{X}^2
	& = \dot{X}^2_\parallel + \dot{X}^2_\perp + 2 \dot{X}_\parallel \dot{X}_\perp   
\end{align}
Using the radial $(R,t,\vec{x})$ coordinates for the embedding space we have:
\begin{align*}
	\dot{X}_\perp & = (\dot{R},0,\vec{0}) \\
	\dot{X}_\parallel & = (0, \, \dot{t},\, \dot{\vec{x}} )
\end{align*}
The wanted curve is a geodesic running on the hyperboloid
$R=R\ads=$ const. and therefore $\dot{R}$ is zero.
We now take $\lambda$ as an affine parameter.
The metric of the embedding space is block diagonal
for all three coordinate sets:
\begin{align}
	G_{MN} = \, 
		\begin{pmatrix}
			\sigma & 0 \\
			0 & g_{\mu \nu}
		\end{pmatrix}
\end{align}
Hence $\dot{X}_\parallel \dot{X}_\perp$ also vanishes.
Along a geodesic we know that the length of the tangent vector is constant:
\begin{align}
	\label{ads_geod07}
	\nabla_{ \! \dot{x}} \, \dot{x} = 0 
	\qquad \quad \rightarrow \qquad \quad
	\nabla _{\! \dot{x}} \, \dot{x}^2 = 0
\end{align}
Thus for a curve in embedding space which is a geodesic in AdS we have
\begin{align}
	\label{ads_geod08}
	\sigma \dot{X}^2 = \, \sigma \dot{X}_\parallel^2
		= \, \sigma \dot{x}^2 = \, c = \text{const.} 
\end{align}
Moreover (\ref{ads_geod07}) also implies $\ddot{X}_\parallel = 0$ and therefore 
\beq
	\ddot{X}^M V_M = \, 0 \qquad \qquad \forall \;\; V \! \in T_X \ADS
\eeq
Together with (\ref{ads_geod03}) this means
that $\ddot{X}$ must be parallel to $X$:
\beq
	\label{ads_geod085}
	\ddot{X}^M \! {\ssty ( \! \lambda \!)} = f {\ssty \! ( \! \lambda \!)} \, X^M \! {\ssty ( \! \lambda %%@
\!)}
\eeq
Plugging (\ref{ads_geod085}) and (\ref{ads_geod08})
in (\ref{ads_geod02}) we find:
\beq
	f {\ssty \! ( \! \lambda \!)} = - \frac{c}{R\ads^2}
\eeq
For timelike geodesics we have $c \, >0$ and therefore obtain:
\beq
	\label{ads_geod12}
	X^M \! {\ssty ( \! \lambda \!)}
	= \, a^M \cos ( \omega \lambda ) + \, b^M \sin ( \omega \lambda )
	\qquad \qquad
	{\ssty \omega = \frac{\sqrt{c}}{ R\ads}}
\eeq
Plugging this into (\ref{ads_geod00})
and then evaluating for $\lambda = 0, \frac{\pi}{2}$
we find that the vectors $a$ and $b$ are constrained by:
\begin{align*}
	a^M b_M & = 0 \\
	a^M a_M = \, b^M b_M & = \, \sigma R^2\ads
\end{align*}
For spacelike geodesics we have $c \, <0$ and find:
\beq
	\label{ads_geod13}
	X^M \! {\ssty ( \! \lambda \!)}
	= \, a^M \cosh ( \omega \lambda ) + \, b^M \sinh ( \omega \lambda )
	\qquad \qquad
	{\ssty \omega = \frac{\sqrt{-c}}{ R\ads}}
\eeq
Plugging into (\ref{ads_geod00})
and evaluating for $\lambda = 0, \pm \infty$ we find the constraints:
\begin{align*}
	a^M b_M & = 0 \\
	a^M a_M = - b^M b_M & = \sigma R^2\ads
\end{align*}
For nullgeodesics we have $c=0$ which leads to $\ddot{X}^M= \, 0$.
Hence lightlike geodesics on AdS are just straight lines in embedding space
and via (\ref{ads_geod08}) they are also (lightlike) geodesics
in the sense of embedding space.
In contrast, timelike (spacelike) geodesics in AdS
remain timelike (spacelike) 
but are not geodesics in embedding space.
\beq
	\label{ads_geod14}
	X^M \! {\ssty ( \! \lambda \!)} = \, a^M \lambda + \, b^M
\eeq
Plugging into (\ref{ads_geod00})
and evaluating for $\lambda = 0, \pm \infty$ this time we encounter the constraints:
\begin{align*}
	b^M b_M & = \sigma R^2\ads \\
	a^M a_M = a^M b_M & = 0
\end{align*}
Planes in embedding space spanned by two vectors $a^M$ and $b^M$
and docked at the point $c^M$ are described by
$X^M {\ssty ( \! \alpha \! ,\beta \!)} = \alpha \, a^M \! + \beta \, b^M \! + c^M$.
Therefore all three equations (\ref{ads_geod12}), (\ref{ads_geod13})
and (\ref{ads_geod14}) say that geodesics of AdS run in planes in embedding space
which contain the origin and are spanned
by the pseudoorthogonal vectors $a^M \!$ and $b^M$.
\\
The AdS geodesics are the intersection lines of these planes and the hyperboloid.
Timelike geodesics form ellipses in embedding space,
lightlike geodesics are lines and spacelike geodesics are hyperbolae.
\\
This section is based on computations taken from \cite{dorn_unpub} by Dorn
and the results are in agreement with Fronsdal \cite{fronsdal}.
\\
The geodesics of AdS are drawn in figure \ref{fig:ads_geod}.
The subfigures on the left half show geodesics running through the AdS origin
and those on the right half show geodesics through an arbitrary point.
Green background indicates timelike geodesics,
orange background marks the spacelike geodesics
running through the origin/arbitrary point
and grey background the spacelike ones through its antipode point.
The black curves running through these points are geodesics 
(i.e. the closed ellipses on green and the hyperbolae on orange/grey,
while other coloured sets of curves are auxiliary lines only).
The two subfigures at the bottom give a view "from below"
(i.e. from a point sitting in the deep on the $X^1 \!$-axis). 
\\
Comparing the subfigures on the right half
with figure \ref{fig:ads_hyp_poincare}
one can recognize that the two regions of the Poincar\'e patch
are the regions covered by the spacelike geodesics
through the point $X_P=(+\infty,+\infty,\ldots,0)$
respectively its antipode point.
These spacelike geodesics correspond to the green curves
of constant $\taubar$ in figure \ref{fig:penrose_ads_poincare}.

		\subsection{Time ordering on AdS}
		\label{sec:ads_timord}
In this section we check the invariance of time ordering in AdS
under the action of SO${\ssty (2,d-1)}$.
First we consider time ordering on the hyperboloid
and continue with the covering space.
The computations again follow Dorn \cite{dorn_unpub}.

			\subsubsection{Time ordering for AdS: Hyperboloid}
			\label{sec:ads_timord_hyp}
Trouble with time ordering on the hyperboloid
is caused by having closed timelike curves.
Therefore time ordering a priori in general cannot be transitive
e.g. point a is later then c is later then b which in turn is later then a.
\begin{figure} [H]
	\begin{center}
	\igx[width=6cm]{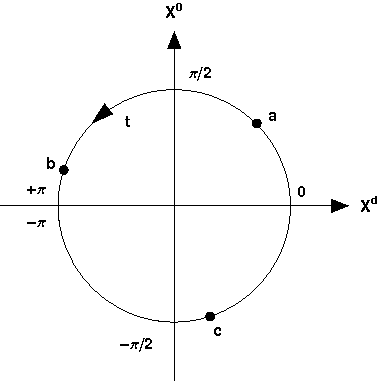}
	\end{center}
	\caption{Time ordering on the AdS hyperboloid: $X^d X^0 \!$-plane}
	\label{fig:ads_timord_hyp}
\end{figure}
Figure \ref{fig:ads_timord_hyp} shows the projection
of three points in AdS into the $(X^d,X^0)$-plane.
With $t_{x,y}$ ranging within $] \!\! - \! \pi, \! +\pi]$ on the hyperboloid
the values of $(t_x \!\! - \! t_y)$ lie within $] \!\! - \! 2\pi, \! +2\pi[$.
We assign a time order to two points $x$ and $y$ in AdS 
by defining that $x$ is 
$^{\; \text{later}}_{\text{earlier}}$ than $y$
if the time variable $t$ $^{\text{decreases}}_{\text{increases}}$
(except at the identified point $t= -\pi = + \pi $
where $t$ jumps discontinuously)
while going from $t_x$ to $t_y$
in the shortest possible way on the circle.
\\
This corresponds to either $0 \lessgtr (t_x \!\! - \! t_y) \lessgtr \pm \pi$
or $\mp 2\pi \lessgtr (t_x \!\! - \! t_y) \lessgtr \mp \pi$
or compactly combined $\sin (t_x \!\! - \! t_y) \gtrless 0$.
This again means that in order to ensure 
the invariance of time ordering on the hyperboloid
we have to show the invariance of
$\; \sign \, \sin (t_x \!\! - \! t_y) \; $ under SO${\ssty (2,d-1)}$. 
Defining
\begin{align}
	\label{eq:ads_timord_hyp01}
	S = \, \frac{X^0 Y^d - X^d Y^0}{R^2\ads} \,
	& = \sqrt{1+\vec{x}^{\, 2}} \sqrt{1+\vec{y}^{\, 2}} \, \sin (t_x \! -t_y)
			\\
	& = \; \cosh \rho_x \, \cosh \rho_y \; \sin (t_x \! -t_y)
			\notag \\
	& = 1 / \, ( \cos \ovl{\rho}_x \cos \ovl{\rho}_y ) \, \sin (t_x \! -t_y)
			\notag	
\end{align}
we see that $\; \sign \, S = \, \sign \sin (t_x \!\! - \! t_y)$.
Therefore we now only need to check the invariance of $\sign \; S$
for causally connected points under boosts in the (0,1)-plane.
Boosts in the other (0,k)-planes and (k,d)-planes are analogue
and rotations in the (0,d)-plane or a (j,k)-plane
trivially leave $(t_x \!\! - \! t_y)$ invariant.
$R^2\ads$ is a scalar product
and thereby SO${\ssty (2,d-1)}$ invariant.
With $X'$ being $X$ boosted by the boost matrix $A$
of rapidity $\chi \, \elof \; [ \! -\infty, \!+\infty]$ we have:
\begin{align}
	\label{eq:ads_timord_hyp02}
	X'^M \, & = \, A^M_{\;\; N} \, X^N
	\qquad \quad
	A^M_{\;\; N} \, = \,
		\begin{pmatrix}
			\; \cosh \chi & \sinh \chi & 0 \\
			\; \sinh \chi & \cosh \chi & 0 \\
			0 & 0 & \; \mathbb{1}_{d \! - \! 1}         
		\end{pmatrix}
		\\
	\label{eq:ads_timord_hyp03}
	\rightarrow \quad	A^0_{\;\; 0} & \geq \; \mid \! A^0_{\;\; 1} \! \mid
	\qquad \quad
	A^0_{\;\; 0} \geq \, 1	 
\end{align} 
Two considered points $X$ and $Y$ on the hyperboloid
surely are causally connected
if there exists a timelike geodesic connection between them.
From section (\ref{sec:ads_geod}) we know
that timelike geodesics are ellipses given by the intersection
of the hyperboloid with a two-dimensional plane
fixed by the origin and the two points $X$ and $Y$.
Hence timelike geodesics fulfill both
\begin{align}
	X^{^2} \, & = \, \sigma R^2\ads
	\\
	\label{eq:ads_timord_hyp05}
	\vec{X} \, & = \, \vec{a} \, X^0 \; + \; \vec{b} \, X^d
\end{align}
wherein the parameter vectors $\vec{a}$ and $\vec{b}$
are fixed by the points $X$ and $Y$.
For a timelike geodesic $X {\ssty \! ( \! \tau \! )}$
with affine contour parameter $\tau$
we know that $\sigma \! \dot{X}^{^2} > 0$ by definition
along all points of the geodesic.
Using (\ref{eq:ads_timord_hyp05}) we find:
\begin{align}
	\label{eq:ads_timord_hyp06}
	\sigma \! \dot{X}^{^2} \! & = \,
	(1 \! - \! \vec{a}^{\, 2}) \, \dot{X}^{0^2} \! + \:
	(1 \! - \! \vec{b}^{\, 2}) \, \dot{X}^{d^2} \! - \:
	2 \, \vec{a}\, \vec{b} \, \dot{X}^0 \dot{X}^d
	\qquad \qquad
	{ \ssty \dot{X} = \frac{d}{d \tau} X }
	\\
	\label{eq:ads_timord_hyp07}
	& \qquad \quad \; \vec{a}^{\, 2} \! < 1
	\qquad \quad \;\;\;
	\vec{b}^{\, 2} \! < 1
\end{align}
Since timelike geodesics are closed ellipses on the hyperboloid,
each of them passes points with $\dot{X}^0 \! =0$
and other points with $\dot{X}^d \! =0$.
Looking at $\sigma \! \dot{X}^{^2} \! > 0$
at a point with $\dot{X}^0 \! =0$
we can read off from (\ref{eq:ads_timord_hyp06})
that $\vec{b}^{\, 2} \! < 1$ 
and looking at points with $\dot{X}^d \! =0$ we find
that $\vec{a}^{\, 2} \! < 1$.
Then for $X$ and $Y$ connected by a timelike geodesic
we can compute for a boost in the (0,1)-plane:
\begin{align}
	\ob{ \, X'^0 Y'^d \! - X'^d Y'^0}^{S'} & 
	\os{(\ref{eq:ads_timord_hyp02})}{=} \,
			 (A^0_{\;\;0} X^0 \! + A^0_{\;\;1} X^1) Y^d
	- X^d (A^0_{\;\;0} Y^0 \! + A^0_{\;\;1} Y^1)
		\notag \\
	& \os{(\ref{eq:ads_timord_hyp05})}{=}
	 \, \ub{(A^0_{\;\;0} \! + a^1 \! A^0_{\;\;1})}_{
			(\ref{eq:ads_timord_hyp03}),(\ref{eq:ads_timord_hyp07})
					\rightarrow \; > \, 0}
			\ub{(X^0 Y^d \! - X^d Y^0)}_S
\end{align}
Thus $\sign \; \snake{S} = \sign \; S$, which shows
that $\sign \; S$ is invariant under SO${\ssty (2,d-1)}$
for points on the hyperboloid
which are connected by a timelike geodesic.
\begin{figure} [H]
	\begin{center}
		\igx[height = 8.6cm]{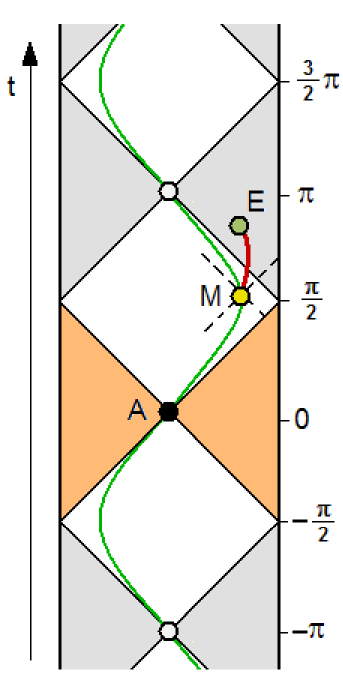}
	\end{center}
	\caption{Penrose diagram of universal covering AdS}
	\label{fig:penrose_ads_uc_nongeocon}
\end{figure}
However, as drawn in figure \ref{fig:penrose_ads_uc_nongeocon},
there also exist causally connected points like $A$ and $E$
which cannot be connected by a timelike geodesic.
Nevertheless we can always connect these points
via two timelike geodesics connected at an intermediate point $M$.
For these the time order is invariant,
and since the relations ${\ssty \, \gtreqqless \,}$ are transitive,
time ordering is also SO${\ssty (2,d-1)}$-invariant
for causally but non-geodesically connected points:
\begin{align*}
	\ub{\, t_E > t_M}_{\text{invariant}}
	\quad \text{and} \quad 
	\ub{\, t_M > t_A}_{\text{invariant}}
	\qquad \longrightarrow \qquad
	\ub{\, t_E > t_M > t_A}_{\text{invariant}}
\end{align*}
Thus the action of the AdS invariant time ordering operator $T_{\text{hyp}}$
for a field $\phi {\ssty \! (\!x\!)}$ reads:
\begin{align}
	T_{\text{hyp}} \, \phi {\ssty (\!x\!)} \, \phi {\ssty (\!y\!)} \,
	& = \, \theta ({\ssty \sin (t_x - t_y)\!}) \,
			\phi {\ssty (\!x\!)} \, \phi {\ssty (\!y\!)}
			\qquad \qquad \quad {\ssty t_{x,y} \, {\sssty \in } \; ]-\pi,+\pi]}
			\notag \\
	\label{eq:ads_timord_hyp09}
	& \quad \; + \, \theta ({\ssty \sin (t_y - t_x)\!}) \, 
			\phi {\ssty (\!y\!)} \, \phi {\ssty (\!x\!)}
\end{align}
This agrees with the results published by Castell \cite{castell_68}
wherein the nontransitivity of this time ordering is already stated:
from
$T \, \phi {\ssty (\!x\!)} \, \phi {\ssty (\!y\!)}
= \phi {\ssty (\!x\!)} \, \phi {\ssty (\!y\!)}$
and
$T \, \phi {\ssty (\!y\!)} \, \phi {\ssty (\!z\!)}
= \phi {\ssty (\!y\!)} \, \phi {\ssty (\!z\!)}$
we cannot conclude that
$T \, \phi {\ssty (\!x\!)} \, \phi {\ssty (\!z\!)}
= \phi {\ssty (\!x\!)} \, \phi {\ssty (\!z\!)}$.
\\
We can also write time ordering in an alternative way
using the intervals discussed above equation(\ref{eq:ads_timord_hyp01}):
\begin{align}
	T_{\text{hyp}} \, \phi {\ssty (\!x\!)} \, \phi {\ssty (\!y\!)} \,
	& = \; \bigl[ \theta {\ssty (t_x \! - t_y)\!} \, \theta {\ssty (t_y \! - t_x+\pi)\!}
				+ \theta {\ssty (t_y \! - t_x-\pi)\!} \bigr] \,
			\phi {\ssty (\!x\!)} \, \phi {\ssty (\!y\!)}
			\qquad \quad {\ssty t_{x,y} \, {\sssty \in } \; ]-\pi,+\pi]}
			\notag \\
	& \quad \, + \! \bigl[ \theta {\ssty (t_y \! - t_x)\!} \, \theta {\ssty (t_x \! - %%@
t_y+\pi)\!}
				+ \theta {\ssty (t_x \! - t_y-\pi)\!} \bigr] \, 
			\phi {\ssty (\!y\!)} \, \phi {\ssty (\!x\!)}
\end{align}
We could freely add a step function
for the condition $(t_x \!\! - \!t_y) \gtrless \mp 2\pi$.
However there is no need for doing so
since for $t_{x,y} \, \elof \; ] \! - \! \pi, \!+\pi ] \; $ it is always equal to one
and it also makes no difference
for the production of delta sources
considered in section \ref{sec:timord_delta}.

			\subsubsection{Time ordering for AdS: Universal cover}
			\label{sec:ads_timord_uc}
The problem with time ordering on the hyperboloid is not caused
by the absence of a direction of time
but by the presence of closed timelike curves.
Since timelike curves are open on the universal cover
we are not faced with this problem here.
\\
In order to demonstrate the invariance of standard
$\sign \, (t_x \! - t_y)$ time ordering we now need to to show,
that for a timelike curve $X {\ssty \! (\! \tau \!)}$ in AdS
the direction of the growing contour parameter $\tau$
is SO${\ssty (2,d-1)}$ invariant.
Here a growing contour parameter can be pointing either in
or against the direction of time $t$.
The sense of the time arrow in the $(X^d,X^0)$-plane
is chosen counterclockwise in the direction of growing $t$.
\begin{figure} [H]
	\begin{center}
	\igx[width=6cm]{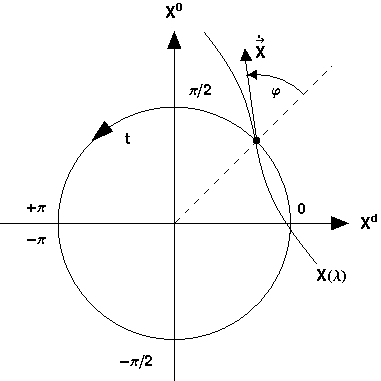}
	\end{center}
	\caption{Time ordering on the universal covering AdS: $X^d X^0 \!$-plane}
	\label{fig:ads_timord_uc}
\end{figure}
For the rest of this section we use the notation $\vec{X}=(X^d,X^0)$.
Figure \ref{fig:ads_timord_uc} shows the projection
of a curve $X {\ssty \! (\! \tau \!)}$ in AdS into the $(X^d,X^0)$-plane.
The dashed line is perpendicular
to the coordinate line of $t$ in the considered point. 
The tangent vector $\dot{\vec{X}}$ is pointing
$^{\;\;\; \text{in}}_{\text{against}}$ the direction of time if
$0 \lessgtr \varphi \lessgtr \pm \pi$ 
i.e. if $\sin \varphi \gtrless 0$.
We find:
\begin{align}
	\ub{ \, X^d \dot{X}^0 - X^0 \dot{X}^d}_{\snake{S}}
	& = \; \mid \! \vec{X} \! \mid \; \mid \! \dot{\vec{X}} \! \mid \sin \varphi
		& & { \ssty \dot{X} \, = \, \frac{d}{d \tau} X }
		 \notag \\
	\label{eq:ads_timord_uc01}
	& = R^2\ads \cosh ^2 \!\! \rho \;\, \dot{t}
		& & { \ssty \dot{t} \; = \, \frac{d}{d \tau} \, t (\tau) }
\end{align}
$\snake{S}$ is the area of the parallelogram
spanned by $\vec{X}$ and $\dot{\vec{X}}$
bearing analogy with the vector product in three dimensions.
\\
For a timelike curve $ t {\ssty (\! \tau \!)}$ must be strictly monotonic
and the contour parameter $\tau$ is pointing 
$^{\;\;\; \text{in}}_{\text{against}}$ the direction of time
if $ t {\ssty (\! \tau \!)}$ is strictly monotonic
$^{\text{increasing}}_{\text{decreasing}}$
i.e. if $\dot{t} \gtrless 0$.
In equation (\ref{eq:ads_timord_uc01}) we see
that the signs of $\dot{t}$, $\sin \varphi$ and $\snake{S}$ are the same.
\\
But the proof that $\sign \, \snake{S}$ is SO${\ssty (2,d-1)}$ invariant
runs parallel to the proof for $S$ in the previous section.
Therefore the growing contour parameter $\tau$
is SO${\ssty (2,d-1)}$-invariantly pointing in the same chronological direction.
\\
We choose $t=\tau$ and with this natural choice the action
of the AdS-invariant time ordering operator $T_{\text{cov}}$
for a field $\phi {\ssty \! (\!x\!)}$ reads:  
\begin{align}
	\label{eq:ads_timord_uc02}
	T_{\text{cov}} \, \phi {\ssty (\!x\!)} \, \phi {\ssty (\!y\!)} \,
	& = \, \theta {\ssty (t_x \! - t_y \! )} \;
			\phi {\ssty (\!x\!)} \, \phi {\ssty (\!y\!)}
			\qquad \qquad \quad {\ssty t_{x,y} \, \in \, ]-\infty,+\infty[}
			\notag \\
	& \quad \; + \, \theta {\ssty (t_y \! - t_x \! )} \:
			\phi {\ssty (\!y\!)} \, \phi {\ssty (\!x\!)}
\end{align}
If we choose $\tau$ to only point in the direction of time
but without being identical with $t$,
then in (\ref{eq:ads_timord_uc02}) we simply have to replace $t$ by $\tau$.
	\clearpage
	\section{Basic properties of de Sitter spacetime}
	\label{sec:basprop_ds}
In this section we consider basic geometric properties
of de Sitter spacetime.
It is held in the same style and sequence as section \ref{sec:basprop_ads}
with the intention being
that the similarities and differences between AdS and dS
shall conveniently be seen.

		\subsection{Coordinate systems}
\begin{figure} [H]
	\centering
		\subfigure[]{\igx[width = 6cm]{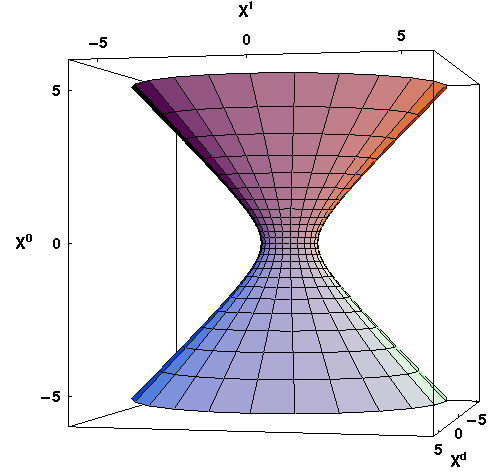}}
		\quad
		\subfigure[]{\igx[width = 5.5cm]{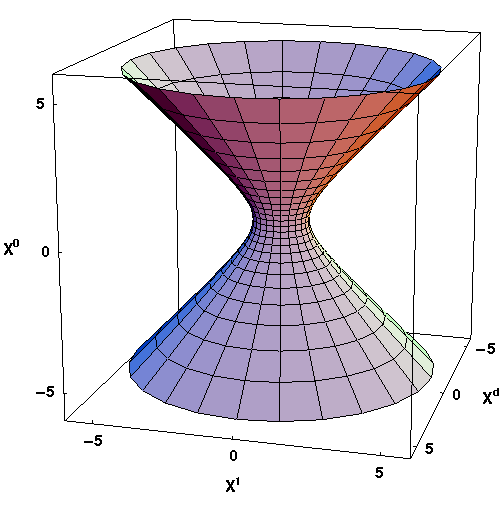}}
	\caption{de Sitter hyperboloid}
	\label{fig:ds_hyp_global}
\end{figure}

A de Sitter space dS is a homogeneous isotropic space
with constant positive curvature. $d$-dimensional de Sitter space dS$_d$
can be realized as a hyperboloid in a $(d+1)$-dimensional embedding space
with the indefinite metric
\begin{align}
	\label{eq:ds_coord01}
	\eta = \sigma \, \text{diag} \,(+,-,\ldots,-) \qquad \qquad \sigma = \pm 1
\end{align}
The overall sign $\sigma$ again is only conventional choice
but will be maintained throughout our considerations.
Cartesian coordinates for a point X in the embedding space are given by
\begin{align}
	X = (X^0 \! , \vec{X}, X^d) \qquad \vec{X} = (X^1,\ldots,X^{d-1})
\end{align}
and the $(d\!+\!1)$-dimensional d'Alembertian in embedding space is
\begin{align}
	\Box_{bulk} = \, \eta^{MN} \del_M \del_N 
	= \, \sigma \, (\del_{0}^2 \! - \del_{d}^2 - \vec{\del} \, ^2)
\end{align}
with indices in Latin upper cases running within $(0,\ldots,d)$.
dS$_d$ then corresponds to a hyperboloid in embedding space with
\beq
	\label{eq:ds_coord04}
	X^{^2} =  \, \eta_{MN} X^M X^N
	= \, \sigma \Bigl( X^{0^2} \!\! - X^{d^2} \! -\vec{X}^{^2} \Bigr) 
	= \, - \sigma R^{^2} \ds = \, const.
\eeq
Figure \ref{fig:ds_hyp_global} shows such a hyperboloid
for the case of dS$_{d=2}$.
For the embedding space we can introduce
several systems of radial coordinates
from which dS emerges by demoting the radial coordinate $R$
to a constant parameter $R \ds$.
In \cite{leshouches} many coordinate systems 
(and also a lot of other basic properties)
of dS spacetime are presented.
Here we consider three of them and later also a fourth one.
\begin{align}
	X^0 \; & = \; R \; \sinh \tau \quad \;\;\,
				= \; R \; \tan T \qquad
	            = \; R \, \sqrt{1-\vec{x} \,^2} \, \sinh t \qquad \quad
				{\ssty -\infty \, < \, t, \tau \, < \, + \infty}
	 			\notag \\
	X^d \; & = \; R \, \xi^d {\ssty \! (\!\vec{\varphi})} \cosh \tau
				= \; R \, \xi^d {\ssty \! (\!\vec{\varphi})} / \cos T
	            = \; R \, \sqrt{1-\vec{x}\,^2} \, \cosh t \qquad \quad
	 			{\ssty -\frac{\pi}{2} \, < \, T \, < \, \frac{\pi}{2}}
				\notag \\
	\label{ds_coord05}
	X^i \; & = \; R \; \xi^i {\ssty \! (\!\vec{\varphi})} \cosh \tau
				= \; R \; \xi^i {\ssty \! (\!\vec{\varphi})} / \cos T
				= \; R \: x^i \qquad 
				 \qquad \qquad \qquad \quad
				 {\ssty \vec{\xi}\, ^{2 \,}=1 }				  
\end{align}
In figure \ref{fig:ds_hyp_global} the circles
running around the hyperboloid
are lines of constant time variables $\tau$ and $T$
and the hyperbolae are lines of constant $\phii^1$. 
Following \cite{leshouches} we refer to $(\tau,\vec{\varphi})$
as global coordinates because they cover the whole hyperboloid
and to $(T,\vec{\varphi})$ as conformal coordinates.
$\vec{\varphi}=(\varphi^1,\ldots,\varphi^{d-1})$ and $\vec{x}$
are spatial coordinates in dS.
We can read off, that the topology of dS$_d$ is 
$ \mathbb{R}^1_{\text{time}} \cross S^{d-1}_{\text{space}}$.
Therefore we are not faced with closed timelike curves.
The conformal boundary of dS$_d$ is formed by all points with
$T=\pm\pi/2$ and its topology is that of two spheres $S^{d-1}$.
\\
The point $X_O=(0,\ldots,0,R)$ in embedding space is the origin
of deSitter spacetime: $\; T=t=\tau=\vec{\varphi}=\vec{x} =r=\rho=0\;$.
The antipodal point $\tilde{X}$ of a point $X$ on the hyperboloid
in embedding space is unique: $\tilde{X}=-X$.
\begin{align}
	X \, & = \, (R,\tau,\vec{\xi}) \quad \;\;\,
		= \, (R,T,\vec{\xi}) \quad \;\;
		= \, (R,t,\vec{x})
	\notag \\
	\label{eq:ds_coord10_5}
	\rightarrow \quad
	\snake{X} \, & = \, (R, -\tau, -\vec{\xi}) \,
		= \, (R, -T, -\vec{\xi})
		= \, (R, t, -\vec{x})
\end{align}
For the
$\vec{\xi} {\ssty \! (\!\vec{\varphi})} = (\xi^1,\ldots,\xi^d)$ coordinates,
we use standard spherical coordinates as in subsection \ref{sec:ads_coord},
but here for a $(d\!-\!1)$-sphere $S^{d-1}$ and in modified sequence: 
\begin{align}
	\xi^d & = \, \cos \varphi^1
	 	& & 		{\ssty 0 \, \leq \; \varphi^k \;\;\;\; \leq \; \pi}
		\notag \\
	\xi^{d-1} & = \, \cos \varphi^2 \, \sin \varphi^1 
	 	& & 		{\ssty 0 \, \leq \; \varphi^{d-1} \, < \; 2\pi}
		\notag \\
	& \, \ldots & & {\ssty k \, = \, 2, \, \ldots, \, (d-1)} \notag \\
	\xi^{2} & = \, \cos \varphi^{d-1} \,
							\sin \varphi^{d-2} \, \ldots  \, \sin \varphi^1 & & 
		\notag \\
	\xi^{1} & = \, \sin \varphi^{d-1} \,
							\sin \varphi^{d-2} \, \ldots  \, \sin \varphi^1 & &
		\notag \\
	\rightarrow \qquad
		-\vec{\xi} {\ssty (\phii^1 \! , \, \ldots, \, \phii^{d-2} \! , \, \phii^{d-1})}
		& = \, \vec{\xi} {\ssty (\pi-\phii^1 \! , \, \ldots, \, \pi-\phii^{d-2} \! , \,
							\phii^{d-1} \! \pm \pi)} & &
\end{align}
Our South Pole is sitting at $(\phii^1,\ldots,\phii^{d-2}) = 0$
and the North pole sits at $(\phii^1,\ldots,\phii^{d-2}) = \pi $.
Since the sequence of the $\vec{\xi}$ is exactly the other way round
in \cite{leshouches}, the South Pole therein sits at $\pi$
and the North Pole at $\pi$.
Except from this definition, our embedding space notation
is the same as the one used therein.
\begin{figure} [H]
	\begin{center}
		\igx[width = 8cm]{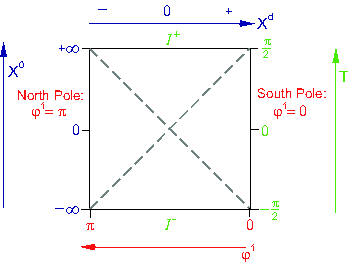}
	\end{center}
	\caption{Penrose diagram of de Sitter spacetime: Conformal coordinates}
	\label{fig:penrose_ds_conf_01}
\end{figure}
Figure \ref{fig:penrose_ds_conf_01} is the Penrose diagram of dS
with the conformal time variable $T$ (linear)
and $(X^0 \! \sim \tan T)$ (nonlinear) at the vertical axis.
The angular variable $\phii^1$ is pointing leftward (linearly)
and thus $(X^d \! \sim \cos \phii^1\! / \cos T )$
points rightward (nonlinearly).
Lightrays travel at angles of $45\degree$ and the dashed grey lines
are the lightcone of a point with $0=T=X^0 \! =X^d$
and $\phii^1\! = \pi \! / 2$.
In contrast to AdS, de Sitter spacetime is globally hyperbolic. 
\\
The spacelike hypersurfaces $I^\pm$ are the conformal
boundaries at future and past infinity.
North and South Pole appear as timelike lines in the diagram
and each correspond to a single point on the sphere
whereas each point in the interior of the diagram
represents a subsphere S$^{d-2}$.
\begin{figure} [H]
	\begin{center}
		\igx[width = 8cm]{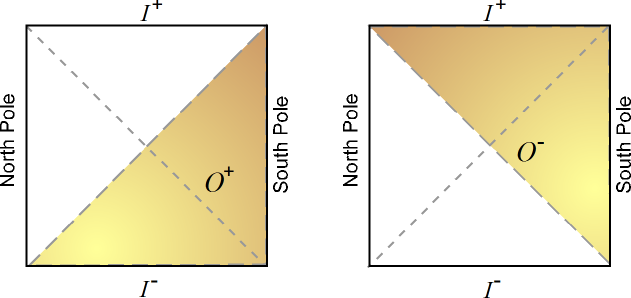}
	\end{center}
	\caption{Past and future lightcone of the South Pole}
	\label{fig:penrose_ds_conf_02}
\end{figure}
Figure \ref{fig:penrose_ds_conf_02} is the same Penrose diagram.
The left diagram shows the past lightcone $O^+$
and the right one the futur lightcone $O^-$ of the South Pole,
their intersection is called southern causal diamond.
\begin{figure} [H]
	\centering
		\subfigure[]{\igx[width = 5cm]{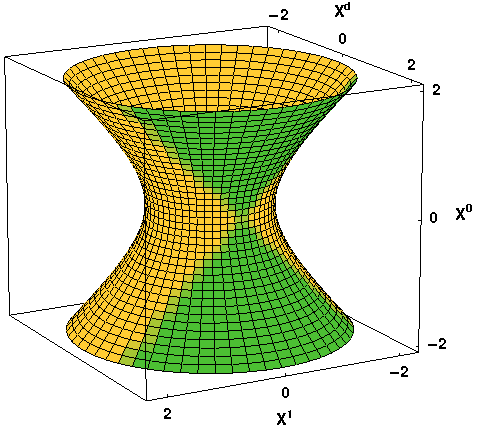}}
		\subfigure[]{\igx[width = 5cm]{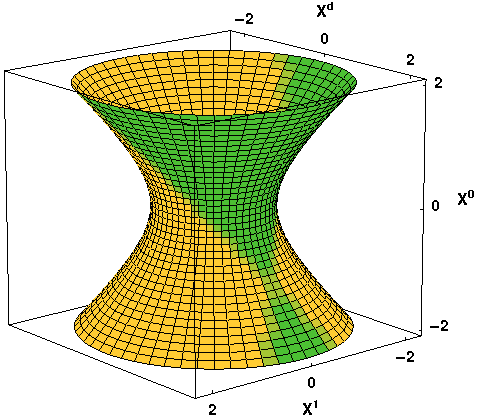}}
		\\
		\subfigure[]{\igx[width = 5cm]{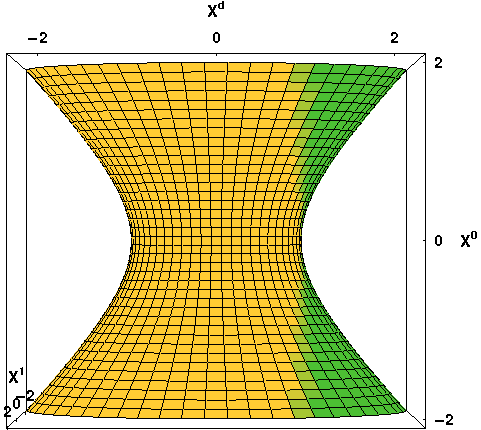}}
		\subfigure[]{\igx[width = 5cm]{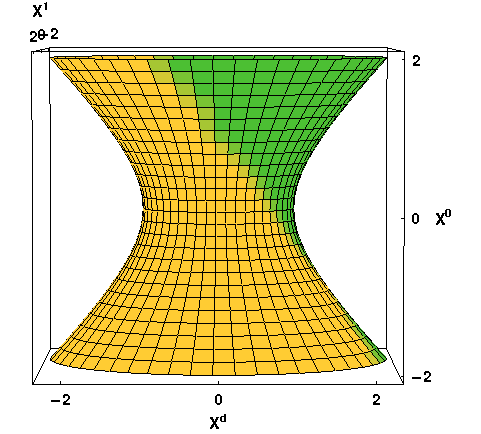}}
		\\
		\subfigure[]{\igx[width = 5cm]{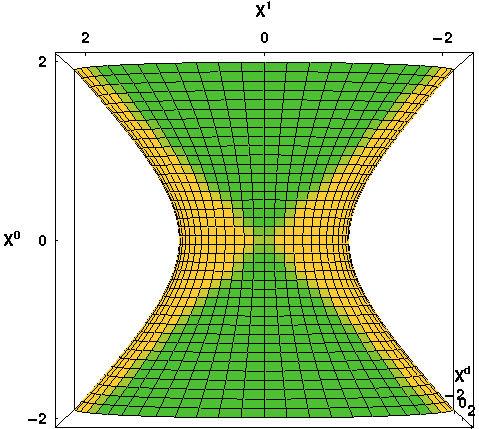}}
		\subfigure[]{\igx[width = 5cm]{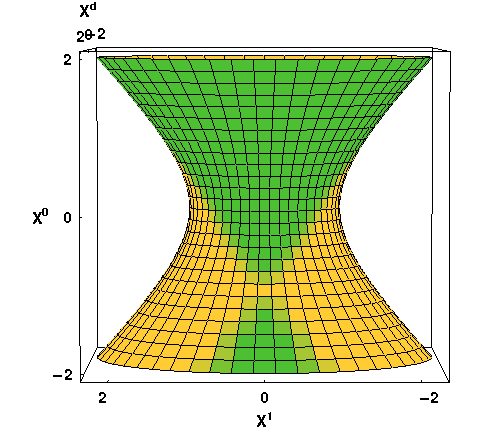}}
	\caption{de Sitter hyperboloid: lightcones}
	\label{fig:ds_hyp_lightcones}
\end{figure}
Figure \ref{fig:ds_hyp_lightcones} illustrates
the form of the lightcones one the de Sitter hyperboloid.
Subfigures (a), (c) and (e) on the left show the lightcones of the AdS origin
while (b), (d) and (f) one the right are plotted for an arbitrary reference point.
The spacelike regions appear in orange again
and the timelike ones in green.
Again the lightcones are cut out of the hyperboloid
by a plane in embedding space containing our reference point
and being perpendicular (with respect to the metric of embedding space)
to the vector pointing from embedding space origin
towards the reference point.
\\
In subsection \ref{sec:ds_geod} we will find that the form
of the dS geodesics on the hyperboloid
is the the same as in AdS with only timelike and spacelike exchanged.
Thus (like AdS) de Sitter spacetime is not geodesically convex.
This can also be read off from the Penrose diagram
for $T$ and $\phii^{d-1} \in [0,2\pi[$
which looks like figure \ref{fig:penrose_ads_hyp}
rotated to the side. 
\\
Turning to the $(t,\vec{x})$ coordinates,
one realizes that they do not cover the whole hyperboloid
but only the region with $X^d > 0$ and $\vec{X}^{2}<R^2$
which is the southern causal diamond \cite{leshouches}.
The northern diamond with $X^d < 0$ and $\vec{X}^{2}<R^2$
can be covered by adding an overall minus sign
to the definitions of $X^0$ and $X^d$ given above.
The $^{\, \text{eastern}}_{\text{western}}$ diamond
with $X^0 \gtrless 0$ and $\vec{X}^{2}>R^2$ can be covered
by changing $\sqrt{1-\vec{x} \,^2}$ into $\sqrt{\vec{x} \,^2 \!- \!1 \,}$
and exchanging $(\cosh t)$ with $(\sinh t)$
in the definitions of $X^0$ and $X^d$
for the $^{\text{southern}}_{\text{northern}}$ diamond.
Hence we have $-\infty < t,x^k < +\infty$.
\\
A priori we can freely choose the signs of each hyperbolic sine function
and thereby the direction of time in each diamond.
We fix the signs by choosing plus for the southern diamond
and requiring that crossing the lines $\vec{X}=R^2\ds$
shall not change the direction of the time coordinate.
Moreover we remark that the static coordinates are not unique:
the two points $(X^0,\vec{X},X^d)$ and $(-X^0,\vec{X},-X^d)$
are mapped to the same coordinates $(R,t,\vec{x})$.
Thus the dS hyperboloid covers the $(t,\vec{x})$-space twice:
southern and eastern diamond together one time,
and northern and western diamond a second time.
\begin{figure} [H]
	\centering
		\subfigure[]{\igx[width = 5cm]{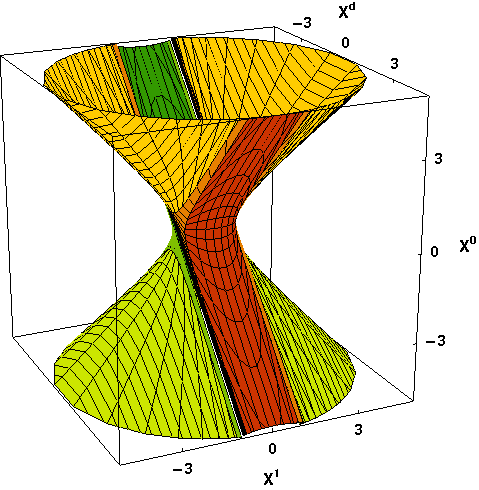}}
		\subfigure[]{\igx[width = 5cm]{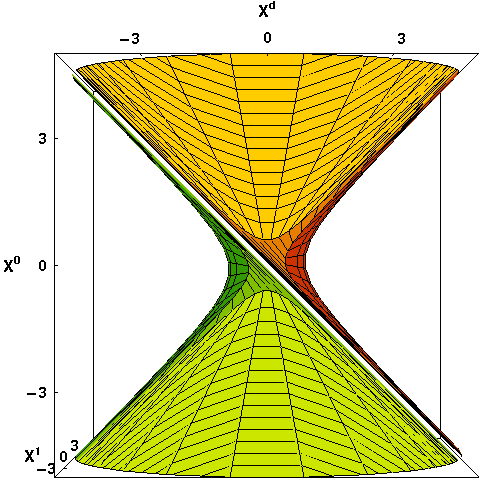}}
		\\
		\subfigure[]{\igx[width = 5cm]{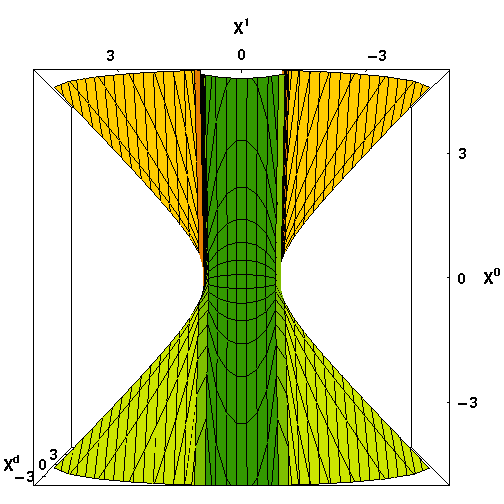}}
		\subfigure[]{\igx[width = 5cm]{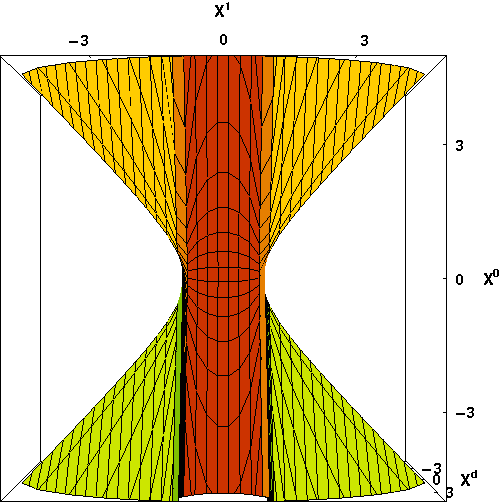}}
	\caption{de Sitter hyperboloid: planar coordinates}
	\label{fig:ds_hyp_static}
\end{figure}
Figure \ref{fig:ds_hyp_static} shows the de Sitter hyperboloid
in static coordinates.
The southern causal diamond is the stripe in strawberry red,
the northern diamond is the one in moss-green,
the eastern one is painted in tangerine and the western one in lime.
		\vspace{-0.5cm}
\begin{figure} [H]
	\begin{center}
		\igx[width=8cm]{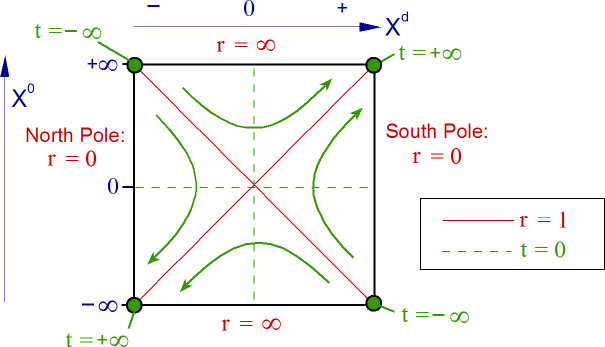}
	\end{center}
		\vspace{-0.5cm}
	\caption{Penrose diagram of de Sitter spacetime for static coordinates}
	\label{fig:penrose_ds_conf_03}
\end{figure}
\vspace{-0.2cm}
Figure \ref{fig:penrose_ds_conf_03} is again the Penrose diagram,
this time with coordinate lines of $t$. 
\\
The inverted relations between coordinates in embedding space
and in de Sitter spacetime are given by:
\vspace{-0.2cm}
\begin{align}
	R & = \sqrt{X^{0^2} \!\! -X^{d^2} \! -\vec{X}^{^2}}
	\phantom{\biggl(  \biggr)} \\
	\sinh \tau = \tan T 
	& = X^{0}/\sqrt{(X^{0^2} \!\! -X^{d^2} \! -\vec{X}^{^2})} \\
	\tanh t & = \theta {\ssty (R^2\ds \! - \! \vec{X}^{2})} \, \frac{X^0}{X^d} \, 
					+ \, \theta {\ssty (\vec{X}^{2} \! - \! R^2\ds)} \, \frac{X^d}{X^0}
	\\
	\vec{x} & = \frac{\vec{X}}{\sqrt{X^{0^2} \!\! - X^{d^2} \! -\vec{X}^{^2}}}
\end{align}
In deSitter spacetime there also exists a patchwise coordinate system
with a particular simple metric.
These so called planar coordinates are given by:
\begin{align}
	X^0 & \; = \; \frac{R\ds}{2 \ovl{\tau}} \, 
					\bigl( \ovl{\tau}^{\,2} \! - \vec{a}^{\,2} \! - 1\bigr)
				&
				{\ssty -\infty \, < \; \ovl{\tau},a^k \, < \; +\infty} &
	 			\notag \\
	X^d & \; = \; \frac{R\ds}{2 \ovl{\tau}} \, 
					\bigl( \ovl{\tau}^{\,2} \! - \vec{a}^{\,2} \! + 1\bigr)
					&
					{\ssty \ovl{\tau} \, \neq \, 0} &
				\notag \\
	X^k & \; =  \; R\ds \, \frac{\, a^k \,}{\ovl{\tau}}
				&
				{\ssty \vec{a}\, = \, (a^k) \qquad k \, = \, 1, \, \ldots, \, (d-1)} &
				\notag				  
\end{align}
Our planar time variable $\ovl{\tau}$
is connected to the planar time variable $t_{SSV}$
of Spradlin, Strominger and Volovich in \cite{leshouches}
via $\ovl{\tau}= e^{t_{SSV}}$. 
The inverted relations are:
\vspace{-0.2cm}
\begin{align}
	\ovl{\tau} \, & = \; \frac{X^d \! - X^0}
										{\sqrt{X^{0^2} \!\! - \!\vec{X}^{^2}}}
			\notag \\
	a^k & = \, X^k \, \frac{X^d \! - X^0}
										{\,(X^{0^2} \!\! - \!\vec{X}^{^2})}
			\notag
\end{align}
For $\ovl{\tau}=0$ the mapping
into embedding space coordinates is ill defined.
The patch with $\ovl{\tau} > 0$
covers the one half of the hyperboloid with $X^d \! > \! X^0$
and the patch with $\ovl{\tau} < 0$ covers the other half $X^d \! < \! X^0$.
\begin{figure} [H]
	\centering
		\subfigure[]{\igx[width = 7cm]{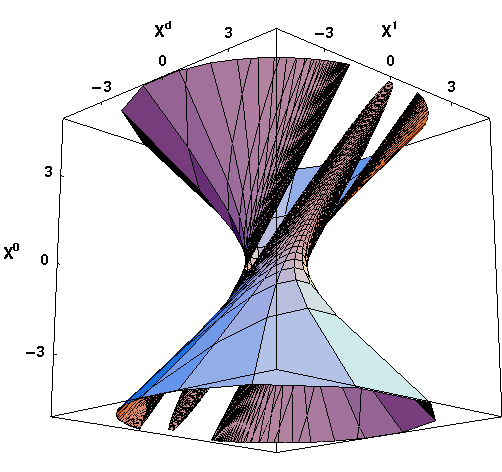}}
		\\
		\subfigure[]{\igx[width = 7cm]{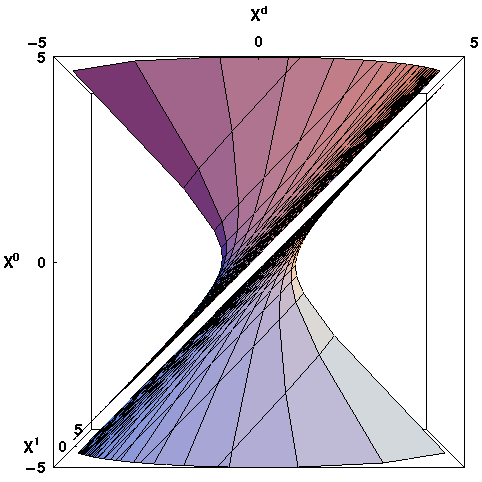}}
		\\
		\subfigure[]{\igx[width = 7cm]{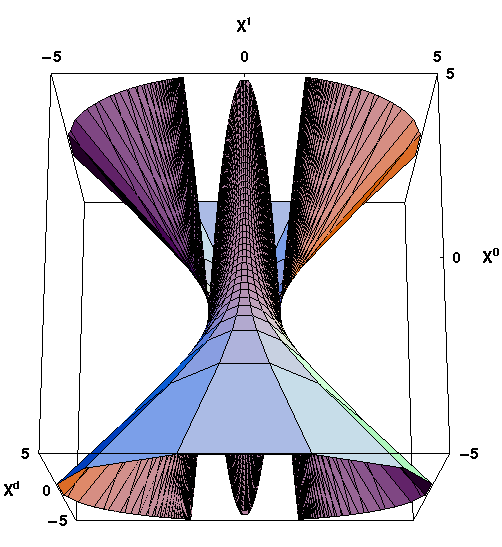}}
	\caption{de Sitter hyperboloid: planar coordinates}
	\label{fig:ds_hyp_planar}
\end{figure}
Figure \ref{fig:ds_hyp_planar} shows the dS hyperboloid
in planar coordinates.
\ref{fig:ds_hyp_planar} (b) demonstrates (again slightly exaggerated)
how the $(X^0 \! = X^d)$ - plane separates the two patches.
\begin{figure} [H]
	\begin{center}
		\igx[width = 12cm]{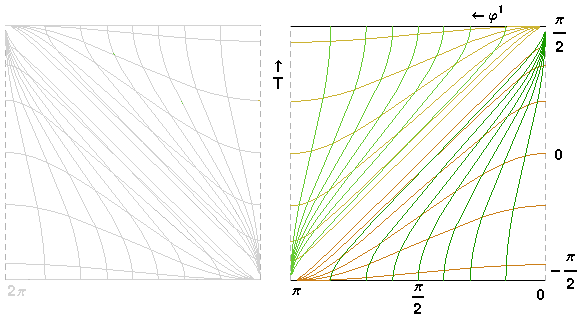}
	\end{center}
	\caption{Penrose diagram of de Sitter spacetime: planar coordinates}
	\label{fig:penrose_ds_planar}
\end{figure}
Figure \ref{fig:penrose_ds_planar} is yet another Penrose diagram
of de Sitter spacetime, now for planar coordinates.
The orange curves are lines of fixed $\taubar$
and the green ones of fixed $a^1$.
We notice that $\taubar$ and $a^1$ exchanged their roles
with regard to AdS.
For $d > 2$ we have $\phii^1$ within $[0,\pi]$
as shown in the coloured diagram,
only for the hyperboloid of dS$_{d=2}$ we have $\phii^1$
within $[0,2\pi[$ and the diagram is then extended
by the grey part on the left hand side.
Thus if for the latter case
we identify the lines of $(\phii^1=0)$ and $(\phii^1=2\pi)$
and then horizontally roll up the diagram into a cylinder, 
then we obtain a flattened version of the hyperboloid
covered by planar coordinates.
\\
In the planar coordinates there is no origin because of $\ovl{\tau} \neq 0$.
We just find that $X_O$ corresponds to $\vec{a}=0$ with $\ovl{\tau}=1$.
In planar coordinates we find for the antipodal point:
\begin{align}
	X \, & = \, (R, \; \ovl{\tau}, \, \vec{a}) \notag \\
			\rightarrow \quad
	\snake{X} \, & = \, (R,\, -\ovl{\tau}, \, \vec{a})
			\quad \notag
\end{align}

			\subsubsection{Chordal distance}
The chordal distance $u$ of two points $x$ and $y$
in deSitter spacetime is the squared distance
of their corresponding points $X$ and $Y$
on the hyperboloid in embedding space,
defined again avoiding a factor of $^1 \!\! / \! _2$ on the right hand side:
%%
%%____________________equation_ds_coord11_____________
\begin{align}
	u {\scriptstyle (X\!,Y)} & \equiv \bigl( X-Y \bigr)^2 \\
	\tilde{u} {\scriptstyle (X\!,Y)}
	\equiv \, u {\scriptstyle (X\!,\tilde{Y})} & = \bigl( X+Y \bigr)^2
\end{align}
$\snake{u}$  is the antipodal chordal distance.
Both chordal distances are unique on the hyperboloid
and SO${\ssty (1,d)}$-invariant.
\\
When written without arguments, the chordal distances
are meant to refer to the arguments $(X,Y)$ as above.
There exist a few simple relations between them for dS:
%%
%%____________________equation_ds_coord11_5_____________
\begin{align}
	\label{eq:ds_chordal03}
	u & = -2 \sigma R^2\ds \! - 2XY
	& u \! + \snake{u} & = -4 \sigma R^2\ds \\
	\label{eq:ds_chordal04}
	\snake{u} & = -2 \sigma R^2\ds \! + 2XY
	& u \snake{u} & = 4R^4\ds - 4 (XY)^2
\end{align}

		\subsection{dS metrics}
The squared infinitesimal length element is
\begin{align}
	ds^2 \ds & = \, \Bigl[ \eta_{MN}dX^MdX^N \Bigr]
	_{ \! R=R \ds=const.}\notag \\
	\label{eq:ds_metrics17_3}
	& = \, \sigma R^2 \ds \,
			\Bigl( d \tau ^2 - \cosh^2 \! \tau \: d \Omega^2_{d-1} \Bigr) \\
	\label{eq:ds_metrics17}
	& = \, \frac{\sigma R^2 \ds}{\cos^2 T} \,
	\Bigl( dT^2 - d \Omega^2_{d-1} \Bigr)
	 \\
	\label{eq:ds_metrics17_5}
	& = \, \sigma R^2 \ds \Biggl[ (1\!-\!\vec{x}\, ^2) \, dt^2 - \biggl( \!
	       	\frac{x^i x^j}{\,(1-\vec{x}\, ^2)}+\delta^{ij} \! \biggr)
			dx^i dx^j \Biggr]
	\\
	\label{eq:ds_metrics17_6}
	& = \, \sigma \frac{\, R\ds^2}{\ovl{\tau}^2} \,
			\Bigl( -d\ovl{\tau}^{\,2} \! + d\vec{a}^{\,2} \Bigr)
\end{align}
wherein $d \Omega _{d-1}$ is the infinitesimal line element
on the unit sphere $S^{d-1}$ and indices in Latin lower cases
are running within $(1,\ldots,d\!-\!1)$.
For the metric on $S^{d-1}$ in standard spherical coordinates one finds:
\beq
	\label{eq:ds_metrics_sphere}
	d\Omega_{d-1}^2 \, = \; d\varphi^{1^2} \!\!
									+ (\sin^2 \varphi^1) \, d\varphi^{2^2} \!\! + \ldots +
								(\sin^2 \varphi^1 \! {\ssty \ldots } \,
								\sin^2 \varphi^{d-2}) \, d\varphi^{d-1^2}  
\eeq
\\
Metric (\ref{eq:ds_metrics17}) is conformally equivalent
to the complete $d$-dimensional Einstein static universe
and thus (with the coordinate ranges also matching)
dS$_d$ can be conformally mapped into the full ESU$_d$
(compare with subsection \ref{sec:ads_metrics}). 
Therefore we refer to the global $(T,\vec{\varphi})$ coordinates
as conformal coordinates and call $T$ conformal time variable.
\\
The planar coordinates lead to a conformally flat metric
and hence $\ovl{\tau}$ is a conformal time variable.
\\
Only metric (\ref{eq:ds_metrics17_5}) is time independent,
which is the reason why we call $(t, \vec{x})$ static coordinates.
Fortunately the metric is the same for all four diamonds.
The static coordinates have the advantage
that (in contrast to the other time variables) $\del_t$ is a Killing vector
fulfilling the Killing equation (\ref{eq:ads_metric_killing})
and therefore can be used to define time evolution and a Hamiltonian.
\\
For the induced metric in this coordinate set
we find with indices in greek lower cases
running within $(0,\ldots,d\!-\!1)$ and $x^0 \! = t$:
\begin{align}
	g_{\mu\nu} \! & = \, -\sigma R^2 \ds \!
		\begin{pmatrix}
			(\vec{x}\,^2 \!-\!1) & 0 & 0 & 0 & \cdots \quad \\
			0 & \frac{x^1x^1}{(1-\vec{x}\,^2)}+1
			   & \frac{x^1x^2}{(1-\vec{x}\,^2)}
			   & \frac{x^1x^3}{(1-\vec{x}\,^2)}	& \cdots \quad \\
			0 & \frac{x^2x^1}{(1-\vec{x}\,^2)}
			   & \frac{x^2x^2}{(1-\vec{x}\,^2)}+1
			   & \frac{x^2x^3}{(1-\vec{x}\,^2)}	& \cdots \quad \\
			0 & \frac{x^3x^1}{(1-\vec{x}\,^2)}
			   & \frac{x^3x^2}{(1-\vec{x}\,^2)}
			   & \frac{x^3x^3}{(1-\vec{x}\,^2)}+1 & \cdots \quad \\
			\vdots & \vdots & \vdots & \vdots & \ddots \quad
		\end{pmatrix} \\
		\notag \\
	g^{\mu\nu} \! & = \, \frac{\sigma}{R^2 \ds}
		\begin{pmatrix}
			\; \, \frac{1}{(1-\vec{x}\,^2)} \;\, & 0 & 0 & 0 & \cdots \quad \\
			0 & x^1 x^1 \! -1 \quad
			   & x^1x^2
			   & x^1x^3	& \cdots \quad \\
			0 & x^2x^1
			   & x^2x^2 \! - 1 \quad
			   & x^2x^3	& \cdots \quad \\
			0 & x^3x^1
			   & x^3x^2
			   & x^3x^3 \! - 1	\quad & \cdots \quad \\
			\vdots & \vdots & \vdots & \vdots & \ddots \quad
		\end{pmatrix}
\end{align}
\begin{align}
	g_{ij} \, & = -\sigma R^2 \ds \biggl[ \frac{x^ix^j}{(1-\vec{x}\,^2)}
				+ \delta^{ij} \biggr] \\
	g^{ij} \, & = \; \frac{\sigma}{R^2 \ds} \;
						\biggl[ \;\; x^ix^j \;\; - \;\; \delta^{ij} \; \biggr] \\
	\notag \\
	\det g_{ij} \,
		& = \, (-\sigma R\ds^2)^{d-\!1} / \, (1 \! - \! \vec{x}\,^2) \\
	\det g_{\mu \nu}  & = \, -(-\sigma R\ds^2)^d
\end{align}
The d'Alembertian on dS then is
\begin{align}
	\Box \ds \, & = \, \frac{1}{\sqrt{g}} \, \del _\mu\sqrt{g} \, g^{\mu \nu}
			\del _\nu
			\qquad \qquad \qquad \qquad \qquad
			{\scriptstyle g \, = \, \mid \det g_{\mu \nu} \mid \, = \, R \ds^{2d}}
			\notag \\
	\label{eq:ds_coord23}
	& = \, \frac{\sigma}{R^2 \ds} \,
				\biggl( \frac{1}{\,(1-\vec{x}\,^2)} \, \del _t^2
			\underbrace {- \, \del _k \del _k + x^j \! x^k \del _j \del _k
								+ 2 x^k \! \del _k} \biggr) \\
	& \qquad \qquad \qquad \qquad \qquad \qquad \qquad \quad
	^{= \, \sigma R^2 \ds \Box_{\vec{x}}} \notag
\end{align}
and the Christoffel symbols (independent of $\sigma$) read:
\begin{align}
	\Gamma ^\lambda _{\alpha \beta} & \equiv \frac{1}{2} \, g^{\lambda \mu}
		\bigl( \del _\alpha g_{\beta \mu} + \del _\beta g_{\mu \alpha}
			- \del _\mu g_{\alpha \beta} \bigr)
	\\
	\Gamma ^t _{tt} & = \Gamma ^t _{\!jk} = \Gamma ^j _{tk}
		=  \Gamma ^j _{kt} = \, 0
	\\	
	\Gamma ^t _{tk} & =  \Gamma ^t _{kt} =
	 	\frac{-x^k}{(1 \! - \! \vec{x}\,^2)}
	\\
	\Gamma ^k _{tt} & = \qquad x^k \, (1 \! - \! \vec{x}\,^2)
		\qquad = \, \sigma \, g_{tt} \, \frac{x^k}{R \ds ^2}
	\\
	\Gamma ^k _{ab} & = \, x^k
		\biggl( \! \frac{x^a x^b}{\, (1 \! - \! \vec{x}\,^2)} + \delta^{ab} \! \biggr) 
		= \, \sigma \, g_{ab} \, \frac{-x^k}{R \ds ^2}		
\end{align}
The metric (\ref{eq:ds_metrics17_3})
of the global $(\tau,\vec{\phii})$ coordinates
with (\ref{eq:ds_metrics_sphere}) is diagonal
and using $\; g = \, \mid \! \det g_{\mu \nu} \! \mid \, = 
R\ds^{2d}(\cosh^2 \! \tau)^{d-1} (\sin^2 \! \phii^1)^{d-2}
\!\! \ldots (\sin^2 \! \phii^{d-1}) \;$
we find the d'Alembertian:
\begin{align}
	\label{eq:ds_metrics30}
	\Box\ds \, & = \, \frac{1}{\sqrt{g}} \, \del_\mu \sqrt{g} \,
			g^{\mu \nu} \del_\nu
			\\
	& = \, \frac{\sigma}{R^2 \ds} \, \bigl( \del^2_\tau
														+ (d \!-\!1) \tanh \tau \, \del_\tau \bigr) + \,
			\ub{ {\ssty \frac{1}{\sqrt{g}}} \, \del _j \sqrt{g} \, g^{jj} \del _j}
			\notag \\
	& \qqqquad \qqqquad \qqqquad \qqqquad \quad \,
			^{\Box_{\vec{\phii}} \;\;
			(\text{summation over} \; j \; = \, 1, \, \ldots, \, (d-1))}
			\notag 
\end{align}

		\subsection{Conformal dimension and mass term for dS spacetime}
		\label{sec:ds_confdim}
In \cite{leshouches} Spradlin, Strominger and Volovich 
for dS$_{d}$ spacetime give a relation between a parameter $\Delta$
and the mass of our scalar field:
%%
%%______________________equation_ds_confdim_01________________
\beq
	\label{eq:ds_confdim01}
	\Delta_\pm = \: \frac{d\!-\!\!1}{2} \pm
							\sqrt{\biggl( \frac{d\!-\!\!1}{2} \biggr)^{\!\! 2} \! 
								  - m^2 \! R \ds^2 \;}
\eeq 
This $\Delta_\pm$ is the conformal dimension (conformal weight)
of a CFT field that via dS/CFT correspondence is related
to a scalar field with mass $m^2$ on dS.
$\Delta_\pm$ are the two solutions of the quadratic equation 
%%
%%______________________equation_ds_confdim_02________________
\beq
	\label{eq:ds_confdim02}
	\Delta\!^2 -(d\!-\!\!1)\Delta + m^2 \! R \ds^2 = \, 0
\eeq 
We see that the conformal weights are real for small masses fulfilling
%%
%%______________________equation_ds_confdim_03________________
\beq
	\label{eq:ds_confdim03}
	m^2 \! R \ds^2 \, \leq \, \biggl( \frac{d\!-\!\!1}{2} \biggr)^{\!\! 2}
\eeq 
and become complex elsewise.
If the mass value is equal to the negative
conformal mass value of AdS (\ref{eq:breitfreed09}) 
we get the same values for the conformal dimensions:
\beq
	\label{eq:ds_confdim_09}
	m_c^2 R \ds^2 \, = \, \frac{d}{2}\biggl( \frac{d\!-\!\!2}{2} \biggr)
	\qquad \longrightarrow \qquad
	^{\Delta_{c+} \, = \, d/2} _{\Delta_{c-} \, = \, d/2 \, -1}
\eeq

		\subsection{Geodesics in dS spacetime}
		\label{sec:ds_geod}
In this section we turn to finding geodesics for dS$_d$ spacetime
following roughly the same road as taken in the section for the AdS case. 
We begin considering an arbitrary curve $X^M \! {\ssty ( \! \lambda \!)}$
with contour parameter $\lambda$ on the hyperboloid (\ref{eq:ds_coord04})
in embedding space:
\begin{align}
	\label{ds_geod00}
	X^M X_M \, & = \, -\sigma R \ds ^2 \\
	\label{ds_geod01}
	\rightarrow \!\! \qquad \qquad \qquad \dot{X}^M X_M \, & = \, 0
	\qquad \qqquad { \ssty \dot{X}^M = \frac{d}{d \lambda} X^M }
	\\
	\label{ds_geod02}
	\rightarrow \quad \ddot{X}^M X_M + \dot{X}^M \dot{X}_M \, & = \, 0
\end{align}
Since (\ref{ds_geod01}) is true for all curves
passing the point $X$ on the hyperboloid
we know that 
\beq
	\label{ds_geod03}
	X^M V_M = \, 0 \qquad \qquad \forall \;\; V \! \in T_X \DS
\eeq
holds for all vectors $V$ in the tangent space of the point $X$ of dS.
By the usual definition a curve in embedding space is called:
\begin{align}
	\text{timelike in } X   \quad
	&  \leftrightarrow \quad \sigma\dot{X}^2 {\ssty \! ( \! X \! )}  > 0 \notag \\
	\label{eq:ds_geod05}
	\text{lightlike in } X  \quad
	&  \leftrightarrow \quad \sigma\dot{X}^2 {\ssty \! ( \! X \! )} = 0 \\
	\text{spacelike in } X \quad
	&  \leftrightarrow \quad \sigma\dot{X}^2 {\ssty \! ( \! X \! )} < 0 \notag	
\end{align}
Likewise to AdS, we can decompose the vector $\dot{X}{\ssty \! ( \! X \! )}$
into a vector $\dot{X}_\parallel$ living in the tangent space $T_X\DS$
and another vector $\dot{X}_\perp$ which does not live there:
\begin{align}
	\dot{X} & = \dot{X}_\parallel + \dot{X}_\perp
	\notag \\
	\rightarrow \quad \dot{X}^2
	& = \dot{X}^2_\parallel + \dot{X}^2_\perp + 2 \dot{X}_\parallel %%@
\dot{X}_\perp   
\end{align}
Using the radial coordinates for the embedding space we have:
\begin{align*}
	\dot{X}_\perp
		& = \, (\dot{R},0,\vec{0}) \\
	\dot{X}_\parallel
		& = \, (0, \dot{t},\dot{\vec{x}} )
			= \, (0, \dot{T},\dot{\vec{\varphi}} \, )
			= \, (0, \dot{\tau},\dot{\vec{\varphi}} \,  )
\end{align*}
The wanted curve is a geodesic running on the hyperboloid
$R=R\ds=$ const. and therefore $\dot{R}$ is zero.
We now take $\lambda$ as an affine parameter.
The metric of the embedding space is block diagonal
for all three coordinate sets:
\begin{align}
	G_{MN} = \, 
		\begin{pmatrix}
			-\sigma & 0 \\
			0 & g_{\mu \nu}
		\end{pmatrix}
\end{align}
Hence $\dot{X}_\parallel \dot{X}_\perp$ also vanishes.
Along a geodesic we know that the length of the tangent vector is constant:
\begin{align}
	\label{ds_geod07}
	\nabla_{ \! \dot{x}} \, \dot{x} = 0 
	\qquad \quad \rightarrow \qquad \quad
	\nabla _{\! \dot{x}} \, \dot{x}^2 = 0
\end{align}
Thus for a curve in embedding space which is a geodesic in dS we have
\begin{align}
	\label{ds_geod08}
	\sigma \dot{X}^2 = \, \sigma \dot{X}_\parallel^2
		= \, \sigma \dot{x}^2 = \, c = \text{const.} 
\end{align}
Moreover (\ref{ds_geod07}) also implies $\ddot{X}_\parallel = 0$ and %%@
therefore 
\beq
	\ddot{X}^M V_M = \, 0 \qquad \qquad \forall \;\; V \! \in T_X \DS
\eeq
Together with (\ref{ds_geod03}) this means
that $\ddot{X}$ must be parallel to $X$:
\beq
	\ddot{X}^M \! {\ssty ( \! \lambda \!)}
		= f {\ssty \! ( \! \lambda \!)} \, X^M \! {\ssty ( \! \lambda \!)}
\eeq
Plugging (\ref{ds_geod08}) in (\ref{ds_geod02}) we find:
\beq
	f {\ssty \! ( \! \lambda \!)} = \frac{c}{R\ds^2}
\eeq
For timelike geodesics we have $ c \, > 0 $ and therefore obtain:
\beq
	\label{ds_geod12}
	X^M \! {\ssty ( \! \lambda \!)}
	= \, a^M \cosh ( \omega \lambda ) + \, b^M \sinh ( \omega \lambda )
	\qquad \qquad
	{\ssty \omega = \frac{\sqrt{c}}{ R\ds}}
\eeq
Plugging this into (\ref{ds_geod00})
and then evaluating for $\lambda = 0, \pm \infty$
we find that the vectors $a$ and $b$ are constrained by:
\begin{align*}
	a^M b_M & = 0 \\
	a^M a_M = \, -b^M b_M & = \, -\sigma R^2\ds
\end{align*}
For spacelike geodesics we have $c \, <0$ and find:
\beq
	\label{ds_geod13}
	X^M \! {\ssty ( \! \lambda \!)}
	= \, a^M \cos ( \omega \lambda ) + \, b^M \sin ( \omega \lambda )
	\qquad \qquad
	{\ssty \omega = \frac{\sqrt{-c}}{ R\ds}}
\eeq
Plugging into (\ref{ds_geod00})
and evaluating for $\lambda = 0, \frac{\pi}{2}$ we find the constraints:
\begin{align*}
	a^M b_M & = 0 \\
	a^M a_M = b^M b_M & = -\sigma R^2\ds
\end{align*}
For nullgeodesics we have $c=0$ which leads to $\ddot{X}^M= \, 0$.
Hence lightlike geodesics on dS are just straight lines in embedding space
and via (\ref{ds_geod08}) they are also lightlike geodesics
in the sense of embedding space.
In contrast, timelike (spacelike) geodesics in dS
remain timelike (spacelike) but are not geodesics in embedding space.
\beq
	\label{ds_geod14}
	X^M \! {\ssty ( \! \lambda \!)} = \, a^M \lambda + \, b^M
\eeq
Plugging into (\ref{ds_geod00})
and evaluating for $\lambda = 0, \pm \infty$ this time we encounter the %%@
constraints:
\begin{align*}
	b^M b_M & = -\sigma R^2\ds \\
	a^M a_M = a^M b_M & = 0
\end{align*}
Planes in embedding space spanned by two vectors $a^M$ and $b^M$
and docked at the point $c^M$ are described by
$X^M {\ssty ( \! \alpha \! ,\beta \!)} = \alpha \, a^M \! + \beta \, b^M \! + c^M$.
Therefore all three equations (\ref{ds_geod12}), (\ref{ds_geod13})
and (\ref{ds_geod14}) say that geodesics of dS run in planes in embedding %%@
space
which contain the origin and are spanned
by the pseudoorthogonal vectors $a^M \!$ and $b^M$.
\\
The dS geodesics are the intersection lines of these planes and the %%@
hyperboloid.
Timelike geodesics are hyperbolae in embedding space,
lightlike geodesics are lines and spacelike geodesics form ellipses.
This is just the contrary of the situation in AdS.
\\
\begin{figure} [H]
	\centering
		\subfigure[]{\igx[width = 5cm]{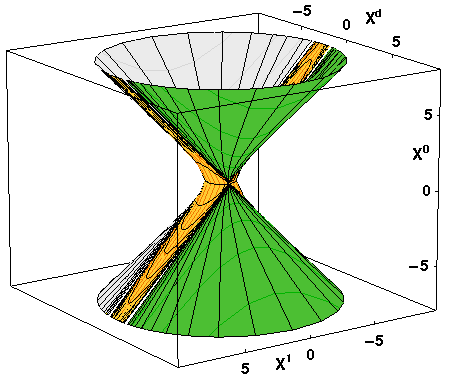}}
		\subfigure[]{\igx[width = 5cm]{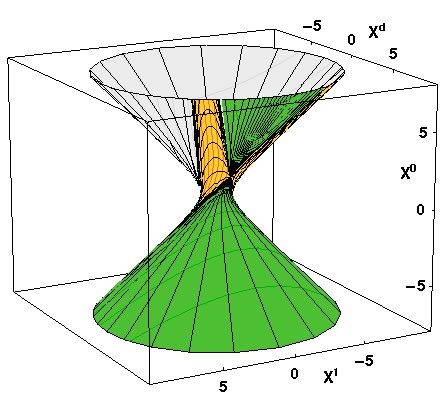}}
		\\
		\vspace{-0.2cm}
		\subfigure[]{\igx[width = 5cm]{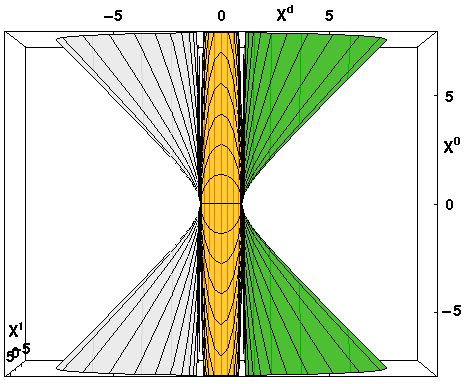}}
		\subfigure[]{\igx[width = 5cm]{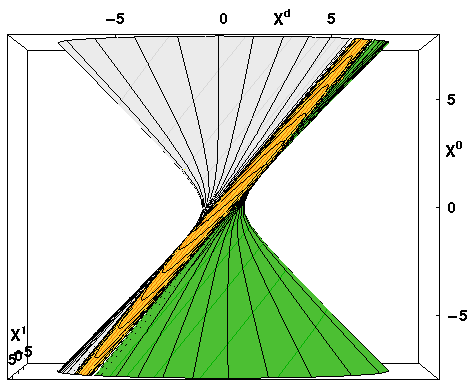}}
		\\
		\vspace{-0.2cm}
		\subfigure[]{\igx[width = 5cm]{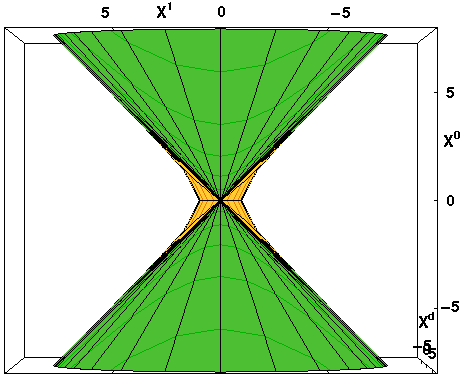}}
		\subfigure[]{\igx[width = 5cm]{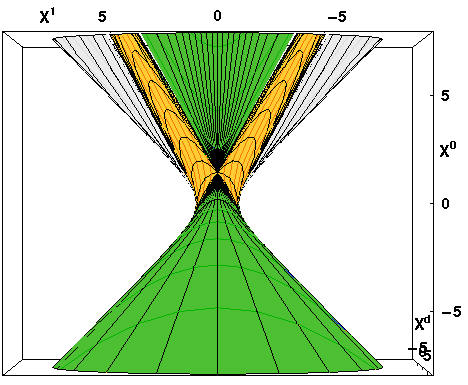}}
		\\
		\vspace{-0.3cm}
		\subfigure[]{\igx[width = 5cm]{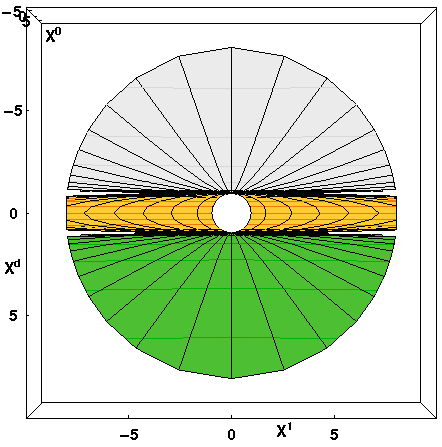}}
		\subfigure[]{\igx[width = 5cm]{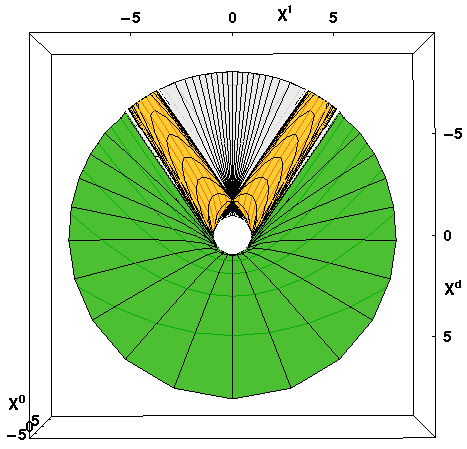}}
	\caption{dS hyperboloid: geodesics}
	\label{fig:ds_geod}
\end{figure}
The geodesics of dS spacetime are drawn in figure \ref{fig:ads_geod}.
The subfigures on the left half show geodesics running through the dS origin
and those on the right half show geodesics through an arbitrary point.
Orange background indicates spacelike geodesics,
green background marks the timelike geodesics,
running through the origin/arbitrary point
and grey background the timelike ones through its antipode point.
The black curves running through these points are geodesics 
(i.e. the closed ellipses on orange and the hyperbolae on green/grey,
while other coloured sets of curves are auxiliary lines only).
The two subfigures at the bottom give a view "from below"
(i.e. from a point sitting in the deep on the $X^0 \!$-axis). 
\\
Comparing the subfigures on the right half
with figure \ref{fig:ds_hyp_planar}
one can recognize that the two regions of the planar coordinates
are the regions covered by the timelike geodesics
through the point $X_P=(+\infty,0,\ldots,+\infty)$
respectively its antipode point.
These timelike geodesics correspond to the green curves
of constant $a^1$ in figure \ref{fig:penrose_ds_planar}.

		\subsection{Time ordering in dS}
In this section we check the invariance of time ordering in dS spacetime
under the action of SO${\ssty (1,d)}$.
First we consider time ordering for the time coordinates $T$ and $\tau$
and take a look at the time $t$ in the second subsection.

			\subsubsection{Time ordering for $T$ and $\tau$}
			\label{sec:ds_timord_ttau}
Only in this subsection indices in Latin lower cases run within (1, \ldots, d)
while Latin upper cases cover the full range of (0, \ldots, d).
Since $R=R\ds$ is constant on the hyperboloid,
both time coordinates $T$ and $\tau$
only depend on the embedding space coordinate $X^0$:
\begin{align*}
	\sinh \tau = \tan T = X^0 \! / \, R
	\qqqquad
	^{\;\;\; -\infty \, < \, \tau \, < \, +\infty}_{-\pi/2 \, < \, T \, < \, +\pi/2}
\end{align*}
Both $T$ and $\tau$ only depend on $X^0$
and grow strictly monotonic,
hence the coordinate lines of $T$ and $\tau$
are just the coordinate lines of $X^0$.
Thus in an $(X^0,X^k)$-plane we see
that a curve $\gamma$ on the hyperboloid in embedding space
with contour parameter $\lambda$
is running $^{\;\;\;\, \text{in}}_{\text{against}}$ the direction of time
if $\dot{X}^0{\ssty \!(\!\lambda\!)} \! = dX^0 \! / d\lambda \gtrless 0$.
\begin{figure} [H]
	\begin{center}
	\igx[width=6cm]{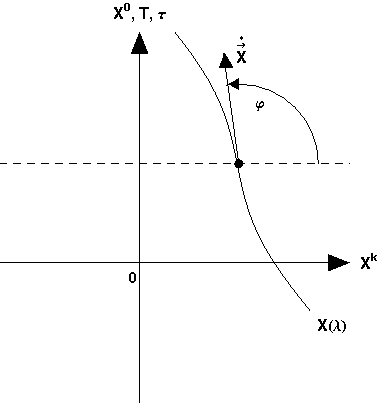}
	\end{center}
	\caption{Time ordering in dS for $T$ and $\tau$: $X^0 X^k \!$-plane}
	\label{fig:ds_timord_ttau}
\end{figure}
Figure \ref{fig:ds_timord_ttau} shows the projection
of a curve $X {\ssty \! (\! \lambda \!)}$ in dS into an $(X^k,X^0)$-plane.
The dashed line is again perpendicular
to the coordinate lines of $T$ and $\tau$ in the considered point. 
Being a scalar $R=R\ds$ already is SO${\ssty (1,d)}$ invariant
and checking the invariance of $\sign \, \dot{X}^0$
runs pretty parallel to what we did for the universal covering AdS
in subsection \ref{sec:ads_timord_hyp} and \ref{sec:ads_timord_uc}.
\\
It is sufficient to do this check for boosts in the (0,1)-plane
since the other (0,k)-planes are analogue
and rotations in the (j,k)-planes do not affect the time coordinates. 
Our boost matrix is again:
\begin{align}
	\label{eq:ds_timord_ttau01}
	\dot{X}'^M \, & = \, A^M_{\;\; N} \, \dot{X}^N
	\qquad \quad
	A^M_{\;\; N} \, = \,
		\begin{pmatrix}
			\; \cosh \alpha & \sinh \alpha & 0 \\
			\; \sinh \alpha & \cosh \alpha & 0 \\
			0 & 0 & \; \mathbb{1}_{d \! - \! 1}         
		\end{pmatrix}
		\\
	\label{eq:ds_timord_hyp02}
	\rightarrow \quad	A^0_{\;\; 0} & \geq \; \mid \! A^0_{\;\; 1} \! \mid
	\qquad \quad
	A^0_{\;\; 0} \geq \, 1	 
\end{align} 
Two points on the curve are causally connected
if they can be joined by a timelike geodesic.
In contrast to AdS, in de Sitter spacetime there
no causally connected points
(but spacelike separated points)
which cannot be joined by a geodesic.
\\
Geodesics again are sections of planes (containing
the embedding space origin) with the hyperboloid
and therefore lie within planes
with parameter vectors $\vec{a}$ and $\vec{b} \,$
fixed by two points connected via the geodesic:
\begin{align}
	\vec{X} \, & = \, \vec{a} \, X^0 \; + \; \vec{b} \, X^1
	\notag \\
	\label{eq:ds_timord_ttau03}
	\rightarrow \qquad
	\dot{\vec{X}} \, & = \, \vec{a} \, \dot{X}^0 \; + \; \vec{b} \, \dot{X}^1
\end{align}
For timelike geodesics we have $\sigma \dot{X}^2 > 0$ and therefore
\begin{align}
	\sigma \! \dot{X}^{^2} \! & = \,
	(1 \! - \! \vec{a}^{\, 2}) \, \dot{X}^{0^2} \! + \:
	(1 \! + \! \vec{b}^{\, 2}) \, \dot{X}^{1^2} \! - \:
	2 \, \vec{a}\, \vec{b} \, \dot{X}^0 \dot{X}^1
	\notag \\
	& = \, \dot{X}^{0^2} \! - \dot{X}^{1^2} \!
		- \ub{\bigl( \vec{a} \dot{X}^0 + \vec{b} \dot{X}^1 \bigr)^{2}}
		_{\;\;\; \geq 0} \; > \, 0
	\notag \\
	\label{eq:ds_timord_ttau04}
	& \rightarrow \qquad
	\mid \! \dot{X}^0 \!\! \mid  \;\, > \; \mid \! \dot{X}^1 \!\! \mid		 
\end{align}
Hence altogether we can compute for timelike curves:
\begin{align}
	\dot{X}'^{0} & = \, A_{00} \dot{X}^{0} + A_{01} \dot{X}^{1}
	\notag \\
	& = \, \ub{A_{00}}_{\; > \, 0} \, \ub{\dot{X}^{0}}_{\; > \, 0} \,
			\pm \, \ub{\mid \! A_{01} \!\mid}_{\; < \, A_{00}} 
			\: \ub{\mid \! \dot{X}^{1} \! \mid}_{\; < \, \dot{X}^0} \,
			\; > \: 0
\end{align} 
This shows that $\sign \, \dot{X}^0$ is SO${\ssty (1,d)}$ invariant
for timelike curves.
Therefore also we know that $\sign (\tau_x \! - \! \tau_y)$ and
$\sign (T_x \! - \! T_y)$ are SO${\ssty (1,d)}$ invariant
and can write the action of the dS time ordering operator $T_{\text{dS}}$
in the standard way:
\begin{align}
	T_{\text{dS}} \, \phi {\ssty (\!x\!)} \, \phi {\ssty (\!y\!)} \,
	& = \, \theta {\ssty (\tau_x \! - \tau_y \! )} \;
			\phi {\ssty (\!x\!)} \, \phi {\ssty (\!y\!)}
			\; + \, \theta {\ssty (\tau_y \! - \tau_x \! )} \:
			\phi {\ssty (\!y\!)} \, \phi {\ssty (\!x\!)}
%%			\qqqquad {\ssty \tau_{x,y} \, \in \, [-\infty,+\infty]}
			\notag \\
	& = \, \theta {\ssty (T_x \! - T_y \! )} \;
			\phi {\ssty (\!x\!)} \, \phi {\ssty (\!y\!)}
			\; + \, \theta {\ssty (T_y \! - T_x \! )} \:
			\phi {\ssty (\!y\!)} \, \phi {\ssty (\!x\!)}
%%			\qqqquad {\ssty T_{x,y} \, \in \, [-\pi/2,+\pi/2]}
			\notag
\end{align}

			\subsubsection{Time ordering for $t$}
			\label{sec:ds_timord_t}
Time ordering for $t$ is different to the previous cases
because here the time coordinate is composed
from two embedding space coordinates
with different signs in the embedding space metric.
A priori we expect difficulties with time ordering in static coordinates
because boosts or rotations in the various planes
can map points from one of the causal diamonds into another diamond.
We return to the notation that indices
 in Latin lower cases run within (1, \ldots, d-1).
\begin{figure} [H]
	\begin{center}
	\igx[width=8cm]{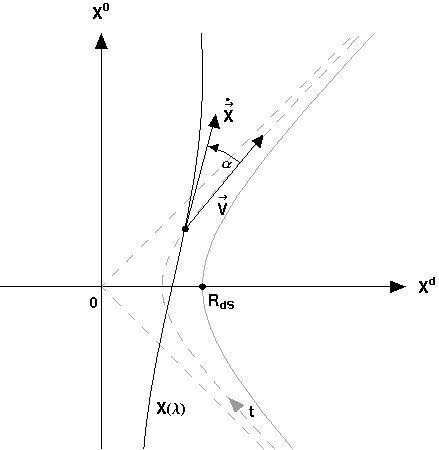}
	\end{center}
	\caption{Time ordering in dS for $t$: $X^0 X^d \!$-plane}
	\label{fig:ds_timord_t}
\end{figure}
Figure \ref{fig:ds_timord_ttau} shows the projection
of a piece of a curve $X {\ssty \! (\! \lambda \!)}$
in the southern diamond of dS into an $(X^k,X^0)$-plane.
The grey lines are coordinate lines of $t$.
The dashed straight ones at the left are the limits of the southern diamond,
the full one at the right is the limit of the dS hyperboloid
and the dashed one between them is a coordinate line
running through the considered point.
Here the notation is $\vec{X}=\bigl(^{X^d}_{X^0} \bigr)$.
$\dot{\vec{X}}= {\ssty \frac{d}{d \lambda}} \vec{X}$
is the tangent vector of the curve in the point
$\vec{X}\! {\ssty( \! \lambda \!)}$ and
\begin{align*}
	\vec{V}
	& = {\ssty \frac{d}{dt}} R\ds \sqrt{1-\vec{x}^{\, 2}} \,
			(\, ^{\sinh t}_{\cosh t})
	\\
	& = \quad R\ds \sqrt{1-\vec{x}^{\, 2}} \,
			(\, _{\sinh t}^{\cosh t})
		\; = \; \bigl(^{X^0}_{X^d} \bigr)
\end{align*}
is the tangent vector of the coordinate line
($\vec{x} {\ssty( \! X \! ( \! \lambda \! ) \! )} =$ constant)
in the same point.
We see that the curve runs $^{\;\;\;\, \text{in}}_{\text{against}}$
the direction of the time $t$ in the considered point
if $\; \bigl( \cos \alpha \gtrless 0 \bigr) \; \leftrightarrow \;
\bigl(\vec{V} \! \dot{\vec{X}} \gtrless 0\bigr) \; \leftrightarrow \;
\bigl( X^0 \dot{X}^d \!\! + \! X^d \dot{X}^0 \gtrless 0\bigr)$
which in static coordinates writes as
$\bigl( \, \dot{t} \, R^2\ds (1 \! - \! \vec{x}^{\,2})
(\sinh^2 \! t + \cosh^2\! t) \gtrless 0 \, \bigr) \,
\leftrightarrow \, \bigl( \, \dot{t} \gtrless 0 \, \bigr)$. 
\\
We can do similar investigations for the other three diamonds
and always obtain the same criterion
$\bigl( X^0 \dot{X}^d \!\! + \! X^d \dot{X}^0 \gtrless 0\bigr)$.
Thus we must check the invariance of
$\sign \, \bigl( X^d \dot{X}^0 + X^0 \dot{X}^d \bigr)$ under SO${\ssty %%@
(1,d)}$.
It is sufficient to check the cases of boosts in the (0,1) and (0,d)-plane
and the rotation in the (d-1,d)-plane since all other cases are analog
and rotations in the (j,k)-planes do affect neither $t$ nor $\vec{x}^{\, 2}$.
\\
We start by giving the corresponding matrices
which transform both $X^M$ and $\dot{X}^M$.
With the rapidity $\chi \, \elof \; [ \! -\infty, \!+\infty]$
the matrices $A$ for a boost in the (0,1)-plane 
and $B$ for the (0,d)-plane read:
\begin{align}
	\label{eq:ds_timord_t01}
	\dot{X}'^M \, & = \, A^M_{\;\; N} \, \dot{X}^N
	&
	A^M_{\;\; N} \, & = \,
		\begin{pmatrix}
			\cosh \chi & \sinh \chi  & 0 \; \\
			\sinh \chi & \cosh \chi & 0 \; \\
			0 & 0 &  \mathbb{1}_{d \! - \! 1}          
		\end{pmatrix}
	\\
	\label{eq:ds_timord_t02}
	\dot{X}'^M \, & = \, B^M_{\;\; N} \, \dot{X}^N
	&
	B^M_{\;\; N} \, & = \,
		\begin{pmatrix}
			\cosh \chi & 0 & \sinh \chi \\
			 0 & \mathbb{1}_{d \! - \! 1} & 0 \\          
			\sinh \chi & 0 & \cosh \chi 
		\end{pmatrix}
\end{align} 
With the angle of rotation $\varphi \, \elof \; ] \! -\pi, \!+\pi]$
the matrix $C$ for a rotation in the (d-1,d)-plane writes:
\begin{align}
	\label{eq:ds_timord_t03}
	\dot{X}'^M \, & = \, C^M_{\;\; N} \, \dot{X}^N
	&
	C^M_{\;\; N} \, & = \,
		\begin{pmatrix}
			\; \mathbb{1}_{d \! - \! 1} & 0 & 0 \; \\          
			0 &  \cos \varphi & \sin \varphi  \; \\
			0 & -\sin \varphi & \cos \varphi \;
		\end{pmatrix}
\end{align} 
We continue by confirming our preoccupation
against the desired SO${\ssty (1,d)}$-invariance
via a counter-example.
We consider a timelike geodesic of the form (\ref{ds_geod12})
\beqs
	X^M \! {\ssty ( \! \lambda \!)}
	= \, a^M \cosh ( \omega \lambda ) + \, b^M \sinh ( \omega \lambda )
	\qquad \qquad
	{\ssty \omega = \frac{\sqrt{c}}{ R\ds} > 0}
\eeqs
running through the point
$X {\ssty \! (\lambda=0)} = (0,\ldots,0,R\ds)$ in embedding space.
Thus we have $a^M \! =(0,\ldots,0,R\ds)$ and with
$b^M \! =(R\ds,0,\ldots,0)$ we can fulfill all three constraints
$a^M b_M \! = 0$ and $a^M a_M \!= -b^M b_M \! = -\sigma R^2\ds$.
\\
Using $\dot{X} {\ssty(\lambda=0)} = \omega b^M$
we can verify that the curve is running in the direction of the time $t$ 
in the point $X \! {\ssty(\lambda=0)}$: 
$\; \bigl( X^0 \dot{X}^d \!\! + \! X^d \dot{X}^0  \bigr) = \, \omega R^2\ds > %%@
0$.
\\
But using the matrix C given above we find that
$\bigl( X'^0 \dot{X}'^d \!\! + \! X'^d \dot{X}'^0  \bigr)
  = \, \omega R^2\ds \cos \varphi$
which becomes smaller then zero for
$\varphi \, \elof \, ]- \! \pi,-\frac{\pi}{2} \, [ \;\;$ or
$ \;\; ] \!+\frac{\pi}{2}, \!+\pi \,[$. 
\\
Thus the rotation in the (d-1,d)-plane
reverses the temporal direction of the curve
as soon as the considered point
is moved into the northern diamond.
Now we could change the direction of the time arrows in figure ??? XX ??? 
by changing the signs of the hyperbolic sine functions in some diamonds.
But no matter how we arrange them,
there will always be some SO${\ssty (1,d)}$-boost or rotation,
which does not change the direction of the curve in embedding space
yet moves it into a diamond with opposite direction of the time $t$.
\\
Hence when using standard $\theta {\ssty (t_x \!-t_y)}$ time ordering
one must be aware 
that it is not generally invariant under SO${\ssty (1,d)}$.

		\subsection{Connection between AdS and dS spacetime}
Anti de Sitter and de Sitter spacetime with the same radius
$R\ads=R\ds=R$ are connected in a simple way.
We start with the $d$-dimensional AdS hyperboloid
$\sigma\ads X^2\ads=+R^2 \, $
in a $(d\!+\!1)$-dimensional embedding space with the metric
$\eta\ads \! =\sigma\ads \text { diag } (+,-,\ldots,-,+) $.
Therefrom we obtain $d$-dimensional de Sitter spacetime 
$\sigma\ds X^2\ds=-R^2 \,$
in a $(d\!+\!1)$-dimensional embedding space with the metric
$\eta\ds=\sigma\ds \text{ diag } (+,-,\ldots,-,-) $
wherein $\sigma\ds = -\sigma\ads$
via the following coordinate transformation in embedding space:
\\
\beq
	X^M\ads = B^M_{\;\,N} X^N\ds \qqqquad
	B^M_{\;\,N} = \text{ diag } (\pm i, \pm i, \ldots, \pm i, \pm 1 )
\eeq
The coordinate transformation naturally
leaves the scalar product in embedding space invariant
$(g_{MN}{\ssty (X\ads)} X^M \ads X^N \ads)
=(g_{AB}{\ssty (X\ds)} X^A \ds X^B \ds)=R^2$ 
while transforming  the metric
$(g_{MN}{\ssty (X\ads)} \rightarrow g_{AB}{\ssty (X\ds)})$
just in the right way for matching definition
(\ref{eq:ds_coord04}) of de Sitter spacetime.
\\
A priori the signs in the B-matrix
can be chosen arbitrarily for each entry.
Choosing $B^M_{\;\,N} = \text{ diag } (-i, \ldots, -i, +1 )$ via the identifactions
$(t\ads=-it\ds)$ and $(\vec{x}\ads=-i\vec{x}\ds)$ 
turns the definitions of the $(t\ads,\vec{x}\ads)$ coordinates (\ref{ads_coord05})
exactly into those of the $(t\ds,\vec{x}\ds)$ coordinates (\ref{ds_coord05})
for the southern causal diamond.
However, we notice that the other diamonds and coordinate systems
do not transform that easily.

\chapter{Propagators}
	\section{Propagators for Minkowski spacetime}
	\label{sec:props_mink}
		\subsection{General properties of the propagators}
		\label{sec:props_mink_gen}
In this section the definitions and relations between different propagators
for a hermitian scalar field $\phi {\ssty(\!x\!)}$ are reviewed
for Minkowski spacetime.
\\
The positive/negative frequency parts of $\phi{\ssty(\!x\!)}$
are written as $\phi^{\ssty +/-}{\ssty(\!x\!)}$
(see chapter \ref{sec:minkfou}). 
$x$ denotes the coordinates in spacetime with the time coordinate $x^0 \!=t$.
The metric used is $g=$ diag $(+,-,-,-)$.
\\
Up to imaginary prefactors the various propagator functions 
according to Birrell and Davies \cite{birrell} (chapter 2.7)
are defined and referred to as:
\begin{alignat}{2}
	\label{eq:minkprop01}
	\text{Wightman functions:} & \quad
		&
		G^+ {\scriptstyle(\!x\!,y\!)}
		& \equiv \, \left\langle 0 \! \mid \phi {\scriptstyle(\!x\!)} \,
		\phi {\scriptstyle(\!y\!)} \mid \! 0 \right\rangle 
		= \, \left\langle 0 \! \mid \phi^{\scriptscriptstyle +} \! {\scriptstyle(\!x\!)} \,
		\phi^{\scriptscriptstyle -} \! {\scriptstyle(\!y\!)} \mid \! 0 \right\rangle \\
	&
		&
		G^- {\scriptstyle(\!x\!,y\!)}
		& \equiv \, \left\langle 0 \! \mid \phi {\scriptstyle(\!y\!)} \,
		\phi {\scriptstyle(\!x\!)} \mid \! 0 \right\rangle
		= \, \left\langle 0 \! \mid \phi^{\scriptscriptstyle +} \! {\scriptstyle(\!y\!)} \,
		\phi^{\scriptscriptstyle -} \! {\scriptstyle(\!x\!)} \mid \! 0 \right\rangle \\
	\notag \\
	\label{eq:minkprop03}
	{\ssty \text{Hadamard's elementary function:}} &
		&
		G^{\scriptscriptstyle(\!1\!)} {\scriptstyle(\!x\!,y\!)}
		& \equiv \, \left\langle 0 \! \mid \left\{ \phi {\scriptstyle(\!x\!)} \, ,
		\phi {\scriptstyle(\!y\!)} \right\} \mid \! 0 \right\rangle \\
		& 
		& & = \, G^+ \! {\scriptstyle(\!x\!,y\!)} \, + \, G^- \! {\scriptstyle(\!x\!,y\!)}
			 \notag \\
	\text{Schwinger function:} &
		&
		G \, {\scriptstyle(\!x\!,y\!)}
		& \equiv \, \left\langle 0 \! \mid \left[ \phi {\scriptstyle(\!x\!)} \, ,
		\phi {\scriptstyle(\!y\!)} \right] \mid \! 0 \right\rangle \\
		& & & = \, G^+ \! {\scriptstyle(\!x\!,y\!)} \, - \, G^- \! {\scriptstyle(\!x\!,y\!)}
	\notag \\
	\notag \\
	\text{retarded Green function:} &
		&
		G \!_R \, {\scriptstyle(\!x\!,y\!)}
		& \equiv \, +\theta {\scriptstyle(\!t_x-t_y\!)} \, G {\scriptstyle(\!x\!,y\!)} \\
	\text{advanced Green function:} &
		&
		G \!_A \, {\scriptstyle(\!x\!,y\!)}
		& \equiv \, -\theta {\scriptstyle(\!t_y-t_x\!)} \, G {\scriptstyle(\!x\!,y\!)} \\
	\notag \\
	\label{eq:minkprop07}
	\text{Feynman propagator:} &
	&
		G \!_F \, {\scriptstyle(\!x\!,y\!)}
		& \equiv \, \left\langle 0 \! \mid T\, \phi {\scriptstyle(\!x\!)} \,
		\phi {\scriptstyle(\!y\!)} \mid \! 0 \right\rangle \\
	\label{eq:minkprop08}
	&
		&
		& = \, \theta {\scriptstyle(\!t_x-t_y\!)} \, G^+ \! {\scriptstyle(\!x\!,y\!)}
		+ \, \theta {\scriptstyle(\!t_y-t_x\!)} \, G^- \! {\scriptstyle(\!x\!,y\!)}
\end{alignat}
The Wightman function $G^+$ propagates the positive
and $G^-$ the negative frequencies of the field $\phi$.
$T$ is the Dyson time ordering operator and sorts earlier operators rightwards. \\
Taking into consideration that the propagator
is supposed to be invariant under translations
the argument ${\scriptstyle(\!x\!,y\!)}$ becomes ${\scriptstyle(\!x-y\!)}$
\; and ${\scriptstyle(\!y\!,x\!)}$ becomes ${\scriptstyle(\!y-x\!)}$.\\
Defined as above the propagators are connected by the following relations:
\begin{align}
	G^\pm {\scriptstyle(\!x\!,y\!)}
	& = {\scriptstyle \frac{1}{2}} \left(
			G^{\scriptscriptstyle(\!1\!)} \! {\scriptstyle(\!x\!,y\!)} \,
			\pm \, G {\scriptstyle(\!x\!,y\!)} \right) \\
	\label{eq:minkprop10}	
	G^\pm {\scriptstyle(\!x\!,y\!)}
	& = G^\mp {\scriptstyle(\!y\!,x\!)} \\
	\label{eq:minkprop11}
	G^\pm {\scriptstyle(\!x\!,y\!)}
	& = G^{\pm*} {\scriptstyle(\!y\!,x\!)}
	\qquad \qquad \;\; {\scriptstyle ( \text{because }\phi \text{ is hermitian} ) } \\
	\notag \\
	\label{eq:minkprop12}
	G^{\scriptscriptstyle(\!1\!)} \! {\scriptstyle(\!x\!,y\!)}
	& = + G^{\scriptscriptstyle(\!1\!)} \! {\scriptstyle(\!y\!,x\!)} \qquad \qquad 
	{\ssty \rightarrow \; G^{\scriptscriptstyle(\!1\!)} \text{ is an even function}}
	\\
	\label{eq:minkprop13}
	G^{\scriptscriptstyle(\!1\!)} \! {\scriptstyle(\!x\!,y\!)}
	& = \, G^{{\scriptscriptstyle(\!1\!)}*} {\scriptstyle(\!y\!,x\!)} \qquad \qquad 
	{\ssty \rightarrow \; G^{\scriptscriptstyle(\!1\!)} \text{ is a real function}}
	\\
	\label{eq:minkprop14}
	G {\scriptstyle(\!x\!,y\!)}
	& = - G \, \! {\scriptstyle(\!y\!,x\!)}
	\qquad \qquad \;\; {\scriptstyle \rightarrow \; G \text{ is an odd function}} \\
	\label{eq:minkprop15}
	G {\scriptstyle(\!x\!,y\!)}
	& = \, G^* \, \! {\scriptstyle(\!y\!,x\!)}
	\qqqquad \;\;\, {\scriptstyle \rightarrow \; G \text{ is purely imaginary}}
	\\
	\notag \\
	G \! _R {\scriptstyle(\!x\!,y\!)} & = G \! _A {\scriptstyle(\!y\!,x\!)} \\
	G _R^* {\scriptstyle(\!x\!,y\!)} & = - G \! _R {\scriptstyle(\!x\!,y\!)}
	\qqqquad {\scriptstyle \rightarrow \; G \! _R \text{ is purely imaginary}}
	\\
	G _A^* {\scriptstyle(\!x\!,y\!)} & = - G \! _A {\scriptstyle(\!x\!,y\!)}
	\qqqquad {\scriptstyle \rightarrow \; G \! _A \text{ is purely imaginary}}
	\\
	\notag \\
	\label{eq:minkprop19}
	G \! _F {\scriptstyle(\!x\!,y\!)}
	& = \, G \! _R {\scriptstyle(\!x\!,y\!)} \, + \, G^- \! {\scriptstyle(\!x\!,y\!)} \\
	\label{eq:minkprop20}
	& = \, G \! _A {\scriptstyle(\!x\!,y\!)} \, + \, G^+ \! {\scriptstyle(\!x\!,y\!)} \\	
	& = {\scriptstyle \frac{1}{2}}
	    \left( G \! _R {\scriptstyle(\!x\!,y\!)} \, + \, G \! _A {\scriptstyle(\!x\!,y\!)} \,
		+ \, G^{\scriptscriptstyle(\!1\!)} \! {\scriptstyle(\!x\!,y\!)} \right) \\
	& = {\scriptstyle \frac{1}{2}}
	    \bigl[ \epsilon {\ssty (\!t_x-t_y\!)} G \! {\scriptstyle(\!x\!,y\!)} \,
		+ \, G^{\scriptscriptstyle(\!1\!)} \! {\scriptstyle(\!x\!,y\!)} \bigr] \\
	\label{eq:minkprop22}
	G \! _F {\scriptstyle(\!x\!,y\!)}
	& = G \! _F {\scriptstyle(\!y\!,x\!)}
	\qqqquad \;\;\; {\scriptstyle \rightarrow \; G \! _F \text{ is an even function}}
\end{align}
With $\phi {\scriptstyle(\!x\!)}$ fulfilling
the Klein-Gordon equation as equation of motion
the Wightman functions $G^\pm$ by definition are homogeneous solutions:
\begin{align}
	\left( \Box _x \!\! + m^{\scriptscriptstyle 2} \right)
			G^\pm {\scriptstyle(\!x\!,y\!)} & = 0 \\
	\rightarrow \quad \left( \Box _x \!\! + m^{\scriptscriptstyle 2} \right)
	G^{\scriptscriptstyle(\!1\!)} {\scriptstyle(\!x\!,y\!)} & = 0 \\
	\rightarrow \quad \left( \Box _x \!\! + m^{\scriptscriptstyle 2} \right) \; 
	G {\scriptstyle(\!x\!,y\!)} \,\; & = 0
\end{align}
Then with (\ref{eq:minkprop19}) and (\ref{eq:minkprop20}) we have
\begin{equation}
	\left( \Box _x \!\! + m^{\scriptscriptstyle 2} \right) G \! _F {\ssty(\!x\!,y\!)}
	\; = \; \left( \Box _x \!\! + m^{\scriptscriptstyle 2} \right)
			G \! _R {\scriptstyle(\!x\!,y\!)}
	\; = \; \left( \Box _x \!\! + m^{\scriptscriptstyle 2} \right)
			G \! _A {\scriptstyle(\!x\!,y\!)}	
\end{equation}
and from (\ref{eq:minkprop08}) using $\Box = \partial _0^2 - \vec{\partial}^2$ 
and the equal time commutation relation
$\left[ \phi {\scriptstyle(\!x\!)} ,\phi {\ssty(\!y\!)}\right] _{t_x=t_y} \! = \, 0 \;$
(see also appendix \ref{sec:timord_delta})
we obtain
\begin{equation}
	\label{eq:minkprop27}
	\left( \Box _x \!\! + m^{\sssty 2} \right) G \! _F {\scriptstyle(\!x\!,y\!)}
	\; = \; \delta {\ssty (\!t_x-t_y\!)} \,
	\bigl( \partial _0 G^+ \! - \partial _0 G^- \bigr) _{t_x=t_y}
\end{equation}

		\subsection{The method of Fourier transformation}
		\label{sec:minkfou}
\begin{figure} [H]
	\begin{center}
	\igx[width=9cm]{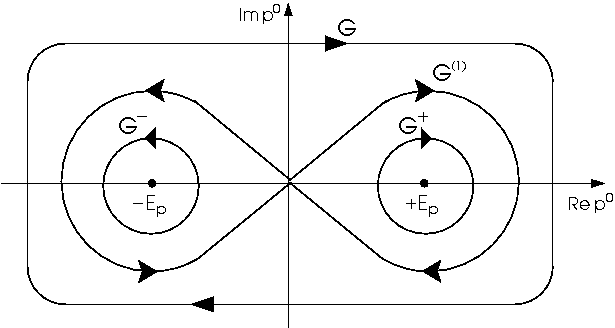}
	\end{center}
	\caption{Wightman, Hadamard and Schwinger function}
	\label{fig:minkprop01}
\end{figure}

\begin{figure} [H]
	\begin{center}
	\igx[width=9cm]{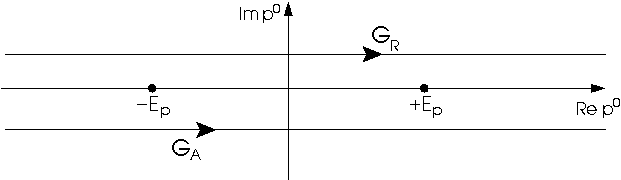}
	\end{center}
	\caption{retarded and advanced Green function}
	\label{fig:minkprop02}
\end{figure}

\begin{figure} [H]
	\begin{center}
	\igx[width=9cm]{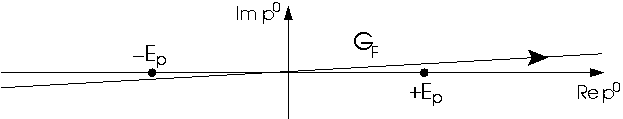}
	\end{center}
	\caption{Feynman propagator}
	\label{fig:minkprop03}
\end{figure}

\begin{figure} [H]
	\begin{center}
	\igx[width=9cm]{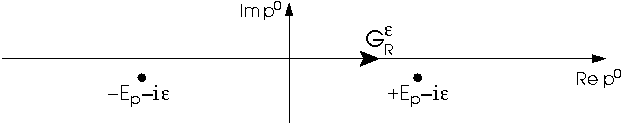}
	\end{center}
	\caption{shifted retarded Green function}
	\label{fig:minkprop04}
\end{figure}

\begin{figure} [H]
	\begin{center}
	\igx[width=9cm]{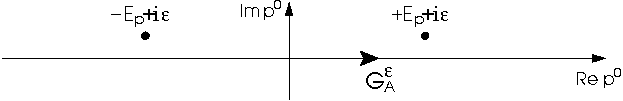}
	\end{center}
	\caption{shifted advanced Green function}
	\label{fig:minkprop05}
\end{figure}

\begin{figure} [H]
	\begin{center}
	\igx[width=9cm]{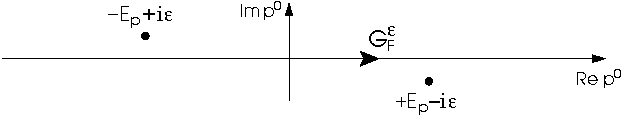}
	\end{center}
	\caption{rotated Feynman propagator}
	\label{fig:minkprop06}
\end{figure}

Conventionally in four dimensional Minkowski spacetime
Fourier transformation is used in order to express fields and propagators. \\
We can use the definitions on pages 21-26 in Peskin, Schroeder \cite{peskin}
for the field $\phi$ and its conjugated impulse field $\pi$:
\begin{align}
	& \quad \;
	\ob{\qquad \qquad \qquad \qquad \qquad \;}^{\phi^+ {\ssty \! (\!x\!)}}
	\;\;\; \ob{\qquad \qquad \qquad \;}^{\phi^- {\ssty \! (\!x\!)}}
	\notag \\
	\phi {\ssty(\!x\!)}
	& = \int \!\! \frac{d^3\!p}{(2\pi\!)^3} \frac{1}{\sqrt{2E_{\vec{p}}}}
	\left( a_{\vec{p}} \, e^{-ipx} \!
	+ a^\dagger_{\vec{p}} \, e^{+ipx} \right) _{\!p^0=E_{\vec{p}}} \\
	\pi {\ssty(\!x\!)}
	& = \, \partial _0 \phi {\ssty(\!x\!)} \qquad \qquad \qquad \qquad 
	E_{\vec{p}} = +\sqrt{\vec{p}^{\sssty \; 2} \! + m^{\sssty 2}}
\end{align}
The integral over the annihilator/creator term
is the positive/negative frequency part of $\phi {\ssty(\!x\!)}$.
The algebra of the creation operators $a^\dagger_{\vec{p}}$
and the annihilation operators $a_{\vec{p}}$ is
\begin{align}
	[ a^{}_{\vec{p}},a^\dagger_{\vec{q}} ]
	& = (2\pi\!)^3 \, \delta^{\sssty\! (3)} {\ssty \! (\!\vec{p}-\vec{q})} \\
	[ a^{}_{\vec{p}},a_{\vec{q}} ]
	& = [ a^\dagger_{\vec{p}},a^\dagger_{\vec{q}} ] = \, 0
\end{align}
which realizes the equal time commutation relations
\begin{align}
	[ \phi {\ssty(\!x\!)} ,\pi {\ssty(\!y\!)}] _{t_x=t_y}
	& = i \, \delta^{\sssty\! (3)} {\ssty \! (\!\vec{x}-\vec{y})} \\
	[ \phi {\ssty(\!x\!)} ,\phi {\ssty(\!y\!)}] _{t_x=t_y}
	& = [ \pi {\ssty(\!x\!)} ,\pi {\ssty(\!y\!)}] _{t_x=t_y} = \, 0
\end{align}
which in turn from (\ref{eq:minkprop27}) generate one delta source term
in the Klein-Gordon equation for the Feynman propagator:
\begin{equation}
	\label{eq:minkprop34_5}
	\left( \Box _x \!\! + m^{\sssty 2} \right) G \! _F {\ssty(\!x\!,y\!)}
	\; = \; -i \, \delta^{\sssty\! (4)} {\ssty \! (\!x-y\!)}
\end{equation}
Evaluating the Wightman functions yields
\begin{equation}
	G^\pm {\ssty(\!x\!,y\!)}
	=  \int \!\! \frac{d^3\!p}{(2\pi\!)^3} \frac{1}{2E_{\vec{p}}} \, e^{\mp ip\, (x-y)}
\end{equation}
from which the other propagators can be assembled according to their definitions given above.
Considering the integral $I$ in the complex $p^{0}$-plane
\begin{align}
	I & = \; i \int \!\! \frac{d^3\!p}{(2\pi\!)^3} \int \!\! \frac{dp^0}{(2\pi\!)}
	\, \frac{e^{-ip\, (x-y)}}{p^{\sssty 2}-m^{\sssty 2}} \notag \\
	& = \; i \int \!\! \frac{d^3\!p}{(2\pi\!)^3} \int \!\! \frac{dp^0}{(2\pi\!)}
	\, \frac{e^{-ip\, (x-y)}}{p^{{\sssty 0}^2}-E^{\sssty 2}_{\vec{p}}} \notag \\
	& = \; i \int \!\! \frac{d^3\!p}{(2\pi\!)^3} \int \!\! \frac{dp^0}{(2\pi\!)}
	\, \frac{e^{-ip\, (x-y)}}{(p^0--E_{\vec{p}})(p^0-+E_{\vec{p}})} 
\end{align}
we find two poles at $\mp E_{\vec{p}}$.
The propagators then correspond to the different contours of integration shown 
in figures \ref{fig:minkprop01}, \ref{fig:minkprop02} and \ref{fig:minkprop03}
which are ajusted for our conventions from
Birrell and Davies \cite{birrell} (Fig. 3 in chapter 2.7).
The contours of $G_F$, $G_R$ and $G_A$ can be closed counterclockwise
above the real $p^0$-axis for $(t_x \! - \! t_y)<0$
and clockwise below for $(t_x \! - \! t_y)>0$ picking up the encircled poles
according to Cauchy's integral formula for holomorphic functions:
\begin{equation}
	\ointctrclockwise \limits_{pole} \! dp \, \frac{f(p)}{p-pole}
	\, = \, 2 \pi i \, f{\ssty (pole)}
\end{equation}
In order to be able to perform the integration exactly on the real $p^0$-axis
we have to shift or rotate the poles.
We remember that $\epsilon$ is infinitesimally small so that
$\epsilon$ times any positive quantity is just $\epsilon$.  
For the retarded Green function the shift is
$\pm E_{\vec{p}} \, \rightarrow \, \pm E_{\vec{p}} - i \epsilon$
and for the advanced Green function $\pm E_{\vec{p}} \, \rightarrow \, \pm E_{\vec{p}} + i %%@
\epsilon$.
The shifted poles and contours are shown in figures \ref{fig:minkprop04} and %%@
\ref{fig:minkprop05}.
We obtain:
\begin{align}
	G^{\epsilon}_R {\ssty(\!x\!,y\!)}
	& = i \int \!\! \frac{d^3\!p}{(2\pi\!)^3} \int\limits^{+\infty}_{-\infty} \!\! %%@
\frac{dp^0}{(2\pi\!)}
	\; \frac{e^{-ip\, (x-y)}}
	        {p^{{\sssty 0}^2} + 2i \epsilon p^{\sssty 0}
			 - \epsilon^{\sssty 2} - E^{\sssty 2}_{\vec{p}}} \notag \\
	G^{\epsilon}_A {\ssty(\!x\!,y\!)}
	& = i \int \!\! \frac{d^3\!p}{(2\pi\!)^3} \int\limits^{+\infty}_{-\infty} \!\! %%@
\frac{dp^0}{(2\pi\!)}
	\; \frac{e^{-ip\, (x-y)}}
	        {p^{{\sssty 0}^2} - 2i\epsilon p^{\sssty 0}
			 - \epsilon^{\sssty 2} - E^{\sssty 2}_{\vec{p}}} \notag
\end{align}
For the Feynman propagator the clockwise rotation is
$\pm E_{\vec{p}} \, \rightarrow \, \pm E_{\vec{p}} \, (1 - i \epsilon)$.
The rotated poles and contour are shown in figure \ref{fig:minkprop06}
and we obtain:
\begin{align}
	G^{\epsilon}_F {\ssty(\!x\!,y\!)}
	& = i \int \!\! \frac{d^3\!p}{(2\pi\!)^3} \int\limits^{+\infty}_{-\infty} \!\!
	 \frac{dp^0}{(2\pi\!)}
	\; \frac{e^{-ip\, (x-y)}}
	{p^{{\sssty 0}^2} -E^{\sssty 2}_{\vec{p}}(1-i\epsilon)^2}
		\notag \\
	\label{eq:minkprop38}
	& = i \int \!\! \frac{d^3\!p}{(2\pi\!)^3} \int\limits^{+\infty}_{-\infty} \!\!
	 \frac{dp^0}{(2\pi\!)}
	\; \frac{e^{-ip\, (x-y)}}
	{p^{\sssty 2}-m^{\sssty 2} + E^{\sssty 2}_{\vec{p}}(2i\epsilon+\epsilon^2)}
\end{align}
This example shows how the explicit method (\ref{eq:minkprop08})
of time ordering in Minkowski spacetime
leads to the $i\epsilon$-prescription (\ref{eq:minkprop38})
in the construction of the Feynman propagator.

		\subsection{The DvB method in Minkowski spacetime}
		\label{sec:dvb_mink}
In \cite{dvb} C. Dullemond and E. van Beveren develop
a configuration space method for finding the various propagators
and apply it to four dimensional Minkowski and AdS$_4$ spacetime.
The advantage of their method is that it delivers in one go
all propagator functions including $i\epsilon$-prescriptions 
without needing to perform a mode summation
as done by Burgess and Lutken in \cite{burgess} for AdS 
and by Bousso, Maloney and Strominger in \cite{bousso}
for de Sitter spacetime.
Analytic continuation from euclidean time obtained via Wick rotation
is not needed as well.
\\
The following sections are devoted to first understanding their method
and then generalizing it to AdS spaces of arbitrary dimension $d$.
The notation used here is connected to the one used by
Dullemond and van Beveren (DvB) by the prefactors
\begin{align*}
	G^\pm & = \, - \; G^\pm \dvb \\
	G & = \, -i \, G \dvb \\
	G^{\sssty(\!1\!)} & = \, - \; \overline{G} \dvb \\
	G_{F,R,A} & = \, -i \, G_{F,R,A}^{\sssty \!D\!v\!B}
\end{align*}
In order to become familiar with the DvB method 
we begin by considering it in Minkowski spacetime
with $x$ denoting a four vector and the metric still being $(+,-,-,-)$. \\
The general strategy of Dullemond and van Beveren consists of
first finding a candidate function $\phi{\ssty(\!x\!,y\!)}$ that is
a solution of the homogeneous Klein-Gordon equation.
\footnote{
	We must not mix up the scalar field $\phi{\ssty(\!x\!)}$ with 
	the candidate function $\phi{\ssty(\!x\!,y\!)} = \phi{\ssty(\!\lambda\!)}$.
	The field is always recognizable by one single latin lower case in the %%@
argument,
	the function by one greek lower case or multiple arguments.}
The second step is modifying it purposively in order to later
identify the modified offspring versions with the various
propagators (\ref{eq:minkprop01})-(\ref{eq:minkprop07}).
\\
First we need to find a solution
$\phi{\ssty(\!x\!,y\!)} = \phi{\ssty(\!\lambda\!)}$
of the homogeneous Klein-Gordon equation
that drops off to zero at infinite distance $\lambda$:
\begin{align}
	\left( \Box \! + m^{\sssty 2} \right) \! \phi {\ssty(\!\lambda\!)}
	& \overset{!}{=} 0
	\qquad \qquad \lambda {\ssty(\!x \! ,y\!)} = \kappa (x \! - \! y)^2
	\qquad \qquad {\ssty \kappa \; > 0} \\
	\phi {\ssty(\!\lambda \rightarrow \pm \infty\!)}
	& \overset{!}{=} 0 \notag
\end{align}
The factor $\kappa$ is introduced in order to take into account
the mass $m$ of our scalar field later. 
In general $\lambda$ shall provide a measure
for the distance of the two points $x$ and $y$,
which is invariant under the symmetry transformations
of the considered spacetime.
When $\lambda$ is written without arguments
then it always denotes $\lambda {\ssty(\!x \! ,y\!)}$. \\
On the light cone $\lambda = 0$
this solution in general will turn out to be ill defined,
thus we apply one of the following two complex shifts
in the time coordinate:
\begin{align}
	\label{eq:dvb_mink_shift_01}
	\text{\ssiz either} & &
	t^\pm_x & = t_x \pm i {\ssty \frac{\epsilon}{2 \kappa}} &
	t^\pm_y & = t_y \qqqquad
	{\ssty ( \text{with }} \epsilon {\ssty  > 0 \text{ and finite})}
	\\
	\label{eq:dvb_mink_shift_02}
	\text{\ssiz or} & &
	t^\pm_x & = t_x \pm i {\ssty \frac{\epsilon}{4 \kappa}} &
	t^\pm_y & = t_x \mp i {\ssty \frac{\epsilon}{4 \kappa}}
\end{align}
In Minkowski spacetime both shifts are equivalent
and do not change the d'Alem-
bertian while shifting $(t_x\!-\!t_y) \rightarrow (t_x \!-\! t_y \pm i
{\ssty \frac{\epsilon}{2 \kappa}})$ and thereby
\begin{align*}
	\lambda {\ssty(\!x\!,y\!)} \quad \rightarrow \quad 
	\lambda _\pm {\ssty(\!x\!,y\!)}
	& \equiv \lambda \pm i \epsilon (t_x \! - \! t_y)
	- {\ssty \frac{\epsilon^2}{4 \kappa}}
	\\
	\rightarrow \quad \lambda^* _\pm {\ssty(\!x\!,y\!)}
	& = \lambda _\mp {\ssty(\!x\!,y\!)}
		= \lambda _\pm {\ssty(\!y\!,x\!)}
\end{align*}
This shift will render $\phi{\ssty(\!\lambda\!)}$
well defined for all real values of the argument.
In the limit of small $\epsilon$ this definition of $\lambda _\pm$
will also render the solutions
$\phi_\pm{\ssty(\!x\!,y\!)} \! = \phi {\ssty(\!\lambda_\pm\!(\!x\!,y\!))}$
invariant under orthochronous Lorentz transformations,
i.e. those that do not change the direction of time
i.e. the sign of $(t_x \! - \! t_y)$.
Now we dispose of two solutions
of the homogenous Klein-Gordon equation:
\begin{align}
	\left( \Box _x \!\! + m^{\sssty 2} \right) \!
	\phi_\pm {\ssty(\!\lambda_\pm\!)} & = 0
\end{align}
We intend to soon identify $G^\pm {\ssty(\!x\!,y\!)}$
with $\phi_\mp {\ssty(\!\lambda_\mp\!)}$
and notice that the therefor required relation (\ref{eq:minkprop10})
$\phi_\pm{\ssty(\!x\!,y\!)} \! =\phi_\mp{\ssty(\!y\!,x\!)}$
is already fulfilled via the definition of $\lambda_\pm$. 
The second necessary relation (\ref{eq:minkprop11})
$\phi_\pm{\ssty(\!x\!,y\!)} \! =\phi_\pm^*{\ssty(\!y\!,x\!)}$
is also accomplished because of 
$\phi_\pm{\ssty(\!\lambda^*_\pm\!)} \! %%@
=\phi_\pm^*{\ssty(\!\lambda_\pm\!)}$. \\
In order to bestow a spatial delta function upon equation %%@
(\ref{eq:minkprop27})
\beqs
	\left( \Box _x \!\! + m^{\sssty 2} \right) G \! _F {\ssty(\!x\!,y\!)}
	\; = \; \delta {\ssty(\!x\!^0-y\!^0\!)}
	\left( \partial _0 G^+ \! - \partial _0 G^- \right) _{\! t_x=t_y}
\eeqs
and thereby turning it into (\ref{eq:minkprop34_5})
\beqs
	\left( \Box _x \!\! + m^{\sssty 2} \right) G \! _F {\ssty(\!x\!,y\!)}
	\; = \; -i \, \delta^{\sssty\! (4)} {\ssty \! (\!x-y\!)}
\eeqs
we additionally require $\phi_\pm$ to satisfy
\begin{equation}
\label{eq:dvb07}
	\underset{\epsilon \rightarrow 0}{\lim} \;
	\bigl[ \partial _{t_x} \phi_\pm {\ssty(\!x\!,y\!)} \bigr]_{t_x=t_y} \,
	= \, \pm i \, \beta \, \delta^{\sssty\! (3)} \!
	{\ssty(\!\vec{x}-\vec{y}\!)} 
\end{equation}
with $\beta$ being a normalization constant.
Then we can enact the following identifications
fulfilling equations (\ref{eq:minkprop03} - \ref{eq:minkprop22}):
\begin{align}
	\label{eq:dvb08}
	G^\pm {\ssty(\!x\!,y\!)}
	& = \, \frac{\ssty 1}{\ssty 2 \beta} \;
		\underset{\epsilon \rightarrow 0}{\lim} \;
	    \phi_\mp {\ssty(\!\lambda_\mp\!)} \\
	\longrightarrow \;\;\;
	G^{\sssty(\!1\!)} {\ssty(\!x\!,y\!)}
	& = \, G^+ {\ssty \! (\!x\!,y\!)} + G^- {\ssty \! (\!x\!,y\!)}
		= \phantom{\mp}\frac{\ssty 1}{\ssty \beta} \;\;
	    \text{Re} \; \underset{\epsilon \rightarrow 0}{\lim} \;
		 \phi_\pm {\ssty(\!\lambda_\pm\!)} \\
	\longrightarrow \quad \;\:
	G {\ssty(\!x\!,y\!)}
	& = \, G^+ {\ssty \! (\!x\!,y\!)} - G^- {\ssty \! (\!x\!,y\!)}
		= \mp \frac{\ssty 1}{\ssty \beta} \;
	    i \, \text{Im} \; \underset{\epsilon \rightarrow 0}{\lim} \;
		\phi_\pm {\ssty(\!\lambda_\pm\!)} \\
	\longrightarrow \quad G \!_R {\ssty(\!x\!,y\!)}
	& = \,+ \theta {\ssty(\!t_x\!-\!t_y\!)} \, G {\ssty(\!x\!,y\!)} \;\;\;
	  = \mp \theta {\ssty(\!t_x\!-\!t_y\!)}
	  	\frac{\ssty 1}{\ssty \beta} \;
	    i \, \text{Im} \; \underset{\epsilon \rightarrow 0}{\lim} \;
		\phi_\pm {\ssty(\!\lambda_\pm\!)} \\
	\longrightarrow \quad G \!_A {\ssty(\!x\!,y\!)}
	& = \, -\theta {\ssty(\!t_y\!-\!t_x\!)} \, G {\ssty(\!x\!,y\!)} \;\;\;
	  = \pm \theta {\ssty(\!t_y\!-\!t_x\!)}
	  \frac{\ssty 1}{\ssty \beta} \;
	    i \, \text{Im} \; \underset{\epsilon \rightarrow 0}{\lim} \;
		\phi_\pm {\ssty(\!\lambda_\pm\!)} \\
	\longrightarrow \quad G \!_F {\ssty(\!x\!,y\!)}
	& = \, \theta {\ssty(\!t_x\!-\!t_y\!)} \, G^+ \! {\ssty(\!x\!,y\!)}
		+ \, \theta {\ssty(\!t_y\!-\!t_x\!)} \, G^- \! {\ssty(\!x\!,y\!)}
		\notag \\
	& = \frac{\ssty 1}{\ssty 2\beta} \;
		\underset{\epsilon \rightarrow 0}{\lim} \;
	    \bigl[ \theta {\ssty(\!t_x\!-\!t_y\!)}
		\phi {\ssty(\!\lambda - i \epsilon (\!t_x\!-\!t_y\!) -
		\frac{\epsilon ^2}{4 \kappa}\!)}
		+ \theta {\ssty(\!t_y\!-\!t_x\!)} 
		\phi {\ssty(\!\lambda + i \epsilon (\!t_x\!-\!t_y\!) -
		\frac{\epsilon ^2}{4 \kappa}\!)} \bigr]
		\notag \\
	& = \frac{\ssty 1}{\ssty 2\beta} \;
		\underset{\epsilon \rightarrow 0}{\lim} \;
	    \bigl[ \theta {\ssty(\!t_x\!-\!t_y\!)}
		\phi {\ssty(\!\lambda - i \epsilon \mid t_x\!-\!t_y\! \mid 
		-\frac{\epsilon ^2}{4 \kappa})}
		+ \theta {\ssty(\!t_y\!-\!t_x\!)} 
		\phi {\ssty(\!\lambda - i \epsilon \mid t_x\!-\!t_y\! \mid 
			- \frac{\epsilon ^2}{4 \kappa})} \bigr]
		\notag \\
	\label{eq:dvb13}
	& = \frac{\ssty 1}{\ssty 2\beta} \;
		\underset{\epsilon \rightarrow 0}{\lim} \;
	    \phi {\ssty(\!\lambda 
			- i \epsilon \ub{\ssty \mid t_x\!-\!t_y\! \mid}_{\geq 0}
			- \frac{\epsilon ^2}{4 \kappa}\!)}
\end{align}
In the Feynman propagator the time ordering leads to an i$\epsilon$-term
which is zero for $t_x=t_y$ and positive otherwise,
this time in configuration space.
Due to our way of construction we find that the Feynman propagator
is indeed a solution of the inhomogeneous Klein-Gordon equation:
\begin{align}
	\label{eq:dvb14}
	\left( \Box_x \! + m^{\sssty 2} \right) G \! _F {\ssty(\!x\!,y\!)} \; 
	& = -i \, \delta^{\sssty \! (4)} \! {\ssty(\!x\!-\!y\!)}
\end{align}

				\subsubsection{Generalized version}
				\label{sec:genver}
In order to apply the DvB method to spacetimes with nonvanishing
constant curvature we need to generalize it a bit.
As a generalized version of (\ref{eq:dvb07}) we will encounter
in $d$-dimensional constantly curved spacetime
\begin{align}
	\label{eq:dvb_genver_delta_x}
	\limepszero	\Bigl( \del_{t_x} 
			\phi^\pm{\ssty \! ( \! \lambda_{\! \pm} \!)} \Bigr)_{\! t_x=t_y}
	& = \; \pm i \, \beta_\phi \, f {\ssty \! (\vec{x})} \;
			\delta^{(\!d\!-\!1\!)} {\ssty (\vec{x}-\vec{y})}
\end{align}
with some function $f {\ssty \! (\vec{x})}$.
As a generalization of (\ref{eq:minkprop27}) we will meet
(see appendix \ref{sec:timord_delta}):
\begin{align}
	\label{eq:dvb_genver_delta_t}
	\left( \sigma \Box_x \!\! + m^{\scriptscriptstyle 2} \right)
		G \! _F {\scriptstyle(\!x\!,y\!)} \;
	& = \; \frac{\beta_G}{f \! {\ssty (\vec{x})}} \,
	\delta {\scriptstyle(\!t_x\!-t_y\!)} \,
		\bigl[ \partial _{t_x} \! G^+ \! - \partial _{t_x} \! G^- \bigr] _{t_x\!=t_y}
\end{align}
with the sign $\; \sigma \!= \!\pm 1$ introduced in chapter \ref{chap:intro}.
Therefore in order to fulfill the generalized version of the
inhomogeneous Klein-Gordon equation (\ref{eq:dvb14})
\begin{align}
	\label{eq:dvb_genver_kg}
	\left( \sigma \Box_x \!\! + m^{\scriptscriptstyle 2} \right)
	G \! _F {\scriptstyle(\!x\!,y\!)} \; & = \;
	\frac{-i}{\sqrt{g}} \; \delta^{\scriptscriptstyle (\!d)} {\scriptstyle \! %%@
(\!x-y\!)}
\end{align}
we have to perform a modified version of the identification (\ref{eq:dvb08}):
\begin{align}
	\label{eq:dvb_gen_ident}
	G^\pm {\scriptstyle(\!x\!,y\!)}
	& = \, \frac{\scriptstyle 1}{\scriptstyle 2 \sqrt{g} \, \beta_\phi \beta_G} \;
		\limepszero \, \phi_\mp {\scriptstyle(\!\lambda_\mp\!)}
\end{align}
For the invariance of the propagators
under the SO$\sssty(., \, .)$
symmetry group of the considered spacetime
it is therein necessary that $ \beta_\phi$ and $\beta_G$ are constants.
$\sqrt{g}$ must either be constant
or fit properly into the integral of normalization
as we will find in subsection \ref{sec:dvb_ds_delnorm}
for de Sitter spacetime.

				\subsubsection{Massless scalar field in Minkowski spacetime}
In order to get used to this method, we first employ it
in four dimensional Minkowski spacetime.
We start with the massless case $m=0$ and choose $\kappa = 1$.
Our homogeneous solution of the massless Klein-Gordon equation is:
\begin{align*}
	\phi     {\scriptstyle(\!\lambda\!)} & = \, \frac{1}{\lambda} \notag \\
	\rightarrow \phi_\pm {\scriptstyle(\!\lambda\!)}
	& = \, \frac{1}{\lambda \pm i \epsilon (\!t_x\!-\!t_y\!)
	- \frac{\epsilon ^2}{4}} \notag
\end{align*}
On page 649 in Bogoliubov, Shirkov \cite{bogshir} we find:
\begin{align}
	\underset{\epsilon \rightarrow 0}{\lim} \; \frac{1}{\alpha \pm i \epsilon} \,
	& = \, \mp i \, \pi \, \delta {\scriptstyle(\!\alpha\!)} \; +
	    \; P \left( \frac{1}{\alpha}\right) \\ 
	\rightarrow \qquad \underset{\epsilon \rightarrow 0}{\lim} \;
		\phi_\pm {\scriptstyle(\!\lambda\!)}
	= \, \underset{\epsilon \rightarrow 0}{\lim} \; 
	  \frac{1}{\lambda \pm i \epsilon \, \epsilon {\scriptstyle(\!t_x\!-\!t_y\!)}} \,
	& = \, \mp i \, \pi \, \epsilon {\scriptstyle(\!t_x\!-\!t_y\!)} \,
		\delta {\scriptstyle(\!\lambda\!)} \; +
	    \; P \left( \frac{1}{\lambda}\right)
\end{align}
We see the required properties (\ref{eq:minkprop10})
and (\ref{eq:minkprop11}) fullfilled. Moreover we have
\begin{align*}
	\int \!\! d^3 \! x \, \bigl[ \partial _0
		\phi_\pm {\scriptstyle(\!x\!,y\!)} \bigr]_{t_x=t_y}
	& = \int \!\! d^3 \! x \,
	    \frac{\mp i \, \epsilon}
		     {\left( -(\vec{x}-\vec{y})^{\scriptscriptstyle \, 2}-\frac{\epsilon ^2}{4}
			  \right)^2} \\
	& = \mp 8 \pi i \!
	    \underbrace { \int \limits_0^\infty \!\! dr \, \frac{r^2} {\left( r^2 + 1 %%@
\right)^2}} \\
	& = \mp 2 \pi^2 i \qquad ^{\pi/4}
\end{align*}
wherein we have used equation 56 on page 157 in Bronstein \cite{bronstein}:
\begin{align*}
	\int \!\! dy \, \frac{y^2} {\left( y^2 + a^2 \right)^2}
	& = \frac {-y}{2\left( y^2 + a^2 \right)} \, + \, \frac{1}{2a} \arctan
			\Bigl( \frac{y}{a} \Bigr)
	\qquad\qquad {\scriptstyle a>0}
\end{align*}
Hence with
\begin{equation*}
	\underset{\epsilon \rightarrow 0}{\lim} \;
	\bigl[ \partial _{t_x} \phi_\pm {\scriptstyle(\!x\!)} \bigr]_{t_x=t_y} \,
	= \, 0 \qquad {\scriptstyle \forall \; \vec{x} \neq \vec{y}} 
\end{equation*}
we have reproduced property (\ref{eq:dvb07}) with $\beta = -2 \pi^2$:
\begin{equation*}
	\underset{\epsilon \rightarrow 0}{\lim} \;
	\bigl[ \partial _{t_x} \phi_\pm {\scriptstyle(\!x\!)} \bigr]_{t_x=t_y} \,
	= \, \mp i \, 2 \pi^2 \, \delta^{\scriptscriptstyle \! (3)} \!
		 {\scriptstyle(\!\vec{x}-\vec{y}\!)} 
\end{equation*}
Inserting everything in relations (\ref{eq:dvb08})-(\ref{eq:dvb13})
we arrive for the massless case at
\begin{align}
	G^\pm {\scriptstyle(\!x\!,y\!)}
	& = \frac{\scriptstyle -1}{\scriptstyle 4 \pi^2} \;
	\underset{\epsilon \rightarrow 0}{\lim} \;
	    \frac{1}{\lambda \mp i \epsilon (\!t_x\!-\!t_y\!)
		- {\ssty \frac{\epsilon ^2}{4}}} \\
	G^{\scriptscriptstyle(\!1\!)} {\scriptstyle(\!x\!,y\!)}
	& = \frac{\scriptstyle -1}{\scriptstyle 2 \pi^2} \; P(1/\lambda) \\
	G {\scriptstyle(\!x\!,y\!)}
	& = \frac{\scriptstyle -i}{\scriptstyle 2 \pi} \;
	    \epsilon {\scriptstyle(\!t_x\!-\!t_y\!)} \, \delta {\scriptstyle(\!\lambda\!)}
		 \\
	G \!_R {\scriptstyle(\!x\!,y\!)}
	& = \frac{\scriptstyle -i}{\scriptstyle 2 \pi} \;
	    \theta {\scriptstyle(\!t_x\!-\!t_y\!)} \, \delta {\scriptstyle(\!\lambda\!)} \\
	G \!_A {\scriptstyle(\!x\!,y\!)}
	& = \frac{\scriptstyle -i}{\scriptstyle 2 \pi} \;
	    \theta {\scriptstyle(\!t_y\!-\!t_x\!)} \, \delta {\scriptstyle(\!\lambda\!)} \\
	G \!_F {\scriptstyle(\!x\!,y\!)}
	& = \frac{\scriptstyle -1}{\scriptstyle 4 \pi^2} \;
		\underset{\epsilon \rightarrow 0}{\lim} \;
	    \frac{1}{\lambda - i \epsilon \! \mid \! t_x \! - \! t_y \! \mid
		 - {\ssty \frac{\epsilon ^2}{4}}}
\end{align}
From (\ref{eq:dvb14}) we know that
$\Box_x \, G \! _F {\scriptstyle(\!x\!,y\!)}
	= \, -i \, \delta^{\scriptscriptstyle\! (4)} \! {\scriptstyle(\!x-y\!)}$
which is confirmed by (\ref{eq:delta12}).
We notice that the propagators show singularities on the whole lightcone
yet only generate one single delta source
in the inhomogeneous Klein-Gordon equation.

				\subsubsection{Massive scalar field  in Minkowski spacetime}			
In the massive case $m \neq 0$ finding the solution
$\phi {\scriptstyle(\!\lambda\!)}$ is a little bit more difficult.
We choose $\kappa=m^2$ and the shift
$x^0 \rightarrow x^0 \pm i \frac{\epsilon}{2 m^2}$ leads to
$\lambda \rightarrow \lambda _\pm \equiv \lambda \pm i \epsilon %%@
(\!t_x\!-\!t_y\!)
- \frac{\epsilon^2}{4 m^2}$.
In terms of $\lambda$ the Klein-Gordon equation reads
% -------------------- equation dvb23--------------------------------------
\begin{equation}
	\label{eq:dvb23}
	0 = \bigl( 1 + 8 \partial _\lambda + \, 4 \lambda \partial^2 _\lambda \bigr) %%@
\,
	    \phi {\scriptstyle(\!\lambda\!)}
\end{equation}
We define $\eta=\sqrt{+\lambda}$ for $\lambda > 0$
and $\sigma=\sqrt{-\lambda}$  for $\lambda < 0$. 
Considering positive $\lambda$ equation (\ref{eq:dvb23}) turns into
\begin{align*}
	0 & = \bigl( 1 + \frac{3}{\eta} \partial _\eta + \, \partial^2 _\eta \bigr) \,
	      \phi {\scriptstyle(\!\eta\!)} \\
	\rightarrow \quad 0 
	& = \bigl(\eta^2 - 1 + \eta \partial _\eta + \, \eta^2 \partial^2 _\eta \bigr) \,
	    \eta  \phi {\scriptstyle(\!\eta\!)}
\end{align*}
This is the Bessel equation with index 1.
(eq. 9.1.1 on p. 358 in Abramowitz, Stegun \cite{astegun})
So if $U_1{\scriptstyle(\!\eta\!)}$ is a solution of this Bessel equation,
then $\phi{\scriptstyle(\!\eta\!)} = U_1{\scriptstyle(\!\eta\!)} \! / \eta$
is a solution of the Klein-Gordon equation for positive $\lambda$. \\
With negative $\lambda$ equation (\ref{eq:dvb23}) becomes
\begin{align*}
	0 & = \bigl( 1 - \frac{3}{\sigma} \partial _\sigma - \, \partial^2 _\sigma %%@
\bigr) \,
	      \phi {\scriptstyle(\!\sigma\!)} \\
	\rightarrow \quad 0 
	& = \bigl(-\sigma^2 - 1 + \sigma \partial _\sigma
		+ \, \sigma^2 \partial^2 _\sigma \bigr) \,
	    \sigma  \phi {\scriptstyle(\!\sigma\!)}
\end{align*}
This is the modified Bessel equation with index 1
(eq. 9.6.1, p.374, Abramowitz, Stegun \cite{astegun}).
So if $V_1{\scriptstyle(\!\sigma\!)}$ is a solution of this modified Bessel %%@
equation,
then $\phi{\scriptstyle(\!\sigma\!)} = V_1{\scriptstyle(\!\sigma\!)} \! / \sigma$
is a solution of the Klein-Gordon equation for negative $\lambda$. \\
The solution that falls off to zero for $\sigma \rightarrow \infty$ is
$\phi{\scriptstyle(\!\sigma\!)} = K_1{\scriptstyle(\!\sigma\!)} \! / \sigma$.
$K_1$ is the modified Hankel function with index 1 and
$H^{\scriptstyle (^1_2)}_1$
are the Hankel functions of first and second kind with index 1.
With $K_1{\scriptstyle(\!\sigma\!)} =
- \frac{\pi}{2}H^{\scriptstyle (^1_2)}_1{\scriptstyle(\!\pm i\sigma\!)}$
(eq. 9.6.4, p.375, Abramowitz, Stegun \cite{astegun})
and $\sqrt{-\lambda_\pm} = \mp i \sqrt{\lambda_\pm}$ this leads to
% -------------------- equation dvb24--------------------------------------
\begin{equation}
	\label{eq:dvb24}
	\phi_\pm {\scriptstyle(\!\lambda\!)} \,
	= \, \frac{\mp i \pi /2}{\sqrt{\lambda _\pm}}
	  H^{\scriptstyle (^1_2)}_1 ({\scriptstyle\!\sqrt{\lambda _\pm}})
\end{equation}
derived here for negative $\lambda$.
With the Hankel functions being solutions of the Bessel equation
this expression is also valid for positive $\lambda$.
Using equations 9.1.3,4,11 in Abramowitz, Stegun \cite{astegun}
we can write equation (\ref{eq:dvb24}) as
% -------------------- equation dvb25--------------------------------------
\begin{align}
	\phi_\pm {\scriptstyle(\!\lambda\!)} \,
	& = \, \frac{J_1({\scriptstyle \! \sqrt{\lambda}})}{2{\scriptstyle \! %%@
\sqrt{\lambda}}}
	    \, \Bigl[ \ln (\lambda _\pm) \mp i \pi \Bigr] - \frac{1}{\lambda _\pm}
		+ \chi {\scriptstyle (\!\lambda\!)} \\
	\underset{\epsilon \rightarrow 0}{\lim} \, \phi_\pm %%@
{\scriptstyle(\!\lambda\!)} \,
	& = \, \frac{J_1({\scriptstyle \! \sqrt{\lambda}})}{2{\scriptstyle \! %%@
\sqrt{\lambda}}}
	    \ln \mid \!  \lambda \! \mid - P(1/ \lambda) + \chi {\ssty (\!\lambda\!)} %%@
\notag \\
	& \;\;\;\;\; \mp i \pi \epsilon {\scriptstyle (\!t_x\!-\!t_y\!)}
	    \biggl[ \theta {\scriptstyle (\!\lambda\!)}
		\frac{J_1({\scriptstyle \! \sqrt{\lambda}})}{2{\scriptstyle \! %%@
\sqrt{\lambda}}}
		- \delta {\scriptstyle (\!\lambda\!)} \biggr]
\end{align}
wherein $\chi$ is a real function given in
equation 9.1.11 in Abramowitz, Stegun \cite{astegun}.
When taking the derivative, only the contribution of the pole term survives.
Likewise to the massless case we thereby reproduce property %%@
(\ref{eq:dvb07}):
\begin{equation*}
	\underset{\epsilon \rightarrow 0}{\lim} \;
	\bigl[ \partial _0 \phi_\pm {\scriptstyle(\!x\!)} \bigr]_{t_x=t_y} \,
	= \, \pm i \, \underbrace{ \, \frac{2\pi^2}{m^2}}_\beta \,
		\delta^{\scriptscriptstyle \! (3)} \! {\scriptstyle(\!\vec{x}-\vec{y}\!)}
\end{equation*}
Applying again relations (\ref{eq:dvb08})-(\ref{eq:dvb13})
in the massive case we find:
% -------------------- equation dvb27--------------------------------------
\begin{align}
	G^\pm {\scriptstyle(\!x\!,y\!)}
	& = \frac{\scriptstyle m^2}{\scriptstyle 4 \pi^2} \;
		\underset{\epsilon \rightarrow 0}{\lim} \;
	    \phi_\mp {\scriptstyle(\!\lambda\!)} \\
	G^{\scriptscriptstyle(\!1\!)} {\scriptstyle(\!x\!,y\!)}
	& = \frac{\scriptstyle m^2}{\scriptstyle 2 \pi^2} \, \biggl[ 
		\frac{J_1({\scriptstyle \! \sqrt{\lambda}})}{2{\scriptstyle \! %%@
\sqrt{\lambda}}}
		\ln \mid \!  \lambda \! \mid - P(1/ \lambda)
		+ \chi {\scriptstyle (\!\lambda\!)} \biggr] \\
	G {\scriptstyle(\!x\!,y\!)}
	& = \frac{\scriptstyle im^2}{\scriptstyle 2 \pi} \; \epsilon
		 {\scriptstyle(\!t_x\!-\!t_y\!)}
	    \biggl[ \theta {\scriptstyle (\!\lambda\!)}
		\frac{J_1({\scriptstyle \! \sqrt{\lambda}})}{2{\scriptstyle \! %%@
\sqrt{\lambda}}}
		- \delta {\scriptstyle (\!\lambda\!)} \biggr] \\
	G \!_R {\scriptstyle(\!x\!,y\!)}
	& = \frac{\scriptstyle im^2}{\scriptstyle 2 \pi} \; \theta %%@
{\scriptstyle(\!t_x\!-\!t_y\!)}
	    \biggl[ \theta {\scriptstyle (\!\lambda\!)}
		\frac{J_1({\scriptstyle \! \sqrt{\lambda}})}{2{\scriptstyle \! %%@
\sqrt{\lambda}}}
		- \delta {\scriptstyle (\!\lambda\!)} \biggr] \\
	G \!_A {\scriptstyle(\!x\!,y\!)}
	& = \frac{\scriptstyle im^2}{\scriptstyle 2 \pi} \; \theta %%@
{\scriptstyle(\!t_y\!-\!t_x\!)}
	    \biggl[ \theta {\scriptstyle (\!\lambda\!)}
		\frac{J_1({\scriptstyle \! \sqrt{\lambda}})}{2{\scriptstyle \! %%@
\sqrt{\lambda}}}
		- \delta {\scriptstyle (\!\lambda\!)} \biggr] \\
	G \!_F {\scriptstyle(\!x\!,y\!)}
	& = \frac{\scriptstyle m^2}{\scriptstyle 4 \pi^2} \;
		\underset{\epsilon \rightarrow 0}{\lim} \;
	    \phi{\scriptstyle (\!\lambda - i \epsilon\!)} \notag \\
	& = \frac{\scriptstyle m^2}{\scriptstyle 4 \pi^2} \;
		\underset{\epsilon \rightarrow 0}{\lim} \;
	    \frac{\mp i \pi /2}
		{\sqrt{\lambda \! - \! i \epsilon \! \mid \! t_x \!\! - \! t_y \!\! \mid
		 - \! {\ssty \frac{\epsilon ^2}{4m^2}}} \,} \,
		H^{\ssty (^1_2)}_1 \! \bigl( {\ssty \!
			\sqrt{\lambda -i \epsilon \mid t_x \! - \! t_y \! \mid
		 - {\ssty \frac{\epsilon ^2}{4}}}} \, \bigr)
\end{align}
These results found by Dullemond and van Beveren
are in agreement with those listed
in appendix I.B p.649 in Bogoliubov, Shirkov \cite{bogshir}. \\
From (\ref{eq:dvb14}) we know again that
$(\Box_x \! + m^{\scriptscriptstyle 2}) \, G \! _F {\scriptstyle(\!x\!,y\!)}
 = \, -i \, \delta^{\scriptscriptstyle\! (4)} \! {\scriptstyle(\!x-y\!)}$.
As in the massless case the propagators exhibit singularities
on the whole lightcone $\lambda = 0$ but only lead to one single delta source
at the coincident point $x=y$. \\
We finally note that in Minkowski spacetime the $i\epsilon$-term
is always time independent, a property that is subject to change in AdS %%@
spacetime.
	\section{Propagators for AdS spacetime}
	\label{sec:props_ads}
		\subsection{General properties of propagators in AdS}
		\label{ads_genprops}
In this section we turn to investigate propagators in AdS,
still on the traces of Dullemond and van Beveren \cite{dvb}.
\\
The definitions and relations between the different propagators
for hermitian scalar fields $\phi {\ssty(\!x\!)}$ in AdS
are the same as in Minkowski spacetime
(\ref{eq:minkprop01}-\ref{eq:minkprop22}).
Now $x=(t,\vec{x})$ is a point in AdS
with $t \equiv x^0$ serving as time coordinate.
\begin{alignat}{2}
	\label{eq:dvb_ads01}
	\text{Wightman functions:} & \;\;\,
		&
		G^+ {\ssty(\!x\!,y\!)}
		& \equiv \, \left\langle 0 \! \mid \phi {\ssty(\!x\!)} \,
		\phi {\ssty(\!y\!)} \mid \! 0 \right\rangle 
		=  \left\langle 0 \! \mid \phi^{\sssty +} \! {\ssty(\!x\!)} \,
		\phi^{\sssty -} \! {\ssty(\!y\!)} \mid \! 0 \right\rangle \\
	&
		&
		G^- {\ssty(\!x\!,y\!)}
		& \equiv \, \left\langle 0 \! \mid \phi {\ssty(\!y\!)} \,
		\phi {\ssty(\!x\!)} \mid \! 0 \right\rangle
		= \left\langle 0 \! \mid \phi^{\sssty +} \! {\ssty(\!y\!)} \,
		\phi^{\sssty -} \! {\ssty(\!x\!)} \mid \! 0 \right\rangle \\
	\notag \\
	\label{eq:dvb_ads03}
	{\ssty \text{Hadamard's elementary function:}} &
		&
		G^{\sssty(\!1\!)} {\ssty(\!x\!,y\!)}
		& \equiv \, \left\langle 0 \! \mid \left\{ \phi {\ssty(\!x\!)} \, ,
		\phi {\ssty(\!y\!)} \right\} \mid \! 0 \right\rangle \\
		& & & = \, G^+ \! {\ssty(\!x\!,y\!)} \, + \, G^- \! {\ssty(\!x\!,y\!)}
		\notag \\
	\text{Schwinger function:} &
		&
		G \, {\ssty(\!x\!,y\!)}
		& \equiv \, \left\langle 0 \! \mid \left[ \phi {\ssty(\!x\!)} \, ,
		\phi {\ssty(\!y\!)} \right] \mid \! 0 \right\rangle \\
		& & & = \, G^+ \! {\ssty(\!x\!,y\!)} \, - \, G^- \! {\ssty(\!x\!,y\!)}
		\notag \\
	\notag \\
	\text{retarded Green function:} &
		&
		G \!_R \, {\ssty(\!x\!,y\!)}
		& \equiv \, +\theta {\ssty(\!t_x\!-t_y\!)} \, G {\ssty(\!x\!,y\!)} \\
	\text{advanced Green function:} &
		&
		G \!_A \, {\ssty(\!x\!,y\!)}
		& \equiv \, -\theta {\ssty(\!t_y\!-t_x\!)} \, G {\ssty(\!x\!,y\!)} \\
	\notag \\
	\text{Feynman propagator:} &
		&
		G \!_F \, {\ssty(\!x\!,y\!)}
		& \equiv \, \left\langle 0 \! \mid T_{\text{cov}} \, \phi {\ssty(\!x\!)} \,
		\phi {\ssty(\!y\!)} \mid \! 0 \right\rangle \\
	\label{eq:dvb_ads08}
	&
		&
		& = \, \theta {\ssty(\!t_x\!-t_y\!)} \, G^+ \! {\ssty(\!x\!,y\!)}
		+ \, \theta {\ssty(\!t_y\!-t_x\!)} \, G^- \! {\ssty(\!x\!,y\!)}
\end{alignat}
$T_{\text{cov}}$ is the time ordering operator for the universal covering space
(see subsection \ref{sec:ads_timord_uc}).
We will briefly comment the case of the hyperboloid in a subsection further %%@
below.
\\  
Defined this way, the propagator functions again are connected by the %%@
relations:
\begin{align}
	G^\pm {\ssty(\!x\!,y\!)}
	& = {\ssty \frac{1}{2}} \left( G^{\sssty(\!1\!)} \! {\ssty(\!x\!,y\!)} \,
		\pm \, G {\ssty(\!x\!,y\!)} \right) \\
	\label{eq:dvb_ads10}	
	G^\pm {\ssty(\!x\!,y\!)}
	& = G^\mp {\ssty(\!y\!,x\!)} \\
	\label{eq:dvb_ads11}
	G^\pm {\ssty(\!x\!,y\!)}
	& = G^{\pm*} {\ssty(\!y\!,x\!)}
	\qquad \qquad \;\; {\ssty ( \text{because }\phi \text{ is hermitian} ) } \\
	\notag \\
	\label{eq:dvb_ads12}
	G^{\sssty(\!1\!)} \! {\ssty(\!x\!,y\!)}
	& = + G^{\sssty(\!1\!)} \! {\ssty(\!y\!,x\!)} \qquad \qquad 
	{\ssty \rightarrow \; G^{\sssty(\!1\!)}
	\text{ is a symmetric function}} \\
	\label{eq:dvb_ads13}
	G^{\sssty(\!1\!)} \! {\ssty(\!x\!,y\!)}
	& = \, G^{{\sssty(\!1\!)}*} {\ssty(\!y\!,x\!)} \qquad \qquad 
	{\ssty \rightarrow \; G^{\sssty(\!1\!)}
	\text{ is a real function}} \\
	\label{eq:dvb_ads14}
	G {\ssty(\!x\!,y\!)}
	& = - G \, \! {\ssty(\!y\!,x\!)}
	\qquad \qquad \;\; {\ssty \rightarrow \;
	G \text{ is an antisymmetric function}} \\
	\label{eq:dvb_ads15}
	G {\ssty(\!x\!,y\!)}
	& = \, G^* \, \! {\ssty(\!y\!,x\!)}
	\qquad \qquad \;\;\, {\ssty \rightarrow \;
	G \text{ is purely imaginary}} \\
	\notag \\
	G \! _R {\ssty(\!x\!,y\!)} & = G \! _A {\ssty(\!y\!,x\!)} \\
	G _R^* {\ssty(\!x\!,y\!)} & = - G \! _R {\ssty(\!x\!,y\!)}
	\qquad \qquad {\ssty \rightarrow \; G \! _R
	\text{ is purely imaginary}} \\
	G _A^* {\ssty(\!x\!,y\!)} & = - G \! _A {\ssty(\!x\!,y\!)}
	\qquad \qquad {\ssty \rightarrow \; G \! _A
	\text{ is purely imaginary}} \\
	\notag \\
	\label{eq:dvb_ads19}
	G \! _F {\ssty(\!x\!,y\!)}
	& = \, G \! _R {\ssty(\!x\!,y\!)} \, + \, G^- \! {\ssty(\!x\!,y\!)} \\
	\label{eq:dvb_ads20}
	& = \, G \! _A {\ssty(\!x\!,y\!)} \, + \, G^+ \! {\ssty(\!x\!,y\!)} \\	
	& = {\ssty \frac{1}{2}}
	    \left( G \! _R {\ssty(\!x\!,y\!)} \, + \, G \! _A {\ssty(\!x\!,y\!)} \,
		+ \, G^{\sssty(\!1\!)} \! {\ssty(\!x\!,y\!)} \right) \\
	& = {\ssty \frac{1}{2}}
	    \bigl[ \epsilon {\ssty (\!t_x\!-t_y\!)} G \! {\ssty(\!x\!,y\!)} \,
		+ \, G^{\sssty(\!1\!)} \! {\ssty(\!x\!,y\!)} \bigr]
		\\
	\label{eq:dvb_ads201}
	G \! _F {\ssty(\!x\!,y\!)}
	& = G \! _F {\ssty(\!y\!,x\!)}
	\qquad \qquad \;\;\; {\ssty \rightarrow \; G \! _F
	\text{ is a symmetric function}}
\end{align}
Now referring back to to the overall sign $\sigma = \pm 1$
of the embedding space's metric introduced in equation %%@
(\ref{eq:ads_coord01}),
we note that the Klein-Gordon operator on AdS reads $(\sigma \Box \ads + %%@
m^2)$.
\\
Our field $\phi {\ssty(\!x\!)}$ again fulfills
the Klein-Gordon equation as its equation of motion.
So the Wightman functions $G^\pm$ by definition are homogeneous solutions:
\begin{align}
	\left( \sigma \Box \ADS _x \!\! + m^{\sssty 2} \right)
		G^\pm {\ssty(\!x\!,y\!)} = \;\; & 0 \, = \,
		\left( \Box \ADS _x \!\! + \sigma m^{\sssty 2} \right)
		G^\pm {\ssty(\!x\!,y\!)} \\
	\rightarrow \quad \left( \sigma \Box \ADS _x \!\!
	+ m^{\sssty 2} \right)
		G^{\sssty(\!1\!)} {\ssty(\!x\!,y\!)} = \;\; & 0 \, = \, 
		\left( \Box \ADS _x \!\! +\sigma  m^{\sssty 2} \right)
		G^{\sssty(\!1\!)} {\ssty(\!x\!,y\!)} \\
	\rightarrow \quad \left( \sigma \Box \ADS _x \!\!
	+ m^{\sssty 2} \right) \; 
		G {\ssty(\!x\!,y\!)} \;\, = \;\; & 0 \, = \,
		\left( \Box \ADS _x \!\! + \sigma m^{\sssty 2} \right) \; 
		G {\ssty(\!x\!,y\!)}
\end{align}
Then with (\ref{eq:dvb_ads19}) and (\ref{eq:dvb_ads20}) we have again
\begin{equation}
	\left( \sigma \Box \ADS _x \!\! + m^{\sssty 2} \right)
	G \! _F {\ssty(\!x\!,y\!)}
	\; = \; \left( \sigma \Box \ADS _x \!\! + m^{\sssty 2} \right)
	G \! _R {\ssty(\!x\!,y\!)}
	\; = \; \left( \sigma \Box \ADS _x \!\! + m^{\sssty 2} \right)
	G \! _A {\ssty(\!x\!,y\!)}	
\end{equation}
Starting from the definition (\ref{eq:dvb_ads08})
of the Feynman propagator 
and using equation (\ref{eq:ads_coord23})
\beq
	\sigma \Box \ADS _x = \frac{1}{R\ads^2(1\! + \! \vec{x}\,^2)} \,
		\partial _{t_x}^2 + \sigma \Box \ADS _{\vec{x}} 
\eeq
together with the equal time commutation relation
$\; \left[ \phi {\ssty(\!x\!)}
,\phi {\ssty(\!y\!)}\right] _{\, t_x=t_y} = \, 0 \;$
we obtain (see appendix \ref{sec:timord_delta}):
\begin{align}
	\label{eq:dvb_ads29}
	\left( \sigma \Box \ADS _x \!\! + m^{\sssty 2} \right)
		G \! _F {\ssty(\!x\!,y\!)} \;
	& = \; \frac{\delta {\ssty(\!t_x\!-t_y\!)}}{R\ads^2(1+\vec{x}\,^2)} \,
		\bigl[ \partial _{t_x} \! G^+ \! - \partial _{t_x} \! G^- \bigr] _{t_x\!=t_y} \\
	& = \; \frac{\delta {\ssty(\!t_x\!-t_y\!)}}
		{\underbrace{R\ads^2}_{1/ \beta_G}
		\underbrace{(1+\vec{x}\,^2)}_{f {\ssty \! (\vec{x})}}} \,
		\bigl[ \partial _{t_x} \! G {\ssty(\!x\!,y\!)} \bigr] _{t_x\!=t_y} \notag
\end{align}
This expression already delivers a delta function for the time variables
of the spacetime points $x$ and $y$. \\
Defining the various propagators as in equations
(\ref{eq:dvb_ads01})-(\ref{eq:dvb_ads08})
corresponds to define as Feynman propagator the function
fulfilling the inhomogeneous Klein-Gordon equation
\begin{align}
	\label{eq:dvb_ads29_5}
	\left( \sigma \Box \ADS _x \!\! + m^{\sssty 2} \right)
	G \! _F {\ssty(\!x\!,y\!)} \; & = \;\,
	\frac{-i}{\sqrt{g}} \; \delta^{\sssty\! (\!d)} {\ssty \! (\!x-y\!)}
	\; = \; \frac{-i}{R\ads^d} \;
	\delta^{\sssty\! (\!d)} {\ssty \! (\!x-y\!)} \\
	\label{eq:dvb_ads30}
	\left(\Box \ADS _x \!\! + \sigma m^{\sssty 2} \right)
	G \! _F {\ssty(\!x\!,y\!)} \; & = \;\frac{-i \sigma}{\sqrt{g}} \;
	 \delta^{\sssty\! (\!d)} {\ssty \! (\!x-y\!)}
	= \; \frac{-i \sigma}{R\ads^d} \;
		\delta^{\sssty\! (\!d)} {\ssty \! (\!x-y\!)}
\end{align}
This definition is consistent with \cite{arx0201253, arx0307229, dvb},
because in continuation $t \rightarrow t_E=-it$ to Euclidean space
the factor $i$ in front of the delta source becomes $-1$.

				\subsubsection
						{Solving the inhomogeneous Klein-Gordon equation}
				\label{sec:solvinhom}
Once found, we can use retarded, advanced and Feynman propagator
for finding the solution $\phi {\ssty (\!x\!)}$
of the inhomogeneous Klein-Gordon equation
for any given source $\rho {\ssty (\!x\!)}$ on the right hand side.
Letting the inverse Klein-Gordon operator act from the left on
(\ref{eq:dvb_ads29_5}) we obtain:
\begin{align}
	G \! _F {\ssty(\!x\!,y\!)} \,
	& = \, \left( \sigma \Box _x \!\! + m^{\sssty 2} \right)^{\! -1}
			\frac{-i}{\sqrt{g}} \; \delta^{\sssty\! (\!d)}
			{\ssty \! (\!x-y\!)} \notag
\end{align}
This equation shows to be very useful in the following calculation
\begin{align}
	\left( \sigma \Box _x \!\! + m^{\sssty 2} \right)
			\phi {\ssty(\!x)} \,
	& \overset{!}{=} \, \rho {\ssty(\!x\!)} \\ 
		\notag \\
	\rightarrow \quad \phi {\ssty(\!x\!)} \,
	& = \, \phi_0 {\ssty(\!x\!)} \, + \,
			\left( \sigma \Box _x \!\! + m^{\sssty 2} \right)^{\! -1}
			\rho {\ssty(\!x\!)}
		\notag \\
	& = \, \phi_0 {\ssty(\!x\!)} \, + \,
			\left( \sigma \Box _x \!\! + m^{\sssty 2} \right)^{\! -1}
			\!\! \int \!\! d^{d} \!y \, \rho {\ssty(\!y\!)} \,
			\delta^{\sssty\! (\!d)} {\ssty \! (\!x-y\!)}
		\notag \\
	& = \, \phi_0 {\ssty(\!x\!)} \, + \,
			\!\! \int \!\! d^{d} \!y \, \rho {\ssty(\!y\!)} \,
			\left( \sigma \Box _x \!\! + m^{\sssty 2} \right)^{\! -1}
			\delta^{\sssty\! (\!d)} {\ssty \! (\!x-y\!)} \,
			\frac{\ssty \! \sqrt{g}}{\ssty \! \sqrt{g}}
		\notag \\
	\label{eq:solvinhom_integral}
	& = \, \phi_0 {\ssty(\!x\!)} \, + \,
			i \!\! \int \!\! d^{d} \!y \, \sqrt{g} \;
			G \! _{F,R,A} {\ssty(\!x\!,y\!)} \, \rho {\ssty(\!y\!)}
\end{align}
wherein $\phi_0 {\ssty(\!x\!)}$ is a solution
of the free (homogeneous) Klein-Gordon equation.

		\subsection{From Klein-Gordon to the hypergeometric equation}
Because of the designated SO$\ssty (2,d-1)$ invariance
(up to time ordering subtleties) of the propagators
it is clear that our candidate function
can depend only on the chordal distance:
$\phi=\phi{\ssty ( \! \lambda {\sssty(\!} u {\sssty \!)} \!)}$.
\begin{align}
	u {\ssty ( \! X \!,Y \!)}\, & = \, (X-Y)^2 \, = \, invar. \notag \\
	X^2 \, & = \, \sigma \! R_X^2 \, = \,invar. \notag \\
	Y^2 \, & = \, \sigma \! R_Y^2 \, = \,invar. \notag \\	
	XY \, & = \, {\ssty \frac{1}{2}}
		(\sigma \! R_X^2 + \sigma \! R_Y^2 - u)	= \, invar. \notag
\end{align}
We make the following ansatz
\begin{align}
	\label{eq:dvb_ads31}
	\lambda \, = \, 1 - \, \frac{(XY)^2}{X^2Y^2} \, 
	= \, 1- \biggl( \! \frac{\sigma}{2}
			\Bigl[ \! {\ssty \frac{R_X}{R_Y} + \frac{R_Y}{R_X} } \! \Bigr]
	 		- \frac{u}{\, 2R_X \! R_Y} \biggr)^{\!\!2}
\end{align}
and will utilize it soon. But for the moment we need to turn to finding
a relation between the d'Alembertians in embedding space and AdS.
We denote with $z$ the coordinates on AdS.
Setting $X\!^N=R \, \omega^N \! {\ssty (z)}$,
which is realized by all given coordinate sets,
we get $\omega^2 \! \equiv \eta_{M \! N} \omega^M \! \omega^N = \sigma$
therefrom $\eta_{M \! N} \omega^M(\del_\nu\omega^N) = 0$
and further on
\begin{align*}
	ds^2 & = \eta_{M \! N} dX^M dX^N \\
		 & = \sigma \, dR^2 + R^2
		     \underbrace{\eta_{M \! N} (\partial_\mu \omega^M)(\partial_\nu \omega^N)} \,
			 dz^\mu dz^\nu \\
		 & \qquad \qquad \qquad \qquad \quad ^{h_{\mu \nu} {\ssty (z)}}
\end{align*}
So the hyperbolical form $G \! _{M \! N}$ of the metric in the embedding space is:
\begin{equation*}
	G \! _{M \! N}{\ssty (R,z)} = 
	\begin{pmatrix}
		\sigma & 0 \\
		0         & g_{\mu \nu} {\ssty (R,z)}
	\end{pmatrix}
	\quad \quad {\ssty \text{with}}
	\qquad g_{\mu \nu} {\ssty (R,z)} = R^2 h_{\mu \nu}{\ssty (z)} 
\end{equation*}
$g_{\mu \nu}$ is the induced metric on a hyperboloid with $R^2 = const. > 0$
and therefore the metric of an AdS space. Defining
\begin{align*}
	G{\ssty (\!R,z)} \, \equiv \, \mid \det G \! _{M \! N} \mid \qquad \quad
	g{\ssty (\!R,z)} \, \equiv \, \mid \det g_{\mu \nu} \mid \qquad \quad
	h{\ssty (z)} \, \equiv \, \mid \det h_{\mu \nu} \mid
\end{align*}
we have $G{\ssty (\!R,z)} = g{\ssty (\!R,z)} = R^{2d} h{\ssty (z)}$.
We find for the d'Alembertians $\Box _{bulk}$ in the embedding space and 
$\Box \ads$ on the hyperbolic hypersurfaces with constant $R$:
\begin{align}
	\Box _{bulk}
	& \equiv \frac{1}{\sqrt{G}} \, \partial _M \sqrt{G} \, G^{M \! N} \partial _N \notag \\
	& = \frac{\sigma}{\sqrt{g}} \, \partial _R \sqrt{g} \, \partial _R + \Box \ads
	\qquad \qquad \quad \Box \ads \equiv \frac{1}{\sqrt{g}} \,
	\partial _\mu \sqrt{g} \, g^{\mu \nu} \partial _\nu \notag \\	
	\label{eq:dvb_ads32}
	& = \sigma \partial _R^2 + \frac{\sigma d}{R} \, \partial _R + \Box \ads
\end{align}
Introducing the generalized angular momentum operators
\begin{equation*}
	M_{M \! N} \equiv i \bigl( X_M \partial _N -  X_N \partial _M \bigr)
	\; \qquad \quad \quad \longrightarrow \;\, M_{M \! N}  R = 0 
\end{equation*}
and the generalized squared total angular momentum operator
\begin{align*}
	M^2 
	& \equiv \frac{1}{2} M_{M \! N} M^{M \! N}
	\qquad \qquad \qquad \quad \longrightarrow \;\, M^2  R = 0 \\
	& = dR \, \partial _R + R^2 \partial _R^2 - \sigma R^2 \Box_{bulk}
\end{align*}
we find:
\begin{equation}
	\label{eq:dvb_ads33}
	\Box_{bulk} = \sigma \partial _R^2 + \frac{\sigma d}{R} \, \partial _R 
					- \sigma \frac{M^2}{R^2}
\end{equation}
Comparing (\ref{eq:dvb_ads32}) and (\ref{eq:dvb_ads33}) shows the equivalence
of the squared angular momentum operator and the d'Alembertian on the hyperboloid:
\begin{equation}
	\label{eq:dvb_ads34}
	- \sigma \frac{M^2}{R^2} = \Box \ads
\end{equation}
We will derive our candidate function
$\phi {\ssty (\!x\!, y\!)}$ which lives on AdS
from a function $\Psi {\ssty (\!X\!, Y\!)}$
which lives in embedding space.
Because of SO$\ssty (2,d-1)$ invariance $\Psi$
can only depend on $R_X$, $R_Y$, $\lambda$
and a distance winding number $n$. 
We make the following separation ansatz
introducing a new parameter $\Delta$:
\begin{equation}
	\Psi \! {\ssty (\!X\!,Y\!)}
	= \Psi {\ssty (\! R_X\!, R_Y\!, \lambda ,n \!)}
	= R_X^{-\Delta} \, R_Y^{-\Delta} \,
		\phi _{\! \Delta}{\ssty \! ( \! \lambda ,n \!)}
\end{equation}
This parameter $\Delta$ is $-m$ in the notation
of Dullemond and van Beveren: $-\Delta =m \dvb$.
Using (\ref{eq:dvb_ads33}) we  find with $R_X\!=\!R\ads$:
\begin{align}
	\label{eq:dvb_ads37}
	0 = \Box ^{(\!X\!)} _{bulk} \Psi \! {\ssty (\! X\!,Y\!)}
			\quad \Longleftrightarrow \quad 0 \, 
	& = \, \biggl[ \frac{-M^2}{R_X^2} +
			\frac{\Delta(\Delta \! - \! d \! + \! 1)}{R_X^2} \biggr] \,
			\phi _{\! \Delta}{\ssty \! ( \! \lambda ,n \!)} \notag \\
	& = \, \biggl[ \sigma \Box  \ads \, + \underbrace{
			\frac{\Delta(\Delta \! - \! d \! + \! 1)}{R\ads^2}}_{m^2} \biggr] \,
			\phi _{\! \Delta}{\ssty \! ( \! \lambda ,n \!)} 
\end{align}
Moreover we have:
\begin{align}
	\label{eq:dvb_ads38}
	0 & = \Box _{bulk} \Psi \,  
	      = \, \partial _M \partial ^M \Psi \! {\ssty (\! X\!,Y \!)} \notag \\ 
	\Longleftrightarrow \quad 0  
	& = \biggl[ \lambda (1 - \lambda ) \, \partial _\lambda ^2 \,
	                   + \biggl( \! \frac{d}{2} - \lambda \, \frac{\, d \!+\!1}{2} \biggr)
					   \partial _\lambda \underbrace{ -
	                   \frac{\Delta}{2} \frac{(-\Delta \! + d \! - \! 1)}{2}}_{\, \frac{1}{4}
					   \, m^2 R\ads^2}
					    \, \biggr] \,
						\phi_{\! \Delta}{\ssty \! ( \! \lambda ,n \!)}
\end{align}
So  being a solution of equation (\ref{eq:dvb_ads38})  
and obeying the homogeneous Klein-Gordon equation (\ref{eq:dvb_ads37})
in AdS$_d$ with the squared mass $m^2=\Delta(\Delta-d+1)/R\ads^2$
is equivalent for $\phi _{\! \Delta}{\ssty \! ( \! \lambda ,n \!)}$.
This mass term exactly matches relation (\ref{eq:breitfreed01}). \\
Equation (\ref{eq:dvb_ads38}) is of the hypergeometric type
defined on p. 562 in Abramowitz, Stegun \cite{astegun} as:
\begin{align}
	\label{eq:dvb_ads38_3}
	0 \, = \, \biggl[ \lambda (1 - \lambda ) \, \partial _\lambda ^2 \,
	                        + \Bigl[ c - \lambda (a \! + \! b \! + \! 1) \Bigr] \partial _\lambda
							- ab \biggr] \,
	\phi _{\! \Delta}{\ssty \! ( \! \lambda ,n \!)}
\end{align}
In our case the parameters are
\begin{equation}
	a = \frac{\Delta}{2} \quad \quad 
	b = \frac{-\Delta \! + d \! - \! 1}{2} \quad \quad
	c = \frac{d}{2}
\end{equation}
The regular solution of the hypergeometric differential equation
is the hypergeometric function $F {\ssty (a,b,c, \lambda)}$.
Some useful properties of this hypergeometric function
are listed in appendix \ref{sec:hypergeo}.
In particular $F {\ssty (a,b,c, \lambda)}$
is is symmetric under the exchange of $a$ and $b$
i.e. under $\Delta \rightarrow (-\Delta \! + \! d \! - \! 1)$
i.e. under $\Delta _+ \! \leftrightarrow \Delta _-$
(see section \ref{sec:ads_breitfreed}).
We have found a first candidate function:
\beq
	F (a,b,c, \lambda) \,
	= \, F \Bigl(\frac{\, \Delta_+}{2},\frac{\, \Delta_-}{2},\frac{d}{2},
		 \, \lambda \Bigr)
\eeq
This function is regular at $\lambda = 1$
because Re $(c \! - \! a \!- \! b) = \frac{1}{2} > 0$. 
\begin{figure} [H]
	\begin{center}
	\igx[width=5cm]{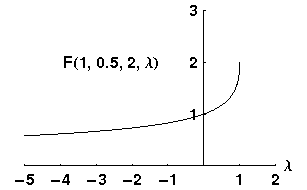}
	\end{center}
	\caption{First candidate function}
	\label{fig:hypergeo_ads_summe}
\end{figure}
In figure \ref{fig:hypergeo_ads_summe} the function is plotted
for $(d=4)$ with the conformal values \eqref{eq:breitfreed09}
of $\Delta_\pm$.
Looking at (\ref{eq:dvb_ads38_3}) we find that substituting
$\ovl{\lambda} = (1 \! - \! \lambda)$ yields the same differential equation
with the same parameters $a,b$ and $\ovl{c}=\frac{1}{2}$.
Therefore we have found a second candidate function: 
\beq
	F (a,b,\ovl{c}, \ovl{\lambda}) \,
	= \, F \Bigl(\frac{\, \Delta_+}{2},\frac{\, \Delta_-}{2},\frac{1}{2},
		 \, \ovl{\lambda} \Bigr)
\eeq
This function is divergent in $\ovl{\lambda} = 1$
for $d \geq 4$ because Re $(c \! - \! a \!- \! b) = 1-\frac{d}{2} \leq -1$.
In figure \ref{fig:hypergeo_ads_summe_bar} the function is plotted
with the same parameters as in the plot above.
\begin{figure} [H]
	\begin{center}
	\igx[width=6cm]{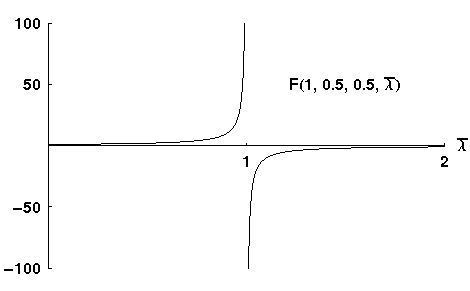}
	\end{center}
	\caption{Second candidate function}
	\label{fig:hypergeo_ads_summe_bar}
\end{figure}

		\subsection{Inspection of the candidate functions}
\begin{figure} [H]
	\begin{center}
	\igx[width=7cm]{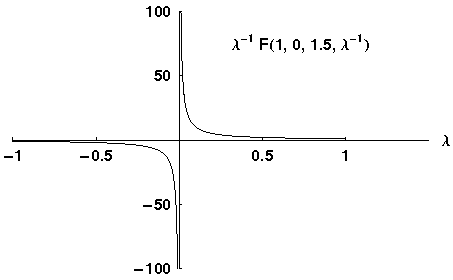}
	\end{center}
	\caption{First summand $\phi_\Delta^{(a)}$}
	\label{fig:hypergeo_ads_prop}
\end{figure}

What we are looking for is a function that converges to zero
if the chordal distance $u$ goes to (positive or negative) infinity.
When working with hypergeometric functions
we have two standard quantities going to zero for large chordal distances:
\begin{align*}
	u \rightarrow \pm \infty
	\quad \Longrightarrow \quad
	\lambda \rightarrow - \infty \quad
	& \Longrightarrow \quad
	\; \;\; \frac{1}{\lambda} \;\;\; \; \rightarrow \, 0 \\
	& \Longrightarrow \quad
	\frac{1}{1- \lambda} \; \rightarrow \, 0
\end{align*}
Dullemond and van Beveren work with $\lambda^{-\!1}$ in \cite{dvb}.
$(1 \! - \! \lambda)$ is the quantity used by Burgess and Lutken
in their calculation of the Feynman propagator
via a mode summation \cite{burgess}
(${\ssty (1 \! - \! \lambda)}$ {\scriptsize therein is denoted as}
${\ssty z^{-2}}$).
Their result is cited often e.g. \cite{arx0201253}
wherein the euclidean version of the propagator is given
({\ssiz and} ${\ssty (1 \! - \! \lambda)}$ {\ssiz is denoted as} ${\ssty \xi^2}$).
\\
Using these two quantities we can split up our solution in two different ways
according to equations (\ref{eq:hypergeo07}) and (\ref{eq:hypergeo08}):
\begin{align}
	F {\scriptstyle (a,b,c, \lambda)} \,
	& = \, \frac{\Gamma{\ssty (\!c\!)} \,\Gamma{\ssty (\!b-a\!)}}
					 {\Gamma{\ssty (\!b\!)} \,\Gamma{\ssty (\!c-a\!)}} \,
					 \biggl( \! \frac{1}{-\lambda} \! \biggr)^{\!\!a} \,
					 F \Bigl( a,a\!-\!c\!+\!1,a\!-\!b\!+\!1, \frac{\ssty 1}
					 			{\ssty \lambda} \Bigr)
					  \notag \\
	\label{eq:dvb_ads40}
	& \qquad + \; \frac{\Gamma{\ssty (\!c\!)} \,\Gamma{\ssty (\!a-b\!)}}
					 {\Gamma{\ssty (\!a\!)} \,\Gamma{\ssty (\!c-b\!)}} \,
					 \biggl( \! \frac{1}{-\lambda} \! \biggr)^{\!\!b} \,
					 F \Bigl( b,b\!-\!c\!+\!1,b\!-\!a\!+\!1, \frac{\ssty 1}
					 {\ssty \lambda} \Bigr) \\
	\notag \\
	& = \, \frac{\Gamma{\ssty (\!c\!)} \,\Gamma{\ssty (\!b-a\!)}}
					 {\Gamma{\ssty (\!b\!)} \,\Gamma{\ssty (\!c-a\!)}} \,
					 \biggl( \! \frac{1}{1 \! - \! \lambda} \! \biggr)^{\!\!a} \,
					 F \Bigl( a,c\!-\!b,a\!-\!b\!+\!1, \frac{\ssty 1}{\ssty 1-\lambda} \Bigr)
					  \notag \\
	\label{eq:dvb_ads41}
	& \qquad + \; \frac{\Gamma{\ssty (\!c\!)} \,\Gamma{\ssty (\!a-b\!)}}
					 {\Gamma{\ssty (\!a\!)} \,\Gamma{\ssty (\!c-b\!)}} \,
					 \biggl( \! \frac{1}{1 \! - \! \lambda} \! \biggr)^{\!\!b} \,
					 F \Bigl( b,c\!-\!a,b\!-\!a\!+\!1, \frac{\ssty 1}{\ssty 1-\lambda} \Bigr)
\end{align}
However using equations (\ref{eq:hypergeo05}) and (\ref{eq:hypergeo06})
we see
that the first summand in (\ref{eq:dvb_ads40}) is equal to the first summand
of (\ref{eq:dvb_ads41}) and that the second summands also match.
In the following calculations we will make use of (\ref{eq:dvb_ads40}). \\
Each of the two summands in (\ref{eq:dvb_ads40}) is a solution
of the hypergeometric differential equation (see \cite{astegun} 15.5.A).
Hence in both cases we can chose either part of (\ref{eq:dvb_ads40})
as a solution of (\ref{eq:dvb_ads38}) and consider them separately,
the prefactors with the four Gamma functions dropped:
\begin{align}
	\label{eq:dvb_ads42}
	\phi^{\ssty (\!a\!)}_{\! \Delta}{\scriptstyle \! ( \! \lambda \!)} \,
	& = \, \biggl( \! \frac{1}{-\lambda} \! \biggr)^{\!a} \,
			 F \Bigl( a,a\!-\!c\!+\!1,a\!-\!b\!+\!1, \frac{\ssty 1}{\ssty \lambda} \Bigr)
			  \\
	\label{eq:dvb_ads43}
	\phi^{\ssty (\!b\!)}_{\! \Delta}{\scriptstyle \! ( \! \lambda \!)} \,
	& = \, \biggl( \! \frac{1}{-\lambda} \! \biggr)^{\!b} \,
			 F \Bigl( b,b\!-\!c\!+\!1,b\!-\!a\!+\!1, \frac{\ssty 1}{\ssty \lambda} \Bigr)
\end{align}
These two functions satisfy the homogeneous Klein-Gordon equation
on AdS and therefore are candidates for the construction
of our various propagators.
In figure \ref{fig:hypergeo_ads_prop} the function $\phi_\Delta^{(a)}$
is plotted with the same parameters as in the plots before. 
If we perform such a splitting using (\ref{eq:hypergeo08})
with the second candidate function $F(a,b,\ovl{c},\ovl{\lambda})$
end resubstitute $\ovl{\lambda}$ by $(1-\lambda)$
we end up with the same two functions (\ref{eq:dvb_ads42})
and (\ref{eq:dvb_ads43}).
Solely the fractions in front of $F$ become $(1/\lambda)^{a,b}$.
Hence by factoring out $(-1)^{a,b}$ we obtain exactly the same functions.
This factor later would be devoured by the normalization,
thus we do not need to consider this case further.
\\
With the conditions (\ref{eq:breitfreed07}) and (\ref{eq:breitfreed08})
by Breitenlohner and Freedman and using 
$\cos \overline{\rho}= (1/ \! -\! \lambda)^{\! 1/2}$
for $t_y \!\! = \! \vec{y} \! = \! 0$
we find that the propagator must vanish for large $\lambda$ faster than 
\beqs
	\biggl( \! \frac{1}{-\lambda} \! \biggr)^{\! \frac{1}{2}\frac{d-\!1}{2}}
\eeqs
for the regular definition of the energy functional respectively faster than
\beqs
	\biggl( \! \frac{1}{-\lambda} \! \biggr)^{\! \frac{1}{2}\frac{d-\!3}{2}}
\eeqs
for its improved version. Thus we find for the regular energy definition
that $\phi^{\ssty (\!a\!)}_{\! \Delta}$ is allowed for $\Delta > (d-\!1)\!/2$
which is fulfilled by $\Delta_+$ only.
$\phi^{\ssty (\!b\!)}_{\! \Delta}$ is admitted for $\Delta < (d-\!1)\!/2$
which is fulfilled by $\Delta_-$ only.
This is the reason why we have to treat $\phi^{\ssty (\!a\!)}_{\! \Delta}$
and $\phi^{\ssty (\!b\!)}_{\! \Delta}$ separately
and cannot use the original candidate function which is 
(up to the dropped $\Gamma$-functions) the sum of both.
\\
For the improved energy definition $\phi^{\ssty (\!a\!)}_{\! \Delta}$
is allowed for $\Delta > (d-\!3)\!/2$
which always is fulfilled by $\Delta_+$ but also by $\Delta_-$ if
\beq
	m^2 R\ads^2 \, < \, 1 - \biggl( \frac{d-\!1}{2} \biggr)^{\! 2}
\eeq
 $\phi^{\ssty (\!b\!)}_{\! \Delta}$ is then permitted for $\Delta < (d+\!1)\!/2$
which is always fulfilled by $\Delta_-$ and also by $\Delta_+$
if the same mass condition as above is satisfied.
We will not consider this case in detail here. \\
We just note that since both candidate functions
are allowed for the satisfied mass condition,
also both propagators soon constructed from them will be allowed.
Thus adjusting their normalization with the dropped $\Gamma$-functions
one could use the entire unsplitted candidate function as propagator.
So far we have found:
\begin{align}
	\label{eq:dvb_ads42_5}
	\phi^{\ssty (\!a\!)}_{\! \Delta_+}{\scriptstyle \! ( \! \lambda \!)} \,
	& = \, \biggl( \!\! \frac{1}{-\lambda} \! \biggr)^{\!\! \frac{\, \Delta_+}{2}} \,
			 F \Biggl( \frac{\, \Delta_+}{2}, \frac{\, \Delta_+\!\!-\!d\!+\!2}{2}, \, 
			 \Delta_+\! - \! \frac{\,d\!-\!3}{2}, \frac{1}{\lambda} \Biggr) \\
	\label{eq:dvb_ads43_5}
	\phi^{\ssty (\!b\!)}_{\! \Delta_-}{\scriptstyle \! ( \! \lambda \!)} \,
	& = \, \biggl( \!\! \frac{1}{-\lambda} \! \biggr)^{\!\!
			\frac{-\Delta_- \! +d- \! 1}{2}}
			 F \Biggl( \frac{-\Delta_- \!\!+\!d\!-\!1}{2}, \frac{\,1\!-\!\Delta_-}{2}, 
			 -\Delta_-\! + \! \frac{\,d\!+\!1}{2}, \frac{1}{\lambda} \Biggr)
\end{align}
We can check that $\phi^{\ssty (\!b\!)}_{\! \Delta_-}{\ssty \! ( \! \lambda \!)}
=\phi^{\ssty (\!a\!)}_{-\Delta_- \! +d- \! 1}{\ssty \! ( \! \lambda \!)}
=\phi^{\ssty (\!a\!)}_{\! \Delta_+}{\ssty \! ( \! \lambda \!)}$.
Hence we can always switch from
$\phi^{\ssty (\!a\!)}_{\! \Delta_+}{\ssty \! ( \! \lambda \!)}$
to $\phi^{\ssty (\!b\!)}_{\! \Delta_-}{\ssty \! ( \! \lambda \!)}$
by replacing $\Delta_+ \rightarrow (-\Delta_- \!\!+\!d\!-\!1)$.
Therefore in the following we will only study the case of
$\phi^{\ssty (\!a\!)}_{\! \Delta_+}{\ssty \! ( \! \lambda \!)}$. \\
Now we must have a closer look on $\lambda$.
Evaluating ansatz (\ref{eq:dvb_ads31}) for AdS we find
\begin{align}
	\lambda \, = \, 1 - \, \frac{(XY)^2}{R\ads^4} \, 
	= \, 1- \biggl( \! \sigma - \frac{u}{2R\ads^2} \biggr)^{\!2}
\end{align}
and obtain for the $(t,\vec{x})$ coordinate set:
\begin{align}
	\lambda & = \, 1 - \Bigl[ \sqrt{1 \! +\vec{x}^{\,2}}\sqrt{1 \! +\vec{y}^{\,2}} %%@
\,
										 \cos \, (t_x\!\!-\!t_y) - \vec{x} \vec{y} \,\Bigr]^2 \notag \\
	& = \, 1 - \Bigl[ (1 \! +\vec{x}^{\,2})(1 \! +\vec{y}^{\,2}) \cos ^2 %%@
(t_x\!\!-\!t_y)
			 + (\vec{x} \vec{y})^2 \notag \\ 
	& \qquad \qquad \quad - \, 2 \,  \vec{x} \vec{y} \,
   \sqrt{1 \! +\vec{x}^{\,2}}\sqrt{1 \! +\vec{y}^{\,2}} \,
   \cos \, (t_x\!\!-\!t_y) \Bigr]
\end{align}
For the origin $t_y \!\! = \! \vec{y} \! = \! 0$ as reference point this reduces %%@
to:
\begin{align*}
	\lambda & = \, 1 - (1 \! +\vec{x}^{\,2}) \cos ^2 (t_x)
\end{align*}
We see that for certain values of $t_x, \vec{x},t_y,\vec{y}$
our invariant quantity $\lambda$ becomes zero %% or one
which is pathological because $\phi^{\ssty (\!a\!)}_{\! \Delta_+}$
%% and $\phi^{\ssty (\!b\!)}_{\! \Delta}$
is singular in $\lambda = 0$.
%% and the hypergeometric differential equation has singular points
%% at $\lambda = 0$ and $\lambda = 1$.
%%

				\subsubsection{Infinitesimal shift in complex time plane}
In order to remove this flaw we apply again a shift in the complex time plane
\beqs
	t_x \; \longrightarrow \; t_x \pm \, i \, \frac{\epsilon}{2}  
\eeqs 
which results in complex coordinates $X^0_\pm$ and $X^d_\pm$:
\beq
	\begin{pmatrix}
		X^0_\pm \\
		X^d_\pm
	\end{pmatrix}
	\, = \,
	\underbrace{
	\begin{pmatrix}
		\;\;\;\; \cosh {\ssty \frac{\epsilon}{2}}
		& \pm i \sinh {\ssty \frac{\epsilon}{2}}
		\\
		\mp i \sinh {\ssty \frac{\epsilon}{2}}
		& \;\;\;\; \cosh {\ssty \frac{\epsilon}{2}}
	\end{pmatrix}}_{A\ads}
	\begin{pmatrix}
		X^0 \\
		X^d
	\end{pmatrix}
\eeq
We can check that $A\ads^T A\ads = \mathbb{1}$, $\det A\ads = 1$
and therefore $A\ads \in SO(2,{\mathbb C})$.
Altogether the complex time shift results in a complex rotation
in the $X^0 \! X^d$ plane
generated by the generalized angular momentum operator $M_{0d}$
which commutes with $R^2\!$, $\Box_{bulk}$ and $M^2$ \cite{dvb}.
In the limit of small $\epsilon$ we find:
\begin{align*}
	X^0_\pm \, & \approx \, R_X \sqrt{1+\vec{x}^{\,2}} \;
							\Bigl[ (1\!+\!{\ssty \frac{\epsilon^2}{8}} ) \sin t_x 
							\pm \, i {\ssty \frac{\epsilon}{2}} \cos t_x \Bigr] \\
					& = \, (1\!+\!{\ssty \frac{\epsilon^2}{8}} ) X^0
							\pm \, i {\ssty \frac{\epsilon}{2}} X^d \\							
	X^d_\pm \, & \approx \, R_X \sqrt{1+\vec{x}^{\,2}} \;
							\Bigl[ (1\!+\!{\ssty \frac{\epsilon^2}{8}} ) \cos t_x 
							\mp \, i {\ssty \frac{\epsilon}{2}} \sin t_x \Bigr] \\
					& = \, (1\!+\!{\ssty \frac{\epsilon^2}{8}} ) X^d
							\mp \, i {\ssty \frac{\epsilon}{2}} X^{0}
\end{align*}
We leave $t_y$ and thereby $Y$ unchanged.
Next we define
\begin{align}
	\lambda_\pm \, & \equiv \,\; 1 - \, \frac{(X_\pm Y)^2}{R\ads^4} \quad
			= \, 1- \biggl( \! \sigma - \frac{u _\pm}{2R\ads^2} \biggr)^{\!2}
			\qquad \qquad {\ssty u_\pm \equiv (X_{\!\pm} \!-\!Y)^2} \\
			\notag \\
	\gamma \, & \equiv \qquad \frac{XY}{R\ads^2} \qquad \quad \,
			= \, \sqrt{1 \! +\vec{x}^{\,2}}\sqrt{1 \! +\vec{y}^{\,2}} \,
			 \cos \, (t_x\!\!-\!t_y) - \vec{x} \vec{y} \\
			\notag \\
	S \, & \equiv \, \frac{X^0Y^d \! - \! X^dY^{0}}{R\ads^2} \,
			= \, \sqrt{1 \! +\vec{x}^{\,2}}\sqrt{1 \! +\vec{y}^{\,2}} \,
			 \sin \, (t_x\!\!-\!t_y) \\
			 \notag \\
	\rightarrow \quad
	\lambda \, & = \, 1 \! - \! \gamma^2 \qquad \qquad
	\xi \, = \, 1 \! / \gamma \notag
\end{align}
$\gamma$ is SO$\ssty (2,d-1)$ invariant
and sign$(S)=\,$sign$\, \sin\, (t_x\!\!-\!t_y)$ 
is invariant under orthochronous SO$\ssty (2,d-1)$.
For small $\epsilon$ this definition of $\lambda_\pm$ yields
\begin{align}
	\lambda_\pm & \approx \, 1 - \Bigl[ (1\!+\!{\ssty \frac{\epsilon^2}{8}} )
								\sqrt{1 \! +\vec{x}^{\,2}}\sqrt{1 \! +\vec{y}^{\,2}} \,
								 \cos \, (t_x\!\!-\!t_y) - \vec{x} \vec{y} \notag \\
								 & \qquad \qquad \qquad \qquad \qquad
								 \mp i  {\ssty \frac{\epsilon}{2}} \, \sqrt{1 \! +\vec{x}^{\,2}}
								 \sqrt{1 \! +\vec{y}^{\,2}} \, \sin \, (t_x\!\!-\!t_y) \Bigr]^2
	\\
	\label{eq:dvb_ads50}
	& \approx \, \ub{1 - \Bigl[ \!
			\sqrt{1 \!\! +\! \vec{x}^{\,2}}\sqrt{1 \!\!	+ \!\vec{y}^{\,2}} 
			\cos \, (t_x\!\!-\!t_y) - \vec{x} \vec{y} \, \Bigr]^{2} }
			_{\lambda \leq 1} 
			\\
			& \qquad 
			-{\ssty \frac{\epsilon^2}{4}} \ub{\Bigl[ \!(1 \!\! +\!\vec{x}^{\,2})
			(1 \!\! +\!\vec{y}^{\,2}) \cos 2 (t_x\!\!-\!t_y) -
			\vec{x} \vec{y} \,\sqrt{1 \!\! +\! \vec{x}^{\,2}}
			\sqrt{1 \!\! + \!\vec{y}^{\,2}} \, 
			\cos \, (t_x\!\!-\!t_y) \Bigr] }_{\;\;\; \gtreqless \,0}
			\notag \\
			& \qquad  
			\pm i \epsilon \ub{ 
			\sqrt{1 \!\! + \! \vec{x}^{\,2}} \sqrt{1 \!\! + \!\vec{y}^{\,2}}
			\sin \, (t_x\!\!-\!t_y)}_S 
			\ub{\Bigl[ \sqrt{1 \!\! + \! \vec{x}^{\,2}}	\sqrt{1 \!\! + \!\vec{y}^{\,2}}
					 \cos (t_x\!\!-\!t_y) -	\vec{x} \vec{y} \,
					 \Bigr]}_\gamma
			\notag \\
	\label{eq:dvb_ads51}
	& \, = \, \underbrace{1 - \gamma^2}_{\lambda}
			 - \, {\ssty \frac{\epsilon^2}{4}}{ \ssty (\ldots)}
			\pm i \epsilon \, \gamma S
\end{align}
so that again we have
$\lambda_\pm {\ssty (\!x\!,y\!)} = \lambda^*_\mp {\ssty (\!x\!,y\!)}
  = \lambda_\mp {\ssty (\!y\!,x\!)}$.
\begin{figure} [H]
	\begin{center}
	\igx[width=10.5cm]{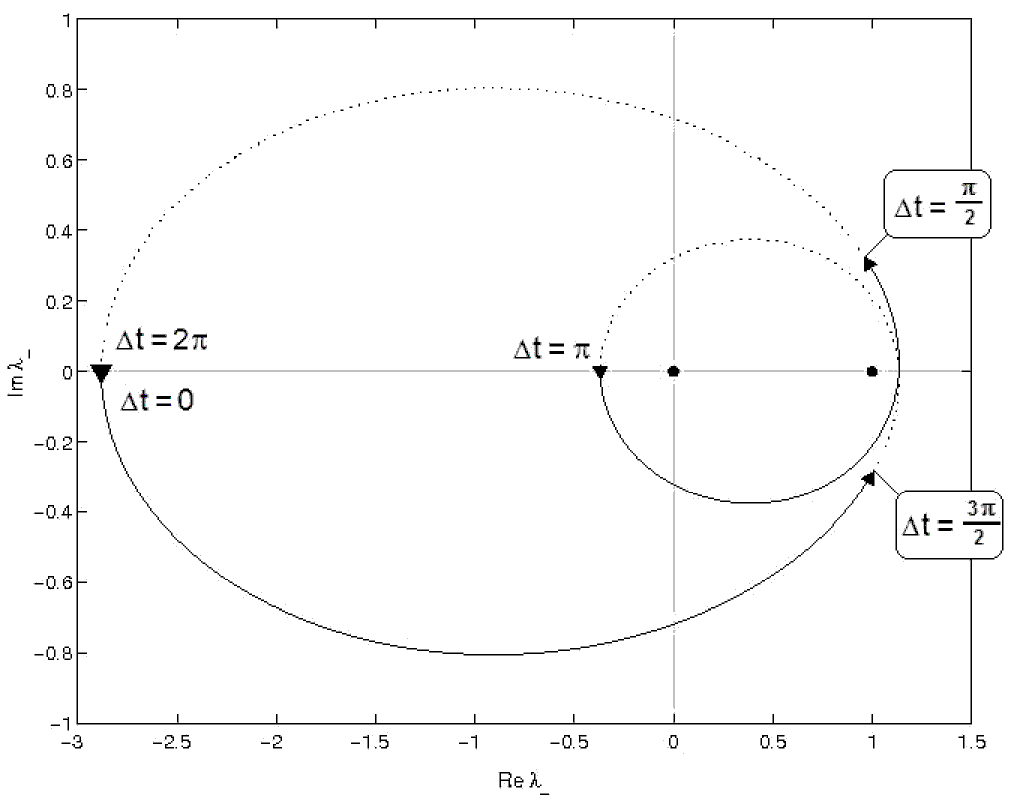}	
	\end{center}
	\caption{time dependence of $\lambda_-$}
	\label{fig:dvb_ads01}
\end{figure}
Figure \ref{fig:dvb_ads01} shows the time dependence of $\lambda _-$
in the complex plane.
It is plotted with purpose of illustration only for the convenient values
$\vec{x}=(1,\vec{0} \,)$, $\vec{y}=(-0.4,\vec{0} \,)$ and $\epsilon = 0.1$. \\
The graph starts at the larger triangle pointing downwards
at the very left hand side where $\sDelta t \equiv (t_x \! - \! t_y)=0$.
It continues counterclockwise with
$\Delta t = \frac{\pi}{2}, \, \pi, \, \frac{3\pi}{2}$ at the smaller triangles.
$\lambda_+$ starts at the same triangle as $\lambda_-$
but always moves in clockwise direction.
In the limit of small $\epsilon$ the triangles for
$\Delta t = \frac{\pi}{2}, \frac{3\pi}{2}$
move close to the real axis. 
\\
We see $\lambda_\pm$ running on one smaller and one larger ellipse by turns
and having a period of $2\pi$ in $\sDelta t$.  
The points $\lambda \! = 0$ and $\lambda \! = 1$
are always enclosed and never crossed for finite $\epsilon$.
Therefore we obtain the well defined function
\begin{align}
	\label{eq:dvb_ads52}
	\phi^{\ssty (\!a\!)}_{\! \Delta_+ \pm}  {\ssty \! ( \! \lambda_{\! \pm} \!)} \,
	& = \, \biggl( \! \frac{1}{-\lambda_\pm} \! \biggr)^{\!a} \,
			 F \Bigl( a,a\!-\!c\!+\!1,a\!-\!b\!+\!1,
			 \frac{\ssty 1}{\ssty \lambda_\pm} \Bigr)
\end{align}

				\subsubsection{Distance winding number}
Now we define a distance winding number $n$ in the following 
SO$\ssty (2,d-1)$-invariant way:
\begin{align}
	\text{(i)} \quad n {\ssty ( \! x \! ,y \!)} & = 0 \quad \text{ if } (t_x,\vec{x})
	\text{ can be continuously moved to } (t_y,\vec{y}) \notag \\
	\label{eq:dvb_ads57}
	& \qquad \quad \:
	\text{without changing the sign of } \gamma \text{ and} \\
	\text{(ii)} \quad \;\;
	\sDelta n & = \pm 1 \; \text{ whenever } \gamma
	\text{ changes sign and } \sDelta t \equiv (t_x\!-\!t_y) \gtrless 0
		\notag	
\end{align}
Whenever $\gamma$ changes sign
then $\lambda_\pm$ is just passing the point 1 in the complex plane. 
In general the distance winding number $n$ depends not only
on the difference of the times but also on the $\vec{x}$ and $\vec{y}$ coordinates.
When below $n$ is written without arguments
then it is always to be understood as $n {\ssty ( \! x \! ,y \!)}$.
\\
Only for $\!\vec{x}\,\vec{y}\!=\!0$ and therefore also
for the special reference point $t_y \!\! = \! \vec{y} \! = \! 0$
both ellipses coincide (which changes the period of $\lambda_\pm$ to $\pi$). 
For this case our definition of $n$ yields the simpler expression:
\beq
	\label{eq:dvb_ads54}
	(n\!-\!{\ssty \frac{1}{2}}) \, \pi \, < \, (t_x \! - \! t_y) \, 
			< \, (n\!+\!{\ssty \frac{1}{2}}) \, \pi
\eeq  
But we will consider the general case.
We first note that in the limit of small $\epsilon$
the ellipses move infinitesimally close to the real axis
(with the critical points 0 and 1 still being encircled).
Thereby the phases $\varphi$ of $\lambda_\pm$
become integer multiples of $\pi$.
\\
When $t_x$ is changed with constant $\vec{x}$ and $\vec{y}$
then $\lambda_\pm$ moves on the ellipses
and sooner or later returns to its starting point.
Thereby it picks up a change in distance winding number
and in phase: $\sDelta \varphi = \mp 2\pi n$.\\
Our function $\phi^{\ssty (\!a\!)}_{\! \Delta_+ \pm}$
%% and $\phi^{\ssty (\!b\!)}_{\! \Delta \pm}$ 
then also undergoes a change of phase
which is independent of the contour as long as it is deformed continuously
without passing the critical points $\lambda_\pm \! = 0; \, 1$.
For the case of large $\lambda$
we can read off from equation (\ref{eq:dvb_ads52}):
%% or rather (\ref{eq:dvb_ads53}).
%%
\begin{align*}
	\phi^{\ssty (\!a\!)}_{\! \Delta_+ \pm}  {\ssty \! ( \! \lambda_{\! \pm} \!)} \,
	& \approx \, \biggl( \! \frac{1}{-\lambda_\pm} \! \biggr)^{\!\!a}
%%	\phi^{\ssty (\!b\!)}_{\! \Delta \pm}  {\ssty \! ( \! \lambda_{\! \pm} \!)} \,
%%	& \approx \, \biggl( \! \frac{1}{-\lambda_\pm} \! \biggr)^{\!b} \,
\end{align*}
Therefrom we obtain the following additional relation:
\begin{align}
	\label{eq:dvb_ads55}
	\phi^{\ssty (\!a\!)}_{\! \Delta_+ \pm}  {\ssty \! (\lambda_{\! \pm}
			e^{\mp 2 \pi i n}\!)} \, 
	& = \, e^{\pm i \pi n \Delta_+} \,
			\phi^{\ssty (\!a\!)}_{\! \Delta_+ \pm}  {\ssty \! ( \!\lambda_{\! \pm} \!)}
		\qqqquad \quad
		{\ssty \lambda \gg 1}
%%	\label{eq:dvb_ads56}
%%	\phi^{\ssty (\!b\!)}_{\! \Delta \pm}  {\ssty \! ( \!n, \lambda_{\! \pm} \!)} \, 
%%	& = \, e^{\pm i \pi n (-\Delta+d-\!1)} \,
%%			\phi^{\ssty (\!b\!)}_{\! \Delta \pm}  {\ssty \! ( \!n=0, \lambda_{\! \pm} \!)}
\end{align}

				\subsubsection{Delta source and normalization}
In order to construct the Feynman propagator fulfilling the inhomogeneous
Klein-Gordon equation (\ref{eq:dvb_ads30}), we now want to show
that applying the Klein-Gordon operator to our candidate functions
yields the desired delta source. \\
According to (\ref{eq:dvb_ads29}) we therefor need to examine
$\bigl( \del_{t_x} \phi_\pm \bigr)_{\!t_x=t_y}$
with the intention of showing that it delivers a delta function
for the spatial variables of the spacetime points $x$ and $y$. \\
Starting from equation (\ref{eq:dvb_ads52})
%% and (\ref{eq:dvb_ads53})
and employing (\ref{eq:hypergeo07}) we get 
\begin{align*}
	\phi^{\ssty (\!a\!)}_{\! \Delta_+ \pm}  {\ssty \! ( \! \lambda_{\! \pm} \!)} \,
	& = \, \underbrace{ \frac{\Gamma{\ssty (\!a-b+1\!)} \,\Gamma{\ssty %%@
(\!1-c\!)}}
					 			{\Gamma{\ssty (\!a-c+1\!)} \,\Gamma{\ssty (\!1-b\!)}}}
					  			_{\;\;\; \alpha_{1a}} \,
			 F \bigl( a,b,c, \lambda_\pm \bigr) \\
	& \qquad + \, 
		\underbrace{\frac{\Gamma{\ssty (\!a-b+1\!)} \,\Gamma{\ssty (\!c-1\!)}}
					 	{\Gamma{\ssty (\!a\!)} \,\Gamma{\ssty (\!c-b\!)}}}
					  			_{\;\;\; \alpha_{2a}} \,
	\biggl( \! \frac{1}{-\lambda_\pm} \! \biggr)^{\!\!\! \frac{d-2}{2}} \!
			 F \bigl( a\!-\!c\!+\!1,b\!-\!c\!+\!1,2\!-\!c, \lambda_\pm \bigr)
\end{align*}
Next we put to work the chain rule of differentiation
$(\del_{t_x} \phi_\pm) = (\del_{t_x} \lambda_\pm)(\del_{\lambda_\pm} %%@
\phi_\pm)$
wherein for the last factor we can make use of equation (\ref{eq:hypergeo04}).
\begin{align}
	\label{eq:dvb_ads58}
	\Bigl( \del_{t_x}
		\phi^{\ssty (\!a\!)}_{\! \Delta_+ \pm} {\ssty \! ( \! \lambda_{\! \pm} \!)} \,
		\Bigr)_{\! t_x=t_y} \!\!\!
	& = \bigl( \del_{t_x} \lambda_\pm \bigr)_{\! t_x=t_y}
			\Bigl[ {\ssty \frac{ab}{c}} \, \alpha_{1a}
					F {\ssty ( a+1, \, b+1, \, c+1, \, \lambda_\pm)} \\
	& \qquad \qquad \qquad \quad + 
					{\ssty \frac{(\!a-c+1\!)(\!b-c+1\!)}{(\!2-c\!)}} \, \alpha_{2a} \,
					\bigl( {\ssty \frac{1}{-\lambda_\pm}} \bigr)^{\!\! \frac{d-2}{2}}
					 F {\ssty ( a-c+2, \, b-c+2, \, 3-c, \, \lambda_\pm )} \notag \\
	& \qquad \qquad \qquad \quad + 
					{\ssty \frac{d-2}{2}} \, \alpha_{2a} \,
					\bigl( {\ssty \frac{1}{-\lambda_\pm}} \bigr)^{\! \frac{d}{2}}
					 F {\ssty ( a-c+1, \, b-c+1, \, 2-c, \, \lambda_\pm )} \Bigr]_{t_x=t_y}
	\notag
	%%
%%	\label{eq:dvb_ads59}
%%	\Bigl( \del_{t_x}
%%		\phi^{\ssty (\!b\!)}_{\! \Delta \pm}  {\ssty \! ( \! \lambda_{\! \pm} \!)} \,
%%		\Bigr)_{\! t_x=t_y} \!\!\!
%%	& = \bigl( \del_{t_x} \lambda_\pm \bigr)_{\! t_x=t_y}
%%			\Bigl[ {\ssty \frac{ab}{c}} \, \alpha_{1b}
%%					F {\ssty ( a+1, \, b+1, \, c+1, \, \lambda_\pm)} \\
%%	& \qquad \qquad \qquad \quad + 
%%					{\ssty \frac{(\!a-c+1\!)(\!b-c+1\!)}{(\!2-c\!)}} \, \alpha_{2b} \,
%%					\bigl( {\ssty \frac{1}{-\lambda_\pm}} \bigr)^{\!\! \frac{d-2}{2}}
%%					 F {\ssty ( a-c+2, \, b-c+2, \, 3-c, \, \lambda_\pm )} \notag \\
%%	& \qquad \qquad \qquad \quad + 
%%					{\ssty \frac{d-2}{2}} \, \alpha_{2b} \,
%%					\bigl( {\ssty \frac{1}{-\lambda_\pm}} \bigr)^{\! \frac{d}{2}}
%%					 F {\ssty ( a-c+1, \, b-c+1, \, 2-c, \, \lambda_\pm )} %%@
\Bigr]_{t_x=t_y}
%%		\notag
\end{align}
From equation (\ref{eq:dvb_ads50}) we can read off that for small $\epsilon$ %%@
we have
\begin{align}
	\label{eq:dvb_ads60}
	\Bigl( \del_{t_x} \lambda_\pm \! \Bigr)_{\! t_x=t_y} \!
	& \approx \, \pm i \epsilon \Bigl[ (1 \!\! +\!\vec{x}^{\,2}) (1 \!\! %%@
+\!\vec{y}^{\,2})
			- \vec{x} \vec{y} \,\sqrt{1 \!\! +\! \vec{x}^{\,2}}\sqrt{1 \!\! + %%@
\!\vec{y}^{\,2}}
			\: \Bigr] \\
	\Bigl( \lambda_\pm \! \Bigr)_{\! t_x=t_y} \!
	& \approx \, 1 - \Bigl[ (1\!+\!{\ssty \frac{\epsilon^2}{8}} )
								\sqrt{1 \! +\vec{x}^{\,2}}\sqrt{1 \! +\vec{y}^{\,2}} \,
								  - \vec{x} \vec{y} \: \Bigr]^2 \notag \\
	& \approx \, 1 - \underbrace{\Bigl[ \!\sqrt{1 \! +\vec{x}^{\,2}}
							\sqrt{1 \! +\vec{y}^{\,2}} - \vec{x} \vec{y} \: \Bigr]^{\!2}}
							_{\gamma^2_{(t_x=t_y)}}	\notag \\
	& \qquad - {\ssty \frac{\epsilon^2}{4}}
		\underbrace{
		\Bigl[ (1 \!\! +\!\vec{x}^{\,2}) (1 \!\! +\!\vec{y}^{\,2}) - \vec{x} \vec{y} \,
		\sqrt{1 \!\! +\! \vec{x}^{\,2}}\sqrt{1 \!\! + \!\vec{y}^{\,2}} \: \Bigr] }
		_{\;\;\;\;\;\;\;\; \quad > 0 \;\; \forall \, \vec{x},\vec{y}}
\end{align}
which for our special reference point $t_y \!\! = \! \vec{y} \! = \! 0$
reduces to the simple expression
\begin{align}
	\Bigl( \lambda_\pm \! \Bigr)_{\! t_x=t_y} \!
	& \approx \, - \Bigl( \vec{x}^{\,2} \! + {\ssty \frac{\epsilon^2}{4}} \Bigr)
\end{align}
We note that
$\gamma^2_{(t_x=t_y \! )}
	= \bigl[ \! \sqrt{1 \! +\vec{x}^{\,2}}\sqrt{1 \! +\vec{y}^{\,2}} \,
	- \vec{x} \vec{y} \: \bigr]^2 \, \geq \, 1 \quad \forall \,\vec{x}, \vec{y} %%@
\quad$
and is equal to $1$ if and only if $\vec{x} = \vec{y}$.
This special feature renders $(\lambda_\pm)_{\! t_x=t_y}$
negative and thus nonzero for all $\vec{x} \neq \vec{y}$.
This in turn signifies that indeed we can let our $\epsilon$
run to zero whenever $\vec{x} \neq \vec{y}$
so that $(\del_{t_x} \phi^{\ssty (\!a\!)}_{\! \Delta_+ \pm})_{t_x=t_y}$
vanishes for this case.
\\
In order to determine suitable normalisation constants we need to integrate
$(\del_{t_x} \phi^{\ssty (\!a\!)}_{\! \Delta_+ \pm})_{t_x=t_y}$
over $(d\!-\!1)$ dimensional $\vec{x}$-space.
Doing so, we find that the first two summands in equation %%@
(\ref{eq:dvb_ads58})
%% and (\ref{eq:dvb_ads59})
yield contributions proportional to $\epsilon$ respectively $\epsilon^2$
which therefore vanish in the limit of small $\epsilon$. \\
When turning to the third summand, we first compute the $\vec{x}$-space %%@
integral
over $\epsilon\, (1\! / \!-\!\lambda_\pm)^{-d/2}_{t_x=t_y}$
for the case of our reference point $t_y \!\! = \! \vec{y} \! = \! 0$ and find
using integral (\ref{eq:delta155})
\begin{align}
	\int \!\! d^{d-\!1}x \,
	\frac{\epsilon}
			{\bigl( \vec{x}^{\, 2} + {\ssty \frac{\epsilon^2}{4}} \bigr)^{\! %%@
\frac{d}{2}}} \;
	& = \; \frac{4 \pi^{\frac{d\!-\!1}{2}}}{\Gamma({\ssty \frac{d-1}{2}})} \,
			\int \limits_0^\infty \!\! dr \, \frac{r^{d-2}}
			{\bigl( r^2 + 1 \bigr)^{\! \frac{d}{2}}} \,
	= \, \frac{2 \pi^{\! \frac{d}{2}}}{\Gamma({\ssty \frac{d}{2}})} \\
	\Longrightarrow \quad \limepszero 
		\frac{\epsilon} {(-\lambda_\pm)_{t_x=t_y}^{\frac{d}{2}}} \;
	& = \; \frac{2 \pi^{\! \frac{d}{2}}}{\Gamma({\ssty \frac{d}{2}})} \;
		\delta^{(\!d\!-\!1\!)}{\ssty (\vec{x}-\vec{y})}
\end{align}
Looking at (\ref{eq:dvb_ads60}) we see that for $\vec{y} = \vec{x}$ we have
$\bigl( \del_{t_x} \lambda_\pm \! \bigr)_{\! t_x=t_y} \!
			\approx \, \pm i \epsilon (1 \!\! +\!\vec{x}^{\,2})$
and therefore arrive at the desired spatial delta function:
\begin{align}
	\limepszero	\Bigl( \del_{t_x} 
		\phi^{\ssty (\!a\!)}_{\! \Delta_+ \pm}{\ssty \! ( \! \lambda_{\! \pm} \!)}
		\Bigr)_{\! t_x=t_y}
	& = \; \pm i \, \underbrace{\alpha_{2a} \, 
		\frac{2 \pi^{\! \frac{d}{2}}}{\Gamma({\ssty \frac{d}{2}-1})}}_{\beta_{\phi %%@
+}} \;
		\underbrace{(1 \!\! +\!\vec{x}^{\,2})}_{f {\ssty \! (\vec{x})}}
		\; \delta^{(\!d\!-\!1\!)}{\ssty (\vec{x}-\vec{y})}
%%	\limepszero	\Bigl( \del_{t_x}
%%		\phi^{\ssty (\!b\!)}_{\! \Delta \pm} {\ssty \! ( \! \lambda_{\! \pm} \!)}
%%		\Bigr)_{\! t_x=t_y}
%%	& = \; \pm i \, \underbrace{\alpha_{2b} \, 
%%		\frac{2 \pi^{\! \frac{d}{2}}}{\Gamma({\ssty \frac{d}{2}-1})}}_{\beta_b} %%@
\;
%%		(1 \!\! +\!\vec{x}^{\,2}) \; \delta^{(\!d\!-\!1\!)}{\ssty (\vec{x}-\vec{y})} %%@
\;
\end{align}
With this relation fixed we can now identify following section \ref{sec:genver}:
\begin{align}
	\label{eq:dvb_ads67}
	G_{\! \Delta_+}^\pm {\ssty (\!x\!,y\!)} \,
	& = \, \frac{R\ads^{2-d}}{\, 2 \beta_{\phi +}} \;\;
			\limepszero \,
			\phi^{\ssty (\!a\!)}_{\! \Delta_+ \mp}{\ssty \! ( \! \lambda_{\! \mp})} \\
	\notag \\
	\label{eq:dvb_ads68}
	G_{\! \Delta_-}^\pm {\ssty (\!x\!,y\!)} \,
	& = \, \frac{R\ads^{2-d}}{\, 2 \beta_{\phi -}} \;\;
			\limepszero \,
			\phi^{\ssty (\!b\!)}_{\! \Delta_- \mp}{\ssty \! ( \! \lambda_{\! \mp})}
\end{align}
and construct the other propagators according to
(\ref{eq:dvb_ads03}-\ref{eq:dvb_ads08}).
Now we can step back and evaluate equation (\ref{eq:dvb_ads29}):
\begin{align}
	\label{eq:dvb_ads67_5}
	\left( \sigma \Box _x \!\! + m^{\scriptscriptstyle 2} \right)
		G \! _F^{\,\Delta_\pm} {\scriptstyle(\!x\!,y\!)} \;
	& = \; \frac{\delta {\scriptstyle(\!t_x\!-t_y\!)}}{R\ads^2(1+\vec{x}\,^2)} \,
		\bigl[ \partial _{t_x} \! G^+_{\! \Delta_\pm} \!
			- \partial _{t_x} \! G^-_{\! \Delta_\pm} \bigr] _{t_x\!=t_y}
			\\
	& = \; \frac{\delta {\scriptstyle(\!t_x\!-t_y\!)}}{R\ads^2(1+\vec{x}\,^2)} \,
		(-2i) \, \frac{R\ads^{2-d}}{\, 2 \beta_{\phi \pm}} \; \beta_{\phi \pm} \,
		(1 \!\! +\!\vec{x}^{\,2}) \; \delta^{(\!d\!-\!1\!)}{\ssty (\vec{x}-\vec{y})}
		\notag \\
	& = \; \frac{-i}{R\ads^d} \;
		\delta^{\scriptscriptstyle\! (\!d\!)} {\scriptstyle \! (\!x-y\!)}
\end{align}
We see that the Feynman propagator fulfills just the inhomogeneous
Klein-Gordon equation (\ref{eq:dvb_ads30}) with one delta source
on its right hand side:
\begin{align*}
	\left( \sigma \Box _x \!\! + m^{\scriptscriptstyle 2} \right)
	G \! _F^{\,\Delta_\pm} {\scriptstyle(\!x\!,y\!)} \; & = \;\,
	\frac{-i}{\sqrt{g}} \; \delta^{\scriptscriptstyle\! (\!d\!)} {\scriptstyle \! %%@
(\!x-y\!)}
	\; = \; \frac{-i}{R\ads^d} \;
	\delta^{\scriptscriptstyle\! (\!d\!)} {\scriptstyle \! (\!x-y\!)}
	\notag \\
	\left(\Box _x \!\! + \sigma m^{\scriptscriptstyle 2} \right)
	G \! _F^{\,\Delta_\pm} {\scriptstyle(\!x\!,y\!)} \; & = \;\frac{-i %%@
\sigma}{\sqrt{g}} \;
	 \delta^{\scriptscriptstyle\! (\!d\!)} {\scriptstyle \! (\!x-y\!)}
	= \; \frac{-i \sigma}{R\ads^d} \;
		\delta^{\scriptscriptstyle\! (\!d\!)} {\scriptstyle \! (\!x-y\!)}
\end{align*}
After a little cleanup the normalization constants condense to:
\begin{align}
	\beta_{\phi +} & = \, 2^{\Delta_+} \pi^{\! \frac{d-1}{2}} \,
		\frac{\Gamma({\ssty \Delta_+-\frac{d-3}{2}})}{\Gamma({\ssty \Delta_+})} %%@
\\
	\beta_{\phi -} & = \, 2^{-\! \Delta_- + d - \! 1} \pi^{\! \frac{d-1}{2}} \,
	\frac{\Gamma({\ssty -\Delta_-+\frac{d+1}{2}})}{\Gamma({\ssty %%@
-\Delta_-+d-1})}
\end{align}
Of course we do not want the $\Gamma{\ssty (\ldots)}$ functions
in our normalization constants to run to infinity.
One could expect this requirement to lead to restrictions on $\Delta_\pm$,
but these are already fulfilled by the range of validity
assigned to the candidate functions with labels (a) and (b)
in the discussion below equation (\ref{eq:dvb_ads43})
or equivalently by the range of values
naturally assigned to $\Delta_\pm$ by their definition. \\
%%
%%From now on we premise that in the limit of small $\epsilon$ we can use 
%%$F \bigl( {\ssty \ldots \, , \, \frac{1}{\lambda_\pm}} \bigr)
%%  \approx F \bigl( {\ssty \ldots \, , \, \frac{1}{\lambda}}\bigr)$.
%%

				\subsubsection{Comment on the AdS case of the hyperboloid}
				\label{sec:ads_comhyp}
As we can see in appendix \ref{sec:timord_delta},
for the case of the hyperboloid instead of equation (\ref{eq:dvb_ads29})
we obtain (\ref{timord_delta04}):
\begin{align}
	\left( \sigma \Box \ADS _x \!\! + m^{\sssty 2} \right) \, G \! _F {\ssty(\!x\!,y\!)}
	& = \, {\ssty \frac{1}{R\ads^2(1 \! + \! \vec{x}\,^2)}}
			\Bigl( \partial _{t_x} \! G^+ {\scriptstyle(\!x\!,y\!)}
			- \partial _{t_x} \! G^- {\scriptstyle(\!x\!,y\!)} \Bigr)  \\
	& \qqqquad	\qqqquad
		\Bigl[ \delta {\ssty (t_x - t_y)}
			- \delta {\ssty (t_x - t_y-\pi)} - \delta {\ssty (t_x - t_y+\pi)} \Bigr]
		\notag
\end{align}
The $\delta {\ssty (t_x - t_y)}$ term behaves exactly as in the universal covering %%@
case
and leads to the delta source at the coincident point $x=y$.
\\
Turning to the $\delta {\ssty (t_x - t_y \pm \pi)}$ terms
we regard equation (\ref{eq:dvb_ads50}).
We recognize that with $(t_x \! - t_y) = \pm \pi$
we need to insert $-\vec{x} \,$ instead of $\, \vec{x} \,$
in order to obtain the same $\lambda_\pm$
and $(\del_{t_x} \lambda_\pm)$
which we had for $(t_x \! - t_y) = 0$
and which together generated the spatial delta function
as shown in the previous subsection.
Therefore both terms generate a delta source
at the antipodal point $x= \snake{y}$
and instead of (\ref{eq:dvb_ads67_5}) we obtain:
\begin{align}
	\label{eq:dvb_ads_comhyp}
	\left( \sigma \Box _x \!\! + m^{\scriptscriptstyle 2} \right)
		G \! _F^{\,\Delta_\pm} {\scriptstyle(\!x\!,y\!)} \;
	& = \; \frac{-i}{R\ads^d} \,
		\bigl(
		\delta^{\scriptscriptstyle\! (\!d\!)} {\scriptstyle \! (\!x-y\!)}
		- 2 \, \delta^{\scriptscriptstyle\! (\!d\!)} {\scriptstyle \! (\!x- \snake{y})}
		\bigr)
\end{align}
The delta source at the antipodal point
is a direct consequence of the unphysical closed timelike curves
and the thereby induced time ordering prescription.
\\
Unfortunately because of (\ref{eq:dvb_ads_comhyp})
we cannot use the Feynman propagator on the hyperboloid
in order to solve the inhomogeneous Klein-Gordon equation
in the way shown in subsection \ref{sec:solvinhom}.

		\subsection{Listings of the propagators}
			\subsubsection{Propagators in $^1 \! / _\lambda$\,-\,form}
Below we list the  various propagators for AdS$_d$
in the $\frac{1}{\lambda}$ form.
Herein $\lambda_\pm$ is meant as in \eqref{eq:dvb_ads50} 
carrying its full phase (acquired via its position in the complex plane
and the distance winding number $n$)
as discussed around equation (\ref{eq:dvb_ads57}).
Likewise to the candidate functions, we can obtain the $\Delta_-$ propagators
from the $\Delta_+$ propagators via the replacement
$\Delta_+ \! \rightarrow (-\Delta_-\!\!+\!d\!-\!1)$. \\
Having used the method of Dullemond and van Beveren,
our results naturally agree with theirs for the case of $d=4$
which was studied by them.
\begin{align}
	%% ---    G+-   --- 1/Lambda
	G_{\! \Delta_+}^\pm \! {\ssty (\!x\!,y\!)}
	& = \, \frac{R\ads^{2-d}}{\, 2 \beta_{\phi +} } \: \limepszero \:
				 F \bigl( {\ssty \! \frac{\, \Delta_+}{2},\frac{\, \Delta_+\!-d+2}{2}, \,
			  			\Delta_+\!-\frac{d-3}{2}, \, \frac{1}{\lambda_\mp}}\bigr) \, 
	\biggl( \! \frac{1}{-\lambda_\mp} \! \biggr)^{\!\! \frac{\, \Delta_+}{2}} \\
	G_{\! \Delta_-}^\pm \! {\ssty (\!x\!,y\!)}
	& = \,\frac{R\ads^{2-d}}{\, 2 \beta_{\phi -}} \: \limepszero \:
			F \bigl( {\ssty \! \frac{-\Delta_-\!+d-1}{2},\frac{\,1-\Delta_-}{2}, \,
			  			-\Delta_-\!+\frac{d+1}{2}, \, \frac{1}{\lambda_\mp}}\bigr) \,
	\biggl( \! \frac{1}{-\lambda_\mp} \! \biggr)
	^{\!\!\! \frac{-\Delta_-\!+d-\!1}{2}}
	\\ \notag
\end{align}
\begin{align}
	%% --- G^(1) 1/Lambda
	\label{eq:dvb_ads72}
	G_{\! \Delta_+}^{\sssty (\!1\!)} \! {\ssty (\!x\!,y\!)} \, 
	& = \, \frac{R\ads^{2-d}}{\, \beta_{\phi +}} \: \limepszero \: \text{Re} \:
			 F \bigl( {\ssty \! \frac{\,\Delta_+}{2},\frac{\,\Delta_+\!-d+2}{2}, \,
			  			\Delta_+\!-\frac{d-3}{2}, \, \frac{1}{\lambda_-}}\bigr) \;
			\biggl( \! \frac{1}{-\lambda_-} \!  \biggr)^{\!\! \frac{\,\Delta_+}{2}} \\
	%%
%%	G_{\! \Delta (\!b\!)}^{\sssty (\!1\!)} {\ssty (\!x\!,y\!)} \, 
%%	& = \,\frac{R\ads^{2-d}}{\, \beta_{\phi -}} \;
%%			F \bigl( {\ssty \! \frac{-\Delta+d-1}{2},\frac{1-\Delta}{2}, \,
%%			  			-\Delta+\frac{d+1}{2}, \, \frac{1}{\lambda_-}}\bigr) \;
%%			\text{Re} \;	\limepszero 
%%			\biggl( \! \frac{1}{-\lambda_-} \!  \biggr)^{\!\!\! \frac{-\Delta+d-1}{2}} 
%%\end{align}
%%
%%\begin{align}
	%% --- G ---- 1/Lambda
	\label{eq:dvb_ads74}
	G_{\! \Delta_+} \! {\ssty (\!x\!,y\!)} \, 
	& = \, \frac{R\ads^{2-d}}{\, \beta_{\phi +} } \: \limepszero \: i \, \text{Im} \,
			 F \bigl( {\ssty \! \frac{\,\Delta_+}{2},\frac{\,\Delta_+\!-d+2}{2}, \,
			  			\Delta_+\!-\frac{d-3}{2}, \, \frac{1}{\lambda_-}}\bigr) \;
			\biggl( \! \frac{1}{-\lambda_-} \!  \biggr)^{\!\! \frac{\,\Delta_+}{2}}
	%%
%%	G_{\! \Delta (\!b\!)} {\ssty (\!x\!,y\!)} \, 
%%	& = \,\frac{R\ads^{2-d}}{\, \beta_{\phi -}} \;
%%			F \bigl( {\ssty \! \frac{-\Delta+d-1}{2},\frac{1-\Delta}{2}, \,
%%			  			-\Delta+\frac{d+1}{2}, \, \frac{1}{\lambda_-}}\bigr) \;
%%	 i \, \text{Im} \;  \limepszero 
%%			\biggl( \! \frac{1}{-\lambda_-} \!  \biggr)^{\!\!\! \frac{-\Delta+d-1}{2}} 
%%	\\ \notag
\end{align}
\begin{align}
%% --- G_R ---- 1/Lambda
	G^{ \Delta_+}_{\!R} \! {\ssty (\!x\!,y\!)}
	& = \, \theta {\ssty (\!t_x\!-t_y\!)} \,
			\frac{R\ads^{2-d}}{\, \beta_{\phi +} } \, \limepszero i \, \text{Im} \,
			 F \bigl( {\ssty \! \frac{\,\Delta_+}{2},\frac{\,\Delta_+\!-d+2}{2}, \,
			  			\Delta_+\!-\frac{d-3}{2}, \, \frac{1}{\lambda_-}}\bigr)
			\biggl( \! \frac{1}{-\lambda_-} \!  \biggr)^{\!\!\! \frac{\,\Delta_+}{2}} \\
	%%
%%	G^{ \Delta (\!b\!)}_{\!R} {\ssty (\!x\!,y\!)}
%%	& = \, \theta {\ssty (\!t_x\!-t_y\!)} \,
%%			\frac{R\ads^{2-d}}{\, \beta_{\phi -} } \,
%%			F \bigl( {\ssty \! \frac{-\Delta+d-1}{2},\frac{1-\Delta}{2}, \,
%%			  			-\Delta+\frac{d+1}{2}, \, \frac{1}{\lambda_-}}\bigr) \,
%%			i \, \text{Im} \;  \limepszero
%%				\biggl( \! \frac{1}{-\lambda_-} \!  \biggr)^{\!\!\!
%% 		 \frac{-\Delta+d-1}{2}}
%%\end{align}
%%
%%\begin{align}
%% --- G_A ---- 1/Lambda
	G^{ \Delta_+}_{\!A} \! {\ssty (\!x\!,y\!)}
	& = -\theta {\ssty (\!t_y\!-t_x\!)}
			\frac{R\ads^{2-d}}{\, \beta_{\phi +}  } \, \limepszero i \, \text{Im} \,
			 F \bigl( {\ssty \! \frac{\,\Delta_+}{2},\frac{\,\Delta_+\!-d+2}{2}, \,
			  			\Delta_+\!-\frac{d-3}{2}, \, \frac{1}{\lambda_-}}\bigr) \!
			\biggl( \! \frac{1}{-\lambda_-} \!  \biggr)^{\!\!\! \frac{\,\Delta_+}{2}}
	%%
%%	G^{ \Delta (\!b\!)}_{\!A} {\ssty (\!x\!,y\!)} \, 
%%	& = \, -\theta {\ssty (\!t_y\!-t_x\!)} \,
%%			\frac{R\ads^{2-d}}{\, \beta_{\phi -} } \;
%%			F \bigl( {\ssty \! \frac{-\Delta+d-1}{2},\frac{1-\Delta}{2}, \,
%%			  			-\Delta+\frac{d+1}{2}, \, \frac{1}{\lambda_-}}\bigr) \;
%%		i \, \text{Im} \;\limepszero 
%%			\biggl( \! \frac{1}{-\lambda_-} \!  \biggr)^{\!\!\! \frac{-\Delta+d-1}{2}} 
\end{align}
\begin{align}
	%% ---    G_F   --- 1/Lambda
	G^{\,\Delta_+}_{\! F} \! {\ssty (\!x\!,y\!)} \, 
	& = \, \frac{R\ads^{2-d}}{\, 2 \beta_{\phi +}  } \: \limepszero \:
			 F \bigl( {\ssty \! \frac{\,\Delta_+}{2},\frac{\,\Delta_+\!-d+2}{2}, \,
			  			\Delta_+\!-\frac{d-3}{2}, \, \frac{1}{\lambda_F}}\bigr) \,
	\biggl( \! \frac{1}{-\lambda_F} \! \biggr)^{\!\! \frac{\,\Delta_+}{2}} \\
	G^{\,\Delta_-}_{\! F} \! {\ssty (\!x\!,y\!)} \, 
	& = \, \frac{R\ads^{2-d}}{\, 2 \beta_{\phi -} } \: \limepszero \:
			F \bigl( {\ssty \! \frac{-\Delta_-\!+d-1}{2},\frac{\,1-\Delta_-}{2}, \,
			  			-\Delta_-\!+\frac{d+1}{2}, \, \frac{1}{\lambda_F}}\bigr) \, 
	\biggl( \! \frac{1}{-\lambda_F} \! \biggr)^{\!\!\! \frac{-\Delta_-\!+d-1}{2}}
\end{align}
\\
Therein $\lambda_F$ (with index F for Feynman)
is defined for the limit of small $\epsilon$ as
\begin{align}
	\lambda_F \!
	& \approx \, \underbrace{1 - \Bigl[ \!
			\sqrt{1 \!\! +\! \vec{x}^{\,2}}\sqrt{1 \!\!	+ \!\vec{y}^{\,2}} 
			\cos \, (t_x\!\!-\!t_y) - \vec{x} \vec{y} \, \Bigr]^{2} }_\lambda
			\\
			& \qquad - {\ssty \frac{\epsilon^2}{4}} \Bigl[ \!(1 \!\! +\!\vec{x}^{\,2})
			(1 \!\! +\!\vec{y}^{\,2}) \cos 2 (t_x\!\!-\!t_y) -
			\vec{x} \vec{y} \,\sqrt{1 \!\! +\! \vec{x}^{\,2}}\sqrt{1 \!\! +
			 \!\vec{y}^{\,2}} \, 
			\cos \, (t_x\!\!-\!t_y) \Bigr]
			\notag \\
			& \qquad - i \epsilon \ub{
			\sqrt{1 \!\! + \! \vec{x}^{\,2}} \sqrt{1 \!\! + \!\vec{y}^{\,2}}
			\sin \! \mid \! t_x\!\!-\!t_y \!\! \mid \!}_{S_F} 
			\ub{ \Bigl[ \sqrt{1 \!\! + \! \vec{x}^{\,2}} \sqrt{1 \!\! + \!\vec{y}^{\,2}}
					 \cos (t_x\!\!-\!t_y) -	\vec{x} \vec{y} \, \Bigr]}_\gamma
			\notag
\end{align}
so that the Feynman propagator is indeed symmetric:
$\, G_{\! F}^{\,\Delta_\pm} \! {\ssty (\!x\!,y\!)}
= G_{\! F}^{\,\Delta_\pm} \! {\ssty (\!y\!,x\!)}$.
The absolute values in the argument of the sine
arise in the course of the calculation as in equation (\ref{eq:dvb13}). 
\\
We observe that the $i \epsilon$-term in the Feynman propagator
in AdS spacetime is no longer time independent
as it was the case in Minkowski spacetime.
\begin{figure} [H]
	\begin{center}
	\igx[height = 15cm]{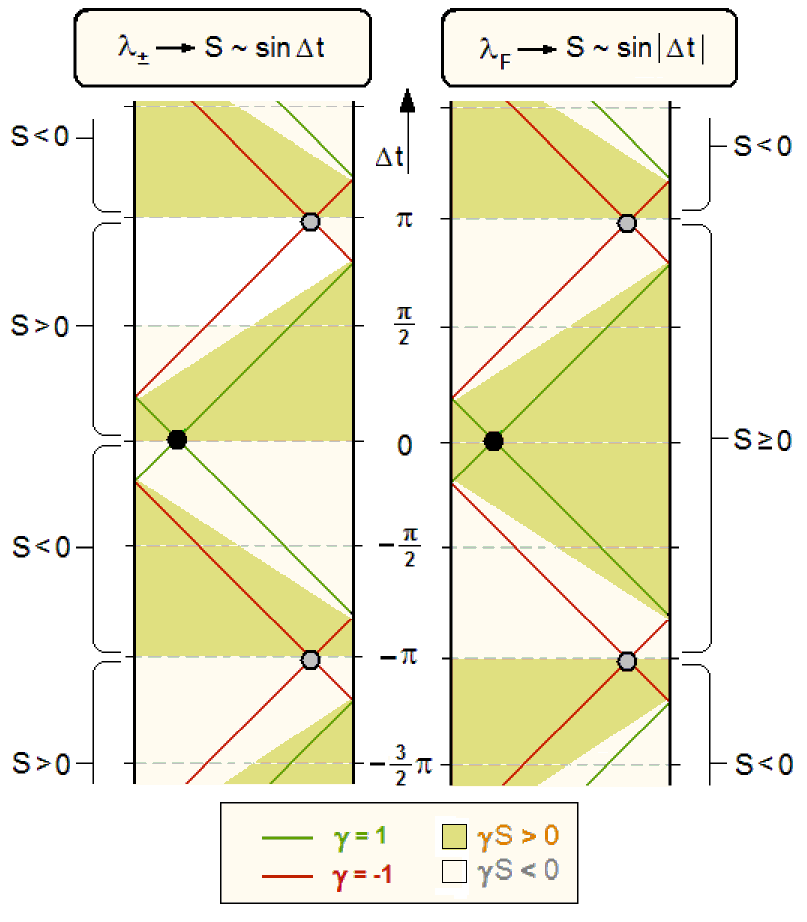}
	\end{center}
	\caption{Penrose diagram of the universal covering AdS}
	\label{fig:penrose_ads_gamma_s}
\end{figure}
Figure \ref{fig:penrose_ads_gamma_s} shows the sign of the
$i\epsilon$-term for $\lambda_\pm$ on the left
and for $\lambda_F$ on the right hand side
in a Penrose diagram of universal covering AdS
for an arbitrary reference point.
The green lightcones of points which 
are mapped to the black reference point
via \eqref{ads_coord05} have $\gamma = 1$
while the red lightcones of points 
which are mapped to its antipodal point
have $\gamma = -1$.
Orange wedges are regions with a positive product $\gamma S$
and white ones with negative $\gamma S$. 
\\
We observe that for $\lambda_\pm$ which is used
in the Wightman functions (homogeneous propagators)
the sign of the $i\epsilon$-term is changing in each of such points.
Applying the Klein-Gordon operator $(\sigma\Box\ADS_x + m^2)$
to these functions yields zero and thus no delta source is generated.
\\
In contrast for $\lambda_F$ used in the Feynman propagator 
the sign of the $i\epsilon$-term is not changing
in the coincident point.
This effect is caused by the step functions of time ordering. 
Letting the Klein-Gordon operator act on the Feynman propagator
generates a delta source at the coincident point.
\\
Figure \ref{fig:penrose_ads_gamma_s} clarifies
that on each single lightcone the sign of the $i\epsilon$-term
depends only on $S$ and in this sense is time dependent only.
If however we consider points at a fixed time $t_{\text{fix}}$
then the sign of the $i\epsilon$-term changes
when moving in space from one lightcone to another one.
The sign of the $i\epsilon$-term
then is determined by both $\gamma$ and $S$
and in this sense is also position depending. 
For further remarks we repeat the definition of $\lambda_\pm$:
\beqs
	\lambda_\pm \, = \, \underbrace{1 - \gamma^2}_{\lambda}
			 - \, {\ssty \frac{\epsilon^2}{4}}{ \ssty (\ldots)}
			\pm i \epsilon \, \gamma S
\eeqs
For the propagators above the hypergeometric function
is defined by its convergent Taylor series
for all $\lambda$ with $\mid \! \lambda \! \mid \, \geq 1$.
The form of the propagators for $\mid \! \lambda \! \mid \, \leq 1$
can be obtained therefrom using (\ref{eq:hypergeo07}). \\
All hypergeometric functions $F{\ssty(A,B,C,z)}$ involved
also converge for $\mid \! z \! \mid \, = 1$ because
Re $(C\!-\!A\!-\!B) = \, \frac{1}{2} \, > 0$. \\
However in all cases $\mid \! \lambda \! \mid {\ssty \gtreqless} \, 1$
for $\Delta_+$ and $\Delta_-$ there appears the term $(1 \! / \lambda_\pm)$
to positive powers only.
In the limit of small $\epsilon$ this causes the propagators 
to become infinitely large whenever it occurs that
$\gamma \equiv \sqrt{1 \! +\vec{x}^{\,2}}\sqrt{1 \! +\vec{y}^{\,2}}
   \cos \, (t_x\!-\!t_y) - \vec{x} \vec{y} \, = \pm 1$.
For constant $\vec{x} , \vec{y} , t_y$ this happens periodically in $t_x$.
As we will see later, this happens every time
the point $x$ is situated on the lightcones of certain points $y'{\ssty (\!y\!)}$.
There are no conspicuities when $\gamma = 0 \, $ or 
$\gamma \rightarrow \pm \infty$. \\
%%
%%	\phi^{\ssty (\!a\!)}_{\! \Delta_+ \pm}  {\ssty \! (\lambda_{\! \pm}
%%			e^{\mp 2 \pi i n}\!)} \, 
%%	& = \, e^{\pm i \pi n \Delta_+} \,
%%			\phi^{\ssty (\!a\!)}_{\! \Delta_+ \pm}  {\ssty \! ( \!\lambda_{\! \pm} \!)}
%%
From (\ref{eq:dvb_ads55}) %%and (\ref{eq:dvb_ads56})
we can extract the additional relations:
\begin{align}
	\label{eq:dvb_ads811}
	G_{\! \Delta_+}^\pm {\ssty (\!\lambda_\mp e^{\pm 2 \pi i n}\!)} \,
	& = \, e^{\mp i \pi n \Delta_+}
			\, G_{\! \Delta_+}^\pm {\ssty (\!\lambda_\mp\!)}
		\qqqquad \qqqquad {\ssty \lambda \gg 1}
		\\
	\label{eq:dvb_ads812}
	G_{\! \Delta_-}^\pm {\ssty (\!\lambda_\mp e^{\pm 2 \pi i n}\!)} \,
	& = \, e^{\mp i \pi n (-\Delta_-\!+d-\!1)}
			\, G_{\! \Delta_-}^\pm {\ssty (\!\lambda_\mp\!)}
		\qqqquad \quad {\ssty \lambda \gg 1}
%%		\\
%%	\notag \\
%%	\label{eq:dvb_ads813}
%%	\rightarrow \quad G^{\Delta_+}_{\! F} {\ssty (\!n,\lambda_F\!)} \,
%%	& = \, e^{-i \pi \mid n \mid \Delta_+}
%%			\, G^{\Delta_+}_{\! F} {\ssty (\!n=0,\lambda_F\!)} \\
%%	\label{eq:dvb_ads814}
%%	\rightarrow \quad G^{\Delta_-}_{\! F} {\ssty (\!n,\lambda_F\!)} \,
%%	& = \, e^{-i \pi \mid n \mid (-\Delta_-\!+d-\!1)}
%%			\, G^{\Delta_-}_{\! F} {\ssty (\!n=0,\lambda_F\!)}
\end{align}

			\subsubsection{Propagators in $\xi$\,-\,form}
Now it is time to remember that long long ago we had found 
two equivalent ways of splitting up our one candidate function
using either $1/ \lambda$ or $\xi = 1/(1 \! - \! \lambda)$.
Therefore an equivalent form of writing the propagator functions exists. 
Using the variable $\xi_\pm = \frac{1}{1-\lambda_\pm}$
the propagators for AdS$_d$ read:
\begin{align}
	%% ---    G+-   --- Xi
	G_{\! \Delta_+}^\pm \! {\ssty (\!x\!,y\!)} \, 
	& = \, \frac{R\ads^{2-d}}{\, 2 \beta_{\phi +}  } \, \limepszero \,
			 F \bigl( {\ssty \! \frac{\,\Delta_+}{2},\frac{\,\Delta_+\!+1}{2}, \,
			  			\Delta_+\!-\frac{d-3}{2}, \, \xi_\mp}\bigr) \; 
			 \xi_\mp^{\Delta_+ \! /2} \; \\
	G_{\! \Delta_-}^\pm \! {\ssty (\!x\!,y\!)} \, 
	& = \,\frac{R\ads^{2-d}}{\, 2 \beta_{\phi -} } \, \limepszero
			F \bigl( {\ssty \! \frac{-\Delta_-\!+d-1}{2},\frac{\,d-\Delta_-}{2}, \,
			  			-\Delta_-\!+\frac{d+1}{2}, \, \xi_\mp} \bigr) \;
			\, \xi_\mp^{( \! -\Delta_-\!+d-1)/2}
	\\ \notag
\end{align}
\begin{align}
	%% --- G^(1) Xi
	G_{\! \Delta_+}^{\sssty (\!1\!)} \! {\ssty (\!x\!,y\!)} \, 
	& = \, \frac{R\ads^{2-d}}{\, \beta_{\phi +}  } \, \text{Re} \, \limepszero \, 
			 F \bigl( {\ssty \! \frac{\,\Delta_+}{2},\frac{\,\Delta_+\!+1}{2}, \,
			  			\Delta_+\!-\frac{d-3}{2}, \, \xi_-}\bigr) \,
			\xi_-^{\Delta_+ \! /2} \\
	%%
%%	G_{\! \Delta (\!b\!)}^{\sssty (\!1\!)} {\ssty (\!x\!,y\!)} \, 
%%	& = \,\frac{R\ads^{2-d}}{\, \beta_{\phi -} } \;
%%			F \bigl( {\ssty \! \frac{-\Delta+d-1}{2},\frac{d-\Delta}{2}, \,
%%			  			-\Delta+\frac{d+1}{2}, \, \xi_\-}\bigr) \;
%%			\text{Re} \, \limepszero \, \xi_-^{-\Delta+d-1}
%%	\\ \notag
%%\end{align}
%%
%%\begin{align}
	%% --- G ---- Xi
	G_{\! \Delta_+} \! {\ssty (\!x\!,y\!)} \, 
	& = \, \frac{R\ads^{2-d}}{\, \beta_{\phi +}  } \, i \, \text{Im} \,
			 F \bigl( {\ssty \! \frac{\,\Delta_+}{2},\frac{\,\Delta_+\!+1}{2}, \,
			  			\Delta_+\!-\frac{d-3}{2}, \, \xi_-}\bigr) \,
			\limepszero \, \xi_-^{\Delta_+ \! /2}
	%%
%%	G_{\! \Delta (\!b\!)} {\ssty (\!x\!,y\!)} \, 
%%	& = \,\frac{R\ads^{2-d}}{\, \beta_{\phi -} } \;
%%			F \bigl( {\ssty \! \frac{-\Delta+d-1}{2},\frac{d-\Delta}{2}, \,
%%			  			-\Delta+\frac{d+1}{2}, \, \xi_\-}\bigr) \;
%%	i \, \text{Im} \, \limepszero \, \xi_-^{-\Delta+d-1}
	\\ \notag
\end{align}
\begin{align}
%% --- G_R ---- Xi
	G^{ \Delta_+}_{\!R} \! {\ssty (\!x\!,y\!)} \, 
	& = \, \theta {\ssty (\!t_x\!-t_y\!)} \,
			\frac{R\ads^{2-d}}{\, \beta_{\phi +}  } \, i \, \text{Im} \, \limepszero \,
			 F \bigl( {\ssty \! \frac{\,\Delta_+}{2},\frac{\,\Delta_+\!+1}{2}, \,
			  			\Delta_+\!-\frac{d-3}{2}, \, \xi_-}\bigr) \,
			\xi_-^{\Delta_+ \! /2} \, \\
	%%
%%	G^{ \Delta (\!b\!)}_{\!R} {\ssty (\!x\!,y\!)} \, 
%%	& = \, \theta {\ssty (\!t_x\!-t_y\!)} \,
%%			\frac{R\ads^{2-d}}{\, \beta_{\phi -} } \;
%%			F \bigl( {\ssty \! \frac{-\Delta+d-1}{2},\frac{d-\Delta}{2}, \,
%%			  			-\Delta+\frac{d+1}{2}, \, \xi_\-}\bigr) \;
%%	i \, \text{Im} \, \limepszero \, \xi_-^{-\Delta+d-1}
%%	\\ \notag
%%\end{align}
%%
%%\begin{align}
%% --- G_A ---- Xi
	G^{ \Delta_+}_{\!A} \! {\ssty (\!x\!,y\!)} \, 
	& = -\theta {\ssty (\!t_y\!-t_x\!)} \,
			\frac{R\ads^{2-d}}{\, \beta_{\phi +}  } \, i \, \text{Im} \, \limepszero \,
			 F \bigl( {\ssty \! \frac{\,\Delta_+}{2},\frac{\,\Delta_+\!+1}{2}, \,
			  			\Delta_+\!-\frac{d-3}{2}, \, \xi_-}\bigr) \,
			\xi_-^{\Delta_+ \! /2}
	%%
%%	G^{ \Delta (\!b\!)}_{\!A} {\ssty (\!x\!,y\!)} \, 
%%	& = \, -\theta {\ssty (\!t_y\!-t_x\!)} \,
%%			\frac{R\ads^{2-d}}{\, \beta_{\phi -} } \;
%%			F \bigl( {\ssty \! \frac{-\Delta+d-1}{2},\frac{d-\Delta}{2}, \,
%%			  			-\Delta+\frac{d+1}{2}, \, \xi_\-}\bigr) \;
%%	i \, \text{Im} \, \limepszero \, \xi_-^{-\Delta+d-1}
	\\ \notag
\end{align}
\begin{align}
	%% ---    G_F   --- Xi
	\label{eq:dvb_ads92} 
	G^{ \Delta_+}_{\! F} \! {\ssty (\!x\!,y\!)} \, 
	& = \, \frac{R\ads^{2-d}}{\, 2 \beta_{\phi +}  } \, \limepszero \,
			 F \bigl( {\ssty \! \frac{\,\Delta_+}{2},\frac{\,\Delta_+\!+1}{2}, \,
			  			\Delta_+\!-\frac{d-3}{2}, \, \xi_F } \bigr) \,
			\xi_F^{\Delta_+ \! /2} \\
	G^{ \Delta_-}_{\! F} \! {\ssty (\!x\!,y\!)} \, 
	& = \, \frac{R\ads^{2-d}}{\, 2 \beta_{\phi -} } \, \limepszero \,
			F \bigl( {\ssty \! \frac{-\Delta_-+d-1}{2},\frac{d-\Delta_-}{2}, \,
			  			-\Delta_-\!+\frac{d+1}{2}, \, \xi_F } \bigr) \, 
			\xi_F^{( \! -\Delta_-\!+d-1)/2}
\end{align}
Again we obtain the $\Delta_-$ propagators
from the $\Delta_+$ propagators through the replacement
$\Delta_+ \! \rightarrow (-\Delta_-\!\!+\!d\!-\!1)$. \\
In the $\xi$-form the hypergeometric functions are of the type
$F ( {\ssty \! k, \, l=k+\frac{1}{2}, \,
			  			k+l-\frac{d-2}{2}, \, z} )$
and thus for even AdS-dimension $d$ can be chopped up
using (15.3.12) on page 560 in Abramowitz, Stegun \cite{astegun}
as it is done by Dullemond and van Beveren in \cite{dvb}.
For the Feynman propagator we have used
\begin{align*}
	\xi_F \, 
	& = \, \frac{1}{1-\lambda_F} \\
	& = \, \biggl[ \!(1 \!\! +\!\vec{x}^{\,2})(1 \!\! +\!\vec{y}^{\,2})
			 \cos ^2  ( t_x\!\!-\!t_y ) 
			 + (\vec{x} \vec{y})^2  - 2 \vec{x} \vec{y} \,
		    \sqrt{1 \!\! +\! \vec{x}^{\,2}}\sqrt{1 \!\! + \!\vec{y}^{\,2}} \, 
			\cos (t_x\!\!-\!t_y ) \notag \\
			& \qquad \;
			+{\ssty \frac{\epsilon^2}{4}} \Bigl[ \!(1 \!\! +\!\vec{x}^{\,2})
			(1 \!\! +\!\vec{y}^{\,2}) \cos 2 (t_x\!\!-\!t_y ) -
			\vec{x} \vec{y} \,\sqrt{1 \!\! +\! \vec{x}^{\,2}}\sqrt{1 \!\! + \!\vec{y}^{\,2}} \, 
			\cos ( t_x\!\!-\!t_y ) \Bigr] \notag \\
			& \qquad \;
			+i {\ssty \frac{\epsilon}{2}} \Bigl[ \!(1 \!\! +\!\vec{x}^{\,2})
			(1 \!\! +\!\vec{y}^{\,2}) \sin 2 \mid \! t_x\!\!-\!t_y \!\! \mid - 2
			\vec{x} \vec{y} \,\sqrt{1 \!\! +\! \vec{x}^{\,2}}\sqrt{1 \!\! + \!\vec{y}^{\,2}} \, 
			\sin \mid \! t_x\!\!-\!t_y \!\! \mid \Bigr] \biggr]^{-1}
\end{align*}
For the propagators above the hypergeometric function
is defined by its convergent Taylor series
for all $\xi$ with $\mid \! \xi \! \mid \, < 1$.
The form of the propagators for $\mid \! \xi \! \mid \, > 1$
can be obtained again using (\ref{eq:hypergeo07}).
For further remarks we remember the definition of $\xi_\pm$:
\beqs
	\xi_\pm 
	= \biggl[ \gamma^2 + {\ssty \frac{\epsilon^2}{4}}({ \ssty \ldots})
			\mp i \epsilon \, \gamma S \,\biggr]^{-1}
\eeqs
For $\mid \! \xi \! \mid \, < 1$
we find that $\xi$ appears in positive power 
which implies that then $\mid \! \gamma \! \mid \, > 1$
has negative power. Hence there is no problem
when $\gamma \rightarrow \pm \infty$
because the propagators simply vanish. \\
On the other hand in the form for  $\mid \! \xi \! \mid \, > 1$
$\xi$ has negative power and therefore $\mid \! \gamma \! \mid \, < 1$
appears with positive power.
Thus whenever $\gamma = 0$
then the propagators vanish in the limit of small $\epsilon$
even though the chordal distance can be finite.
\\
All hypergeometric functions $F{\ssty(A,B,C,z)}$
involved in the $\xi$ forms for $d \geq 4$
diverge for $\mid \! z \! \mid \, = 1$ because we have 
Re $(C\!-\!A\!-\!B) = \, \frac{2-d}{2} \, \leq -1$.
\\ 
The Feynman propagator (\ref{eq:dvb_ads92})
is in agreement with the one found by
Burgess and Lutken in \cite{burgess} that is also cited 
by d'Hoker and Freedman in \cite{arx0201253}
(wherein our $\xi$ is denoted as $\xi^2$).
Finally we give a plot of the unnormalized Feynman propagator in $\xi$-form
for the same parameters as for figure \ref{fig:hypergeo_ads_summe}.
\begin{figure} [H]
	\begin{center}
	\igx[width=5cm]{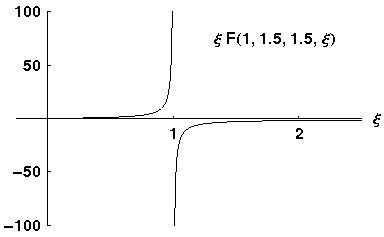}
	\end{center}
	\caption{Feynman propagator in $\xi$-form}
	\label{fig:hypergeo_ads_xi}
\end{figure}

			\subsubsection{Singularities of the propagators}
We conclude with a remark concerning antipodal points.
Using equations (\ref{eq:ads_chordal03}) and (\ref{eq:ads_chordal04})
we can write $\lambda$ and $\xi$
using both chordal and antipodal chordal distance:
%%
%% -------------------- equation dvb_ads94--------------------------------------
\begin{align}
	\quad \lambda {\ssty (\!x\!,y\!)} \,
	& = \, \frac{ \, u {\ssty (\!x\!,y\!)} \, \snake{u} {\ssty (\!x\!,y\!)} }{4R^4\ads}
	= \, \frac{ \, u {\ssty (\!x\!,y\!)} \, u {\ssty (\!x\!, \snake{y} )} }{4R^4\ads}
	& \rightarrow \qquad
	\lambda{\ssty (\!x\!,y\!)} & = \,\lambda{\ssty (\!x\!,\snake{y})} \qquad \\
	\quad \xi{\ssty (\!x\!,y\!)} \, & = \, \frac{1}{1-\frac{u \, \snake{u}}{\, 4R^4\ads}}
	& \rightarrow \qquad
	\xi{\ssty (\!x\!,y\!)} & = \, \xi{\ssty (\!x\!,\snake{y})} \qquad
	\\
	& & {\ssty \text{with } \; u (\!x\!,y\!) \;}
		& {\ssty = \, u \, ( \! X} {\sssty \! (x)} {\ssty , Y} {\sssty \! (y)} {\ssty )} 
	\notag
\end{align}
We recognize that likewise to Minkowski spacetime
the propagators are singular  on the whole lightcone
$u {\ssty (\!x\!,y\!)}=0$ of the reference point $y$
and in addition also on the lightcone $u {\ssty (\!x\!, \snake{y} )}=0$
of the antipodal point $\snake{y}$.
Moreover this periodically happens on the lightcones
$u {\ssty (\!x\!,y')}=0$ of all points $y'$ satisfying
either $ Y \! {\ssty (y')} = Y  \! {\ssty (y)} $ 
or $ Y \! {\ssty (y')} = Y  \! {\ssty (\snake{y})}$
when wrapping the universal covering AdS
again and again around the hyperboloid.
On all these lightcones we have $\lambda=0$ and $\xi=1$.
\\
However, as we had already seen above
and also likewise to Minkowski spacetime,
this solely generates one single delta source 
sitting at the coincidence point $x=y$  
when letting the Klein-Gordon operator act on the Feynman propagator.
This is due to the standard time ordering via $\theta {\ssty (t_x - t_y)}$
on the universal covering AdS.
%%

%%
\begin{comment}
	\section{Propagators for AdS x S  spacetime}
	\label{sec:props_ads_x_s}
		\subsection{General properties of propagators in AdS x S}
		\label{ads_x_s_genprops}
		\input{dvb_ads_x_s_genprop}
		\subsection{From Klein-Gordon to the hypergeometric equation}
		\input{dvb_ads_x_s_kghyper}
		\subsection{Inspection of the candidate functions}
		\input{dvb_ads_x_s_excand}
				\subsubsection{Infinitesimal shift in complex time plane}
				\input{dvb_ads_x_s_infshift}
				\subsubsection{Distance winding number}
				\input{dvb_ads_x_s_windnumb}
				\subsubsection{Delta source and normalization}
				\input{dvb_ads_x_s_delnorm}
		\subsection{Listings of the propagators}
			\subsubsection{Propagators in $^1 \! / _\lambda$\,-\,form}
			\input{dvb_ads_x_s_proplist_lambda}
			\subsubsection{Singularities of the propagators}
			\input{dvb_ads_x_s_proplist_sing}
\end{comment}
%%
	\cleardoublepage
	\section{Propagators for dS spacetime}
	\label{sec:props_ds}
		\subsection{General properties of propagators in dS}
In this section we apply the method of Dullemond and van Beveren
to deSitter spacetime.
The definitions and relations between the different propagators
for hermitian scalar fields $\phi {\scriptstyle(\!x\!)}$ in dS
are the same as the ones listed for AdS in section \ref{ads_genprops}.
Here $x=(\tau,\vec{\phii})$ is a point in dS in global coordinates
with $\tau \equiv x^0$ serving as time coordinate.
With the overall sign $\sigma = \pm 1$ of the embedding space's metric
introduced in equation (\ref{eq:ds_coord01}),
the Klein-Gordon operator on dS reads $(\sigma \Box \ds + m^2)$. \\
Because our field $\phi {\scriptstyle(\!x\!)}$ again fulfills
the Klein-Gordon equation, we have
\begin{align}
	\left( \sigma \Box \DS _x \!\! + m^{\scriptscriptstyle 2} \right)
		G^\pm {\scriptstyle(\!x\!,y\!)} = \;\; & 0 \, = \,
		\left( \Box \DS _x \!\! + \sigma m^{\scriptscriptstyle 2} \right)
		G^\pm {\scriptstyle(\!x\!,y\!)} \\
	\rightarrow \quad \left( \sigma \Box \DS _x \!\! + m^{\scriptscriptstyle 2} %%@
\right)
		G^{\scriptscriptstyle(\!1\!)} {\scriptstyle(\!x\!,y\!)} = \;\; & 0 \, = \, 
		\left( \Box \DS _x \!\! +\sigma  m^{\scriptscriptstyle 2} \right)
		G^{\scriptscriptstyle(\!1\!)} {\scriptstyle(\!x\!,y\!)} \\
	\rightarrow \quad \left( \sigma \Box \DS _x \!\! + m^{\scriptscriptstyle 2} %%@
\right) \; 
		G {\scriptstyle(\!x\!,y\!)} \;\, = \;\; & 0 \, = \,
		\left( \Box \DS _x \!\! + \sigma m^{\scriptscriptstyle 2} \right) \; 
		G {\scriptstyle(\!x\!,y\!)}
\end{align}
and
\begin{equation}
	\left( \sigma \Box \DS _x \!\! + m^{\scriptscriptstyle 2} \right)
	G \! _F {\scriptstyle(\!x\!,y\!)}
	\; = \; \left( \sigma \Box \DS _x \!\! + m^{\scriptscriptstyle 2} \right)
	G \! _R {\scriptstyle(\!x\!,y\!)}
	\; = \; \left( \sigma \Box \DS _x \!\! + m^{\scriptscriptstyle 2} \right)
	G \! _A {\scriptstyle(\!x\!,y\!)}	
\end{equation}
Starting from the definition (\ref{eq:dvb_ads08})
of the Feynman propagator 
and using equation (\ref{eq:ds_metrics30})
\beq
	\sigma \Box\ds \, = \, \frac{1}{R^2 \ds} \, \bigl( \del^2_\tau
									+ (d \!-\!1) \tanh \tau \, \del_\tau \bigr) + \,
										\sigma \Box_{\vec{\phii}}
\eeq
in combination with the equal time commutation relation
$\; \left[ \phi {\scriptstyle(\!x\!)}
,\phi {\scriptstyle(\!y\!)}\right] _{\, \tau_x=\tau_y} = \, 0 \;$
we obtain (see appendix \ref{sec:timord_delta}):
\begin{align}
	\label{eq:dvb_ds29}
	\left( \sigma \Box \DS _x \!\! + m^{\scriptscriptstyle 2} \right)
		G \! _F {\scriptstyle(\!x\!,y\!)} \;
	& = \; \frac{\, \delta {\scriptstyle(\!\tau_x\!-\tau_y\!)}}
		{\ub{R\ds^2}_{\rightarrow \beta_G = R\ds^{-2}}} \,
		\bigl[ \partial _{\tau_x} \! G^+ \! - \partial _{\tau_x} \! G^- \bigr] %%@
_{\tau_x\!=\tau_y}
\end{align}
This expression contributes a delta function for the time variables
of the spacetime points $x$ and $y$.
(Thus the function $f{\ssty(\vec{x})}$ respectively $f{\ssty(\vec{\phii})}$
 introduced in subsection \ref{sec:genver}
 is identically one for the global coordinates.)
\\
Defining the various propagators as in equations
(\ref{eq:dvb_ads01})-(\ref{eq:dvb_ads08})
corresponds to define as Feynman propagator the function
fulfilling the inhomogeneous Klein-Gordon equation
%%
%%_____________________equation_dvb_ds_30___________________
\begin{align}
	\label{eq:dvb_ds29_5}
	\left( \sigma \Box \DS _x \!\! + m^{\scriptscriptstyle 2} \right)
		G \! _F {\scriptstyle(\!x\!,y\!)} \;
	& = \;	\frac{-i}{\sqrt{g}} \;
		\delta^{\scriptscriptstyle\! (\!d)} {\scriptstyle \! (\!x-y\!)}
	\\
	\left(\Box \DS _x \!\! + \sigma m^{\scriptscriptstyle 2} \right)
		G \! _F {\scriptstyle(\!x\!,y\!)} \;
	& = \; \frac{-i \sigma}{\sqrt{g}} \;
		\delta^{\scriptscriptstyle\! (\!d)} {\scriptstyle \! (\!x-y\!)}
	\qqquad
	{\ssty \sqrt{g} \; = \, R\ds^d \, \mid (\cosh^{d-1} \! \tau)
		(\sin^{d-2} \! \phii^1)	\, \ldots \, (\sin \phii^{d-1})	\mid}
		\notag
\end{align}

		\subsection{From Klein-Gordon to the hypergeometric equation}
		\label{sec:dvb_ds_kghyper}
\begin{figure} [H]
	\begin{center}
	\igx[width=7cm]{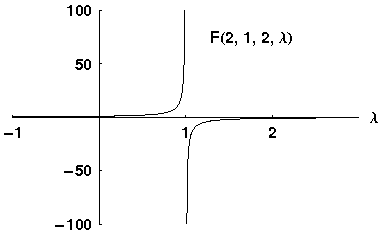}
	\end{center}
	\caption{First candidate function}
	\label{fig:hypergeo_ds_prop}
\end{figure}

Because of the designated SO$\ssty (1,d)$ invariance
(up to time ordering subtleties) of the propagators
it is clear that our candidate function
again can depend only on the chordal distance:
$\phi=\phi{\ssty ( \! \lambda {\sssty(\!} u {\sssty \!)} \!)}$.
\begin{align}
	u {\scriptstyle ( \! X \!,Y \!)}\, & = \, (X-Y)^2 \, = \, invar. \notag \\
	X^2 \, & = \, -\sigma \! R_X^2 \, = \,invar. \notag \\
	Y^2 \, & = \, -\sigma \! R_Y^2 \, = \,invar. \notag \\	
	XY \, & = \, - {\ssty \frac{1}{2}}
		(\sigma \! R_X^2 + \sigma \! R_Y^2 + u)	= \, invar. \notag
\end{align}
We make the following ansatz
\begin{align}
	\label{eq:dvb_ds31}
	\lambda \; = \; \frac{1}{2} + \, \frac{XY}{2 \sqrt{X^2Y^2}} \; 
	= \; \frac{1}{2} - \frac{\sigma}{4} \Bigl( \! {\ssty \frac{R_X}{R_Y}+
			\frac{R_Y}{R_X} } \! \Bigr) - \frac{u}{\, 4 R_X \! R_Y} 
\end{align}
and will utilize it shortly.
For the moment we need to turn to finding
a relation between the d'Alembertians in embedding space and dS.
We denote with $z$ the coordinates on dS.
Setting $X\!^N=R \, \omega^N \! {\scriptstyle (z)}$,
which is realized by all given coordinate sets,
we get $\omega^2 \! = \eta_{M \! N} \omega^M \! \omega^N = -\sigma$
therefrom $\eta_{M \! N} \omega^M(\del_\nu\omega^N) = 0$
and further on
\begin{align*}
	ds^2 & = \eta_{M \! N} dX^M dX^N \\
		 & = -\sigma \, dR^2 + R^2
		     \ub{\eta_{M \! N} (\partial_\mu \omega^M)(\partial_\nu \omega^N)} \,
			 dz^\mu dz^\nu \\
		 & \qquad \qquad \qquad \qquad \quad ^{h_{\mu \nu} {\scriptstyle (z)}}
\end{align*}
The metric in the embedding space using radial coordinates reads:
\begin{equation*}
	G \! _{M \! N}{\scriptstyle (R,z)} = 
	\begin{pmatrix}
		-\sigma & 0 \\
		0         & g_{\mu \nu} {\scriptstyle (R,z)}
	\end{pmatrix}
	\quad \quad {\scriptstyle \text{with}}
	\qquad g_{\mu \nu} {\scriptstyle (R,z)} = R^2 h_{\mu \nu}{\scriptstyle (z)} 
\end{equation*}
$g_{\mu \nu}$ is the induced metric
on a hyperboloid with $R^2 = const. > 0$
and therefore the metric of a deSitter space. Defining
\begin{align*}
	G{\scriptstyle (\!R,z)} \, = \, \mid \det G \! _{M \! N} \mid \qquad \quad
	g{\scriptstyle (\!R,z)} \, = \, \mid \det g_{\mu \nu} \mid \qquad \quad
	h{\scriptstyle (z)} \, = \, \mid \det h_{\mu \nu} \mid
\end{align*}
we have $G{\scriptstyle (\!R,z)} = g{\scriptstyle (\!R,z)} = R^{2d} %%@
h{\scriptstyle (z)}$.
We find for the d'Alembertians $\Box _{bulk}$ in the embedding space and 
$\Box \ds$ on the hyperbolic hypersurfaces with constant $R$:
\begin{align}
	\Box _{bulk}
	& = \frac{1}{\sqrt{G}} \, \partial _M \sqrt{G} \, G^{M \! N} \partial _N
		 \notag \\
	& = \frac{-\sigma}{\sqrt{g}} \, \partial _R \sqrt{g} \, \partial _R + \Box \ds
	\qquad \qquad \quad \Box \ds = \frac{1}{\sqrt{g}} \,
	\partial _\mu \sqrt{g} \, g^{\mu \nu} \partial _\nu \notag \\	
	\label{eq:dvb_ds32}
	& = -\sigma \partial _R^2 - \frac{\sigma d}{R} \, \partial _R + \Box \ds
\end{align}
Introducing the generalized angular momentum operators
\begin{equation*}
	M_{M \! N} = i \bigl( X_M \partial _N -  X_N \partial _M \bigr)
	\; \qquad \quad \quad \longrightarrow \;\, M_{M \! N}  R = 0 
\end{equation*}
and the generalized squared total angular momentum operator
\begin{align*}
	M^2 
	& = \frac{1}{2} M_{M \! N} M^{M \! N}
	\qquad \qquad \qquad \quad \longrightarrow \;\, M^2  R = 0 \\
	& = R^2 \partial _R^2 + d \, R \, \partial _R  + \sigma R^2 \Box_{bulk}
\end{align*}
we find:
\begin{equation}
	\label{eq:dvb_ds33}
	\Box_{bulk} = -\sigma \partial _R^2 -\frac{\sigma d}{R} \, \partial _R 
					+ \sigma \frac{M^2}{R^2}
\end{equation}
Comparing (\ref{eq:dvb_ds32}) and (\ref{eq:dvb_ds33})
shows the equivalence
of the squared angular momentum operator
and the d'Alembertian on the hyperboloid:
\begin{equation}
	\label{eq:dvb_ds34}
	\sigma \frac{M^2}{R^2} = \Box \ds
\end{equation}
We will derive our candidate function
$\phi {\ssty (\!x\!, y\!)}$ which lives on dS
from a function $\Psi {\ssty (\!X\!, Y\!)}$
which lives in embedding space.
Because of SO$\ssty (1,d)$ invariance $\Psi$
can only depend on $R_X$, $R_Y$, $\lambda$.
We make the following separation ansatz
introducing a new parameter $\Delta$:
\begin{equation}
	\Psi \! {\scriptstyle (\!X\!,Y\!)}
	= \Psi {\scriptstyle (\! R_X\!, R_Y\!, \lambda \!)}
	= R_X^{-\Delta} \, R_Y^{-\Delta} \,
		\phi _{\! \Delta}{\scriptstyle \! ( \! \lambda \!)}
\end{equation}
Using (\ref{eq:dvb_ds33}) we  find with $R_X\!=\!R\ds$:
\begin{align}
	\label{eq:dvb_ds37}
	0 = \Box ^{(\!X\!)} _{bulk} \Psi \! {\scriptstyle (\! X\!,Y\!)}
			\quad \Longleftrightarrow \quad 0 \, 
	& = \, \biggl[ \;\frac{M^2}{R_X^2} \, +
			\frac{\Delta(-\Delta \! + \! d \! - \! 1)}{R_X^2} \biggr] \,
			\phi _{\! \Delta}{\scriptstyle \! ( \! \lambda \!)} \notag \\
	& = \, \biggl[ \sigma \Box  \ds \, + \underbrace{
			\frac{\Delta(-\Delta \! + \! d \! - \! 1)}{R\ds^2}}_{m^2} \biggr] \,
			\phi _{\! \Delta}{\scriptstyle \! ( \! \lambda \!)} 
\end{align}
Moreover we have:
\begin{align}
	\label{eq:dvb_ds38}
	0 & = \Box _{bulk} \Psi \,  
	      = \, \partial _M \partial ^M \Psi \! {\scriptstyle (\! X\!,Y \!)} \notag \\ 
	\Longleftrightarrow \quad 0  
	& = \biggl[ \lambda (1 - \lambda ) \, \partial _\lambda ^2 \,
	                   + \biggl( \! \frac{d}{2} - \lambda \, d \biggr)
					   \partial _\lambda
	                   - \ub{\Delta (-\Delta \! + d \! - \! 1)}_{m^2 R\ds^2} \biggr] \,
						\phi_{\! \Delta}{\scriptstyle \! ( \! \lambda ,n \!)}
\end{align}
So  being a solution of equation (\ref{eq:dvb_ds38})  
and obeying the homogeneous Klein-Gordon equation (\ref{eq:dvb_ds37})
in dS$_d$ with the squared mass $m^2=\Delta(-\Delta+d-1)/R\ds^2$
is equivalent for $\phi _{\! \Delta}{\scriptstyle \! ( \! \lambda \!)}$.
This mass term exactly matches relation (\ref{eq:ds_confdim02}).
\\
Equation (\ref{eq:dvb_ds38}) is of the hypergeometric type:
\begin{align}
	\label{eq:dvb_ds38_5}
	0 \, = \, \biggl[ \lambda (1 - \lambda ) \, \partial _\lambda ^2 \,
	                        + \Bigl[ c - \lambda (a \! + \! b \! + \! 1) \Bigr]
							\partial _\lambda
							- ab \biggr] \,
	\phi _{\! \Delta}{\scriptstyle \! ( \! \lambda \!)}
\end{align}
In our case the parameters are
\begin{equation}
	a = \Delta \quad \quad 
	b = -\Delta \! + d \! - \! 1 \quad \quad
	c = \frac{d}{2}
\end{equation}
The regular solution of the hypergeometric differential equation
is the hypergeometric function $F {\scriptstyle (a,b,c, \lambda)}$.
Again $F {\scriptstyle (a,b,c, \lambda)}$
is symmetric under the exchange of $a$ and $b$
i.e. under $\Delta \rightarrow (-\Delta \! + \! d \! - \! 1)$
i.e. under $\Delta _+ \! \leftrightarrow \Delta _-$
(see section \ref{sec:ds_confdim}). \\
We have found a first candidate function:
%%
%%\beq
%%	\alpha \, = \, \frac{\, \Gamma {\ssty (\!a\!)} \Gamma {\ssty (\!b\!)}}
%%				{ (4\pi)^{d/2} \, \Gamma {\ssty (c)}} \, 
%%				= \, \frac{\, \Gamma {\ssty (\!\Delta\!)} \Gamma {\ssty
%%  (\!-\Delta+d-1)}}
%%				{ (4\pi)^{d/2} \, \Gamma {\ssty (d/2)}}
%%\eeq
%%
\beq
	\phi_{\! \Delta}{\scriptstyle \! ( \! \lambda \!)} \, 
	= \, F (a,b,c, \lambda) \,
	= \, F \Bigl(\Delta_+, \Delta_-,\frac{d}{2},
		 \, \lambda \Bigr)
\eeq
In figure \ref{fig:hypergeo_ds_prop} this function is plotted
for $(d=4)$ and the values \eqref{eq:ds_confdim_09}
of $\Delta_\pm$.
Looking at (\ref{eq:dvb_ds38_5}) we find that substituting
$\ovl{\lambda} = 1-\lambda$ yields the same differential equation
with the same parameters $a,b,c$.
Therefore we have found a second candidate function 
\beq
	\phi_{\! \Delta}{\scriptstyle \! ( \! \ovl{\lambda} \!)} \, 
	= \, F (a,b,c, \ovl{\lambda}) \,
	= \, F \Bigl(\Delta_+, \Delta_-,\frac{d}{2},
		 \, \ovl{\lambda} \Bigr)
\eeq
with
\begin{align}
	\lambda \; = \; \frac{1}{2} + \, \frac{XY}{2 \sqrt{X^2Y^2}} \; 
	= \; \frac{1}{2} - \frac{\sigma}{4} \Bigl( \! {\ssty \frac{R_X}{R_Y}+
			\frac{R_Y}{R_X} } \! \Bigr) - \frac{u}{\, 4 R_X \! R_Y} \\
	\ovl{\lambda} \; = \; \frac{1}{2} - \, \frac{XY}{2 \sqrt{X^2Y^2}} \; 
	= \; \frac{1}{2} + \frac{\sigma}{4} \Bigl( \! {\ssty \frac{R_X}{R_Y}+
			\frac{R_Y}{R_X} } \! \Bigr) + \frac{u}{\, 4 R_X \! R_Y} 
\end{align}
We note that $F (a,b,c, \lambda)$ is regular in $\lambda=0$
but  for $d \geq 4$ is divergent in $\lambda=1$
because Re$(c-a-b)=1-\frac{d}{2} \leq -1 $.

		\subsection{Inspection of the candidate functions}
		\label{sec:dvb_ds_excand}
What we are looking for is a function that converges to zero
if the chordal distance $u$ goes to (positive or negative) infinity.
We have the following quantities going to zero for large chordal distances:
\begin{align*}
	u \rightarrow \pm \infty
	\quad \Longrightarrow \quad
	& \lambda \rightarrow \mp \infty
	\quad & \Longrightarrow \quad
	\frac{\,1\,}{\lambda} \; \rightarrow \, 0 \\
	\Longrightarrow \quad
	& \ovl{\lambda} \rightarrow \pm \infty
	\quad & \Longrightarrow \quad
	\frac{\,1\,}{\ovl{\lambda}} \; \rightarrow \, 0
\end{align*}
The following discussion is done for $\lambda$ only,
but is the same for $\ovl{\lambda}$.
Likewise to the AdS case we can split up our solution
according to equation (\ref{eq:hypergeo07}):
%%and (\ref{eq:hypergeo08})
%%
\begin{align}
	F {\scriptstyle (a,b,c, \lambda)} \,
	& = \, \frac{\Gamma{\ssty (\!c\!)} \,\Gamma{\ssty (\!b-a\!)}}
					 {\Gamma{\ssty (\!b\!)} \,\Gamma{\ssty (\!c-a\!)}} \,
					 \biggl( \! \frac{1}{-\lambda} \! \biggr)^{\!\!a} \,
					 F \Bigl( a,a\!-\!c\!+\!1,a\!-\!b\!+\!1, \frac{\ssty 1}{\ssty \lambda} %%@
\Bigr)
					  \notag \\
	\label{eq:dvb_ds40}
	& \qquad + \; \frac{\Gamma{\ssty (\!c\!)} \,\Gamma{\ssty (\!a-b\!)}}
					 {\Gamma{\ssty (\!a\!)} \,\Gamma{\ssty (\!c-b\!)}} \,
					 \biggl( \! \frac{1}{-\lambda} \! \biggr)^{\!\!b} \,
					 F \Bigl( b,b\!-\!c\!+\!1,b\!-\!a\!+\!1, \frac{\ssty 1}{\ssty \lambda} %%@
\Bigr)
\end{align}
Each of the two summands in (\ref{eq:dvb_ads40}) is a solution
of the hypergeometric differential equation (see \cite{astegun} 15.5.A).
In the AdS case we took each summand as a self-contained candidate function
because Breitenlohner and Freedman found the condition
that the propagator has to vanish faster than
$\! \bigl( \! \frac{1}{-\lambda} \! \bigr)^{\! \frac{1}{2}\frac{d-\!1}{2}} \!$
for large $\lambda$.
This could not be satisfied by both summands for the same $\Delta$.
\\
For the dS case the corresponding condition would be 
that the propagator would have to vanish faster than
$\bigl( \! \frac{1}{-\lambda} \! \bigr)^{\! \frac{d-\!1}{2}}$
for large $\lambda$.
However no such condition is known.
\\
For $m^2R^2\ds>({\ssty \frac{d-1}{2}})^2$ the conformal weights
$\Delta_\pm$ are complex with positive real part.
Therefore both summands converge to zero for large $\lambda$.
In the case of $0 < m^2R^2\ds \leq ({\ssty \frac{d-1}{2}})^2$
we find $\Delta_\pm > 0$ 
and therefore both summands in (\ref{eq:dvb_ds40})
again converge to zero for large $\lambda$.
However for $m^2R^2\ds \leq 0$ it happens
that while still $\Delta_+ > 0$ we now face $\Delta_- \leq 0$
and hence the second summand remains nonzero 
or even grows for large $\lambda$.
In this case, which we will not consider in detail here,
one can only use the first summand as a candidate function.
\\
Thus for $m^2R^2\ds > 0$
we do not need to split up our solution
and can continue working with the full candidate functions.  
\\
Now we have a closer look on $\lambda$.
Evaluating ansatz (\ref{eq:dvb_ads31}) for dS we find:
\begin{align}
	\label{eq:dvb_ds40_5}
	\lambda \, & = \; \frac{1}{2} \! + \frac{XY}{2 R\ds^2} \; 
	= \; \frac{\,1\!-\!\sigma}{2} \! - \! \frac{u}{\,4 R\ds^2} \;
	= \; \frac{\,1\!+\!\sigma}{2} \! + \! \frac{\snake{u}}{\,4 R\ds^2}	
	\\
	\label{eq:dvb_ds40_6}
	\ovl{\lambda} \, & = \; \frac{1}{2} \! - \frac{XY}{2 R\ds^2} \; 
	= \; \frac{\,1\!+\!\sigma}{2} \! + \! \frac{u}{\,4 R\ds^2} \;
	= \; \frac{\,1\!-\!\sigma}{2} \! - \! \frac{\snake{u}}{\,4 R\ds^2}	
\end{align}
For the $(\tau,\vec{\phii})$ coordinate set we obtain:
\begin{align}
	\label{eq:dvb_ds40_6a}
	\lambda & = \, \frac{1}{2} + \frac{\sigma}{4}
		\Bigl[ (1 \! - \! \vec{\xi}_x \vec{\xi}_y) \cosh (\tau_x \!\! + \! \tau_y) \, - \,
				 (1 \! + \! \vec{\xi}_x \vec{\xi}_y) \cosh (\tau_x \!\! - \! \tau_y) \Bigr]
\end{align}
For the origin $\tau_y \!\! = \! \vec{\phii}_y \! = \! 0$
as reference point this expression condenses into:
\begin{align}
	\label{eq:dvb_ds40_6b}
	\lambda & = \, \frac{1}{2} - \frac{\sigma}{2} \cos \phii^1 \cosh \tau_x
\end{align}
We see that for certain values of $\tau_x, \vec{\phii}_x,\tau_y,\vec{\phii}_y$
our invariant quantities $\lambda, \ovl{\lambda}$ can become one
which is pathologic because $\phi_{\! \Delta}$
%% and $\phi^{\ssty (\!b\!)}_{\! \Delta}$
is divergent in $\lambda = 1$.
%%and the hypergeometric differential equation has singular points
%%at $\lambda = 0$ and $\lambda = 1$.

				\subsubsection{Infinitesimal shift in complex time plane}
				\label{sec:dvb_ds_infshift}
In order to remove this we apply again shifts in the complex $X^0$-plane
\begin{align}
	X^0 \; & \rightarrow \; X^0_\pm = X^0 \pm \, i {\ssty \frac{\epsilon}{2}}  \\
	Y^0 \; & \rightarrow \; Y^0_\pm = Y^0 \mp \, i {\ssty \frac{\epsilon}{2}}
		\notag
\end{align}
which do not change the d'Alembertian.
Now based on \eqref{eq:dvb_ds40_5} 
there are two possibilities of defining $\lambda_\pm$:
\begin{align}
	\lambda_\pm^{\sssty \! ( \! u \! )}
	& = \, \frac{\,1\!-\!\sigma}{2} - \frac{u_\pm}{\,4 R\ds^2}
		& {\ssty u_\pm } & {\ssty = \, (X_{\!\pm} \!-\!Y_{\!\pm})^2} \\
	\lambda_\pm^{\sssty \! ( \! XY \! )} 
	& = \;\,  \frac{1}{2} \, + \, \frac{X_\pm Y_\pm}{\,2 R\ds^2}
		& {\ssty X_\pm} & {\ssty = \, (X^0_{\!\pm}, \vec{X}, X^d)}
\end{align}
A priori it is not clear which version we should choose.
We will use definition $\lambda_\pm^{\sssty \! ( \! u \! )}$
and call it $\lambda_\pm$ from here on,
simply because it yields the same results as given in \cite{leshouches}.
We just note that using definition $\lambda_\pm^{\sssty \! ( \! XY \! )}$
yields the same signs for the terms involving $i\epsilon$ and $\epsilon^2$,
however instead of $\beta$ in \eqref{eq:dvb_ds67_13} one obtains
a different constant of normalization: 
$\beta^{\sssty ( \! XY \! )} = \beta / \sqrt{2}$.
Definition $\lambda_\pm^{\sssty \! ( \! u \! )}$ yields:
\begin{align}
	\label{eq:dvb_ds_57_2}
	\lambda_\pm = \, \frac{1}{2} + \frac{\sigma}{4}
		\biggl[  (1 \! - \! \vec{\xi}_x \vec{\xi}_y) \cosh (\tau_x \!\! + \! \tau_y)
			  & - (1 \! + \! \vec{\xi}_x \vec{\xi}_y) \cosh (\tau_x \!\! - \! \tau_y)
			  \quad \\
		& \quad
		\mp i {\ssty \frac{2 \epsilon}{R^2\ds}} \bigl( X^0 \! - \! Y^0 \bigr) 
		+ {\ssty \, \frac{\epsilon^2}{R^2\ds} \,} \biggr]
		\notag
\end{align}
so that we have again 
$\lambda_\pm {\ssty (\!x\!,y\!)}
= \lambda^*_\mp {\ssty (\!x\!,y\!)}
= \lambda_\mp {\ssty (\!y\!,x\!)}$.
$\lambda$ is SO$\ssty (1,d)$ invariant
and sign$(X^0 \! - \! Y^0)$ is invariant under orthochronous SO$\ssty (1,d)$
just as in Minkowski spacetime.
Moreover in the limit of small $\epsilon$ we can replace
$\epsilon (X^0 \! - \! Y^0)$ by $\epsilon R\ds (\tau_x \! - \tau_y)$
(see subsection \ref{sec:ds_timord_ttau}).  
\\
While the $i\epsilon$ term is present, the point 1 is always evaded.
We observe that the $i\epsilon$ term vanishes only
iff $X^0 = Y^0$ i.e. iff $\tau_x\!\!=\!\tau_y$.
In this case we have
\begin{align*}
	\lambda_{\pm (\tau_x=\tau_y\!)} = \, \frac{1}{2} + \frac{\sigma}{4}
	\biggl[  (1 \! - \! \vec{\xi}_x \vec{\xi}_y) \cosh 2\tau_x
			   - (1 \! + \! \vec{\xi}_x \vec{\xi}_y)
			   + {\ssty \, \frac{\epsilon^2}{R^2\ds} \,} \biggr]
	\\
	\ovl{\lambda}_{\pm (\tau_x=\tau_y\!)} = \, \frac{1}{2} - \frac{\sigma}{4}
	\biggl[  (1 \! - \! \vec{\xi}_x \vec{\xi}_y) \cosh 2\tau_x
			   - (1 \! + \! \vec{\xi}_x \vec{\xi}_y)
			   + {\ssty \, \frac{\epsilon^2}{R^2\ds} \,} \biggr]
\end{align*}
and thus
\begin{align}
	\label{eq:dvb_ds57_45}
	\lambda^{\sigma=-1}_{\pm (\tau_x=\tau_y\!)} \!
	= \; \ovl{\lambda}^{\sigma=+1}_{\pm (\tau_x=\tau_y\!)} \,
	\leq \, 1 - {\ssty \frac{\epsilon^2}{4R^2\ds}}
	\\
	\lambda^{\sigma=+1}_{\pm (\tau_x=\tau_y\!)}
	= \; \ovl{\lambda}^{\sigma=-1}_{\pm (\tau_x=\tau_y\!)} \,
	\geq \, 0 + {\ssty \frac{\epsilon^2}{4R^2\ds}}
\end{align}
We see that in the upper line
$\lambda_\pm$ is always separated from 1 by $\epsilon^2$ 
while in the lower line $\lambda_\pm$ crosses 1.
Thereby it is clear that $\lambda$ is the right quantity for $\sigma=-1$
and $\ovl{\lambda}$ is the right one for $\sigma=+1$.
This assignment is confirmed by the observation
that for vanishing chordal distance $u=0$
we obtain $\lambda^{\sigma=-1} \! =\ovl{\lambda}^{\sigma=+1} \! =1$
which lets the candidate functions become singular
exactly on the lightcone of the coincident point  
$(X=Y)$ in embedding space i.e. $(x=y)$ in dS.
The opposite assignment would let them become singular
on the lightcone of the antipode point $x=\snake{y}$.
\\
The singularity of our assignment is of delta function type
and realized in the limit of small $\epsilon$
while the singularity of the opposite assignment
is not of delta type since the point 1 is directly crossed.
Moreover in (\ref{eq:dvb_ds57_7}) we can see
that for $d>4$ the singularity is of order $(1-{\ssty \frac{d}{2}})<1$.
Hence $\lambda_\pm=1$ is no residual
and does not contribute in integrals like (\ref{eq:solvinhom_integral}).
\\
Therefore we will only consider the case (\ref{eq:dvb_ds57_45})
and from here on simplify our notation to
$\lambda_\pm=\lambda^{\sigma=-1}_\pm=\ovl{\lambda}^{\sigma=+1}_\pm$.
\\
We remark that $\lambda_{\pm (\tau_x=\tau_y\!)} = 1 - \epsilon^2$
iff $\vec{\xi}_x = \vec{\xi}_y$.
After all we have found a well defined candidate function:
\beq
	\label{eq:dvb_ds_57_5}
	\phi^\pm_{\! \Delta}{\scriptstyle \! ( \! \lambda_\pm \!)} \, 
	= \, F (a,b,c, \lambda_\pm) \,
	= \, F \Bigl(\Delta_+, \Delta_-,\frac{d}{2},
		 \, \lambda_\pm \Bigr)
\eeq

				\subsubsection{Delta source and normalization}
				\label{sec:dvb_ds_delnorm}
In order to construct the Feynman propagator fulfilling the inhomogeneous
Klein-Gordon equation (\ref{eq:dvb_ds29_5}), we again need to show
that applying the Klein-Gordon operator to our candidate functions
yields the desired delta source.
\\
According to (\ref{eq:dvb_ds29}) we therefor shall examine
$\bigl( \del_{\tau_x} \phi_\Delta^\pm \bigr)_{\!\tau_x=\tau_y}$ \!\!
in order to show that it delivers a delta function
for the spatial variables $\vec{\phii}_{x,y}$
of the spacetime points $x$ and $y$.
Likewise to the AdS case we plan to identify
$\phi_\Delta^\pm$ with $G^{\mp}_\Delta$. \\
Starting from equation (\ref{eq:dvb_ds_57_5})
and employing (\ref{eq:hypergeo04_5}) we get 
\begin{align}
	\label{eq:dvb_ds57_7}
	\phi^\pm_{\! \Delta}{\ssty \! ( \! \lambda_\pm \!)} \, 
	= \, \bigl(1\!-\! \lambda_\pm \bigr)^{1-d/2}
	F \bigl( c\!-\!a, \, c\!-\!b, \, c, \, \lambda_\pm \bigr)
\end{align}
Next we put to work the chain rule of differentiation
$(\del_{\tau_x} \phi_\Delta^\pm) = (\del_{\tau_x} \lambda_\pm)
(\del_{\lambda_\pm} \phi_\Delta^\pm)$
wherein for the last factor we can make use of equation (\ref{eq:hypergeo04}).
\begin{align}
	\label{eq:dvb_ds58}
	\Bigl( \del_{\tau_x}
	\phi^\pm_\Delta {\ssty \! ( \! \lambda_{\! \pm} \!)} \! \Bigr)
	_{\! \tau_x=\tau_y} \!\!\!
	& = \, \bigl( \del_{\tau_x} \lambda_\pm \bigr)_{\! \tau_x=\tau_y}
			\Bigl[
			({\ssty \frac{d}{2}}\!-\!1) (1\!-\! \lambda_\pm )^{-\frac{d}{2}}
			F { \ssty( c\!-\!a, \, c\!-\!b, \, c, \, \lambda_\pm)} \\
			& \qquad \qquad \qquad \quad \;\;
			+ {\ssty \frac{(c-a)(c-b)}{c}} (1\!-\! \lambda_\pm )^{1-\frac{d}{2}}
			F {\ssty (c-a+1,c-b+1,c+1, \lambda_\pm)}
			\Bigr]_{\! \tau_x=\tau_y} \notag
\end{align}
From equation (\ref{eq:dvb_ds_57_2}) we can read off
\begin{align}
	\label{eq:dvb_ds60}
	\Bigl( \del_{\tau_x} \lambda_\pm \! \Bigr)_{\! \tau_x=\tau_y} \!
	& = \,\epsilon \Bigl( \pm \, {\ssty \frac{i}{2R\ds}}
			- {\ssty \frac{1}{4\epsilon} \, }
			(1 \! - \vec{\xi}_x \vec{\xi}_y) \sinh 2 \tau_x \Bigr)
			\\
	\Bigl( \lambda_\pm \! \Bigr)_{\! \tau_x=\tau_y} \!
	& = \, {\ssty \frac{1}{2}} - {\ssty \frac{1}{4}}
	\Bigl[  (1 \! - \! \vec{\xi}_x \vec{\xi}_y) \cosh 2\tau_x
			   - (1 \! + \! \vec{\xi}_x \vec{\xi}_y)
			   + {\ssty \, \frac{\epsilon^2}{R^2\ds} \,} \Bigr] 
\end{align}
which for the reference point $\! \vec{\phii}_y \! = \! 0$
(with $\tau_y$ not necessarily zero)
reduces to the simpler expression
\begin{align}
	\Bigl( \lambda_\pm \! \Bigr)_{\! \tau_x=\tau_y} \!
	& = \, {\ssty \frac{1}{2}} - {\ssty \frac{1}{4}}
	\Bigl[  (1 \! - \! \cos \phii^1_x) \cosh 2\tau_x
				- (1 \! + \! \cos \phii^1_x)
				+ {\ssty \, \frac{\epsilon^2}{R^2\ds} \,} \Bigr]
		\notag \\ 
	& = \, 1 - {\ssty \frac{1}{2}} (1 \! - \! \cos \phii^1_x) \cosh^2 \! \tau_x 
				- {\ssty \frac{\epsilon^2}{4R^2\ds} \,}
\end{align}
We already found that $(1 \! - \! \lambda_\pm)=\epsilon^2$
iff $(\vec{\xi}_x \! =\vec{\xi}_y)$ i.e. iff $(\vec{\phii}_x \! =\vec{\phii}_y)$.
This means that for all $\vec{\phii}_x \neq \vec{\phii}_y$
we can let $\epsilon$ run to zero so that $( \del_{\tau_x} \! \phi^\pm_\Delta
{\ssty \! ( \! \lambda_{\! \pm} \!)} )_{\tau_x=\tau_y}$ vanishes in this case.
\\
In order to determine suitable normalization constants we need to integrate
$(\del_{\tau_x} \phi_\Delta^\pm )_{\tau_x=\tau_y}$ over S$^{d-1}$.
Doing this, we find that the second summand in equation (\ref{eq:dvb_ds58})
yields a contribution proportional to $\epsilon^2$ 
which therefore vanishes in the limit of small $\epsilon$.
\\
When turning to the first summand,
we first compute the S$^{d- \!1}$-integral
over $\epsilon\, \sqrt{g} \, (1\! / \!-\!\lambda_\pm)^{-d/2}_{\tau_x=\tau_y}$
for the case of our reference point $\vec{\phii}_y \! = \! 0$
and find using integral (\ref{eq:delta155}):
\begin{align}
	\int \limits_{S^{d-1}} \!\!\!\! d^{d-\!1} \phii \,
	\frac{\epsilon \, \sqrt{g}}
			{\bigl( 1 \! - \! \lambda_\pm \bigr)^{\! \frac{d}{2}}_{\tau_x=\tau_y}} \,
	& = \, \frac{2 \pi^{\frac{d\!-\!1}{2}}}{\Gamma({\ssty \frac{d-1}{2}})}
			\, R\ds^d \! \int \limits_0^\pi \!\! d\phii_x^1 
			\, \frac{\, \epsilon \, \cosh^{d-1} \! \tau_x \, \sin^{d-2} \! \phii_x^1}
			{\Bigl[{\ssty \frac{1}{2}} (1 \! - \! \cos \phii_x^1) \cosh^2 \! \tau_x 
			+ {\ssty \frac{\epsilon^2}{4R^2\ds}} \, \Bigr]^{\! \frac{d}{2}}}
	\notag \\
	& = \, \frac{2 \pi^{\frac{d\!-\!1}{2}}}{\Gamma({\ssty \frac{d-1}{2}})}
			\, R\ds^d \!\!\! \int \limits_0^{\mu \ll 1} \!\!\!\! d\phii_x^1
			\, \frac{\, \epsilon \, \cosh^{d-1} \! \tau_x \, \sin^{d-2} \! \phii_x^1}
			{\Bigl[{\ssty \frac{1}{2}} (1 \! - \! \cos \phii_x^1) \cosh^2 \! \tau_x \!
			+ {\ssty \frac{\epsilon^2}{4R^2\ds}} \, \Bigr]^{\! \frac{d}{2}}}
	\notag \\
	& \qquad \quad
		 	+ \,  \frac{2 \pi^{\frac{d\!-\!1}{2}}}{\Gamma({\ssty \frac{d-1}{2}})}
			\, R\ds^d \!\!\! 
			\ub{ \int \limits_{\mu \ll 1}^\pi \!\!\!\! d\phii_x^1
			\, \frac{\, \epsilon \, \cosh^{d-1} \! \tau_x \, \sin^{d-2} \! \phii_x^1}
			{\Bigl[{\ssty \frac{1}{2}} (1 \! - \! \cos \phii_x^1) \cosh^2 \! \tau_x \! 
			+ {\ssty \frac{\epsilon^2}{4R^2\ds}} \, \Bigr]^{\! \frac{d}{2}}}
			}_{\limepszero \text{exists:} \,  \longrightarrow \, 0}
	\notag \\
	& \!\! \overset{\epsilon \rightarrow 0}{=} 
			\, \frac{2 \pi^{\frac{d\!-\!1}{2}}}{\Gamma({\ssty \frac{d-1}{2}})}
			\, R\ds^{d+1} 2^d \!
			\int \limits_0^\infty \!\! d \phii^1_x \, \frac{\phii_x^{1^{d-2}}}
			{\bigl( \phii_x^{1^2} \!\! + 1  \bigr)^{\! \frac{d}{2}}}
	\notag \\
	& = \; \frac{\, R\ds^{d+1} \, (4 \pi)^{\frac{d}{2}}}
					{\Gamma({\ssty \frac{d}{2}})}
	\\
\end{align}
In the first line of the integration above we already used $\sqrt{g}$.
\begin{align}	
	\Longrightarrow \quad \limepszero 
		\frac{\, \epsilon \, \sqrt{g} }
		{(1 \! - \! \lambda_\pm)_{\tau_x=\tau_y}^{\frac{d}{2}}} \;
	& = \; \frac{\, R\ds^{d+1} \, (4 \pi)^{\frac{d}{2}}}
					{\Gamma({\ssty \frac{d}{2}})} \,
			\delta^{(\!d\!-\!1\!)}{\ssty (\vec{\phii}_x-\vec{\phii}_y)}
\end{align}
Looking at (\ref{eq:dvb_ds60}) we see
that for $(\vec{\phii}_y = \vec{\phii}_x)$
i.e. $(\vec{\xi}_y = \vec{\xi}_x)$ we have
$\bigl( \del_{\tau_x} \lambda_\pm \! \bigr)_{\! \tau_x=\tau_y} \!
			= \, \pm i {\ssty \frac{\epsilon}{2R\ds}}$.
Therefore using (\ref{eq:hypergeo03_5})
we arrive at the desired spatial delta function:
\begin{align}
	 \sqrt{g} \; \limepszero	\Bigl( \del_{\tau_x} 
			\phi^\pm_\Delta{\ssty \! ( \! \lambda_{\! \pm} \!)} \Bigr)
			_{\! \tau_x=\tau_y}
	& = \; \pm i \, \ub{ \frac{\, (4\pi)^{\! \frac{d}{2}} R\ds^d\;
		\Gamma({\ssty \frac{d}{2})}}
		{\, \Gamma({\ssty \Delta_+}) \, \Gamma({\ssty \Delta_-})}
		\frac{1}{2}}_{\beta_\phi} \,		
		\delta^{(\!d\!-\!1\!)} {\ssty (\vec{\phii}_x-\vec{\phii}_y)}
\end{align}
With this relation fixed we can now identify
(up to $\sqrt{g}$ which is already included)
according to (\ref{eq:dvb_gen_ident}):
\begin{align}
	\label{eq:dvb_ds67}
	G_{\! \Delta}^\pm {\ssty (\!x\!,y\!)} \,
	& = \, \frac{1}{\, 2 \beta_\phi \beta_G} \;\;
			\limepszero \, \phi^\mp_\Delta{\ssty \! ( \! \lambda_{\! \mp})}
	\\
	& = \, \ub{\frac{\, \Gamma({\ssty \Delta_+}) \, \Gamma({\ssty \Delta_-})}
					{\, (4\pi)^{\! \frac{d}{2}} R\ds^{d-2}\; \Gamma({\ssty %%@
\frac{d}{2}})}}
					_{1/(2\beta)}
			 \;	\limepszero \, \phi^\mp_\Delta{\ssty \! ( \! \lambda_{\! \mp})}
	\notag
\end{align}
and construct the other propagators according to
(\ref{eq:dvb_ads03}-\ref{eq:dvb_ads08}).
Evaluating equation (\ref{eq:dvb_ds29})
we find that the Feynman propagator fulfills
the inhomogeneous Klein-Gordon equation (\ref{eq:dvb_ds29_5})
with one delta source on its right hand side:
\begin{align*}
	\left( \sigma \Box _x \!\! + m^{ 2} \right)
	G \! _F^{\,\Delta} {\ssty(\!x\!,y\!)} \; & = \;\,
	\frac{-i}{\sqrt{g}} \;
	\delta^{\! (\!d\!)} {\ssty \! (\!x-y\!)}
	\notag \\
	\left(\Box _x \!\! + \sigma m^{ 2} \right)
	G \! _F^{\,\Delta} {\ssty(\!x\!,y\!)} \; 
	& = \;\frac{ -i \sigma }{\sqrt{g}} \;
	 \delta^{\! (\!d\!)} {\ssty \! (\!x-y\!)}
\end{align*}
The normalization constant is:
\begin{align}
	\label{eq:dvb_ds67_13}
	\beta & = \, \frac{\, (4\pi)^{\! \frac{d}{2}} R\ds^{d-2}\;
					 \Gamma({\ssty \frac{d}{2}}) }
					 {2 \Gamma({\ssty \Delta_+}) \, \Gamma({\ssty \Delta_-})}
\end{align}
Again we do not want the $\Gamma{\ssty (\ldots)}$ functions
in our normalization constant to run to infinity.
This requirement is fulfilled for all $m^2R^2\ds > 0$.
In this case $\Delta_\pm$ are either positive or complex
and thus the Gamma functions remain finite
(see discussion above equation \ref{eq:dvb_ds40_5}).

		\subsection{Listing of propagators}
Below we list the various propagators for dS$_d$.
Herein $\lambda_\pm$ is meant to carry its phase
acquired via its position in the complex plane:
\begin{align}
	\lambda_\pm = \, \frac{1}{2} - \frac{1}{4}
		\biggl[  (1 \! - \! \vec{\xi}_x \vec{\xi}_y) \cosh (\tau_x \!\! + \! \tau_y)
			  & - (1 \! + \! \vec{\xi}_x \vec{\xi}_y) \cosh (\tau_x \!\! - \! \tau_y)
			  \quad \\
		& \quad
		\mp i {\ssty \frac{2 \epsilon}{R\ds}} \, (\tau_x \!\! - \! \tau_y) \,  
		+ {\ssty \, \frac{\epsilon^2}{R^2\ds} \,} \biggr]
		\notag
\end{align}
Our Wightman function $G_\Delta^+$
agrees with the Wightman function which is given by
Spradlin, Strominger and Volovich in \cite{leshouches}.
\\
Moreover our $i \epsilon$-prescription is almost the same as theirs.
Spradlin, Strominger and Volovich use the complex shift
$(X^0-Y^0) \rightarrow (X^0-Y^0 - i \epsilon)$ in embedding space
for the $G^+ {\ssty (x,y)}$ propagator,
(but without giving further information about it).
This is the same prescription as in equation (\ref{eq:dvb_mink_shift_01})
for Minkowski spacetime.
It works well when using the chordal distance $u$-terms
in equations (\ref{eq:dvb_ds40_5}) and (\ref{eq:dvb_ds40_5})
but does not produce meaningful results if the $XY$-terms need to be used.  
\\
Our prescription for the propagators $G^\mp {\ssty (x,y)}$
is $X^0 \rightarrow (X^0 \! \pm i {\ssty \frac{\epsilon}{2}})$
together with $Y^0 \rightarrow (Y^0 \! \mp i {\ssty \frac{\epsilon}{2}})$.
This is the same as in equation (\ref{eq:dvb_mink_shift_02})
for Minkowski spacetime
and produces meaningful results for both $u$- and $XY$-terms.
However, as already mentioned above, using the $u$-version
yields a different constant of normalization compared to the $XY$-version. 
Our prescription agrees with the one of Spradlin, Strominger and Volovich
because for $G^+ {\ssty (x,y)}$ ours leads to
$(X^0 \! - \! Y^0) \rightarrow (X^0 \! - \! Y^0 \! - \! i \epsilon)$.
\begin{align}
	G_{\! \Delta}^\pm \! {\ssty (\!x\!,y\!)}
	& = \, \frac{1}{\, 2 \beta} \: \limepszero \,
				 F (\Delta_+, \Delta_-, {\ssty \frac{d}{2}}, \, \lambda_\mp) 
\end{align}
\begin{align}
	G_{\! \Delta}^{\sssty (\!1\!)} \! {\ssty (\!x\!,y\!)} \, 
	& = \, \frac{1}{\, \beta} \;\;\: \text{Re} \: \limepszero \,
				 F (\Delta_+, \Delta_-, {\ssty \frac{d}{2}}, \, \lambda_-) \\
	G_{\! \Delta} \! {\ssty (\!x\!,y\!)} \, 
	& = \, \frac{1}{\, \beta} \; i \, \text{Im} \: \limepszero \,
				 F (\Delta_+, \Delta_-, {\ssty \frac{d}{2}}, \, \lambda_-) 
\end{align}
\begin{align}
	G^{ \Delta}_{\!R} \! {\ssty (\!x\!,y\!)}
	& = \;\;\; \theta {\ssty (\!t_x\!-t_y\!)} \,
			\frac{1}{\, \beta} \, i \, \text{Im} \: \limepszero \,
				 F (\Delta_+, \Delta_-, {\ssty \frac{d}{2}}, \, \lambda_-) \\
	G^{ \Delta}_{\!A} \! {\ssty (\!x\!,y\!)}
	& = -\theta {\ssty (\!t_y\!-t_x\!)} \,
			\frac{1}{\, \beta} \, i \, \text{Im} \: \limepszero \,
				 F (\Delta_+, \Delta_-, {\ssty \frac{d}{2}}, \, \lambda_-) 
\end{align}
\begin{align}
	G^{\,\Delta}_{\! F} \! {\ssty (\!x\!,y\!)} \, 
	& = \, \frac{1}{\, 2 \beta} \; \limepszero \,
				 F (\Delta_+, \Delta_-, {\ssty \frac{d}{2}}, \, \lambda_F) 
\end{align}
\\
Therein $\lambda_F$ (with index F for Feynman)
is defined for the limit of small $\epsilon$ as
\begin{align}
	\lambda_F \! & = \ub{\, \frac{1}{2} - \frac{1}{4} \Bigl[
			  (1 \! - \! \vec{\xi}_x \vec{\xi}_y) \cosh (\tau_x \!\! + \! \tau_y)
			  - (1 \! + \! \vec{\xi}_x \vec{\xi}_y) \cosh (\tau_x \!\! - \! \tau_y)
			  \Bigr] } _{\lambda}
		- \, i {\ssty \frac{\epsilon}{4R\ds}} \! 
			\mid \! \tau_x \!\! - \!\! \tau_y \!\! \mid \!
		- {\ssty \frac{\epsilon^2}{8R^2\ds} \,} 
		\notag
\end{align}
so that the Feynman propagator is indeed symmetric:
$\, G_{\! F}^{\Delta} \! {\ssty (\!x\!,y\!)}
= G_{\! F}^{\Delta} \! {\ssty (\!y\!,x\!)}$.
We observe that in contrast to AdS the sign of the $i \epsilon$-term
in the Feynman propagator
in dS spacetime is always positive except for $\tau_x \! = \tau_y$
as in Minkowski spacetime.
\\
For the propagators above the hypergeometric function
is defined by its convergent Taylor series
for all $\lambda$ with $\mid \! \lambda \! \mid \, < 1$.
The form of the propagators for $\mid \! \lambda \! \mid \, > 1$
can be obtained therefrom using (\ref{eq:hypergeo07}).
For $\mid \! \lambda \! \mid \, < 1$ we can use
(\ref{eq:dvb_ds57_7}).\\
Likewise to Minkowski and AdS spacetime
we find that the propagators become singular
one the whole embedding space lightcone $\lambda=1$
while generating only one delta source at the coincidence point $x=y$.
\\
In appendix \ref{sec:props_ds_t} we also apply the DvB method
for static coordinates.
Again we find agreement in function and normalization
with the results of Spradlin, Strominger and Volovich in \cite{leshouches}.
The only flaw of these coordinates is the non-invariance
of time ordering with respect to $t$ (see subsection \ref{sec:ds_timord_t}).
\\
In both \cite{strominger} (see eqs. A.7 and A.10 therein) 
and \cite {leshouches} the authors
use the same method as Dullemond and van Beveren
in order to derive a hypergeometric differential equation,
but without giving many details on the $i\epsilon$-prescription.
In contrast to this the authors of \cite{bousso}
make extensive use of mode summation,
but do not give an $i\epsilon$-prescription.
\chapter{Summary and Conclusions}
We have carried together an overview of the basic geometric properties
of AdS and dS spacetimes in chapter \ref{chap:intro}.
All coordinate systems are cast into radial form
such that the AdS and dS internal coordinates are independent
of the radii of curvature $R\ads$ and $R\ds$.
The corresponding metrics are listed thereafter.
\\
Geodesics through a point $X$
on the hyperboloid in embedding space are examined then 
and found to be cuts of planes
(containing both $X$ and the origin of embedding space)
with the hyperboloid.
For AdS timelike geodesics are found to run on closed ellipses
in embedding space while lightlike geodesics are straight lines
and spacelike ones are hyperbolae.
For dS spacetime we find the same forms,
just with inverse causal classification.
\\
For AdS spacetimes time ordering
with respect to the time variable $t$ is found being
SO$\ssty(2,d\ads-1)$ invariant for causally connected points.
For dS spacetimes time ordering via the time variables $\tau$ and $T$
is found being SO$\ssty(1,d\ds-1)$ invariant
while time ordering via $t$ is not invariant
under SO$\ssty(1,d\ds-1)$ transformations moving
a point into another causal diamond.
\\
It would be nice to know also whether the time ordering
via $\theta {\ssty (\ovl{\tau}_x - \ovl{\tau}_y)}$
in Poincar\'e and planar coordinates for causally connected points
is invariant under the action of SO$\ssty(2,d\ads-1)$
respectively SO$\ssty(1,d\s)$,
however for reasons of time this question was not investigated.
\\
We have reviewed the method of Dullemond and van Beveren
and applying it we found all scalar propagators
including $i\epsilon$-prescriptions
for AdS and dS spacetimes of arbitrary dimension.
Depending (up to time ordering subtleties) only on 
an SO$\ssty(2,d\ads-1)$ respectively SO$\ssty(1,d\s)$
invariant quantity $\lambda$,
the propagators are invariant as well. 
\\
For AdS the propagator functions are found to be of the type
$\lambda^{-1} F {\ssty (a,b,c,\lambda^{-1})}$
with the parameters $a,b,c$ depending on the dimension $d\ads$
of AdS and the mass $m$ of the scalar field,
which is in agreement with the literature.
\\
For dS spacetime the propagator functions are found to be of the form
$F {\ssty (a,b,c,\lambda)}$
with the parameters $a,b,c$ also depending on the dimension $d\ds$
of dS and the mass $m$ of the scalar field
which agrees with the literature as well.
\\
A convenient advantage of the DvB method is
that it permits finding all propagator functions
including $i\epsilon$-prescriptions in one fell swoop
without having to perform a mode summation
or needing analytic continuation from euclidean time.
Moreover the conformal dimension $\Delta$
of a corresponding CFT field is delivered as a bonus.
A slight disadvantage lies in the integrals of normalization,
which become rather complicated for general reference points
and therefore are calculated for an especially simple reference point only.
\\
We expect the DvB method to work
for all spacetimes ST$_d$ of dimension $d$
which can be embedded in a $(d\!+\!1)$-dimensional
embedding space with pseudo-euclidean metric
and moreover represent a hypersurface $(R=const.)$
for a radial coordinate $R$ of embedding space.  
It is worth investigating whether and how the DvB method
can be extended and modified
in order to find propagators for gauge fields and matter fields.
%%

%%
%%_____________Suffix__________________
%%
\appendix
\chapter{Special functions}
	\section{Dirac's delta function}
In this section concrete realizations of the delta function are constructed
using the d'Alembertian operator, which generally is defined as
\begin{equation}
	\Box \equiv  \frac{1}{\sqrt{g}} \; \partial_{\mu} \, g^{\mu \nu} \! \sqrt{g} \; %%@
\partial_{\nu}
	\qquad \quad g \equiv \;\; \mid \det (g_{\mu \nu}) \mid
\end{equation}
In an $n$-dimensional coordinate space V with points $x=(x_1,\ldots,x_n)$
the $n$-dimensional delta function is characterized by the following two %%@
properties,
which any realization has to fulfill:
\begin{align}
	\label{eq:delta02}
	\delta^{\sssty (\!n\!)} {\ssty \! (x-y)} & = \, 0 
	\qquad {\ssty \forall \; x \neq y} \\
	\label{eq:delta03}
	\int\limits_V \!\! d^{n} \! x \: \delta^{\ssty (\!n\!)} {\ssty \! (x-y)} \: & = \: 1
\end{align}
Because of equation (\ref{eq:delta02}) the product 
$\delta^{\sssty (\!n\!)} {\ssty \! (x-y)} \, f{\ssty \! (x)}$
is vanishing at any point except at $x=y$ and thus can be replaced by 
$\delta^{\sssty (\!n\!)} {\ssty \! (x-y)} \, f{\ssty \! (y)}$. 
Integrating this product, $f{\ssty \! (y)}$
can be pulled outside the integral and applying equation (\ref{eq:delta03})
then yields the useful so-called shifting property:
\begin{equation}
	\label{eq:delta04}
	\int \limits_V \!\! d^{n} \! x \: \delta^{\ssty (\!n\!)} {\ssty \! (x-y)} \,
	f{\ssty \! (x)} \: = \: f{\ssty \! (y)}
	\qquad {\ssty \forall \; f{\sssty \! (x)}}
\end{equation}
The $n$-dimensional $\delta$-function is the product
of $n$ one-dimensional $\delta$-functions
\beq
	\delta^{\ssty (\!n\!)} {\ssty (x-y)} =
	\delta {\ssty (x_1-y_1)} \, \cdot \, \ldots \, \cdot \, \delta {\ssty (x_n-y_n)}
\eeq
with each of them fulfilling (\ref{eq:delta02}), (\ref{eq:delta03})
and therefore also (\ref{eq:delta04}) for $n \! = \! 1$.
Moreover we have the following relations \cite{mathworld}
for one-dimensional $\delta$-functions:
\begin{align}
	\del_{x_1} \theta {\ssty (x_1-y_1)}
		& = \, \delta {\ssty (x-y)}
		\\
	\label{eq:delta04_05}
	\delta ( f {\ssty (x_1)} )
		& = \, \sum \limits_k
			\frac{\delta {\ssty (x_1-x_{1k})}}
					{\; \mid \!\! (\del_{x_1} f){\ssty (x_{1k})} \!\! \mid \,}
		\\
	\label{eq:delta04_06}
	(x_1 \! - \! y_1) \, \del_{x_1} \, \delta {\ssty (x_1-y_1)}
		& = \, - \, \delta {\ssty (x_1-y_1)} \\
	\label{eq:delta04_07}
	(x_1 \! - \! y_1) \, \del_{x_1} \, \delta {\ssty (y_1-x_1)}
		& = \, + \, \delta {\ssty (x_1-y_1)}
\end{align}
with $\theta {\ssty (x_1-y_1)}$ being Heaviside's step function
and $x_{1k}$ the simple roots i.e. the zeros with power one of $f {\ssty %%@
(x_1)}$.
\\
The first considered space is ${\mathbb R}^n$
with the euclidean metric $g={\boldsymbol 1}_n$
which from the general definition of the d'Alembertian
returns the $n$-dimensional Laplacian
$\Delta \equiv \vec{\partial}^{\, 2}$.
If we feed the function
$((\vec{x}-\vec{y}) \, ^2+\epsilon^2)^{-\alpha}$ to the Laplacian,
we find for $\alpha={\textstyle \frac{n-2}{2}}$ that
\begin{equation*} 
	\Delta \frac{1}{((\vec{x}-\vec{y}) \, ^2+\epsilon^2)^{\frac{\sssty n-2}{\sssty %%@
2}}}\,
	= \,\frac{\epsilon^2}{((\vec{x}-\vec{y}) \, ^2+\epsilon^2)^{\frac{\sssty %%@
n+2}{\sssty 2}}} \; n(2-n)
\end{equation*}
For all points $\vec{x} \neq \vec{y}$ our $\epsilon$ can be sent to zero,
so the left hand side of the next equation is a candidate for the delta function %%@
in $R^n$.
\begin{equation*} 
	\underset{\epsilon \rightarrow 0}{\lim} \;
	\Delta \frac{1}{((\vec{x}-\vec{y}) \, ^2+\epsilon^2)^{\frac{\sssty n-2}{\sssty %%@
2}}}\,=\, 0
	\qquad {\ssty \forall \; \vec{x} \neq \vec{y}}
\end{equation*}
The next task is to compute the integral over ${\mathbb R}^n$ of the %%@
candidate function:
\begin{align*}
	\int \limits_{R^n} \!\! d^{n} \! x \: 
	\Delta \frac{1}{((\vec{x}-\vec{y}) \, ^2+\epsilon^2)^{\frac{\sssty n-2}{\sssty %%@
2}}}\,
	& = n(2-n)\int \limits_{R^n} \!\! d^{n} \! x \:
	\frac{\epsilon^2}{(\vec{x}\,^2+\epsilon^2)^{\frac{\sssty n+2}{\sssty 2}}} \\
	& = n(2-n)\int \limits_{R^n} \!\! d^{n} \! \left( \frac{\ssty x}{\ssty \epsilon} %%@
\right) \:
	\frac{\epsilon^{n+2}}{\epsilon^{n+2} \left( \frac{\vec{x} \, ^2}{\epsilon^2} %%@
+ 1 \right)
	^{\frac{\sssty n+2}{\sssty 2}}} \\
	& = n(2-n)\int \limits_{R^n} \!\! d^{n} \! x \:
	\frac{1}{(\vec{x} \, ^2 + 1)^{\frac{\sssty n+2}{\sssty 2}}} \\
	& = n(2-n) \underbrace{\int \limits_{S^n} \!\! d \Omega _n}
	\underbrace{ \int \limits_{0}^{\infty} \!\! dr
	\frac{r^{n-1}}{(r^2 + 1)^{\frac{\sssty n+2}{\sssty 2}}}} \\
	& = n(2-n) \;\; \frac{2 \pi \! ^{\frac{n}{2}}}{\Gamma (\frac{\sssty n}{\sssty %%@
2})}
	\qquad \;\;\; \frac{1}{n} \\
	& = (2-n) \; \frac{2 \pi \! ^{\frac{n}{2}}}{\Gamma (\frac{\sssty n}{\sssty %%@
2})}
\end{align*}
The entire integral has turned out to be independent of $\epsilon$
(the radial integration will be computed at the end of this section).
Thus a realization of the delta function for the ${\mathbb R}^n$ with $n \geq %%@
3$ has been found
which is well defined for real $\vec{x}$.
\begin{equation} 
	\underset{\epsilon \rightarrow 0}{\lim} \;
	\Delta \frac{1}{((\vec{x}-\vec{y}) \, ^2+\epsilon^2)^{\frac{\sssty n-2}{\sssty %%@
2}}}\,
	= \, (2-n) \; \frac{2 \pi \! ^{\frac{n}{2}}}{\Gamma (\frac{\sssty n}{\sssty %%@
2})} \;\;
	\delta^{\sssty (\!n\!)} {\ssty \! (\vec{x}-\vec{y})}
\end{equation}
Besides there are solutions of the homogeneous Laplace equation in ${\mathbb %%@
R}^n$ with $n \geq 3$ 
and $\epsilon$ not necessarily infinitesimal.
We can build them by replacing $x^1 \rightarrow x^1 \pm i^\frac{1}{2} %%@
\frac{\epsilon}{2}$
which then changes $\vec{x} \, ^2 \rightarrow (\vec{x} \, ^2 \pm i^\frac{1}{2} %%@
\epsilon x^1 + i \frac{\epsilon^2}{4})$.
They are also well defined for real $\vec{x}$.
\begin{equation*} 
	\Delta \frac{1}{(\vec{x} \, ^2 \pm i^\frac{1}{2} \epsilon x^1 +i %%@
\frac{\epsilon^2}{4})
	^{\frac{\sssty n-2}{\sssty 2}}}\,
	= \, 0
\end{equation*}
Now $n$-dimensional Minkowski spacetime is considered.
The notation is $g_+ \equiv \diag (+,-,\ldots,-)$
and $g_- \equiv \diag(-,+,\ldots,+)$.
${\mathbb M}^n_+$ is $n$-dimensional spacetime
equipped with $g_+$ and ${\mathbb M}^n_-$ is spacetime with $g_-$. \\
This leads to the d'Alembertians
$\Box _+ \equiv + \, \partial _0^2 - \vec{\partial} ^2$ for ${\mathbb M}^n_+$
and $\Box _- \equiv - \partial _0^2 + \vec{\partial} ^2$ for ${\mathbb %%@
M}^n_-$.
$x_\pm^2$ stands for the Minkowski square of the $n$-vector
$x = (x^0, x^1, \ldots, x^{n-1})$ formed with $g_\pm$. \\
For reasons of brevity we suppress the parameter $y$
in the argument of the delta function in the following calculations,
nevertheless the variable $x$ can simply be replaced by $(x-y)$,
with the d'Alembertian acting always on $x$. \\
We can find the following homogeneous solutions
of the d'Alembert equation in ${\mathbb M}^n_\pm$ with $n \geq 3$
and $\epsilon$ not necessarily infinitesimal.
They are well defined for real $x^\mu$.
\begin{align} 
	\Box_+ \frac{1}{(x_+^2 \pm i \epsilon x^0 - %%@
\frac{\epsilon^2}{4})^{\frac{\sssty n-2}{\sssty 2}}}\,
	& = 0 \\
	\Box_- \frac{1}{(x_-^2 \pm i \epsilon x^0 + %%@
\frac{\epsilon^2}{4})^{\frac{\sssty n-2}{\sssty 2}}}\,
	& = 0
\end{align}
They can be obtained by the shift $x^0 \rightarrow x^0 \pm i %%@
\frac{\epsilon}{2}$
which changes the Minkowski products $x^2_+ \rightarrow (x^2_+ \pm i %%@
\epsilon x^0 - \frac{\epsilon^2}{4})$
and $x^2_- \rightarrow (x^2_- \mp i \epsilon x^0 + \frac{\epsilon^2}{4})$.
Searching candidates for the delta function, one can check that
\begin{align}
	\label{eq:delta08}
	\Box_+ \frac{1}{(x_+^2 \pm i \epsilon^2)^{\frac{\sssty n-2}{\sssty 2}}}\,
	& = \,(\pm i)\frac{\epsilon^2}{(x_+^2 \pm i \epsilon^2)^{\frac{\sssty %%@
n+2}{\sssty 2}}} \; n(2-n) \\
	\label{eq:delta09}
	\Box_- \frac{1}{(x_-^2 \pm i \epsilon^2)^{\frac{\sssty n-2}{\sssty 2}}}\,
	& = \,(\pm i)\frac{\epsilon^2}{(x_-^2 \pm i \epsilon^2)^{\frac{\sssty %%@
n+2}{\sssty 2}}} \; n(2-n) 
\end{align}
For $x^2 \neq 0$ the $\epsilon$ can be send to zero giving two candidates for %%@
delta functions
$\delta^{\sssty (\!n\!)} {\ssty \! (x^2)}$.
The next step is showing they actually are candidates for $\delta^{\sssty %%@
(\!n\!)} {\ssty \! (x)}$.
This is done in ${\mathbb M}^n_+$ with a similar argumentation holding for %%@
${\mathbb M}^n_-$.
First the integral below is carried out on the hyperplane $x^0= \pm a$:
\begin{equation*}
\begin{split}
	& \int \!\! d^{n-1} \! x \; \partial_0 \!
	\left( \frac{1}{(x_+^2 \pm i \epsilon^2)^{\frac{\sssty n-2}{\sssty 2}}} \right)
	_{\!\! x^0 = \pm a} \\
	& = \int \!\! d^{n-1} \! x \;
	\frac{\pm a}{(-\vec{x} \, ^2 + a^2 \pm i \epsilon^2)^{\frac{\sssty n}{\sssty %%@
2}}} \; (2-n)
	\qquad \qquad \ssty{a>0} \\
	& = \pm \int \!\! d^{n-1} \! \left( \frac{\ssty x}{\ssty a} \right)
	\frac{a^{1+(n-1)-n}}{(-\frac{\vec{x}^2}{a^2} + 1 \pm i %%@
\frac{\epsilon^2}{a^2})^{\frac{\sssty n}{\sssty 2}}} \; (2-n)
	\qquad \ssty{a<\infty} \\
	& = \pm \underbrace{ \int \!\! d^{n-1} \! x \;
	\frac{1}{(-\vec{x} \, ^2 + 1 \pm i \epsilon^2)^{\frac{\sssty n}{\sssty 2}}} \; %%@
(2-n)} \\
	& = \pm \; A \qquad \qquad \qquad \; ^A
\end{split}
\end{equation*}
The value of the integral depends only on the sign of the $x^0$-value of the %%@
hyperplane.
Using the theorem of Gauss
\begin{equation*}
	\int \limits_V \!\! d^n \! x \; \vec{\nabla} \vec{j}
	= \int \limits_{\partial V} \!\! d\vec{f} \; \vec{j}
	= \int \limits_{\partial V} \!\! df \; \vec{n}\! _f \vec{j}
\end{equation*}
we can compute the integral of the candidate functions over a slice in %%@
Minkowski spacetime 
between $x^0=a$ and $x^0=b$ with $b>a>0$, so the slice does not contain %%@
the origin.
\begin{equation*}
\begin{split}
	& \int \limits_{slice} \!\! d^n \! x \; \Box_+ \frac{1}{(x_+^2 \pm i %%@
\epsilon^2)^{\frac{\sssty n-2}{\sssty 2}}}
	\qquad \qquad \ssty{\vec{\nabla} \vec{\nabla} \: = \: \partial_\mu g^{\mu %%@
\nu} \partial_\nu} \: = \: \Box \\
	& = \int \limits_{slice} \!\! d^n \! x \; \vec{\nabla} \vec{\nabla}
	\frac{1}{(x_+^2 \pm i \epsilon^2)^{\frac{\sssty n-2}{\sssty 2}}}
	= \int \limits_{\partial slice} \!\! df \; \vec{n}\! _f \vec{\nabla}
	\frac{1}{(x_+^2 \pm i \epsilon^2)^{\frac{\sssty n-2}{\sssty 2}}} \\
	& = \int \limits_{x^0=b} \!\! d^{n-1} \! x \;
	\begin{pmatrix} \ssty{1}          \\ \ssty{0}          \\ \vdots \end{pmatrix}
	\begin{pmatrix} \ssty{\partial_0} \\ \ssty{\partial_1} \\ \vdots \end{pmatrix} 
	\frac{1}{(x_+^2 \pm i \epsilon^2)^{\frac{\sssty n-2}{\sssty 2}}} \\
	& \quad + \int \limits_{x^0=a} \!\! d^{n-1} \! x \;
	\begin{pmatrix} \ssty{-1}         \\ \ssty{0}          \\ \vdots \end{pmatrix}
	\begin{pmatrix} \ssty{\partial_0} \\ \ssty{\partial_1} \\ \vdots \end{pmatrix} 
	\frac{1}{(x_+^2 \pm i \epsilon^2)^{\frac{\sssty n-2}{\sssty 2}}} \\
	& \quad + \quad \ssty{\text{vanishing boundary terms with one } x^i= \; \pm %%@
\infty \quad i=1,\ldots,n-1} \\
	& = \int \limits_{x^0=b} \!\! d^{n-1} \! x \; \partial_0
	\frac{1}{(x_+^2 \pm i \epsilon^2)^{\frac{\sssty n-2}{\sssty 2}}} \;\;
	- \int \limits_{x^0=a} \!\! d^{n-1} \! x \; \partial_0
	\frac{1}{(x_+^2 \pm i \epsilon^2)^{\frac{\sssty n-2}{\sssty 2}}} \\
	& = \; A \; - \; A \; = \; 0
\end{split}
\end{equation*}
For $b<a<0$ we have the same result.
As was found before, we can send $\epsilon$ to zero for $x^2 \neq 0$,
hence contributions to the integral just calculated
can come from the lightcone $x^2 = 0$ only. \\
The integrand is symmetric under sign change of each coordinate,
so there is no possibility for a cancellation of nonzero contributions.
With the integral vanishing for arbitrary $a$ and $b$,
i.e. to position and thickness of the slice,
we can conclude that the integrand must indeed be zero
on and off the lightcone except the origin. \\
In the end of this section we will calculate in detail that
\begin{align}
	\label{eq:delta10}
	\int \limits_{M^n_+} \!\! d^{n} \! x \: 
	\Box_+ \frac{1}{(x_+^2 \pm i \epsilon^2)^{\frac{\sssty n-2}{\sssty 2}}}\,
	& = (\mp i)^{n+1} \; (n-2) \; \frac{2 \pi \! ^{\frac{n}{2}}}{\Gamma %%@
(\frac{\sssty n}{\sssty 2})} \\
	\label{eq:delta11}
	\int \limits_{M^n_-} \!\! d^{n} \! x \: 
	\Box_- \frac{1}{(x_-^2 \pm i \epsilon^2)^{\frac{\sssty n-2}{\sssty 2}}}\,
	& = (\pm i) \; (n-2) \; \frac{2 \pi \! ^{\frac{n}{2}}}{\Gamma (\frac{\sssty %%@
n}{\sssty 2})}
\end{align}	
Thus two realizations of the delta function in $n$-dimensional Minkowski %%@
spacetime with $n \geq 3$ are found.
They are well defined for real $x^\mu$.
\begin{align}
	\label{eq:delta12}
    \underset{\epsilon \rightarrow 0}{\lim} \;
	\Box_+ \frac{1}{((x-y)_+^2 \pm i \epsilon^2)^{\frac{\sssty n-2}{\sssty 2}}}\,
	& = (\mp i)^{n+1} \; (n-2) \; \frac{2 \pi \! ^{\frac{n}{2}}}{\Gamma %%@
(\frac{\sssty n}{\sssty 2})}
	\;\; \delta^{\sssty (\!n\!)} {\ssty \! (x-y)}  \\ 
	\underset{\epsilon \rightarrow 0}{\lim} \;
	\Box_- \frac{1}{((x-y)_-^2 \pm i \epsilon^2)^{\frac{\sssty n-2}{\sssty 2}}}\,
	& = (\pm i) \; (n-2) \; \frac{2 \pi \! ^{\frac{n}{2}}}{\Gamma (\frac{\sssty %%@
n}{\sssty 2})}
    \;\; \delta^{\sssty (\!n\!)} {\ssty \! (x-y)}
\end{align}	
Equation (\ref{eq:delta12}) is in agreement with Dullemond and van Beveren
\cite{dvb} wherein the case ${\mathbb M}^4_+$ is considered. \\ 
What remains is to compute the integrals used above. The integral
\begin{equation}
	\label{eq:delta14}
	\int \limits_{0}^{\infty} \!\! dr \;
	\frac{r^{n-1}}{(r^2 + 1)^{\frac{\sssty n+2}{\sssty 2}}} = \frac{1}{n}
\end{equation}
can be calculated using Euler's Beta function defined in Bronstein %%@
\cite{bronstein} as
\beqs
	B{\ssty (\!a\!,b\!)} \, = \,
		\frac{\, \Gamma{\ssty (\!a\!)} \, \Gamma{\ssty (\!b\!)}}{\Gamma{\ssty %%@
(\!a+b\!)}}
		\, = \, \int \limits_0^1 \!\! ds \, s^{a\!-\!1} \, (1\!-\!s)^{b\!-\!1}
		\qquad \qquad ^{a>0}_{b>0}
\eeqs
Using the substitutions $t=(1\!-\!s) \! /s$ and $r^2 \! =t$ we find
\beq
	\label{eq:delta155}
	\int \limits_0^\infty \!\! dr \, \frac{r^\alpha}{(r^2\!\!+\!1)^\beta}
	\, = \, 
	\frac{\Gamma {\ssty (\frac{\alpha+1}{2})} \, 
			\Gamma {\ssty (\beta-\frac{\alpha+1}{2})}}
			{2 \, \Gamma {\ssty (\beta)}}
	\qquad \qquad {\ssty \beta \, > \, \frac{\alpha \! + \! 1}{2} \, > \, 0}
\eeq
which for $\alpha = (n\!-\!1)$ and $\beta = (n \! + \! 2) \! /2$
gives just ${ \frac{1}{n}}$. \\
The integrand in (\ref{eq:delta14}) vanishes for large values of
$\mid \!\! r \!\! \mid$ and possesses poles at $\pm i$.
Thus the path of integration can be rotated
by $\pm 45^\circ$ in the complex $r$-plane
without changing the value of the integral:
\begin{equation}
	\label{eq:delta15}
	\int \limits_{0}^{\infty} \!\! dr \;
	\frac{r^{n-1}}{(r^2 + 1)^{\frac{\sssty n+2}{\sssty 2}}}
	= \int \limits_{0}^{i^\frac{1}{2}\infty} \!\! dr \;
	\frac{r^{n-1}}{(r^2 + 1)^{\frac{\sssty n+2}{\sssty 2}}}
	= \int \limits_{0}^{(-i)^\frac{1}{2}\infty} \!\! dr \;
	\frac{r^{n-1}}{(r^2 + 1)^{\frac{\sssty n+2}{\sssty 2}}}
	= \frac{1}{n}
\end{equation}
This property is useful for the calculation of the $+i \epsilon^2$ integral in %%@
${\mathbb M}^n_+$:
\begin{equation*}
\begin{split}
	& \int \limits_{M^n_+} \!\! d^{n} \! x \: 
	\Box_+ \frac{1}{(x_+^2 + i \epsilon^2)^{\frac{\sssty n-2}{\sssty 2}}}\,
	= \int \limits_{M^n_+} \!\! d^{n} \! x \:
	\frac{i \epsilon^2}{(x_+^2 + i \epsilon^2)^{\frac{\sssty n+2}{\sssty 2}}} \; %%@
n(2-n) \\
	& = \int \limits_{M^n_+} \!\! d^{n} \! \left( \frac{\ssty x}{\ssty \epsilon} %%@
\right) \:
	\frac{i \epsilon^{n+2}}{\epsilon^{n+2} \left( \frac{x \, ^2_+}{\epsilon^2} + %%@
i \right)
	^{\frac{\sssty n+2}{\sssty 2}}} n(2-n)
	= \int \limits_{M^n_+} \!\! d^{n} \! x \:
	\frac{i}{(x_+^2 + i)^{\frac{\sssty n+2}{\sssty 2}}} \; n(2-n) \\
	& = \int \limits_{R^{n-1}} \!\! d^{n-1} \! x \: \int \limits_{-\infty}^{+\infty} %%@
\!\! dx^0 \:
	\frac{i}{(x^{0^2} - \vec{x} \,^2 + i)^{\frac{\sssty n+2}{\sssty 2}}} \; n(2-n)
	\qquad ^{\text{now clockwise rotation of path of integration}} _{\text{so %%@
that the poles are not crossed}}\\
	& = \int \limits_{R^{n-1}} \!\! d^{n-1} \! x \: \int \limits_{-i\infty}^{+i\infty} %%@
\!\! dx^0 \:
	\frac{i}{(x^{0^2} - \vec{x} \,^2 + i)^{\frac{\sssty n+2}{\sssty 2}}} \; n(2-n)
	\qquad \qquad x^0=iz \\
	& = \int \limits_{R^{n-1}} \!\! d^{n-1} \! x \: \int \limits_{-\infty}^{+\infty} %%@
\!\! dz \:
	\frac{-1}{(-z^2 - \vec{x} \,^2 + i)^{\frac{\sssty n+2}{\sssty 2}}} \; n(2-n) \\
	& = \int \limits_{R^{n}} \!\! d^{n} \! x \:
	\frac{-1}{(-\vec{x} \, ^2 + i)^{\frac{\sssty n+2}{\sssty 2}}} \; n(2-n) \\
	& = \int \limits_0^{\infty} \!\! dr \:
	\frac{-r^{n-1}}{(-r^2 + i)^{\frac{\sssty n+2}{\sssty 2}}} \; n(2-n)
	\; \frac{2 \pi \! ^{\frac{n}{2}}}{\Gamma (\frac{\sssty n}{\sssty 2})}
	\qquad \qquad \qquad \qquad r = u \, i^{-\frac{1}{2}} \quad \rightarrow %%@
\quad r^2=-iu^2\\
	& = \int \limits_0^{i^{\frac{1}{2}}\infty} \!\! du \:
	\frac{-u^{n-1}}{(u^2 + 1)^{\frac{\sssty n+2}{\sssty 2}}} \; n(2-n)
	\; \frac{2 \pi \! ^{\frac{n}{2}}}{\Gamma (\frac{\sssty n}{\sssty 2})}
	\; i^{\frac{-1-(n-1)-(n+2)}{2}}\\
	& = (-i)^{n+1} \; (n-2) \; \frac{2 \pi \! ^{\frac{n}{2}}}{\Gamma (\frac{\sssty %%@
n}{\sssty 2})}
\end{split}	
\end{equation*}	
This is just the result given in (\ref{eq:delta10}).
The other integrals are solved in the same way,
first employing a rotation of the path of integration
in the direction which prevents from crossing the poles,
and then suitably substituting in order to obtain
one of the rotated integrals in (\ref{eq:delta15}).
The values of the four integrals (\ref{eq:delta10}) and (\ref{eq:delta11})
all are obtained independently using this method of integration. \\
When battling with the imaginary roots we can use
$i^{-\frac{1}{2}}=(-i)^{\frac{1}{2}}$ and therefore %%@
$(-i)^{-\frac{1}{2}}=i^{\frac{1}{2}}$.
The root of a complex number means the positive root
which by usual definition always is situated in the right half of the complex %%@
plane. \\
Now the consistency of the prefactors is checked.
For complex numbers $z$ in the upper half of the complex plane we have %%@
$(-z)^\frac{1}{2} = -iz^\frac{1}{2}$
while for complex numbers in the lower half $(-z)^\frac{1}{2} = %%@
iz^\frac{1}{2}$ holds true.
Applying this when comparing (\ref{eq:delta08}) and (\ref{eq:delta09}) we %%@
find
\begin{align*}
	\Box_+ \frac{1}{(x_+^2 \pm i \epsilon^2)^{\frac{\sssty n-2}{\sssty 2}}}\,
	& = \, -(\mp i)^{n+2} \Box_- \frac{1}{(x_-^2 \mp i \epsilon^2)^{\frac{\sssty %%@
n-2}{\sssty 2}}}\,
\end{align*}
Plugging this relation into equation (\ref{eq:delta10}) yields equation %%@
(\ref{eq:delta11})
confirming the correctness of the integration results given above.
	\section{Hypergeometric function}
	\label{sec:hypergeo}
	%\newcommand{\del}{\partial}
%\newcommand{\beq}{\begin{equation}}
%\newcommand{\eeq}{\end{equation}}
%\newcommand{\beqs}{\begin{equation*}}
%\newcommand{\eeqs}{\end{equation*}}
%\newcommand{\ssty}{\scriptstyle}
%\newcommand{\sssty}{\scriptscriptstyle}
%\newcommand{\ads}{_{\sssty \!A\!d\!S}}
%\newcommand{\dvb}{_{\sssty \!D\!v\!B}}

%
%\documentclass[a4paper,10pt]{article}
%
%\usepackage {amsmath}
%\usepackage {amssymb}
%\usepackage {amsxtra}
%\usepackage {epsfig}
%\usepackage {float}
%\usepackage {latexsym}
%
%\begin{document}
%
In this section we list some properties of hypergeometric functions
which can be found in \cite{astegun} by Abramowitz and Stegun 
and in \cite{bateman} by Bateman and Erdelyi.
The hypergeometric differential equation reads:
% -------------------- equation hypergeo 01 --------------------------------------
\begin{align}
	0 \, = \, \biggl[ z (1 - z) \, \partial _z ^2 \,
	                        + \Bigl[ c - z (a \! + \! b \! + \! 1) \Bigr] \partial _z
							- ab \biggr] \,
	F {\scriptstyle (a,b,c, z)}
\end{align}
It has three singular points at $z=0, \, 1, \, \infty$.
The regular solution of the hypergeometric equation
is given by the hypergeometric function,
which is defined as the analytic continuation
of the Gauss hypergeometric series
%
% -------------------- equation hypergeo 02 --------------------------------------
\begin{align}
	F {\scriptstyle (a,b,c, z)} & \equiv \sum _{k=0} ^\infty
	\frac{(a)_k (b)_k}{(c)_k} \, \frac{z^k}{k!} \\
	(a)_k \equiv \frac{\Gamma{\scriptstyle (a+k)}}{\Gamma{\scriptstyle (a)}}
	& = a (a+1) \ldots (a+k-1) \qquad \quad (a)_0 \! = 1
\end{align}
which converges for all $\mid \! z \! \mid < 1$.
For $\mid \! z \! \mid = 1$ it also converges
if Re $(c\!-\!a\!-\!b) > 0$ but diverges if Re $(c\!-\!a\!-\!b) \leq -1$. \\
The hypergeometric function is symmetric under exchange of $a$ and $b$
and is regular at $z\!=0$ with $F {\scriptstyle (a,b,c, 0)} = 1$ and
\beq
	\label{eq:hypergeo03_5}
	F {\scriptstyle (a,b,c, 1)} = \, 
	\frac{\, \Gamma{\scriptstyle (c)} \, \Gamma{\scriptstyle (c-a-b)}}
			{\, \Gamma{\scriptstyle (c-a)} \, \Gamma{\scriptstyle (c-b)}}
	\qquad \qquad
	^{c \, \neq \, 0,-1,-2, \, \ldots}_{\text{Re} \, (c-a-b) \, > \, 0} 
\eeq
The first derivation of the hypergeometric function is given by
%
% -------------------- equation hypergeo 04 --------------------------------------
\beq
	\label{eq:hypergeo04}
	\frac{d}{dz} \, F {\scriptstyle (a,b,c, z)} \, 
	= \,\frac{\, ab \, }{c} \, F {\scriptstyle (a+1,b+1,c+1, z)} 
\eeq
Moreover there exist a number of useful linear transformation formulas
for the hypergeometric function:
% -------------------- equation hypergeo 05 --------------------------------------
\begin{align}
	\label{eq:hypergeo04_5}
	F {\scriptstyle (a,b,c, z)} \,
	& = \, \bigl(1\!-\!z \bigr)^{c-a-b} F \bigl( c\!-\!a, \, c\!-\!b, \, c, \,z \bigr) \\
	\label{eq:hypergeo05}
	& = \, \biggl( \! \frac{1}{1-z} \! \biggr)^{\!a} \,
			F \Bigl( a,c\!-\!b,c, \frac{\ssty z}{\ssty z-1} \Bigr) \\
	\label{eq:hypergeo06}
	& = \, \biggl( \! \frac{1}{1-z} \! \biggr)^{\!b} \,
			F \Bigl( b,c\!-\!a,c, \frac{\ssty z}{\ssty z-1} \Bigr) \\
		\notag \\
	& = \, \frac{\Gamma{\ssty \! (\!c\!)} \,\Gamma{\ssty (\!b-a\!)}}
					 {\Gamma{\ssty \! (\!b\!)} \,\Gamma{\ssty (\!c-a\!)}} \,
					 \biggl( \! \frac{1}{-z} \! \biggr)^{\!a} \,
					 F \Bigl( a,a\!-\!c\!+\!1,a\!-\!b\!+\!1, \frac{\ssty 1}{\ssty z} \Bigr) \notag \\
	\label{eq:hypergeo07}
	& \qquad + \; \frac{\Gamma{\ssty \! (\!c\!)} \,\Gamma{\ssty (\!a-b\!)}}
					 {\Gamma{\ssty \! (\!a\!)} \,\Gamma{\ssty (\!c-b\!)}} \,
					 \biggl( \! \frac{1}{-z} \! \biggr)^{\!b} \,
					 F \Bigl( b,b\!-\!c\!+\!1,b\!-\!a\!+\!1, \frac{\ssty 1}{\ssty z} \Bigr) \\
		\notag \\
	& = \, \frac{\Gamma{\ssty \! (\!c\!)} \,\Gamma{\ssty (\!b-a\!)}}
					 {\Gamma{\ssty \! (\!b\!)} \,\Gamma{\ssty (\!c-a\!)}} \,
					 \biggl( \! \frac{1}{1-z} \! \biggr)^{\!a} \,
					 F \Bigl( a,c\!-\!b,a\!-\!b\!+\!1, \frac{\ssty 1}{\ssty 1-z} \Bigr) \notag \\
	\label{eq:hypergeo08}
	& \qquad + \; \frac{\Gamma{\ssty \! (\!c\!)} \,\Gamma{\ssty (\!a-b\!)}}
					 {\Gamma{\ssty \! (\!a\!)} \,\Gamma{\ssty (\!c-b\!)}} \,
					 \biggl( \! \frac{1}{1-z} \! \biggr)^{\!b} \,
					 F \Bigl( b,c\!-\!a,b\!-\!a\!+\!1, \frac{\ssty 1}{\ssty 1-z} \Bigr)
\end{align}
%
%\end{document}

%%
\chapter{Computations}
	\section{Time ordering and delta sources}
	\label{sec:timord_delta}
In this section we investigate how time ordering prescriptions
generate delta functions in time for the Feynman propagator.
We will take the time ordering on the AdS hyperboloid as an example,
with the results for Minkowski spacetime, the universal covering AdS
and dS spacetime originating from simpler similar computations.
Time ordering on the AdS hyperboloid is given
by equation (\ref{eq:ads_timord_hyp09}):
\begin{align}
	T_{\text{hyp}} \, \phi {\ssty (\!x\!)} \, \phi {\ssty (\!y\!)} \,
	& = \, \theta ({\ssty \sin (t_x - t_y)\!}) \,
			\phi {\ssty (\!x\!)} \, \phi {\ssty (\!y\!)}
			\qquad \qquad \quad {\ssty t_{x,y} \, {\sssty \in } \; ]-\pi,+\pi]}
			\notag \\
	& \quad \; + \, \theta ({\ssty \sin (t_y - t_x)\!}) \, 
			\phi {\ssty (\!y\!)} \, \phi {\ssty (\!x\!)}
\end{align}
Thereby the Feynman propagator reads:
\begin{align}
		G \!_F \, {\scriptstyle(\!x\!,y\!)}
		& \equiv \, \left\langle 0 \! \mid T_{\text{hyp}} \,
			\phi {\ssty(\!x\!)} \,	\phi {\ssty(\!y\!)} \mid \! 0 \right\rangle
			\notag \\
		& = \, \theta ({\ssty \sin (t_x - t_y)\!}) \,
				\left\langle 0 \! \mid \phi {\ssty(\!x\!)} \,
					 \phi {\ssty(\!y\!)} \mid \! 0 \right\rangle
			\notag \\
		& \quad + \, \theta ({\ssty \sin (t_y - t_x)\!}) \,
				\ub{\left\langle 0 \! \mid \phi {\ssty(\!y\!)} \,
					 \phi {\ssty(\!x\!)} \mid \! 0 \right\rangle}
				_{\; \text{shorter} \, = \, \left\langle \phi {\ssty(\!y\!)}
																	\phi {\ssty(\!x\!)} \right\rangle_0}
\end{align}
Applying the Klein-Gordon operator yields:
\begin{align}
	& \left( \sigma \Box \ADS _x \!\! + m^{\sssty 2} \right)
	\, G \! _F {\ssty(\!x\!,y\!)}
		= \Bigl( {\ssty \frac{1}{R\ads^2(1 \! + \! \vec{x}\,^2)}} \, \partial _{t_x}^2  
						\! + \sigma \Box \ADS _{\vec{x}} \! + m^{\sssty 2} \Bigr)
			\notag \\
			& \qquad \qquad \qquad \qquad \qquad
			\Bigl[ \theta ({\ssty \sin (t_x - t_y)\!})
				\bigl\langle \phi {\ssty(\!x\!)} \phi {\ssty(\!y\!)} \bigr\rangle_{\! 0}
				 \! + \, 	\theta ({\ssty \sin (t_y - t_x)\!})
				\bigl\langle \phi {\ssty(\!y\!)} \phi {\ssty(\!x\!)} \bigr\rangle_{\! 0}
			\Bigr]
			\\
	& = \, {\ssty \frac{1}{R\ads^2(1 \! + \! \vec{x}\,^2)}} \, \partial _{t_x} \!
			\Bigl[ \cancel{\delta ({\ssty \sin (t_x - t_y)\!}) \cos(t_x \!\! - \! t_y)
					\bigl\langle \phi {\ssty(\!x\!)} \phi {\ssty(\!y\!)} \bigr\rangle_{\! 0}} %%@
\!
					+ \, \theta ({\ssty \sin (t_x - t_y)\!}) \, \partial _{t_x} \!
					\bigl\langle \phi {\ssty(\!x\!)} \phi {\ssty(\!y\!)} \bigr\rangle_{\! 0}
			\notag \\
			& \qquad \qquad \qquad \quad
					\cancel{- \, \delta ({\ssty \sin (t_y - t_x)\!}) \cos(t_y \!\! - \! t_x)
					\bigl\langle \phi {\ssty(\!y\!)} \phi {\ssty(\!x\!)} \bigr\rangle_{\! 0}} %%@
\!
					+ \, \theta ({\ssty \sin (t_x - t_y)\!}) \, \partial _{t_x} \!
					\bigl\langle \phi {\ssty(\!y\!)} \phi {\ssty(\!x\!)} \bigr\rangle_{\! 0}
			\Bigr]
			\notag \\
			& \qquad
			+ \,\theta ({\ssty \sin (t_x - t_y)\!})
				(\sigma \Box \ADS _{\vec{x}} \! + m^{\sssty 2})
				\bigl\langle \phi {\ssty(\!x\!)} \phi {\ssty(\!y\!)} \bigr\rangle_{\! 0} \!
			+ \,\theta ({\ssty \sin (t_y - t_x)\!})
				(\sigma \Box \ADS _{\vec{x}} \! + m^{\sssty 2})
				\bigl\langle \phi {\ssty(\!y\!)} \phi {\ssty(\!x\!)} \bigr\rangle_{\! 0}
			\notag \\
			\intertext{ \scriptsize The two terms crossed out cancel each other. 
							On the hyperboloid we have
							$ t_{x,y} \, {\sssty \in } $ $ ]-\pi,+\pi]$
							and thereby $(t_x \! -t_y) \, {\sssty \in } \; ]-2\pi,+2\pi[$.
							Thus the delta functions become nonzero only for
							$(t_x \!\!-\! t_y) = 0,\pm \pi$.
							Inserting these values in (\ref{eq:dvb_ads50})
							we find that $\lambda_{\pm}$ becomes real.
							Inserting real $\lambda_{\pm}$ in (\ref{eq:dvb_ads74})
							yields $0 =	G {\ssty (\!x\!,y\!)} = \left\langle [\phi {\ssty(\!x\!)},
							\phi {\ssty(\!y\!)}] \right\rangle_{\! 0}$.
							Therefore we can use
							$\left\langle \phi {\ssty(\!x\!)} \phi {\ssty(\!y\!)} %%@
\right\rangle_{\! 0}
							= \left\langle \phi {\ssty(\!y\!)} \phi {\ssty(\!x\!)} %%@
\right\rangle_{\! 0}$
							and the two terms crossed out cancel each other
							for all $(t_x \!\!-\! t_y \! ) = 0,\pm \pi$
							and vanish for all other values of $t_{x,y}$.}
			\notag \\
	& = \, {\ssty \frac{1}{R\ads^2(1 \! + \! \vec{x}\,^2)}}
			\Bigl( \delta ({\ssty \sin (t_x - t_y)\!}) \cos(t_x \!\! - \! t_y)
					\, \partial _{t_x} \!
					\bigl\langle \phi {\ssty(\!x\!)} \phi {\ssty(\!y\!)} \bigr\rangle_{\! 0} %%@
\!
			\notag \\
					& \qquad \qquad \qquad
					- \delta ({\ssty \sin (t_y - t_x)\!}) \cos(t_y \!\! - \! t_x)
					\, \partial _{t_x} \!
					\bigl\langle \phi {\ssty(\!y\!)} \phi {\ssty(\!x\!)} \bigr\rangle_{\! 0} %%@
\!
			\Bigr)
			\notag \\
			& \qquad
			+ \,\theta ({\ssty \sin (t_x - t_y)\!})
				\ub{(\sigma \Box \ADS _x \! + m^{\sssty 2})
				\bigl\langle \phi {\ssty(\!x\!)} \phi {\ssty(\!y\!)} \bigr\rangle_{\! 0}}_0
			+ \; \theta ({\ssty \sin (t_y - t_x)\!})
				\ub{(\sigma \Box \ADS _x \! + m^{\sssty 2})
				\bigl\langle \phi {\ssty(\!y\!)} \phi {\ssty(\!x\!)} \bigr\rangle_{\! 0}}_0
		\notag \\
	\intertext{\scriptsize The last two terms vanish since the field $\phi %%@
{\ssty(\!x\!)}$
					fulfills the homogeneous Klein-Gordon equation.
					For the next step we use equation (\ref{eq:delta04_05})
					with $(t_x \!\!-\! t_y) = 0,\pm \pi$
					as possible roots of the sine function.}
		\notag \\
	& = \, {\ssty \frac{1}{R\ads^2(1 \! + \! \vec{x}\,^2)}}
			\, \partial _{t_x} \!
			\bigl\langle \phi {\ssty(\!x\!)} \phi {\ssty(\!y\!)} \bigr\rangle_{\! 0}
			\notag \\
			& \qquad \qquad
			\biggl[
			\cos (t_x \!\! - \! t_y) \frac{\, \delta {\ssty (t_x - t_y)}}{\mid \! \cos \, 0  %%@
\! \mid}
			+ \cos (t_x \!\! - \! t_y)
			\frac{\, \delta {\ssty (t_x - t_y-\pi)}}{\mid \! \cos \, \pi  \! \mid}
			+ \cos (t_x \!\! - \! t_y)
			\frac{\, \delta {\ssty (t_x - t_y+\pi)}}{\mid \! \cos \, -\pi  \! \mid}
			\biggr]
			\notag \\
	& \qquad + \, {\ssty \frac{1}{R\ads^2(1 \! + \! \vec{x}\,^2)}}
			\, \partial _{t_x} \!
			\bigl\langle \phi {\ssty(\!y\!)} \phi {\ssty(\!x\!)} \bigr\rangle_{\! 0}
			\notag \\
			& \qquad \qquad
			\biggl[
			\cos (t_y \!\! - \! t_x) \frac{\, \delta {\ssty (t_y - t_x)}}{\mid \! \cos \, 0  %%@
\! \mid}
			+ \cos (t_y \!\! - \! t_x)
			\frac{\, \delta {\ssty (t_y - t_x-\pi)}}{\mid \! \cos \, \pi  \! \mid}
			+ \cos (t_y \!\! - \! t_x)
			\frac{\, \delta {\ssty (t_y - t_x+\pi)}}{\mid \! \cos \, -\pi  \! \mid}
			\biggr]
			\notag \\
			\notag \\
	& = \, {\ssty \frac{1}{R\ads^2(1 \! + \! \vec{x}\,^2)}}
			\Bigl( \partial _{t_x} \!
				\left\langle 0 \! \mid \phi {\ssty(\!x\!)} \,
					 \phi {\ssty(\!y\!)} \mid \! 0 \right\rangle
			- \partial _{t_x} \!
				\left\langle 0 \! \mid \phi {\ssty(\!y\!)} \,
					 \phi {\ssty(\!x\!)} \mid \! 0 \right\rangle \! \Bigr)	
			\notag \\
		& \qquad \qqqquad \qqqquad
			\Bigl[ \delta {\ssty (t_x - t_y)}
			- \delta {\ssty (t_x - t_y-\pi)} - \delta {\ssty (t_x - t_y+\pi)} \Bigr]
			\notag \\
			\notag \\
	\label{timord_delta04}
	& = \, {\ssty \frac{1}{R\ads^2(1 \! + \! \vec{x}\,^2)}}
			\Bigl( \partial _{t_x} \! G^+ {\scriptstyle(\!x\!,y\!)}
			- \partial _{t_x} \! G^- {\scriptstyle(\!x\!,y\!)} \Bigr)
			\Bigl[ \delta {\ssty (t_x - t_y)}
			- \delta {\ssty (t_x - t_y-\pi)} - \delta {\ssty (t_x - t_y+\pi)} \Bigr]
\end{align}
\\
We thereby find that applying the Klein-Gordon operator
to the Feynman propagator generates delta functions in the time variable.
Through the standard time ordering via $\theta {\ssty (t_x \! - t_y)}$
in Minkowski spacetime and the universal covering AdS
and $\theta {\ssty (\tau_x \! - \tau_y)}$ in dS spacetime 
only one single $\delta {\ssty (t_x - t_y)}$ term is generated.
\\
However as a consequence of the special time ordering prescription
via $\theta ({\ssty \sin (t_x - t_y)\!})$
due to the closed timelike curves on the AdS hyperboloid
we encounter three delta functions in time.
We obtain exactly the same result
using the alternative time ordering via
$( \theta {\ssty (t_x \! - t_y)\!} \, \theta {\ssty (t_y \! - t_x+\pi)\!}
+ \theta {\ssty (t_y \! - t_x-\pi)\!})$.
\\
The difference of the time derivatives of the Wightman functions $G^\pm$
contributes delta functions for the spatial variables.
Therefore altogether we find that applying the Klein-Gordon operator
to the Feynman propagator produces delta sources
in certain spacetime points.
\\
In all cases mentioned above there is one delta source
situated at the coincident point $x=y$
while on the AdS hyperboloid there also  is a delta source
(with factor 2 and opposite sign, see subsection \ref{sec:ads_comhyp})
sitting at the antipodal point $x=\snake{y}$.

	\section{DvB method for de Sitter spacetime in static coordinates}
	\label{sec:props_ds_t}
		\subsection{General properties of propagators in dS}
In this section we apply the method of Dullemond and van Beveren
to deSitter spacetime.
The definitions and relations between the different propagators
for hermitian scalar fields $\phi {\ssty(\!x\!)}$ in dS
are the same as the ones listed for AdS in section \ref{ads_genprops}.
Here $x=(t,\vec{x})$ is a point in dS in static coordinates
with $t \equiv x^0$ serving as time coordinate.
With the overall sign $\sigma = \pm 1$ of the embedding space's metric
introduced in equation (\ref{eq:ds_coord01}),
the Klein-Gordon operator on dS reads $(\sigma \Box \ds + m^2)$. \\
Because our field $\phi {\ssty(\!x\!)}$ again fulfills
the Klein-Gordon equation, we have
\begin{align}
	\left( \sigma \Box \DS _x \!\! + m^{\sssty 2} \right)
		G^\pm {\ssty(\!x\!,y\!)} = \;\; & 0 \, = \,
		\left( \Box \DS _x \!\! + \sigma m^{\sssty 2} \right)
		G^\pm {\ssty(\!x\!,y\!)} \\
	\rightarrow \quad \left( \sigma \Box \DS _x \!\! + m^{\sssty 2} \right)
		G^{\sssty(\!1\!)} {\ssty(\!x\!,y\!)} = \;\; & 0 \, = \, 
		\left( \Box \DS _x \!\! +\sigma  m^{\sssty 2} \right)
		G^{\sssty(\!1\!)} {\ssty(\!x\!,y\!)} \\
	\rightarrow \quad \left( \sigma \Box \DS _x \!\! + m^{\sssty 2} \right) \; 
		G {\ssty(\!x\!,y\!)} \;\, = \;\; & 0 \, = \,
		\left( \Box \DS _x \!\! + \sigma m^{\sssty 2} \right) \; 
		G {\ssty(\!x\!,y\!)}
\end{align}
and
\begin{equation}
	\left( \sigma \Box \DS _x \!\! + m^{\sssty 2} \right)
	G \! _F {\ssty(\!x\!,y\!)}
	\; = \; \left( \sigma \Box \DS _x \!\! + m^{\sssty 2} \right)
	G \! _R {\ssty(\!x\!,y\!)}
	\; = \; \left( \sigma \Box \DS _x \!\! + m^{\sssty 2} \right)
	G \! _A {\ssty(\!x\!,y\!)}	
\end{equation}
Starting from the definition (\ref{eq:dvb_ads08})
of the Feynman propagator 
and using equation (\ref{eq:ds_coord23})
\beq
	\sigma \Box \DS _x = \frac{1}{R\ds^2(1-\vec{x}\,^2)} \, \partial _{t_x}^2 
		+ \sigma \Box \DS _{\vec{x}} 
\eeq
in combination with the equal time commutation relation
$\; \left[ \phi {\ssty(\!x\!)}
,\phi {\ssty(\!y\!)}\right] _{\, t_x=t_y} = \, 0 \;$
we obtain (see appendix \ref{sec:timord_delta}):
\begin{align}
	\label{eq:dvb_ds_t_29}
	\left( \sigma \Box \DS _x \!\! + m^{\sssty 2} \right)
		G \! _F {\ssty(\!x\!,y\!)} \;
	& = \; \frac{\delta {\ssty(\!t_x\!-t_y\!)}}
		{\underbrace{R\ds^2}_{1/ \beta_G}
		\underbrace{(1-\vec{x}\,^2)}_{f {\ssty \! (\vec{x})}}} \,
		\bigl[ \partial _{t_x} \! G^+ \! - \partial _{t_x} \! G^- \bigr] _{t_x\!=t_y}
\end{align}
This expression contributes a delta function for the time variables
of the spacetime points $x$ and $y$.
\\
Defining the various propagators as in equations
(\ref{eq:dvb_ads01})-(\ref{eq:dvb_ads08})
corresponds to define as Feynman propagator the function
fulfilling the inhomogeneous Klein-Gordon equation
\begin{align}
	\label{eq:dvb_ds_t_29_5}
	\left( \sigma \Box \DS _x \!\! + m^{\sssty 2} \right)
	G \! _F {\ssty(\!x\!,y\!)} \; & = \;\,
	\frac{-i}{\sqrt{g}} \; \delta^{\sssty\! (\!d)} {\ssty \! (\!x-y\!)}
	\; = \; \frac{-i}{R\ds^d} \;
	\delta^{\sssty\! (\!d)} {\ssty \! (\!x-y\!)} \\
	\label{eq:dvb_ds_t_30}
	\left(\Box \DS _x \!\! + \sigma m^{\sssty 2} \right)
	G \! _F {\ssty(\!x\!,y\!)} \; & = \;\frac{-i \sigma}{\sqrt{g}} \;
	 \delta^{\sssty\! (\!d)} {\ssty \! (\!x-y\!)}
	= \; \frac{-i \sigma}{R\ds^d} \;
		\delta^{\sssty\! (\!d)} {\ssty \! (\!x-y\!)}
\end{align}

		\subsection{From Klein-Gordon to the hypergeometric equation}
		\label{sec:dvb_ds_t_kghyper}
This subsection is independent of the dS internal coordinates
(because the embedding space coordinates are all given in radial form
such that the dS internal coordinates do not depend on the radius $R\ds$)
and therefore is exactly the same as subsection \ref{sec:dvb_ds_kghyper}
for the global coordinates. 

		\subsection{Inspection of the candidate functions}
		\label{sec:dvb_ds_t_excand}
This subsection is also the same as \ref{sec:dvb_ds_excand}
for the global coordinates, only equations
\eqref{eq:dvb_ds40_6a} and \eqref{eq:dvb_ds40_6b}
now look different.
For the $(t,\vec{x})$ coordinate set we obtain
instead of \eqref{eq:dvb_ds40_6a}:
\begin{align}
	\lambda & = \, \frac{1}{2} - \frac{\sigma}{2}
	\Bigl[ \sqrt{1 \! -\vec{x}^{\,2}}\sqrt{1 \! -\vec{y}^{\,2}} \,
			\cosh \, (t_x\!\!-\!t_y) + \vec{x} \vec{y} \,\Bigr]
\end{align}
For the origin $t_y \!\! = \! \vec{y} \! = \! 0$
as reference point instead of \eqref{eq:dvb_ds40_6b} this reduces to:
\begin{align}
	\lambda & = \, \frac{1}{2} -
	\frac{\sigma}{2} \sqrt{1 \! -\vec{x}^{\,2}} \, \cosh t_x
\end{align}
We see again that for certain values of $t_x, \vec{x},t_y,\vec{y}$
our invariant quantities $\lambda, \ovl{\lambda}$ can become one
which is pathologic because $\phi_{\! \Delta}$
%% and $\phi^{\ssty (\!b\!)}_{\! \Delta}$
is divergent in $\lambda = 1$.
%%and the hypergeometric differential equation has singular points
%%at $\lambda = 0$ and $\lambda = 1$.

				\subsubsection{Infinitesimal shift in complex time plane}
				\label{sec:dvb_ds_t_infshift}
In order to remove this we apply again a shift in the complex time plane
\beqs
	t_x \; \longrightarrow \; t_x \pm \, i \, \frac{\epsilon}{2}  
\eeqs 
which results in complex coordinates $X^0_\pm$ and $X^d_\pm$:
\beq
	\begin{pmatrix}
		X^0_\pm \\
		X^d_\pm
	\end{pmatrix}
	\, = \,
	\underbrace{
	\begin{pmatrix}
		\;\;\;\; \cos {\ssty \frac{\epsilon}{2}}	 & \pm i \sin {\ssty %%@
\frac{\epsilon}{2}} \\
		\pm i \sin {\ssty \frac{\epsilon}{2}} & \;\;\;\; \cos {\ssty \frac{\epsilon}{2}}
	\end{pmatrix}}_{A\ds}
	\begin{pmatrix}
		X^0 \\
		X^d
	\end{pmatrix}
\eeq
We can check that $A\ds^\dagger A\ds = \mathbb{1}$, $\det A\ds = 1$
and therefore $A\ds \in SU(2)$.
We leave $t_y$ and thereby $Y$ unchanged.
In the limit of small $\epsilon$ we find
\begin{align*}
	X^0_\pm \, & \approx \, R_X \sqrt{1-\vec{x}^{\,2}} \;
							\Bigl[ (1\!-\!{\ssty \frac{\epsilon^2}{8}} ) \sinh t_x 
							\pm \, i {\ssty \frac{\epsilon}{2}} \cosh t_x \Bigr] \\
					& = \, (1\!-\!{\ssty \frac{\epsilon^2}{8}} ) X^0
							\pm \, i {\ssty \frac{\epsilon}{2}} X^d \\							
	X^d_\pm \, & \approx \, R_X \sqrt{1-\vec{x}^{\,2}} \;
							\Bigl[ (1\!-\!{\ssty \frac{\epsilon^2}{8}} ) \cosh t_x 
							\pm \, i {\ssty \frac{\epsilon}{2}} \sinh t_x \Bigr] \\
					& = \, (1\!-\!{\ssty \frac{\epsilon^2}{8}} ) X^d
							\pm \, i {\ssty \frac{\epsilon}{2}} X^{0}
\end{align*}
Next we define
\begin{align}
	\lambda_\pm \, & \equiv \,\; \frac{1}{2} + \, \frac{X_\pm Y}{2R\ds^2} %%@
\quad
			= \, \frac{\,1\!-\!\sigma}{2} - \frac{u_\pm}{\,4 R\ds^2}
			\qquad \qquad {\ssty u_\pm \equiv (X_{\!\pm} \!-\!Y)^2} \\
	\ovl{\lambda}_\pm \, & \equiv \,\; \frac{1}{2} - \, \frac{X_\pm Y}{2R\ds^2} %%@
\quad
			= \, \frac{\,1\!+\!\sigma}{2} + \frac{u_\pm}{\,4 R\ds^2}
\end{align}
For small $\epsilon$ this definition of $\lambda_\pm$ yields
\begin{align}
	\label{eq:dvb_ds_t_57_2}
	\lambda_\pm & \approx \, \frac{1}{2} - \frac{\sigma}{2}
			\Bigl[ (1\!-\!{\ssty \frac{\epsilon^2}{8}}) \sqrt{1\!-\!\vec{x}^{\, 2}}
			\sqrt{1\!-\!\vec{y}^{\, 2}} \cosh (t_x \!\!-\!t_y) + \vec{x}\vec{y}
			\notag \\
			& \qquad \qquad \qquad \qquad \qquad 
			\pm i {\ssty \frac{\epsilon}{2}} \sqrt{1\!-\!\vec{x}^{\, 2}}		
			\sqrt{1\!-\!\vec{y}^{\, 2}} \sinh (t_x \!\!-\!t_y) \Bigr] \\
	& \approx \, \frac{1}{2} - \frac{\sigma}{2}
			\Bigl[ (1\!-\!{\ssty \frac{\epsilon^2}{8}}) \sqrt{1\!-\!\vec{x}^{\, 2}}
			\sqrt{1\!-\!\vec{y}^{\, 2}} \cosh (t_x \!\!-\!t_y) + \vec{x}\vec{y}
%%			\notag \\
%%			& \qquad \qquad \qquad \qquad \qquad 
			\pm i \epsilon \, \epsilon {\ssty (t_x \!-\!t_y)} \Bigr]
\end{align}
so that again we have
$\lambda_\pm {\ssty (\!x\!,y\!)} = \lambda^*_\mp {\ssty (\!x\!,y\!)}
  = \lambda_\mp {\ssty (\!y\!,x\!)}$. \\ 
$\lambda$ is SO$\ssty (1,d)$ invariant
but unfortunately sign$(t_x \!\!-\!t_y)$ is not generally invariant
under orthochronous SO$\ssty (1,d)$
(see subsection \ref{sec:ds_timord_t})
which is a drawback of this coordinate set.
\\
While the $i\epsilon$ term is present, the point 1 is always evaded.
We observe that the $i\epsilon$ term vanishes only for $t_x\!\!=\!t_y$.
In this case we have
\begin{align*}
	\lambda_{\pm (t_x=t_y\!)} = \, \frac{1}{2} - \frac{\sigma}{2}
			\overbrace{\Bigl( \sqrt{1\!-\!\vec{x}^{\, 2}}	\sqrt{1\!-\!\vec{y}^{\, 2}}
			+ \vec{x}\vec{y} \Bigr)}^\kappa
			+ \; \sigma {\ssty \frac{\epsilon^2}{16}}
			\sqrt{1\!-\!\vec{x}^{\, 2}} \sqrt{1\!-\!\vec{y}^{\, 2}} \\
	\ovl{\lambda}_{\pm (t_x=t_y\!)} = \, \frac{1}{2} + \frac{\sigma}{2}
			\Bigl( \sqrt{1\!-\!\vec{x}^{\, 2}}	\sqrt{1\!-\!\vec{y}^{\, 2}}
			+ \vec{x}\vec{y} \Bigr)
			- \; \sigma {\ssty \frac{\epsilon^2}{16}}
			\sqrt{1\!-\!\vec{x}^{\, 2}} \sqrt{1\!-\!\vec{y}^{\, 2}}
\end{align*}
For $\vec{x}^{\,2} \!, \vec{y}^{\,2} \! < 1$
we find $-1 < \kappa \leq 1$ and further on
\begin{align}
	\label{eq:dvb_ds_t_57_45}
	\lambda^{\sigma=-1}_{\pm (t_x=t_y\!)}
	= \, \ovl{\lambda}^{\sigma=+1}_{\pm (t_x=t_y\!)}
	= \, \overbrace{{\ssty \frac{1}{2}}(1+\kappa)}^{0<\ldots\leq1}
	 \, - \, \epsilon^2 \\
	\lambda^{\sigma=-1}_{\pm (t_x=t_y\!)}
	= \, \ovl{\lambda}^{\sigma=+1}_{\pm (t_x=t_y\!)}
	= \, \underbrace{{\ssty \frac{1}{2}}(1-\kappa)}_{0 \leq \ldots < 1}
	 \, + \, \epsilon^2
\end{align}
We see that in the upper line
$\lambda$ is always separated from 1 by $\epsilon^2$ 
while in the lower line a finite $\epsilon$ for $\kappa \approx -1$
would lead to  $\lambda  \geq 1$.
Thereby it is clear that $\lambda$ is the right quantity for $\sigma=-1$
and $\ovl{\lambda}$ is the right one for $\sigma=+1$.
This assignment is the same found for the global coordinates
in subsection \ref{sec:dvb_ds_infshift}.
Therefore we will only consider the case (\ref{eq:dvb_ds_t_57_45})
and from here on simplify our notation to
$\lambda_\pm=\lambda^{\sigma=-1}_\pm=\ovl{\lambda}^{\sigma=+1}_\pm$.
\\
We remark that $\kappa=1$
and thereby $\lambda_{\pm (t_x=t_y\!)} =1-\epsilon^2$
if and only if $\vec{x}=\vec{y}$.
After all we have found the well defined function
\beq
	\label{eq:dvb_ds_t_57_5}
	\phi^\pm_{\! \Delta}{\scriptstyle \! ( \! \lambda_\pm \!)} \, 
	= \, F (a,b,c, \lambda_\pm) \,
	= \, F \Bigl(\Delta_+, \Delta_-,\frac{d}{2},
		 \, \lambda_\pm \Bigr)
\eeq

				\subsubsection{Delta source and normalization}
In order to construct the Feynman propagator fulfilling the inhomogeneous
Klein-Gordon equation (\ref{eq:dvb_ds_t_29_5}), we now want to show
that applying the Klein-Gordon operator to our candidate functions
yields the desired delta source.
\\
According to (\ref{eq:dvb_ds_t_29}) we therefor need to examine
$\bigl( \del_{t_x} \phi_\Delta^\pm \bigr)_{\!t_x=t_y}$ \!\!
in order to show that it delivers a delta function
for the spatial variables of the spacetime points $x$ and $y$.
Likewise to the AdS case we plan to identify
$\phi_\Delta^\pm$ with $G^{\mp}_\Delta$.
\\
Starting from equation (\ref{eq:dvb_ds_t_57_5})
and employing (\ref{eq:hypergeo04_5}) we get 
\begin{align}
	\label{eq:dvb_ds_t_57_7}
	\phi^\pm_{\! \Delta}{\scriptstyle \! ( \! \lambda_\pm \!)} \, 
	= \, \bigl(1\!-\! \lambda_\pm \bigr)^{1-d/2}
	F \bigl( c\!-\!a, \, c\!-\!b, \, c, \, \lambda_\pm \bigr)
\end{align}
Next we put to work the chain rule of differentiation
$(\del_{t_x} \phi_\Delta^\pm) = (\del_{t_x} \lambda_\pm)
(\del_{\lambda_\pm} \phi_\Delta^\pm)$
wherein for the last factor we can make use of equation (\ref{eq:hypergeo04}).
\begin{align}
	\label{eq:dvb_ds_t_58}
	\Bigl( \del_{t_x}
	\phi^\pm_\Delta {\ssty \! ( \! \lambda_{\! \pm} \!)} \! \Bigr)_{\! t_x=t_y} %%@
\!\!\!
	& = \, \bigl( \del_{t_x} \lambda_\pm \bigr)_{\! t_x=t_y}
			\Bigl[
			({\ssty \frac{d}{2}}\!-\!1) (1\!-\! \lambda_\pm )^{-\frac{d}{2}}
			F { \ssty( c\!-\!a, \, c\!-\!b, \, c, \, \lambda_\pm)} \\
			& \qquad \qquad \qquad \quad \;\;
			+ {\ssty \frac{(c-a)(c-b)}{c}} (1\!-\! \lambda_\pm )^{1-\frac{d}{2}}
			F {\ssty (c-a+1,c-b+1,c+1, \lambda_\pm)}
			\Bigr]_{\! t_x=t_y} \notag
\end{align}
From equation (\ref{eq:dvb_ds_t_57_2})
we can read off that for small $\epsilon$ we have
\begin{align}
	\label{eq:dvb_ds_t_60}
	\Bigl( \del_{t_x} \lambda_\pm \! \Bigr)_{\! t_x=t_y} \!
	& \approx \, \pm \, i {\ssty \frac{\epsilon}{2}} \,
			\sqrt{1 \!\! -\! \vec{x}^{\,2}}\sqrt{1 \!\! - \!\vec{y}^{\,2}} \\
	\Bigl( \lambda_\pm \! \Bigr)_{\! t_x=t_y} \!
	& \approx \, \frac{1}{2} + \frac{1}{2}
			\underbrace{\Bigl( \sqrt{1\!-\!\vec{x}^{\, 2}} \sqrt{1\!-\!\vec{y}^{\, 2}}
			+ \vec{x}\vec{y} \Bigr)}_\kappa
			- {\ssty \frac{\epsilon^2}{16}}
			\sqrt{1\!-\!\vec{x}^{\, 2}} \sqrt{1\!-\!\vec{y}^{\, 2}} 
\end{align}
which for our special reference point $t_y \!\! = \! \vec{y} \! = \! 0$
reduces to the simpler expression
\begin{align}
	\Bigl( \lambda_\pm \! \Bigr)_{\! t_x=t_y} \!
	& \approx \, {\ssty \frac{1}{2}} + {\ssty \frac{1}{2}} \sqrt{1\!-\!\vec{x}^{\, %%@
2}}
			- {\ssty \frac{\epsilon^2}{16}} \sqrt{1\!-\!\vec{x}^{\, 2}} 
\end{align}
We already found that $\kappa=1$ and thereby
$(1-\lambda_\pm)=\epsilon^2$ if and only if $\vec{x}=\vec{y}$.
This means that for all $\vec{x}\neq \vec{y}$ we can let $\epsilon$
run to zero so that
$( \del_{t_x} \! \phi^\pm_\Delta {\ssty \! ( \! \lambda_{\! \pm} \!)} %%@
)_{t_x=t_y}$
vanishes for this case.
\\
In order to determine suitable normalisation constants we need to integrate
$(\del_{t_x} \phi_\Delta^\pm )_{t_x=t_y}$
over $(d\!-\!1)$ dimensional $\vec{x}$-space
with the rectriction $\vec{x}^{\,2}<1$.
Doing so, we find that the second summand
in equation (\ref{eq:dvb_ds_t_58})
yields a contribution proportional to $\epsilon^2$ 
which therefore vanishes in the limit of small $\epsilon$.
\\
When turning to the first summand,
we first compute the $\vec{x}$-space integral
over $\epsilon\, (1\! / \!-\!\lambda_\pm)^{-d/2}_{t_x=t_y}$
for the case of our reference point $t_y \!\! = \! \vec{y} \! = \! 0$ and find
using integral (\ref{eq:delta155}):
\begin{align}
	\int \limits_{\vec{x}^{\,2}<1} \!\!\!\! d^{d-\!1}x \,
	\frac{\epsilon}
			{\bigl( 1 \! - \! \lambda_\pm \bigr)^{\! \frac{d}{2}}_{t_x=t_y}} \,
	& = \, \frac{2 \pi^{\frac{d\!-\!1}{2}}}{\Gamma({\ssty \frac{d-1}{2}})}
			\int \limits_0^1 \!\! dr \, \frac{\epsilon \, r^{d-2}}
			{\bigl({\ssty \frac{1}{2}} \! - \!  {\ssty \frac{1}{2}} \sqrt{1-r^2} 
			+ {\ssty \frac{\epsilon^2}{16}} \sqrt{1-r^2} \,  \bigr)^{\! \frac{d}{2}}}
	\notag \\
	& = \, \frac{2 \pi^{\frac{d\!-\!1}{2}}}{\Gamma({\ssty \frac{d-1}{2}})}
			\int \limits_0^{\xi \ll 1} \!\! dr \, \frac{\epsilon \, r^{d-2}}
			{\bigl({\ssty \frac{1}{2}} \! - \!  {\ssty \frac{1}{2}}
			(1-{\ssty \frac{1}{2}} r^2) 
			+ {\ssty \frac{\epsilon^2}{16}} \sqrt{1-r^2} \,  \bigr)^{\! \frac{d}{2}}}
	\notag \\
	& \qquad \quad
		 	+ \,  \frac{2 \pi^{\frac{d\!-\!1}{2}}}{\Gamma({\ssty \frac{d-1}{2}})}
			\underbrace{\int \limits_\xi^1 \!\! dr \,
			\frac{\epsilon \, r^{d-2}}
			{\bigl( \underbrace{ {\ssty \frac{1}{2}} \! - \!  {\ssty \frac{1}{2}}
			\sqrt{1-r^2} + {\ssty \frac{\epsilon^2}{16}} \sqrt{1-r^2} 
			}_{> \, 0} \,  \bigr)^{\! \frac{d}{2}}}
			}_{\limepszero \text{exists:} \,  \longrightarrow \, 0}
	\notag \\
	& \!\! \overset{\epsilon \rightarrow 0}{=} 
			\, \frac{2^{d+2} \pi^{\frac{d\!-\!1}{2}}}{\Gamma({\ssty \frac{d-1}{2}})}
			\int \limits_0^\infty \!\! dr \, \frac{r^{d-2}}
			{\bigl( r^2 + 1  \bigr)^{\! \frac{d}{2}}}
	\; = \; \frac{2^{d+1} \pi^{\frac{d}{2}}}{\Gamma({\ssty \frac{d}{2}})} \\
	\Longrightarrow \quad \limepszero 
		\frac{\epsilon} {(1 \! - \! \lambda_\pm)_{t_x=t_y}^{\frac{d}{2}}} \;
	& = \; \frac{2^{d+1} \pi^{\! \frac{d}{2}}}{\Gamma({\ssty \frac{d}{2}})} \;
		\delta^{(\!d\!-\!1\!)}{\ssty (\vec{x}-\vec{y})}
\end{align}
Looking at (\ref{eq:dvb_ds_t_60}) we see that for $\vec{y} = \vec{x}$ we have
$\bigl( \del_{t_x} \lambda_\pm \! \bigr)_{\! t_x=t_y} \!
			\approx \, \pm i \epsilon (1 \!\! - \!\vec{x}^{\,2})$.
Therefore using (\ref{eq:hypergeo03_5})
we arrive at the desired spatial delta function:
\begin{align}
	\limepszero	\Bigl( \del_{t_x} 
			\phi^\pm_\Delta{\ssty \! ( \! \lambda_{\! \pm} \!)} \Bigr)_{\! t_x=t_y}
	& = \; \pm i \, \underbrace{  
		\frac{2^d \pi^{\! \frac{d}{2}} \; \Gamma({\ssty \frac{d}{2})}}
		{\, \Gamma({\ssty \Delta_+}) \, \Gamma({\ssty \Delta_-})}}_{\beta_\phi} \;
		(1 \!\! -\!\vec{x}^{\,2}) \; \delta^{(\!d\!-\!1\!)} {\ssty (\vec{x}-\vec{y})}
\end{align}
With this relation fixed we can now identify:
\begin{align}
	\label{eq:dvb_ds_t_67}
	G_{\! \Delta}^\pm {\ssty (\!x\!,y\!)} \,
	& = \, \frac{\, R\ds}{\, 2 \beta_\phi} \;\;
			\limepszero \,
			\phi^\mp_\Delta{\ssty \! ( \! \lambda_{\! \mp})}
\end{align}
and construct the other propagators according to
(\ref{eq:dvb_ads03}-\ref{eq:dvb_ads08}).
Now we can step back and evaluate equation (\ref{eq:dvb_ds_t_29}):
\begin{align*}
	\left( \sigma \Box _x \!\! + m^{\scriptscriptstyle 2} \right)
		G \! _F^{\,\Delta} {\scriptstyle(\!x\!,y\!)} \;
	& = \; \frac{\delta {\scriptstyle(\!t_x\!-t_y\!)}}{R\ds^2(1 \! - \! \vec{x}\,^2)} %%@
\,
		\bigl[ \partial _{t_x} \! G^+_{\! \Delta} \!
			- \partial _{t_x} \! G^-_{\! \Delta} \bigr] _{t_x\!=t_y} \\
	& = \; \frac{\delta {\scriptstyle(\!t_x\!-t_y\!)}}{R\ds^2(1 \! - \! \vec{x}\,^2)} %%@
\,
		(-2i) \, \frac{R\ds}{\, 2 \beta_\phi} \; \beta_\phi \,
		(1 \!\! -\!\vec{x}^{\,2}) \; \delta^{(\!d\!-\!1\!)}{\ssty (\vec{x}-\vec{y})} \\
	& = \; \frac{-i}{R\ds} \;
		\delta^{\scriptscriptstyle\! (\!d\!)} {\scriptstyle \! (\!x-y\!)}
\end{align*}
We see that the Feynman propagator fulfills just the inhomogeneous
Klein-Gordon equation (\ref{eq:dvb_ds_t_29_5})
with one delta source on its right hand side:
\begin{align*}
	\left( \sigma \Box _x \!\! + m^{\scriptscriptstyle 2} \right)
	G \! _F^{\,\Delta} {\scriptstyle(\!x\!,y\!)} \; & = \;\,
	\frac{-i}{\sqrt{g}} \; \delta^{\scriptscriptstyle\! (\!d\!)} {\scriptstyle \! %%@
(\!x-y\!)}
	\; = \; \frac{-i}{R\ds} \;
	\delta^{\scriptscriptstyle\! (\!d\!)} {\scriptstyle \! (\!x-y\!)}
	\notag \\
	\left(\Box _x \!\! + \sigma m^{\scriptscriptstyle 2} \right)
	G \! _F^{\,\Delta} {\scriptstyle(\!x\!,y\!)} \; & = \;\frac{-i \sigma}{\sqrt{g}} %%@
\;
	 \delta^{\scriptscriptstyle\! (\!d\!)} {\scriptstyle \! (\!x-y\!)}
	= \; \frac{-i \sigma}{R\ds} \;
		\delta^{\scriptscriptstyle\! (\!d\!)} {\scriptstyle \! (\!x-y\!)}
\end{align*}
The normalization constant is:
\begin{align}
	\beta_\phi & = \, \frac{(4 \pi)^{\! \frac{d}{2}} \; \Gamma({\ssty %%@
\frac{d}{2}})}
		{\, \Gamma({\ssty \Delta_+ \!}) \, \Gamma({\ssty \Delta_- \!})}
\end{align}
Again we do not want the $\Gamma{\ssty (\ldots)}$ functions
in our normalization constant to run to infinity.
This requirement is fulfilled for all $m^2R^2\ds > 0$.
In this case $\Delta_\pm$ are either positive or complex
and thus the Gamma functions remain finite
(see discussion above equation \ref{eq:dvb_ds40_5}).

		\subsection{Listing of propagators}
Below we list the various propagators for dS$_d$.
Herein $\lambda_\pm$ is meant to carry its phase
acquired via its position in the complex plane:
\begin{align}
	\lambda_\pm & \approx \, \frac{1}{2} + \frac{1}{2}
			\Bigl[ (1\!-\!{\ssty \frac{\epsilon^2}{8}}) \sqrt{1\!-\!\vec{x}^{\, 2}}
			\sqrt{1\!-\!\vec{y}^{\, 2}} \cosh (t_x \!\!-\!t_y) + \vec{x}\vec{y}
			\notag \\
			& \qquad \qquad \qquad \qquad \qquad 
			\pm i {\ssty \frac{\epsilon}{2}} \sqrt{1\!-\!\vec{x}^{\, 2}}		
			\sqrt{1\!-\!\vec{y}^{\, 2}} \sinh (t_x \!\!-\!t_y) \Bigr]
\end{align}
Our Wightman function $G_\Delta^+$ agrees with the Wightman function given %%@
by
Spradlin, Strominger and Volovich in \cite{leshouches}.
\\
However our $i \epsilon$ prescription 
for static coordinates differs from theirs.
They use the complex shift
$X^0-Y^0 \rightarrow X^0-Y^0 - i \epsilon$ in embedding space
(without giving further information).
Following Dullemond and van Beveren we employed a complex shift
for the time variable of deSitter spacetime which turned out
to produce reasonable results, except for the noninvariance
of the time ordering for $t$ under SO$\ssty (1,d)$. 
\begin{align}
	G_{\! \Delta}^\pm \! {\ssty (\!x\!,y\!)}
	& = \, \frac{R\ds}{\, 2 \beta_\phi} \: \limepszero \,
				 F (\Delta_+, \Delta_-, {\ssty \frac{d}{2}}, \, \lambda_\mp) 
\end{align}
\begin{align}
	G_{\! \Delta}^{\sssty (\!1\!)} \! {\ssty (\!x\!,y\!)} \, 
	& = \, \frac{R\ds}{\, \beta_\phi} \;\;\: \text{Re} \: \limepszero \,
				 F (\Delta_+, \Delta_-, {\ssty \frac{d}{2}}, \, \lambda_-) \\
	G_{\! \Delta} \! {\ssty (\!x\!,y\!)} \, 
	& = \, \frac{R\ds}{\, \beta_\phi} \; i \, \text{Im} \: \limepszero \,
				 F (\Delta_+, \Delta_-, {\ssty \frac{d}{2}}, \, \lambda_-) 
\end{align}
\begin{align}
	G^{ \Delta}_{\!R} \! {\ssty (\!x\!,y\!)}
	& = \;\;\; \theta {\ssty (\!t_x\!-t_y\!)} \,
			\frac{R\ds}{\, \beta_\phi} \, i \, \text{Im} \: \limepszero \,
				 F (\Delta_+, \Delta_-, {\ssty \frac{d}{2}}, \, \lambda_-) \\
	G^{ \Delta}_{\!A} \! {\ssty (\!x\!,y\!)}
	& = -\theta {\ssty (\!t_y\!-t_x\!)} \,
			\frac{R\ds}{\, \beta_\phi} \, i \, \text{Im} \: \limepszero \,
				 F (\Delta_+, \Delta_-, {\ssty \frac{d}{2}}, \, \lambda_-) 
\end{align}
\begin{align}
	G^{\,\Delta}_{\! F} \! {\ssty (\!x\!,y\!)} \, 
	& = \, \frac{R\ds}{\, 2 \beta_\phi} \; \limepszero \,
				 F (\Delta_+, \Delta_-, {\ssty \frac{d}{2}}, \, \lambda_F) 
\end{align}
\\
Therein $\lambda_F$ (with index F for Feynman)
is defined for the limit of small $\epsilon$ as
\begin{align*}
	\lambda_F \!
		& \approx \, \frac{1}{2} + \frac{1}{2}
			\Bigl[ (1\!-\!{\ssty \frac{\epsilon^2}{8}}) \sqrt{1\!-\!\vec{x}^{\, 2}}
			\sqrt{1\!-\!\vec{y}^{\, 2}} \cosh (t_x \!\!-\!t_y) + \vec{x}\vec{y}
			\\
		& \qqqquad \qqqquad \qquad
			+ i {\ssty \frac{\epsilon}{2}}
			\sqrt{1\!-\!\vec{x}^{\, 2}}	\sqrt{1\!-\!\vec{y}^{\, 2}} \,
			\sinh \! \mid \! t_x \!\! - \! t_y \!\! \mid \, \Bigr]
\end{align*}
so that the Feynman propagator is indeed symmetric:
$\, G_{\! F}^{\Delta} \! {\ssty (\!x\!,y\!)}
= G_{\! F}^{\Delta} \! {\ssty (\!y\!,x\!)}$.
We observe that in contrast to AdS the sign of the $i \epsilon$-term
in the Feynman propagator
in dS spacetime is always positive except for $\tau_x = \tau_y$
as in Minkowski spacetime.
For the next remark we repeat the structure of $\lambda_\pm$:
\begin{align*}
	\lambda_\pm
	& \approx \, \underbrace{
			{\ssty \frac{1}{2}} + {\ssty \frac{1}{2}}
			\Bigl[ \sqrt{ _{\ldots}} \sqrt{ _{\ldots}} \cosh {\ssty (\ldots)}
			+ \vec{x}\vec{y} \, \Bigr]}_\lambda
			\pm \, i \epsilon {\ssty (\ldots)} - \epsilon^2 {\ssty (\ldots)}
\end{align*}
For the propagators above the hypergeometric function
is defined by its convergent Taylor series
for all $\lambda$ with $\mid \! \lambda \! \mid \, < 1$.
The form of the propagators for $\mid \! \lambda \! \mid \, > 1$
can be obtained therefrom using (\ref{eq:hypergeo07}).
For $\mid \! \lambda \! \mid \, < 1$ we can use (\ref{eq:dvb_ds_t_57_7}).
\\
Likewise to Minkowski and AdS spacetime
we find that the propagators become singular
one the whole embedding space lightcone $\lambda=1$
while generating only one delta source at the coincidence point $x=y$.
%%

%%
%%_______the_bibliography______________________________
%%
%%

%%

%%
\chapter{Acknowledgements}
First of all, at this opportunity I would like to thank 
everyone in my whole family,
especially my parents Hannelore and Peter Dohse,
my brother Paul and my sister Ilka 
for always having supported me throughout my life
and helping me keeping up my studies. 
\\
From the QFT and String Group of the HU at Berlin
first I wish to thank my supervisor Dr. Harald Dorn
for dedicating plenty of time to supporting this piece of work
and both finding and thoroughly counterchecking many arguments.
To Sylvia Richter for her administrative assistance
and to Dr. Hans-J\"org Otto for maintaining our computers
I am also very thankful.
I enjoyed very much working in the QFT and String Theorie Group
of Prof. Jan Plefka and would like to thank
all present and former group members
for the friendly, communicative atmosphere and personal discussions.
Another thank you goes to Prof. Dietmar Ebert
for being devotedly active in teaching
and reminding students to keep in mind regarding
the overall picture and developments in physics, life and society.
\\
I would like to thank my former and present office mates and colleagues
Alexander Hentschel, Andreas Rodigast, Hai Ngo Than, Fabian Spill,
Johannes Vetter and last but certainly not least
Danilo Diaz for the many discussions,
some about physical and lots about other topics in life.
\\
For having survived studying physics all together with me
and the many valuable philosophical contributions
at the Mensa Round Table I would like to express my respect and gratitude
to Aiko, Jens, Julia (also especially for several culinary contributions!)
Christian, Marco, Oli, Hannes, Nico und Martin.
\\
Finally I would like to greet my friends
Paul, Uwe, Lydia, Karsten, Alina, Oli, Nora, Teppi, Heike, Johanna,
Anja, Stefan, Christoph, Mario, Thomas and Ati 
and express my thankfulness to all of you for encouraging me
in personal and professional life and for just being as you are.
A special last thanks to Teppi and Oli for providing
some helpful software solutions! 
\thispagestyle{plain}
\end{document}